\documentclass[acmsmall]{acmart}

\setcopyright{acmlicensed}
\acmJournal{JACM}
\acmYear{2022} 
\acmVolume{70} 
\acmNumber{1} 
\acmArticle{1} 
\acmMonth{12} 
\acmPrice{15.00}
\acmDOI{10.1145/3559103}

\copyrightyear{2022}

\definecolor{Bgreen}{rgb}{0,0.5,0}

\usepackage{eucal}

\renewcommand{\amalg}{\sqcup}

\newcommand{\simplexcategory}{\boldsymbol{\Delta}}
\newcommand{\CC}{\mathcal{C}}
\providecommand{\kat}[1]{\text{\textbf{\textsl{#1}}}}
\newcommand{\Set}{\kat{Set}}
\newcommand{\Gr}{\kat{Gr}}

\newcommand{\Grpd}{\kat{Grpd}}
\newcommand{\elGr}{\kat{elGr}}
\newcommand{\Sh}{\kat{Sh}}
\newcommand{\PrSh}{\kat{PrSh}}
\newcommand{\op}{^{\text{{\rm{op}}}}}
\newcommand{\N}{\mathbb{N}}
\newcommand{\B}{\mathbb{B}}
\newcommand{\F}{\kat{FinSet}}
\newcommand{\Map}{\operatorname{Map}}
\newcommand{\Aut}{\operatorname{Aut}}
\newcommand{\Hom}{\operatorname{Hom}}
\newcommand{\End}{\operatorname{End}}
\newcommand{\Fun}{\operatorname{Fun}}
\newcommand{\id}{\operatorname{id}}
\newcommand{\obj}{\operatorname{obj}}
\newcommand{\triv}{\star}
\newcommand{\coro}[2]{\begin{smallmatrix}#2\\#1\end{smallmatrix}}
\newcommand{\name}[1]{\ulcorner #1\urcorner}
\newcommand{\isopil}{\stackrel{\raisebox{0.1ex}[0ex][0ex]{\(\sim\)}}%
			{\raisebox{-0.15ex}[0.28ex]{\(\rightarrow\)}}}

\DeclareRobustCommand\comma{%
  \mathchoice%
    {\kern1.2pt\raise0pt\hbox{$\displaystyle\downarrow$}\kern1.2pt}
    {\kern1pt\raise0pt\hbox{$\textstyle\downarrow$}\kern1pt}
    {\kern0.4pt\raise0pt\hbox{$\scriptstyle\downarrow$}\kern0.4pt}
    {\kern0.2pt\raise0pt\hbox{$\scriptscriptstyle\downarrow$}\kern0.2pt}
}%
\DeclareRobustCommand\upperstar{%
  \mathchoice%
    {\kern0pt\raise0.55ex\hbox{$\displaystyle *$}\kern0.8pt}
    {\kern0pt\raise0.58ex\hbox{$\textstyle *$}\kern0.8pt}
    {\kern0pt\raise0.45ex\hbox{$\scriptstyle *$}\kern0.4pt}
    {\kern0pt\raise0.4ex\hbox{$\scriptscriptstyle *$}\kern0.2pt}
}%
\DeclareRobustCommand\lowerstar{%
  \mathchoice%
    {\kern0pt\raise-0.65ex\hbox{$\displaystyle *$}\kern0.8pt}
    {\kern0pt\raise-0.68ex\hbox{$\textstyle *$}\kern0.8pt}
    {\kern0pt\raise-0.55ex\hbox{$\scriptstyle *$}\kern0.4pt}
    {\kern0pt\raise-0.5ex\hbox{$\scriptscriptstyle *$}\kern0.2pt}
}%

\newcommand{\lowershriek}{_!}

\newcommand{\Cor}{\kat{Cor}}
\newcommand{\res}{\operatorname{res}}
\newcommand{\iso}{\operatorname{iso}}

\newcommand{\el}{\operatorname{el}}

\newcommand{\un}{\underline}
\newcommand{\ov}{\overline}

\DeclareMathOperator*{\colim}{colim}

\newcommand{\ds}{\mathbf}
\newcommand{\gs}{\mathcal}

\DeclareMathAlphabet {\pnfont}{OT1}{cmss}{m}{n}
\newcommand{\pn}[1]{\scalebox{0.94}{$\pnfont #1$}}
\newcommand{\pnsmall}[1]{\scalebox{0.69}{$\pnfont #1$}}

\usepackage{tikz,tikz-cd}
\usetikzlibrary[arrows,decorations.pathmorphing,backgrounds,positioning,fit,petri]

\usetikzlibrary{shapes, decorations.markings}

\newcommand{\drpullback}{\arrow[phantom]{dr}[very near start,description]{\lrcorner}}
\newcommand{\dlpullback}{\arrow[phantom]{dl}[very near start,description]{\llcorner}}
\newcommand{\urpullback}{\arrow[phantom]{ur}[very near start,description]{\urcorner}}
\newcommand{\ulpullback}{\arrow[phantom]{ul}[very near start,description]{\ulcorner}}

\AtEndPreamble{%

\newenvironment{blanko}[1]%
{%
\refstepcounter{subsection}
\begin{list}{\rm \thesubsection \ \ {\it #1} \ }%
{\setlength{\itemindent}{\parindent}\setlength{\labelsep}{0mm}\setlength{\leftmargin}{0mm}%
\setlength{\labelwidth}{0mm}\setlength{\listparindent}{\parindent}%
\setlength{\parsep}{\parskip}\setlength{\partopsep}{4pt}}%
\item%
}%
{%
\end{list}%
}

\newenvironment{blankothm}[1]%
{%
\refstepcounter{subsection}
\begin{list}{{\sc #1} \thesubsection.\ }%
{\setlength{\itemindent}{\parindent}\setlength{\labelsep}{0mm}\setlength{\leftmargin}{0mm}%
\setlength{\labelwidth}{0mm}\setlength{\listparindent}{\parindent}%
\setlength{\parsep}{\parskip}\setlength{\partopsep}{4pt}}%
\item \em%
}%
{%
\end{list}%
}

}

\newcounter{dummycounter}
\newenvironment{shortlist}%
{%
	\begin{list}%
	{(\arabic{dummycounter})}%
	{\usecounter{dummycounter}%
	\setlength{\itemindent}{\parindent}
	\setlength{\itemsep}{0em}\setlength{\parsep}{0em}\setlength{\topsep}{0em}%
	\setlength{\itemindent}{0em}\setlength{\labelwidth}{1.8em}%
	\setlength{\labelsep}{0.5em}\setlength{\leftmargin}{2.4em}}%
}%
{\end{list}}

\tikzcdset{
	arrow style=tikz
}

\tikzset{
    onto/.style={/tikz/commutative diagrams/twoheadrightarrow}
}
\tikzset{
    into/.style={/tikz/commutative diagrams/rightarrowtail}
}

\tikzset{
	snake it/.style={decorate, decoration=snake}
}

	\tikzstyle directed=[postaction={decorate,decoration={markings,
	  mark=at position .65 with {\arrow[arrowstyle]{stealth}}}}]

	 \tikzstyle smallgrnode=[
	   rectangle, 
	   draw, 
	   fill=black, 
	   inner sep=0mm, 
	   outer sep=0mm, 
	   minimum size=0.7mm
	 ]
	 \tikzstyle grnode=[
	   rectangle, 
	   draw, 
	   fill=black, 
	   inner sep=0mm, 
	   outer sep=0mm, 
	   minimum size=0.9mm
	 ]
	 \tikzstyle grNode=[
	   rectangle, 
	   draw, 
	   fill=black, 
	   inner sep=0mm, 
	   outer sep=0mm, 
	   minimum size=1.1mm
	 ]
	 \tikzstyle port=[
	   inner sep=0mm, 
	   outer sep=0mm, 
	   minimum size=0.0mm
	 ]
	 \tikzstyle import=[
	 below,
	 ]
	 \tikzstyle export=[
	 above
	 ]

\begin{document}

\title{Whole-grain Petri nets and processes}

\author{Joachim Kock}
\email{kock@mat.uab.cat}
\orcid{0000-0003-3358-2812}
\affiliation{%
  \institution{Universitat Aut\`onoma de Barcelona and Centre de Recerca 
  Matem\`atica}
  \streetaddress{Edifici C}
  \city{Cerdanyola}
  \state{Barcelona}
  \country{Catalonia}
  \postcode{08193}
}

\authorsaddresses{%
    This is the author version, not identical to the published version, but 
  adjusted so that the numberings agree, and updated with final 
  publication information.
 \\
 \\
 This work was partially supported by grants MTM2016-80439-P (AEI/FEDER, UE) and 
PID2020-116481GB-I00 (AEI/FEDER, UE) of Spain and 2017-SGR-1725 of Catalonia, and
also through the Severo Ochoa and Mar\'ia de Maeztu Program for Centers and
Units of Excellence in R\&D grant number CEX2020-001084-M.
\\
  Author's address: Joachim Kock, Universitat Aut\`onoma de Barcelona and Centre de Recerca 
  Matem\`atica.
  \\
  Current address: University of Copenhagen.
}
\renewcommand{\shortauthors}{Joachim Kock}

\begin{abstract}
  
  \bigskip
  
    \hrule

  \medskip

  We present a formalism for Petri nets based on polynomial-style
  finite-set configurations and etale maps. The formalism supports
  both a geometric semantics in the style of Goltz and Reisig
  (processes are etale maps from graphs) and an algebraic semantics in
  the style of Meseguer and Montanari, in
  terms of free coloured props, and allows the following unification:
  for $\pn P$ a Petri net, the Segal space of $\pn P$-processes
  is shown to be the free coloured prop-in-groupoids on $\pn P$. There
  is also an unfolding semantics \`a la Winskel, which bypasses the
  classical symmetry problems:  with the new formalism, every Petri net admits
  a universal unfolding, which in turn has associated an event structure and 
  a Scott domain.
   Since everything is encoded with
  explicit sets, Petri nets and their processes have elements. In
  particular, individual-token semantics is native.
  (Collective-token semantics emerges from rather drastic
  quotient constructions \`a la Best--Devillers, involving taking
  $\pi_0$ of the groupoids of states.)
\end{abstract}



\maketitle

\setcounter{section}{-1}


\hrule

\bigskip
\bigskip

\section{Introduction}

\subsection*{Background}

Petri nets are an important framework for describing networks in which
resources interact and transform, such as chemical reaction networks and 
population dynamics, including compartmental models in epidemiology.
In computer science they serve as a widely used model of concurrency.

A Petri net (cf.~Section~\ref{sec:Petri}) has {\em places} holding {\em tokens}, and {\em transitions}
describing how tokens flow and transform. {\em Arcs} connect places and 
transitions to express the interactions. A Petri net is thus a  kind of 
graph, and may look like the following figure. Places are always pictured
as circles, and transitions as squares. Tokens are drawn as bullets:

  
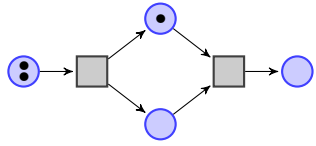
\begin{center}
	\begin{tikzpicture}[scale=0.9, rotate=-90,transform shape]
	  \begin{scope}[
		  node distance=0.9cm,on grid,>=stealth',bend angle=22.5,auto,
		  every place/.style= {minimum size=4mm,thick,draw=blue!75,fill=blue!20}, 
		  every transition/.style={minimum 
		  size=4mm,thick,draw=black!75,fill=black!20}]
		\node [transition] (t1) {};
		\node [place](s1) [below=of t1] {\scriptsize $\bullet\bullet$}; 
		\node [place] (s2) [above=of t1, xshift=-7mm] {\scriptsize $\bullet$};
		\node [place] (s3) [above=of t1, xshift=7mm] {};
		\node [transition] (t2) [above=of s2, xshift=7mm] {};
		\node [place](s4) [above=of t2] {}; 
		\path (s1) edge [post] (t1);
		\path (t1) edge [post] (s2);
		\path (t1) edge [post] (s3);
		\path (s2) edge [post] (t2);
		\path (s3) edge [post] (t2);
		\path (t2) edge [post] (s4);
	  \end{scope}
	\end{tikzpicture}
  \end{center}

The operational semantics of Petri nets is the mathematical formalisation
of their behaviour, expressed informally in the {\em token game}, whereby
firing of transitions move tokens around between places. Two main classes
of operational semantics can be called geometric and algebraic, although
this terminology is not standard in the literature. Geometric semantics
take as starting point the notion of nonsequential processes, formalised as
certain morphisms into a given Petri net from {\em causal nets}, certain
posets or graphs. This viewpoint was pioneered by Petri
himself~\cite{Petri:1977}, and found a very clean formulation in the
seminal 1983 paper of Goltz and Reisig~\cite{Goltz-Reisig:1983}, leading to
many developments and insights regarding reachability, density, safeness,
as expressed for example in the 1988 monograph of Best and
Fern\'andez~\cite{Best-Fernandez}. Algebraic semantics, on the other hand,
focus on the way the transitions of a Petri net generate a formal algebraic
system. After earlier work of Reisig~\cite{DBLP:books/sp/Reisig85a} and
Winskel~\cite{Winskel:1987}, a breakthrough was the 1990 work of Meseguer
and Montanari~\cite{Meseguer-Montanari:monoids}, who observed that Petri
nets can be defined as monoids in certain directed graphs, and that they
freely generate symmetric monoidal categories of certain kinds,
equivalently, free props. The morphisms in such a symmetric monoidal
category are thus built up from the transitions of a given Petri net by
serial and parallel connection (composition and the tensor product), and
are thus certain equivalence classes of firing sequences.

In view of the well-known graphical calculus for monoidal categories in
terms of string diagrams~\cite{Joyal-Street:tensor-calculus}, one could
expect a unification of the geometric and algebraic semantics. This has
turned out to be a difficult problem, though: In the `geometric' frameworks,
there is no way to compose processes in the sense of category theory, and
in the algebraic frameworks there is no synthetic description of processes
--- they rely solely on formal inductive descriptions. Over the past four
decades, a lot of research has been dedicated to this problem. Already in
1987, Best and Devillers~\cite{Best-Devillers:1987} searched for an
algebraic description of the Goltz--Reisig processes, finding that it was
necessary to impose drastic equivalence relations on processes. Starting
from the Meseguer--Montanari formalism, many people, including in
particular Bruni, Meseguer, Montanari, and 
Sassone~\cite{Bruni-Meseguer-Montanari-Sassone:CTCS99},
\cite{Bruni-Meseguer-Montanari-Sassone:2001}, tried to avoid these
quontientings by introducing further bookkeeping on Petri nets. They
arrived at the notion of pre-net, which are Petri nets with numberings. For
the notion of pre-net the semantics problems can be solved, but pre-nets
are not quite a satisfactory substitute for Petri nets. In particular, they
do not allow enough morphisms. The final section of this paper contains
some further discussion of these issues.

Closely related to the geometric-algebraic dichotomy is the distinction
between individual-token and collective-token philosophies, which roughly
asks whether the tokens of a state of a Petri net execution are individual
elements in a set or if they are only the expression of a quantity, a
natural number. (See \cite{Glabbeek-Plotkin:LICS95},
\cite{Glabbeek-Plotkin:2009}, \cite{Bruni-Montanari:2000},
\cite{Glabbeek:2005}, \cite{Bruni-Meseguer-Montanari-Sassone:2001} for more
thorough discussion.) The geometric semantics favour the individual-token
philosophy, since processes are explicit maps from other nets, and in
particular involves mappings of sets. In contrast, the algebraic semantics
are more geared towards the collective-token philosophy, as both their
states and the pre- and post-conditions of a transition are given by
multisets, meaning sets with multiplicities. In its full version (see van
Glabbeek~\cite{Glabbeek:2005}) the individual-token philosophy requires
keeping track of all tokens at all times, which is done by annotating each
token occurrence with the complete history of the transitions that produced
it from a given initial state. (This involves knowing which tokens enter
which input slots of a given transition, and in particular it is necessary
to be able to distinguish such slots, something which is essentially
impossible in the Petri-nets-as-monoids formalism.)

A third main approach to operational semantics is based on the idea of {\em
unfolding}, pioneered in the work of Winskel~\cite{Winskel:thesis} and
Nielsen, Plotkin, and Winskel~\cite{Nielsen-Plotkin-Winskel:1981}. It is
important because it establishes connections to domain theory and
denotational semantics. The key ingredient, the universal unfolding of a
Petri net, was established for {\em safe} Petri nets by Nielsen, Plotkin,
and Winskel, but there are symmetry issues preventing the existence of a
universal unfolding for general Petri nets. Montanari, Meseguer, and
Sassone~\cite{DBLP:journals/mscs/MeseguerMS97},
\cite{DBLP:journals/tcs/MeseguerMS96} worked around the symmetry problems
by considering certain decorated unfoldings, but their construction was not
universal in the categorical sense. Baldan, Bruni, and
Montanari~\cite{Baldan-Bruni-Montanari-DBLP:conf/wadt/BaldanBM02}
established instead the existence of a universal unfolding for pre-nets,
where the symmetries do not come up. Finally Hayman and
Winskel~\cite{Hayman-Winskel:2008}, \cite{Hayman-Winskel:2009} came to
grips with the symmetries themselves and established a weaker form of
universal property, through a notion of adjunction up to symmetry in a
precise technical sense.

\subsection*{Contributions of this paper}

The present paper proposes a rather natural solution to these problems, by
using some elementary homotopy theory to overcome the symmetry issues. The
novelty essentially boils down to one small modification at the
foundational level: to abolish the traditional notion of multisets in
favour of the representable analogue: Instead of sets of multiplicity
functions $S \to \N$, we consider groupoids of $S$-coloured finite sets $A
\to S$; instead of assigning to each $s\in S$ a multiplicity, there is now
assigned an actual set, namely the pre-image $A_s$. The isomorphism classes
of $S$-coloured sets are the traditional multisets on $S$. The benefit is
that $S$-coloured sets have elements, which can be accessed individually,
giving full control over symmetries.

The multiset modification is incorporated in the very definition of
Petri net, following the lead of the analogous formalism for directed
graphs~\cite{Kock:1407.3744} (in turn motivated by the polynomial
formalism in the theory of operads~\cite{Kock:0807}): we define a Petri net 
to be a diagram of finite sets
$$
S \longleftarrow I \longrightarrow T \longleftarrow O \longrightarrow S .
$$
Here $T$ is the set of transitions, $S$ is the set of places, and $I$ and
$O$ are the sets of incoming and outgoing arcs of transitions. In
particular, for a transition $t\in T$, the fibres (i.e.~pre-image sets)
$I_t$ and $O_t$ are explicit sets, and they are not necessarily subsets of
$S$. In this paper, the new Petri nets are called {\em whole-grain Petri
nets}, to distinguish them from the traditional notion of Petri net. A
whole-grain Petri net is thus precisely what we see in the pictures. It is
only a slight modification of the usual definition, but it has important
implications, and the theory develops quite neatly from it, exploiting some
recent insights from algebraic
combinatorics~\cite{Galvez-Kock-Tonks:1602.05082},
\cite{Galvez-Kock-Tonks:1512.07573}, \cite{Galvez-Kock-Tonks:1708.02570},
\cite{Galvez-Kock-Tonks:1612.09225}.

An {\em etale map} is a diagram
  $$
  \begin{tikzcd}
	S' \ar[d, "\alpha"'] & I'\ar[d] \ar[l] \ar[r] \drpullback &  T'\ar[d] & 
	O'\ar[d] \ar[l]\ar[r] \dlpullback & S'\ar[d, "\alpha"]
	\\
	S & I \ar[l]\ar[r]& T & O \ar[l]\ar[r]& S   
  \end{tikzcd}
  $$
(the middle squares being pullbacks). A {\em graph} is a SITOS diagram
where the outer maps $S\leftarrow I$ and $O \to S$ are
injective~\cite{Kock:1407.3744}. A {\em process} is an etale map from an
acyclic graph. The processes of a Petri net $\pn P$ assemble naturally into
a simplicial groupoid $\ds X_\bullet$, shown to be a symmetric monoidal
Segal space. (Segal spaces (recalled below in \S\ref{sec:X}) are a homotopy
version of categories. Appendix~B provides some background.) The following
is the first main result of this work.

\bigskip

{\bf Theorem.} (Cf.~\ref{thm:Xisfree})
{\em $\ds X_\bullet$ is the free prop-in-groupoids on $\pn P$.}

\bigskip

It expresses the reconciliation of the geometric and algebraic semantics for
whole-grain Petri nets. A $1$-categorical analogue of this result is also extracted
(\ref{thm:C}).  Making sense of the free prop on a (whole-grain) Petri net
depends on an explicit fully faithfully embedding $\kat{Petri} \to \PrSh(\elGr)$ of
the category of (whole-grain) Petri nets into the presheaf category on elementary
graphs (\ref{lem:density}, \ref{prop:flat}).  Such presheaves, the structure
underlying coloured props, are called {\em digraphical species}\footnote{Warning: the
word `species' is used here as in combinatorics~\cite{Bergeron-Labelle-Leroux}. It
has nothing to do with the use of the word to mean `places', as occurs sometimes in
Petri net theory motivated by chemical reaction networks (see for
example~\cite{Baez-Pollard:1704.02051} and the references therein).} (or sometimes
{\em bicollections})~\cite{Kock:1407.3744}.

\bigskip

The second main result shows the benefit of whole-grain Petri nets for
unfolding:

\bigskip

{\bf Theorem.} (Cf.~\ref{thm:unfolding}) 
{\em Any (grounded) whole-grain Petri net admits a universal unfolding.}

\bigskip

No safety assumptions are required here, in contrast to the classical result of
Nielsen--Plotkin--Winskel~\cite{Nielsen-Plotkin-Winskel:1981}.
It should be stressed that the proof of this theorem
follows the ideas and arguments employed by Winskel in the safe case --- the only 
novelty is the new whole-grain setting, where the arguments go through even in the 
non-safe case.

\bigskip

The key point of the new formalism, as exploited in the proofs of the main
theorems, is that whole-grain Petri nets are 
configurations of sets, and that these sets have elements that can be
accessed, giving total control over symmetries.
With the systematic use of groupoids, these symmetries are kept around as
they are, giving some advantage over the traditional formalisms of Petri nets,
where the symmetries are difficult to control.

The general importance of groupoids in combinatorics was discovered and
advocated by Joyal~\cite{Joyal:1981} and
Baez--Dolan~\cite{Baez-Dolan:finset-feynman}. The specific insight of
representing algebraic structures by groupoids of configurations of sets
has been found useful in algebraic topology (mainly in the theory of
operads) \cite{Gambino-Kock:0906.4931}, \cite{Kock:0807},
\cite{Joyal-Kock:0908.2675}, \cite{Kock:1407.3744},
\cite{Batanin-Berger:1305.0086}, and in algebraic combinatorics (in
connection with incidence bialgebras and M\"obius
inversion)~\cite{Galvez-Kock-Tonks:1602.05082},
\cite{Galvez-Kock-Tonks:1512.07573}, \cite{Galvez-Kock-Tonks:1612.09225}.
The mathematical tools are thus already available. The whole-grain Petri
nets of the present paper can be seen as an intermediate notion between
traditional Petri nets and
pre-nets~\cite{Bruni-Meseguer-Montanari-Sassone:CTCS99},
\cite{Bruni-Meseguer-Montanari-Sassone:2001}, in the same way as polynomial
monads are intermediate between symmetric operads and non-symmetric
operads~\cite{Kock:0807}.

\bigskip

The basic ideas and results are elementary, relying solely on manipulations with
finite sets, mostly pullbacks and pushouts. Some of the theoretical results and
justifications (Sections~\ref{sec:X}--\ref{sec:1cats}) require some more category
theory, and some elementary homotopy theory.

The core mathematical content could be given more succinctly, but the text
has grown longer for three reasons. Firstly, many examples, figures, and
explanations have been included to illustrate the formalism, where
different from that of the traditional approaches to Petri nets. Secondly,
many arguments of homotopical nature have been accompanied by intuitive
explanations and background material, hopefully making them accessible also
to readers without background in homotopy theory. Finally, an effort has
been made to try to point out origins of ideas and provide comparison with
related developments.

\subsection*{Outline of the paper}

We begin in Section~\ref{sec:graphs} with a brief summary of the
formalism of directed graphs from \cite{Kock:1407.3744}. This is where
the semantics lives, both in geometric and algebraic form, and most of
the work will take place at this level.

In Section~\ref{sec:Petri} we define the whole-grain Petri nets and their
etale maps, and in Section~\ref{sec:processes} we define their processes to
be etale maps from (acyclic) graphs.
Section~\ref{sec:X} sets up the symmetric monoidal Segal space $\ds
X_\bullet$ of processes of a fixed Petri net $\pn P$. 

In Section~\ref{sec:grsp} we fully faithfully embed the category of Petri
nets into the presheaf category on elementary graphs --- these presheaves
are called digraphical species --- and characterise the image as the {\em
flat} digraphical species. This embedding is used to define the free prop
on a Petri net in Section~\ref{sec:props}. We show that the symmetric
monoidal Segal space $\ds X_\bullet$ associated to $\pn P$ is the free
prop-in-groupoids on $\pn P$. This is the first main theorem of the paper,
Theorem~\ref{thm:Xisfree}. In Section~\ref{sec:1cats} we show how to trim
down the symmetric monoidal Segal space to a symmetric monoidal category,
and discuss the homotopy issues involved.

In Section~\ref{sec:hypergraphs} we set up some definitions and results
about certain classes of hypergraphs. This is preparation for
Section~\ref{sec:unfold} where we come to the second main theorem of the paper,
Theorem~\ref{thm:unfolding}, establishing the existence of universal unfoldings for
general Petri nets.

In Section~\ref{sec:generalmap} we look into fancier notions of morphisms
of Petri nets than the basic etale maps, and establish functoriality of
$\ds X_\bullet$ in these more general maps. Modulo the difference in
set-up, this covers the notion of morphisms of Meseguer and
Montanari~\cite{Meseguer-Montanari:monoids} (monoid homomorphisms), as well
as the more general notions given in terms of multi-relations, studied by
Winskel~\cite{Winskel:1987}.

The closing Section~\ref{sec:outlook} makes an attempt at situating this work in a
bigger picture, and in particular provide comparison with the pre-nets of Bruni,
Meseguer, Montanari, and Sassone~\cite{Bruni-Meseguer-Montanari-Sassone:2001}. There
are two short appendices: one with a few basic facts about groupoids and homotopy
pullbacks, and one with simplicial groupoids and Segal spaces.

\subsection*{Related work and outlook}

After this work was first released in preprint form (May 2020), Baez,
Genovese, Master, and
Shulman~\cite{Baez-Genovese-Master-Shulman:2101.04238} have made an
important contribution to the theory, which complements the material in
Sections~\ref{sec:grsp}--\ref{sec:1cats} in various ways, and constitutes a
nice overall picture of monoidal-category theoretic operational semantics.
First of all they propose to work directly with digraphical species (which
they call $\Sigma$-nets), as a substitute for Petri nets. This is a
fascinating idea, going beyond the strong graphical aspect of Petri nets.
Secondly, they give slick categorical proofs of key adjunctions, including
a nice alternative proof of the adjunction described below in
\ref{free-prop-monad}.

Although the field of Petri net theory is very much driven by applications --- in
computer science, natural sciences, and industry --- the present work is motivated by
theoretical interest. Up to the point developed here, I think the theoretical picture
is quite satisfactory and clean, at the price of some homotopy overhead. It is my
opinion that Petri-net theorists and end users can bargain against this price in
various ways (see \cite{Genovese-Gryzlov-Herold-Perone-Post-Videla:1904.12974} for
some explicit bargaining), which I think amounts to falling back on previous
approaches. It is my hope nevertheless that the account given here will resonate with
modern trends in theoretical computer science, and with homotopy type
theory~\cite{HoTT-book} in particular, where specified bijections are part of the
whole set-up in the form of terms of identity types.

At the same time, the whole-grain formalism in itself can be appreciated
without any homotopy overhead. The main contributions, the definitions in
Sections~\ref{sec:graphs}--\ref{sec:processes}, have no other prerequisites
than pullbacks of finite sets. The formalism aligns very well with the way
Petri nets are employed in applications, the main feature being that
whole-grain Petri nets are just configurations of finite sets, lending
themselves to software implementation. Recently, Baas, Fairbanks, Halter,
and Patterson chose the whole-grain formalism as the basis for an
implementation in Julia~\cite{Julia},
\cite{Bezanson-Karpinski-Shah-Edelman:1209.5145} of a Petri net
library~\cite{AlgebraicPetri}
(see~\cite{Patterson-Lynch-Fairbanks:2106.04703} for theoretical
background), and used it in epidemiology, first to give elegant modelling
of COVID-19 data from the UK~\cite{Halter-Patterson:epidem}, and later for
a more general compositional approach to epidemiology,
cf.~Libkind--Baas--Halter--Patterson--Fairbanks~\cite{Libkind-Baas-Halter-Patterson-Fairbanks:2203.16345}.

\section{Graphs (according to \cite{Kock:1407.3744})}
\label{sec:graphs}

The following formalism for directed graphs (and all the results in
this section) are from \cite{Kock:1407.3744}, which in turn was
heavily inspired by the polynomial formalism for
trees~\cite{Kock:0807} and by the formalism for Feynman graphs of
Joyal--Kock~\cite{Joyal-Kock:0908.2675}. Some further comparison is
provided in Section~\ref{sec:outlook}.
The graphs will play the role of what are also called {\em causal
nets} in Petri-net theory (see for example~\cite{Goltz-Reisig:1983}),
their purpose being to define processes.

\begin{blanko}{Graphs (AINOA style).}
  A {\em graph} (meaning directed graph admitting open-ended edges) is a
  diagram of finite sets
  $$
  \begin{tikzcd}
	A & \ar[l, into] I \ar[r] & N & \ar[l] O \ar[r, into] & A
  \end{tikzcd}
  $$
  where the outermost maps are injective. Here $A$ is the set of
  edges, and $N$ is the set of nodes. The set $I$ expresses the
  incidence of edges and nodes from the viewpoint of edges incoming to
  nodes, and $O$ the same for outgoing edges. The injectivity
  condition says that an edge is incoming (or outgoing) for at most
  one node. 
  The elements in $A$ not in the image of $I$ or not in the image of $O$
  are the open-ended edges.
  We shall only consider acyclic
  graphs, meaning having no directed cycles (see \cite{Kock:1407.3744}
  and \ref{levelfunction} below).
\end{blanko}

\begin{blanko}{Example.}\label{ex:gr}
  The graph
  $$
  \begin{tikzcd}
	\{a,b,c,d,e\}  & \{b,c,d\} \ar[l, into] \ar[r, 
	"\begin{array}{c}b\mapsto y\\ c\mapsto z\\ d\mapsto z\end{array}"'] 
	&  \{x,y,z\} & 
	\{c,d, e\}\ar[l, "\begin{array}{c}y\mapsfrom c\\ y\mapsfrom d\\ 
	y\mapsfrom e
  \end{array}"]\ar[r, into]  & \{a,b,c,d,e\}
  \end{tikzcd}
  $$
  can be pictured like this:
  
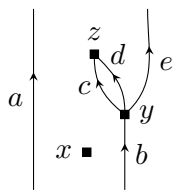
\begin{center}
	  \begin{tikzpicture}[scale=0.9]
	  \begin{scope}
		\tikzstyle arrowstyle=[scale=1]
		\small
		\node [grnode, label=left:$x$] (x) at (0.5,0.5) {};
		\node [grnode, label=right:$y$] (y) at (1.0,1.0) {};
		\node [grnode, label=above:$z$] (z) at (0.6,1.8) {};
		\node [export] (o1) at (-0.2,2.4) {};
		\node [export] (o2) at (1.3,2.4) {};
		\node [import] (i1) at (-0.2,0.0) {};
		\node [import] (i2) at (1.0,0.0) {};
		\draw[directed] (i1) -- node[left] {$a$} (o1);
		\draw[directed] (i2) -- node[right] {$b$} (y);
		\draw[directed] (y) to[out=140, in=270] node[left] {$c$} (z);
		\draw[directed] (y) to[out=90, in=-45] node[above=3pt] {$d$} (z);
		\draw[directed] (y) to[out=45, in=270] node[right] {$e$} (o2);
	  \end{scope}
	  \end{tikzpicture}
  \end{center}
   The picture is a full rendition of the data of the 
  AINOA diagram, except that the sets $I$ and $O$ are not explicit.
  They can be derived from the picture as subsets of $A$.
\end{blanko}

\begin{blanko}{Etale maps.}
  An {\em etale map} of graphs is a diagram
  $$
  \begin{tikzcd}
	A' \ar[d, "\alpha"'] & I'\ar[d] \ar[l] \ar[r] \drpullback &  N'\ar[d] & 
	O'\ar[d] \ar[l]\ar[r] \dlpullback & A'\ar[d, "\alpha"]
	\\
	A & I \ar[l]\ar[r]& N & O \ar[l]\ar[r]& A ,
  \end{tikzcd}
  $$
  where the middle squares are pullbacks.
  The pullback condition expresses that arities of nodes must be
  respected; in other words the map is a `homeomorphism' locally at
  each node. 
  An {\em open map} is an etale map that is furthermore injective on 
  nodes and edges.
\end{blanko}

\begin{blanko}{Remark.}
  The notion of etale map has a clear intuitive content. It also fits
  into the axiomatic notion of classes of etale maps of
  Joyal--Moerdijk~\cite{Joyal-Moerdijk:openmaps}; see also
  \cite{Joyal-Nielsen-Winskel}, \cite{Hayman-Winskel:2008}. There are
  other useful notions of morphisms of AINOA graphs, some of which are
  used in \cite{Kock:1407.3744} (see also~\cite{Kock:1611.10342}),
  but for the present purposes only etale maps are relevant.
\end{blanko}

\begin{blanko}{Sums and connectedness.}
  The category $\Gr$ of acyclic graphs and etale maps has
  categorical sums given by disjoint union of graphs. These are
  calculated pointwise (i.e.~on $A$, $N$, $I$, and $O$ separately).
  The empty graph is neutral for sum.
  A
  graph is {\em connected} if it is non-empty and cannot be written as
  a sum of smaller non-empty graphs.\footnote{Warning: in \cite{Kock:1407.3744}
  the symbol $\Gr$ denotes the category
  of {\em connected} graphs.}
\end{blanko}

\begin{blanko}{Pullbacks.}\label{pbk}
  The category $\Gr$ has pullbacks, computed pointwise (i.e.~for the $A$, $I$, $N$, $O$ components separately).
\end{blanko}

\begin{blanko}{Unit graph and edges.}\label{unit}
  The {\em unit graph} is the graph $1\emptyset\emptyset\emptyset1$.
  An edge in a graph $\pn G = AINOA$
    $$
  \begin{tikzcd}
	1 \ar[d, "\alpha"'] & \emptyset\ar[d] \ar[l] \ar[r]\drpullback&  \emptyset\ar[d] & 
	\emptyset\ar[d] \ar[l]\ar[r]\dlpullback& 1\ar[d, "\alpha"]
	\\
	A & I \ar[l]\ar[r]& N & O \ar[l]\ar[r]& A 
  \end{tikzcd}
  $$
  is \emph{incoming} to $\pn G$ if the right-most square is a pullback, and
  \emph{outgoing} if the left-most square is a pullback (in addition to the standing
  requirement that the two middle squares are pullbacks). Accordingly, the {\em
  in-boundary} of $\pn G$ is defined to be the set $\operatorname{in}(\pn G)$ of
  edges in the complement of $O \rightarrowtail A$  (that is, the complement
  of the image of this injective map) and the {\em out-boundary}
  $\operatorname{out}(\pn G)$ is similarly defined as the complement of $A
  \leftarrowtail I$. An edge is {\em isolated} if it belongs to both the in-boundary
  and the out-boundary. An edge is {\em inner} if it is outgoing to some node {\em
  and} incoming to some node. The set of inner edges is thus the intersection of the
  (images of the) two maps $O \rightarrowtail A \leftarrowtail I$.
  We write $x\lessdot y$ if there is an inner edge from $x$ to $y$.  
  
  In Example~\ref{ex:gr}, the in-boundary is $\{a,b\}$, the 
  out-boundary is $\{a,e\}$, and the set of inner edges is $\{c,d\}$.
\end{blanko}

\begin{blanko}{Elementary graphs.}
  An {\em elementary graph} is a connected graph with no inner edges.
  An elementary graph is thus either a unit graph or a {\em corolla},
  which means one of the form
  $$
  \begin{tikzcd}
	m+n & \ar[l] m \ar[r] & 1 & \ar[l] n \ar[r] & m+n .
  \end{tikzcd}
  $$
  Here $m$ is the set of incoming edges and $n$ is the set of outgoing
  edges.

  Let $\elGr\subset \Gr$ denote the full subcategory of elementary graphs and etale maps
  --- in fact it is practical to work rather with a skeleton of this
  category, as we do henceforth. We denote by $[\triv]$ the unit
  graph, and by $[\coro{m}{n}]$ the corolla with $m$ incoming and $n$
  outgoing edges.  These are pictured, respectively, as
  
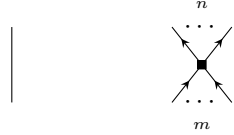
\begin{center}
	\begin{tikzpicture}
	  \begin{scope}
		\tikzstyle arrowstyle=[scale=1]
		\draw[directed] (0,0) -- (0, 1.0);
	  \end{scope}
	  \begin{scope}[shift={(2,-0.0)}]
		\tikzstyle arrowstyle=[scale=1]
		\small
		\node [grnode] (x) at (0.5,0.5) {};
		\node [export] (o1) at (0.0,1.0) {};
		\node [export] (o2) at (1.0,1.0) {};
		\node [import] (i1) at (0.0,0.0) {};
		\node [import] (i2) at (1.0,0.0) {};
		\draw[directed] (i1) --  (x);
		\draw[directed] (i2) --  (x);
		\draw[directed] (x) -- (o1);
		\draw[directed] (x) -- (o2);
		\node at (0.5, -0.0) {\footnotesize $\cdots$};
		\node at (0.5, 1.0) {\footnotesize $\cdots$};
		\node at (0.5, -0.3) {\tiny $m$};
		\node at (0.5, 1.3) {\tiny $n$};
	  \end{scope}
	\end{tikzpicture}	
  \end{center}
  
  Note that the only non-invertible maps in $\elGr$ are the 
  inclusions of 
  $[\triv]$ into a corolla $[\coro{m}{n}]$ (of which there are $m+n$). In addition
  there are the symmetries of the corollas: there are $m!n!$ invertible 
  maps $[\coro{m}{n}]\to [\coro{m}{n}]$.  The etale condition precludes 
  non-invertible maps between corollas.   
  Denote by $\Cor$ the full subcategory consisting of the corollas; it is thus a groupoid.
\end{blanko}

\begin{blanko}{Colimits and gluing.}\label{pushout}
  The category $\Gr$ admits enough colimits to account neatly for 
  gluing~\cite{Kock:1407.3744}.
  In particular, 
  if $\pn M$ is a node-less graph (disjoint union of unit graphs), 
  which embeds into the out-boundary of a graph $\pn G_1$ and 
  embeds into the in-boundary of a graph $\pn G_2$, then the pushout
  
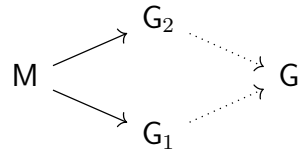
\[\begin{tikzcd}[row sep={22pt,between origins}]
  & \pn G_2 \ar[rd, dotted] & \\
  \pn M \ar[ru] \ar[rd] &&  \pn G  \\
  & \pn G_1 \ar[ru, dotted] &
  \end{tikzcd}
  \]
  exists in the category $\Gr$, and it is calculated pointwise 
  (i.e.~for the $A$, $I$, $N$, $O$ components separately).
  See \cite{Kock:1407.3744} for details. The finite-set pushouts are 
  along injections, and can be computed very explicitly. It gives a 
  clean formalisation of the intuitive idea of gluing parts of the out-boundary 
  of one graph to parts of the in-boundary of another, as exemplified 
  in this picture:
  
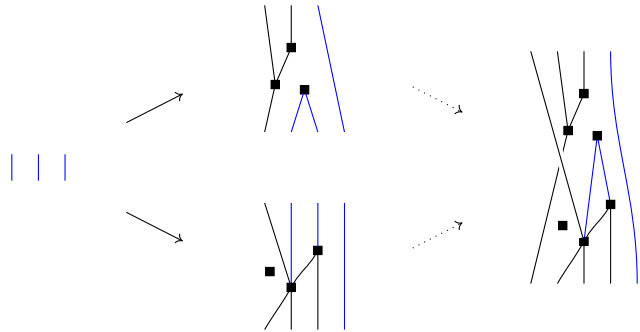
\begin{center}
  \begin{tikzpicture}[scale=0.9]
	\tikzstyle arrowstyle=[scale=1]
	
	\small

	\coordinate (M) at (0,1.8);
	\coordinate (G2) at (3.5,0);
	\coordinate (G1) at (3.5,3.6);
	\coordinate (G) at (7,1.8);	
	
	\draw[-{>[width=4pt]}, shorten >=1.7cm, shorten <=1.2cm] (M) -- (G1);
	\draw[-{>[width=4pt]}, shorten >=1.7cm, shorten <=1.2cm] (M) -- (G2);
	\draw[-{>[width=4pt]}, dotted, shorten >=1.4cm, shorten <=1.4cm] (G1) -- (G);
	\draw[-{>[width=4pt]}, dotted, shorten >=1.4cm, shorten <=1.4cm] (G2) -- (G);

	\begin{scope}[shift={(M)}, scale=0.6]
	\begin{scope}[shift={(-0.5,-0.25)}]
      \draw[blue] (0.0, 0.5) --+ (0.0, -0.5);
	  \draw[blue] (0.5, 0.5) --+ (0.0, -0.5);
	  \draw[blue] (1.0, 0.5) --+ (0.0, -0.5);
	\end{scope}
	\end{scope}

  	\begin{scope}[shift={(G1)}, scale=0.6]
	\begin{scope}[shift={(-0.75,-3.7)}]
	  \node [grnode] (x) at (0.5,3.4) {};
	  \node [grnode] (y) at (0.2,2.7) {};
	  \node [grnode] (z) at (0.75,2.6) {};
	  \node [port] (i0) at (0.0,4.2) {};
	  \node [port] (i1) at (0.5,4.2) {};
	  \node [port] (i2) at (1.0,4.2) {};
	  \node [port] (m1) at (0.0,1.8) {};
	  \node [port] (m2) at (0.5,1.8) {};
	  \node [port] (m3) at (1.0,1.8) {};
	  \node [port] (m4) at (1.5,1.8) {};
	  \draw (i0) -- (y);
	  \draw (i1) -- (x);
	  \draw (x) -- (y);
	  \draw (y) -- (m1);
	  \draw[blue] (z) --  (m2);
	  \draw[blue] (z) --  (m3);
	  \draw[blue] (i2) -- (m4);
	\end{scope}
	\end{scope}
	
  	\begin{scope}[shift={(G2)}, scale=0.6]
	\begin{scope}[shift={(-0.75,-0.3)}]
	  \node [grnode] (u) at (1.0,1.3) {};
	  \node [grnode] (v) at (0.5,0.6) {};
	  \node [grnode] (w) at (0.1,0.9) {};
	  \node [port] (m0) at (0.0,2.2) {};
	  \node [port] (m1) at (0.5,2.2) {};
	  \node [port] (m2) at (1.0,2.2) {};
	  \node [port] (m3) at (1.5,2.2) {};
	  \node [port] (o1) at (0.0,-0.2) {};
	  \node [port] (o2) at (0.5,-0.2) {};
	  \node [port] (o3) at (1.0,-0.2) {};
	  \node [port] (o4) at (1.5,-0.2) {};
	  \draw (m0) -- (v);
	  \draw[blue] (m1) -- (v);
	  \draw[blue] (m2) -- (u);
	  \draw[blue] (m3) -- (o4);
	  \draw (u) to[out=240, in=60] (v);
	  \draw (v) to[out=240, in=60] (o1);
	  \draw (v) -- (o2);
	  \draw (u) --  (o3);
	\end{scope}
	\end{scope}
	
  	\begin{scope}[shift={(G)}, scale=0.6]
	\begin{scope}[shift={(-0.25,-2)}]
	  \node [grnode] (x) at (0.5,3.4) {};
	  \node [grnode] (y) at (0.2,2.7) {};
	  \node [grnode] (z) at (0.75,2.6) {};
	  \node [port] (i0) at (0.0,4.2) {};
	  \node [port] (i1) at (0.5,4.2) {};
	  \node [port] (i2) at (1.0,4.2) {};
	  \draw (i0) -- (y);
	  \draw (i1) -- (x);
	  \draw (x) -- (y);

	  \node [grnode] (u) at (1.0,1.3) {};
	  \node [grnode] (v) at (0.5,0.6) {};
	  \node [grnode] (w) at (0.1,0.9) {};
	  \node [port] (o1) at (0.0,-0.2) {};
	  \node [port] (o2) at (0.5,-0.2) {};
	  \node [port] (o3) at (1.0,-0.2) {};
	  \node [port] (o4) at (1.5,-0.2) {};
	  \draw (u) to[out=240, in=60] (v);
	  \draw (v) to[out=240, in=60] (o1);
	  \draw (v) -- (o2);
	  \draw (u) --  (o3);
	  \node [port] (oo) at (-0.5,-0.2) {};
	  \node [port] (ii) at (-0.5,4.2) {};
	  \draw (y) -- (oo);
	  \draw[draw=white, double=white, very thick]  (ii) -- (v);
	  \draw (ii) -- (v);
	  \draw[blue] (z) --  (u);
	  \draw[blue] (z) --  (v);	  
	  \draw[blue] (i2) to[out=270, in=90] (o4);

	\end{scope}
	\end{scope}

  \end{tikzpicture}
  \end{center}
  The reader is encouraged to write down these graphs in AINOA 
  diagrams and actually compute the pushout.
\end{blanko}

\begin{blankothm}{Lemma}\label{colim-el}
  {\upshape\textrm{(\cite{Kock:1407.3744})}}
  Every graph is canonically the colimit of its elementary
  subgraphs:
  $$
  \pn G \simeq \colim_{\pnsmall E \in \el(\pnsmall G)} \pn E .
  $$
\end{blankothm}
The colimit is over the {\em category of elements}
of the graph $\pn G$, defined as the comma category
$$
\el(\pn G) := \elGr \comma \pn G ,
$$
in turn defined by the
  natural transformation diagram (lax pullback)
  \[
  \begin{tikzcd}[column sep={21mm,between origins}]
  \elGr\comma \pn G \ar[d]\ar[r] & 1 \ar[d, "\name{\pn G}"]  \\
  \elGr \ar[ru, phantom, "\Rightarrow"]
  \ar[r] & \Gr  .
  \end{tikzcd}
  \]
The objects of $\el(\pn G)$ are thus the elementary subgraphs of $\pn G$ (or more
precisely, etale maps from elementary graphs). For details, see
\cite{Kock:1407.3744}.
In concrete terms, the diagram consists of all the edges and all the nodes, with 
each edge mapping into the
corollas it is incident to, as exemplified in this colimit decomposition of a
graph $\pn G$:

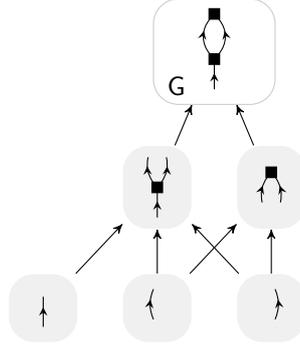
  \begin{center}
	\begin{tikzpicture}[scale=0.8,
	  node distance=0.9cm,on grid,>=stealth',bend angle=22.5,auto,
	  ]
	  
	  \small
			  
	  \node[rectangle, rounded corners=2ex, draw=black!20, 
	    minimum width=13mm, minimum height=11mm]
		(colim) at (0.75,3.4) {};

	  \node[rectangle, rounded corners=2ex, fill=black!06, 
	    minimum width=8mm, minimum height=8mm]
		(deg0graph0) at (-1.5,0.0) {};

	  \node[rectangle, rounded corners=2ex, fill=black!06, 
	    minimum width=8mm, minimum height=8mm]
		(deg0graph1) at (0.0,0.0) {};

	  \node[rectangle, rounded corners=2ex, fill=black!06, 
	    minimum width=8mm, minimum height=8mm]
		(deg0graph2) at (1.5,0.0) {};

	  \node[rectangle, rounded corners=2ex, fill=black!06,
	    minimum width=8mm, minimum height=9mm]
		(deg1graph1) at (0.0,1.6) {};
		
	  \node[rectangle, rounded corners=2ex, fill=black!06,
	    minimum width=8mm, minimum height=9mm]
		(deg1graph2) at (1.5,1.6) {};

	  \draw[->] (deg0graph0) -- (deg1graph1);
	  \draw[->] (deg0graph1) -- (deg1graph1);
	  \draw[->] (deg0graph1) -- (deg1graph2);
	  \draw[->] (deg0graph2) -- (deg1graph1);
	  \draw[->] (deg0graph2) -- (deg1graph2);
	  \draw[->] (deg1graph1) -- (colim);
	  \draw[->] (deg1graph2) -- (colim);

      \begin{scope}[shift={(deg0graph0)}]
      \begin{scope}[shift={(0.0,-0.3)}]
		\tikzstyle arrowstyle=[scale=0.9]
		\node [port] (i1) at (0.0,0.0) {};
		\node [port] (o1) at (0.0,0.45) {};
		\draw[directed] (i1) to (o1);
	  \end{scope}
	  \end{scope}
      \begin{scope}[shift={(deg0graph1)}]
      \begin{scope}[shift={(-0.05,-0.15)}]
		\tikzstyle arrowstyle=[scale=0.9]
		\node [port] (i1) at (0.0,0.0) {};
		\node [port] (o1) at (0.0,0.45) {};
		\draw[directed, bend left] (i1) to (o1);
	  \end{scope}
	  \end{scope}
      \begin{scope}[shift={(deg0graph2)}]
      \begin{scope}[shift={(0.05,-0.15)}]
		\tikzstyle arrowstyle=[scale=0.9]
		\node [port] (i1) at (0.0,0.0) {};
		\node [port] (o1) at (0.0,0.45) {};
		\draw[directed, bend right] (i1) to (o1);
	  \end{scope}
	  \end{scope}
	  
      \begin{scope}[shift={(deg1graph1)}]
      \begin{scope}[shift={(0.0,-0.4)}]
		\tikzstyle arrowstyle=[scale=0.9]
		\node [port] (i1) at (0.0,0.0) {};
		\node [grNode] (t1) at (0.0,0.4) {};
		\node [port] (o1) at (-0.12,0.8) {};
		\node [port] (o2) at (0.12,0.8) {};
		\draw[directed, bend left] (t1) to (o1);
		\draw[directed, bend right] (t1) to (o2);
		\draw[directed] (i1) to (t1);
	  \end{scope}
	  \end{scope}
	  
      \begin{scope}[shift={(deg1graph2)}]
      \begin{scope}[shift={(0.0,-0.5)}]
		\tikzstyle arrowstyle=[scale=0.9]
		\node [port] (i1) at (-0.12,0.33) {};
		\node [port] (i2) at (0.12,0.33) {};
		\node [grNode] (t1) at (0.0,0.8) {};
		\draw[directed, bend left] (i1) to (t1);
		\draw[directed, bend right] (i2) to (t1);
	  \end{scope}
	  \end{scope}

      \begin{scope}[shift={(colim)}]
      \begin{scope}[shift={(0.0,-0.5)}]
		\tikzstyle arrowstyle=[scale=0.9]
		\node [port] (i1) at (0.0,0.0) {};
		\node [grNode] (t1) at (0.0,0.4) {};
		\node [grNode] (t2) at (0.0,1.0) {};
		\draw[directed] (i1) to (t1);
		\draw[directed, bend left=40] (t1) to (t2);
		\draw[directed, bend right=40] (t1) to (t2);
	  \end{scope}
		\node at (-0.5,-0.45) {\footnotesize $\pn G$};
	  \end{scope}
 
\end{tikzpicture} 
\end{center}

\begin{blanko}{Dynamics of graphs.}\label{evolution}
  It is fruitful to view a graph as an interpolation between the
  in-boundary and the out-boundary, an evolution --- not with respect to
  any absolute notion of time, but reflecting the fact that both $N$ and
  $A$ carry a preorder structure. A graph is {\em acyclic} if the preorder
  $N$ is actually a poset (i.e.~is an anti-symmetric
  relation).\footnote{One can also put $N$ and $A$ together in a single
  poset, whose Hasse diagram is then bipartite. Such posets form the
  substrate of the classical poset semantics of Petri nets
  \cite{Petri:1977}, \cite{Genrich-StankiewiczWiechno},
  \cite{Goltz-Reisig:1983}; see for example the
  monograph~\cite{Best-Fernandez}.} Further notions of time can be imposed
  in terms of the following concept:
\end{blanko}

\begin{blanko}{Level functions.}\label{levelfunction}
   A {\em level function} of a graph $\pn G = AINOA$ is a
  monotone map $f: N \to \un k$, where $\un k := \{1,2,\ldots,k\}$ for some $k \in \N$.
  {\em Monotone} means that for every edge from node $x$ to node $y$
  we have $f(x)\leq f(y)$. A level function is {\em strict} if the inequality is 
  strict (that is, $x\lessdot y \Rightarrow f(x)<f(y)$).   A graph is acyclic iff it 
  admits a strict level function.
\end{blanko}

\begin{blanko}{Layers, cuts, and pushout decompositions.}\label{cut}
  Level functions serve in particular to split graphs into layers. For
  example, a level function $f: N \to \un 2$ will partition the node set
  into two sets $N = N_1 + N_2$, namely the pre-images $N_1 := f^{-1}(1)$
  and $N_2 := f^{-1}(2)$. This in turn will induce two open subgraphs $\pn
  G_1$ and $\pn G_2$ called {\em layers}, defined as follows. $\pn G_1$ is
  the open subgraph containing all the nodes in $N_1$ and all their
  incident edges, and also the in-boundary of $\pn G$. Precisely, to obtain
  $\pn G_1$, first take pullbacks as indicated with dotted arrows:
      $$
  \begin{tikzcd}
	\operatorname{in}(\pn G) \cup I_1 \cup O_1 \ar[d, dashed] & 
	I_1\ar[d, dotted] \ar[l, dashed] \ar[r, dotted]\drpullback&  
	N_1\ar[d, into] & 
	O_1\ar[d, dotted] \ar[l, dotted]\ar[r, dashed]\dlpullback& \operatorname{in}(\pn G) 
	\cup I_1 \cup O_1\ar[d, dashed]
	\\
	A & I \ar[l]\ar[r]& N & O \ar[l]\ar[r]& A  ,
  \end{tikzcd}
  $$
  then add the dashed arrows, which are just the inclusions. $\pn G_2$
  is constructed similarly from $N_2$, but with
  $\operatorname{out}(\pn G)$ instead of $\operatorname{in}(\pn G)$.
  The out-boundary of $\pn G_1$ will coincide with the in-boundary of
  $\pn G_2$ (as subsets of $A$). Note that isolated edges will belong
  to both $\pn G_1$ and $\pn G_2$. The intersection $\pn M := \pn G_1
  \cap \pn G_2$ (that is, pullback of the inclusion maps) constitutes
  a disjoint union of units graphs, called a {\em cut}.
  We have
  $$\operatorname{in}(\pn G)=\operatorname{in}(\pn G_1)
  \qquad
  \operatorname{out}(\pn G_1)=\pn M=\operatorname{in}(\pn G_2)
   \qquad
  \operatorname{out}(\pn G_2)=\operatorname{out}(\pn G) .
  $$

  The diagram 
  \[
  \begin{tikzcd}
  \pn M \drpullback \ar[r] \ar[d]& \pn G_2 \ar[d]  \\
  \pn G_1 \ar[r] & \pn G
  \end{tikzcd}
  \]
  is not just a pullback but also a pushout, as in \ref{pushout}.
  
  More generally, a level function $N \to \un k$ provides $k-1$ 
  compatible cuts, 
  splitting $\pn G$ into $k$ subgraphs, giving an iterated-pushout formula
  $$
  \pn G=
  \pn G_1  \amalg_{\pnsmall M_1} \pn G_2 \amalg_{\pnsmall M_2} \cdots 
  \amalg_{\pnsmall M_{k-1}} \pn G_k .
  $$
  
  We stress again that all these constructions take place in the
  category of finite sets. They are at the same time elementary and
  rigourous.
\end{blanko}

\begin{blanko}{Digraphical species~\cite{Kock:1407.3744}.}\label{Sh=PrSh}
  A {\em digraphical species}\footnote{The notion is from 
  \cite{Joyal-Kock:0908.2675}, in the setting of undirected graphs.}
  is a presheaf $\gs F : \elGr\op\to\Set$. 
  An etale map of graphs is called a {\em cover} if it is surjective 
  on nodes and edges. This defines the {\em etale topology} on $\Gr$.
  A {\em sheaf} on $\Gr$ with respect to the etale topology is a presheaf
  $\gs F : \Gr\op\to\Set$ whose value is determined by its values on any cover.
  Since every graph is the colimit of its elementary 
  subgraphs (see \ref{colim-el}), and since these elementary graphs constitute a 
  canonical cover, to give a sheaf it is enough to give its values on elementary 
  graphs.  Altogether, sheaves on 
  $\Gr$ are equivalent to presheaves on $\elGr$:
  \begin{equation}\label{eq:Sh}
  \PrSh(\elGr) \simeq \Sh_{\operatorname{et}}(\Gr) .
  \end{equation}
\end{blanko}

\begin{blanko}{Local structures on graphs.}
  Digraphical species $\gs F$ serve to impose or specify {\em local} 
  structure or property on
  graphs, by considering comma categories $\Gr\comma \gs F$, whose 
  objects $\pn G{\to}\gs F$
  are called {\em $\gs F$-graphs}.  This comma category is defined 
  as the lax pullback
  \[
  \begin{tikzcd}[column sep={41mm,between origins}]
  \Gr\comma \gs F \ar[d]\ar[r] & 1 \ar[d, "\name{\gs F}"]  \\
  \Gr \ar[ru, phantom, "\Rightarrow"]
  \ar[r, "{\tiny G\mapsto \Hom_{\Gr}( - , G)}"'] & \PrSh(\elGr)  .
  \end{tikzcd}
  \]
   An $\gs F$-graph is thus a graph whose 
  edges are decorated with elements in $\gs F[\triv]$, and whose 
  $(m,n)$-nodes are decorated with elements in $\gs F[\coro{m}{n}]$, 
  compatibly with the edge decorations and the projections 
  $\gs F[\coro{m}{n}]\to \gs F[\triv]$.
  
  For example: $k$-regular, polarised, bipartite graphs\footnote{ For
  example, to obtain bipartite graphs, consider the digraphical species
  $\elGr\op\to\Set$ with ${}[\coro{m}{n}] \mapsto \{r,b\}$, (stipulating
  that there are two kinds of nodes, red and blue) and $[\triv] \mapsto
  \{rb, br\}$ (stipulating that there are two kinds of edges, red-to-blue
  and blue-to-red). Now we should say where the arrows of $\elGr$ go: we
  declare of course that for all $[\triv] \to [\coro{m}{n}]$ that hit an
  incoming edge the map should be $\{r,b\} \to \{rb, br\}$ sending $r$ to
  $br$ and $b$ to $rb$ (and the other way around for output). } are $\gs
  F$-graphs for suitable $\gs F$ (whereas non-local notions such as
  connected, strongly regular, distance regular, etc., cannot be encoded
  with digraphical species). We shall see shortly that each Petri net $\pn
  P$ defines a digraphical species $\gs P$, and then define its processes
  to be $\gs P$-graphs.

  The local nature means that the key feature of graphs holds for
  $\gs F$-graphs too: {\em every $\gs F$-graph is canonically the colimit of its
  elementary sub-$\gs F$-graphs, in the category $\Gr\comma \gs F$.}
\end{blanko}

\section{Whole-grain Petri nets}
\label{sec:Petri}

The following definition is possibly the main contribution of this 
work.

\begin{blanko}{Petri nets (SITOS style).}
    A {\em Petri net} $\pn P$ is defined to be a diagram of finite sets
  $$
  \begin{tikzcd}
	S & \ar[l] I \ar[r] & T& \ar[l] O \ar[r] & S 
  \end{tikzcd}
  $$
  without any conditions.
  $T$ is now the set of {\em transitions} (pictured as small squares) and $S$ is the set of 
  {\em places} (pictured as circles). The sets $I$ and $O$ are the sets of {\em arcs},
  expressing the incidences between transitions and places. For $t\in 
  T$, the fibre (i.e.~pre-image set) $I_t$ is called the {\em pre-set} 
  of $t$, and the fibre $O_t$ is 
  called the {\em post-set} of $t$. (Note that these are not 
  necessarily subsets of $S$, so that there can be parallel arcs.)
  
  This notion of Petri net is the only one used in this work. When
  contrast with `traditional' definitions is required, we shall refer to the 
  present notion as {\em whole-grain Petri nets}.
\end{blanko}

\begin{blanko}{Example.}\label{ex:ex}
  The Petri net 
  $$
  \begin{tikzcd}[row sep = 0pt]
  \{s_1,s_2,s_3\}  & \{i_1,i_2,i_3,i_4\} \ar[l]\ar[r]& \{t_1,t_2\} & 
  \{o_1,o_2,o_3,o_4\} \ar[l]\ar[r] &  \{s_1,s_2,s_3\}
  \\
  s_1 & \ar[l, mapsto, shorten <= 20pt, shorten >= 20pt] i_1 \ar[r, mapsto, shorten <= 20pt, shorten >= 20pt] & t_1 & 
  \ar[l, mapsto, shorten <= 20pt, shorten >= 20pt] o_1 \ar[r, mapsto, shorten <= 20pt, shorten >= 20pt ] & s_2
  \\[-4pt]
  s_2 & \ar[l, mapsto, shorten <= 20pt, shorten >= 20pt]i_2 \ar[r, mapsto, shorten <= 20pt, shorten >= 20pt]& t_1 & 
  \ar[l, mapsto, shorten <= 20pt, shorten >= 20pt ] o_2 \ar[r, mapsto, shorten <= 20pt, shorten >= 20pt ] & s_3
  \\[-4pt]
  s_3 & \ar[l, mapsto, shorten <= 20pt, shorten >= 20pt ] i_3 \ar[r, mapsto, shorten <= 20pt, shorten >= 20pt] & t_2 & 
  \ar[l, mapsto, shorten <= 20pt, shorten >= 20pt ] o_3 \ar[r, mapsto, shorten <= 20pt, shorten >= 20pt] & s_1
  \\[-4pt]
  s_3 & \ar[l, mapsto, shorten <= 20pt, shorten >= 20pt] i_4 \ar[r, mapsto, shorten <= 20pt, shorten >= 20pt] & t_2 & 
  \ar[l, mapsto, shorten <= 20pt, shorten >= 20pt ] o_4 \ar[r, mapsto, shorten <= 20pt, shorten >= 20pt] & s_1
  \end{tikzcd}
  $$
  is pictured in the usual way as
  
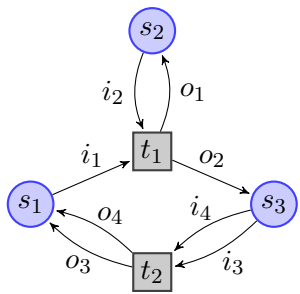
\begin{center}
  \begin{tikzpicture}[scale=0.8,
	 node distance=1.4cm,on grid,>=stealth',bend angle=22.5,auto,
     every place/.style= {minimum size=5mm,thick,draw=blue!75,fill=blue!20}, 
     every transition/.style={minimum size=4mm,thick,draw=black!75,fill=black!20}]

	 \small
	 
	  \node [transition] (t1) {$t_1$};
	  \node [place](s1) [left=of t1, yshift=-7mm]  {$s_1$}; 
	  \node [place] (s2) [above=of t1] {$s_2$};
	  \node [place] (s3) [right=of t1, yshift=-7mm] {$s_3$};
	  \node [transition] (t2) [below=of t1] {$t_2$};

	  \path (s1) edge [post] node[above]  {$i_1$} (t1);
	  \path (t1) edge [post, bend right] node[right]  {$o_1$} (s2);
	  \path (s2) edge [post, bend right] node[left]  {$i_2$} (t1);
	  \path (t1) edge [post] node[above] {$o_2$}(s3);

	  \path (s3) edge [post, bend right=15] node[above, pos=0.6]  {$i_4$} (t2);
	  \path (s3) edge [post, bend left=15] node[below, pos=0.3]  {$i_3$} (t2);
	  \path (t2) edge [post, bend left=15] node[below, pos=0.6]  {$o_3$} (s1);
	  \path (t2) edge [post, bend right=15] node[above, pos=0.3]  {$o_4$} (s1);

  \end{tikzpicture}
  \end{center} 
  There is a one-to-one correspondence between the elements in the 
  SITOS diagram and the elements of the picture. Also the maps of the 
  diagram can be read off the picture.
\end{blanko}

\begin{blanko}{Comparison with traditional definitions of Petri nets.} 
   Traditionally, instead of allowing parallel arcs, arcs can have multiplicities.
   This is usually formalised by equipping two incidence relations $I \subset S\times
   T$ and $O \subset T \times S$ with multiplicity functions. An elegant
   formulation is due to Meseguer and Montanari~\cite{Meseguer-Montanari:monoids} who
   define a Petri net to be a pair of maps $T \rightrightarrows C(S)$, where $C(S)$
   is the free commutative monoid on $S$. 
   
   It is quite curious that the difference
   between parallel arcs and multiplicities does not seem to have been exploited
   before. Apparently, parallel arcs and multiplicities have always been regarded as the
   same thing.\footnote{An early example is Hack (1975)~\cite{Hack:PhD}, who describes
   `generalised Petri nets' as having `bundles of arcs', but when he formalises the
   notion he uses multiplicity functions; see also the influential 1977 survey by
   Peterson~\cite{Peterson:1977}.}
   
   The SITOS diagrams of the whole-grain formalism could be written more economically,
   avoiding the repetition of the set $S$:
   \[
   \begin{tikzcd}[column sep={18mm,between origins}, row sep={6mm,between origins}]
	 & I \ar[ld] \ar[rd] & \\
   T  & & S , \\
   & O \ar[lu] \ar[ru] & 
   \end{tikzcd}
   \]
   emphasising that it is the data of two parallel spans, in analogy with
   the Meseguer--Montanari definition $T \rightrightarrows C(S)$. For
   further comparison, see \ref{whichmonad}--\ref{S}. The SITOS arrangement
   of the diagrams, duplicating the set $S$ and putting it at the ends (as
   we shall always do), turns out to be the most practical in connection
   with morphisms. It is also an important design choice that the diagrams
   extend the polynomial formalism for trees~\cite{Kock:0807} (trees and
   forests are the special case $A \leftarrow M \to N = N \to A$) as well
   as graphs and hypergraphs~\cite{Kock:1407.3744}, as will play a crucial
   role in this work.
\end{blanko}

\begin{blanko}{Morphisms of Petri nets.}
  A Petri net is thought of as a configuration of interacting 
  transitions, in turn prescriptions for computation steps of 
  certain kinds. The main characteristic of a transition is its 
  interface, its input-output typing. Morphisms of Petri nets should
  respect these characteristics (in addition to respecting the 
  incidences expressed by the interaction). For this reason it is 
  natural to stipulate (for the moment) 
  that a {\em morphism of Petri nets} is an etale map of
  diagrams
    $$
  \begin{tikzcd}
	S' \ar[d, "\alpha"'] & I'\ar[d] \ar[l] \ar[r] \drpullback &  T'\ar[d] & 
	O'\ar[d] \ar[l]\ar[r] \dlpullback & S'\ar[d, "\alpha"]
	\\
	S & I \ar[l]\ar[r]& T & O \ar[l]\ar[r]& S   .
  \end{tikzcd}
  $$
  Let $\kat{Petri}$ denote the category of Petri nets and etale maps.
  The etale maps correspond to what Winskel~\cite{Winskel:1984}
  calls {\em folding maps} in the setting of traditional Petri nets 
  (they are also the maps employed in \cite{Baldan-Corradini-Ehrig-Heckel:2005}, 
  \cite{Baez-Master:1808.05415}, \cite{Baez-Genovese-Master-Shulman:2101.04238}).
  Other authors (including Winskel~\cite{Winskel:1987} and 
  Meseguer--Montanari~\cite{Meseguer-Montanari:monoids}) work with more 
  general morphisms.
  We shall come back to these more general morphisms in
  Section~\ref{sec:generalmap}. 
\end{blanko}

\begin{blanko}{Graphs.}
   Graphs (in the sense of Section~\ref{sec:graphs}) are 
  clearly Petri nets,
  interpreting nodes as transitions and edges as places.
  They are the Petri nets with the special property that every place is in the
  pre-set of at most one transition and in the post-set of at most one transition.
  The interplay between graphs and Petri nets is a main ingredient in the
  theory.  In analogy with the category of graphs, we find:
\end{blanko}
\begin{blankothm}{Lemma}
  The category $\kat{Petri}$ has pullbacks, and they are calculated 
  pointwise (i.e.~separately on the $S$, $I$, 
$T$, $O$ components).
\end{blankothm}

\begin{blankothm}{Lemma}\label{lem:Pcolims}
  The category $\kat{Petri}$ has pushouts and coequalisers over unit
  graphs, and they are calculated pointwise. Every Petri net is the
  colimit of elementary graphs over unit graphs (transitions glued
  along places).
\end{blankothm}

\begin{blanko}{Example.} 
   The Petri net $\pn P$ (from Example~\ref{ex:ex}) is the colimit of the
  graphs arranged in the grey diagram:

	
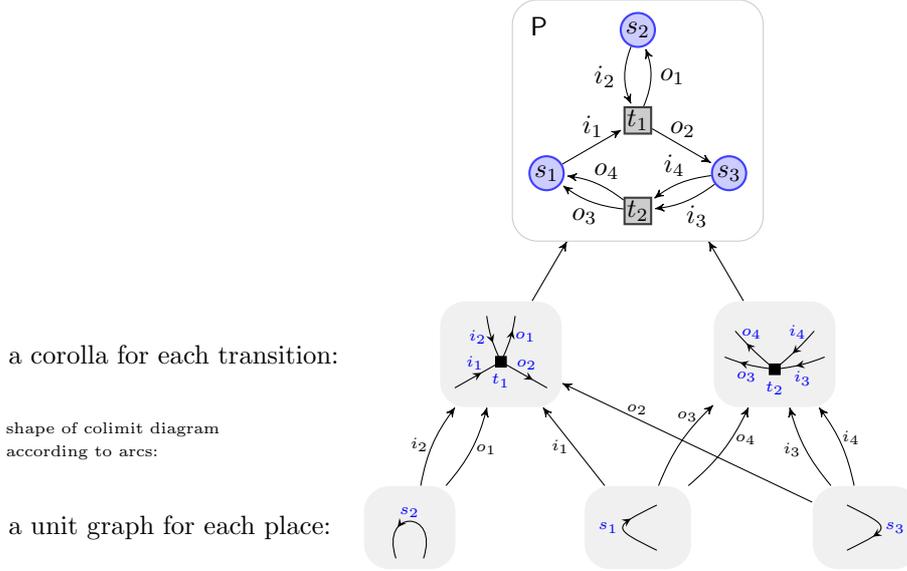
\begin{tikzpicture}[scale=0.85,
	  node distance=0.9cm,on grid,>=stealth',bend angle=22.5,auto,
	  every place/.style= {minimum size=5mm,thick,draw=blue!75,fill=blue!20}, 
	  every transition/.style={minimum size=4mm,thick,draw=black!75,fill=black!20},
	  every token/.style={minimum size=3.5pt, token distance=8pt}
	  ]
	  
	  \small
			  
	  \node[rectangle, rounded corners=2ex, fill=black!06, 
	    minimum width=10mm, minimum height=9mm]
		(edge-s2) at (-3.0,0.0) {};
	  \node[rectangle, rounded corners=2ex, fill=black!06, 
	    minimum width=12mm, minimum height=9mm]
		(edge-s1) at (0.0,0.0) {};
	  \node[rectangle, rounded corners=2ex, fill=black!06, 
	    minimum width=12mm, minimum height=9mm]
		(edge-s3) at (3.0,0.0) {};
	  \node[rectangle, rounded corners=2ex, fill=black!06,
	    minimum width=13mm, minimum height=11mm]
		(graph1) at (-1.8,2.3) {};
	  \node[rectangle, rounded corners=2ex, fill=black!06,
	    minimum width=13mm, minimum height=11mm]
		(graph2) at (1.8,2.3) {};
	  \node[rectangle, rounded corners=2ex, draw=black!20, 
	    minimum width=27mm, minimum height=27mm]
		(colim) at (0.0,5.4) {};
	  \footnotesize
	  \node[anchor=west] at (-8.4,2.3) {a corolla for each transition:};
	  \node[anchor=west] at (-8.4,0) {a unit graph for each place:};
	  \tiny
	  \node[anchor=west] at (-8.4,1.3) {shape of colimit diagram};
	  \node[anchor=west] at (-8.4,1.0) {according to arcs:};

	  \draw[->, bend left=13] (edge-s2) to node[left=-1pt] {\tiny $i_2$} (graph1);
	  \draw[->, bend right=13] (edge-s2) to node[right=-1pt] {\tiny $o_1$}  (graph1);
	  \draw[->] (edge-s1) to node[left=-1pt] {\tiny $i_1$} (graph1);
	  \draw[->, bend left=12] (edge-s1) to node[left=-1pt, pos=0.85] {\tiny $o_3$} (graph2);
	  \draw[->, bend right=12] (edge-s1) to node[right=-1pt, pos=0.65] {\tiny $o_4$} (graph2);
	  \draw[->] (edge-s3) --  node[above=-1pt, pos=0.7] {\tiny $o_2$} (graph1);
	  \draw[->, bend left=12] (edge-s3) to node[left=-1pt, pos=0.5] {\tiny $i_3$} (graph2);
	  \draw[->, bend right=12] (edge-s3) to node[right=-1pt, pos=0.6] {\tiny $i_4$}  (graph2);
	  \draw[->] (graph1) to (colim);
	  \draw[->] (graph2) to (colim);

	  \begin{scope}[shift={(edge-s2)}]
      \begin{scope}[shift={(0.0,-0.1)}]
		\tikzstyle arrowstyle=[scale=0.9]
		\node (o1) at (0.15,-0.4) {};
		\node (i2) at (-0.15,-0.4) {};
		\draw[directed] (o1) .. controls (0.4,0.35) and (-0.4,0.35) ..
		node[above=-1pt] {\color{blue} \tiny $s_2$} 
		(i2);
	  \end{scope}		  
	  \end{scope}	  
	  \begin{scope}[shift={(edge-s1)}]
      \begin{scope}[shift={(-0.15,0)}]
		\tikzstyle arrowstyle=[scale=0.9]
		\draw[directed] 
		(0.4,-0.3) 
		.. controls (-0.2,0.0) and (-0.2,0.0) ..
		node[left=-1pt] {\color{blue} \tiny $s_1$} 
		(0.4,0.3);
	  \end{scope}		  
	  \end{scope}	  
	  \begin{scope}[shift={(edge-s3)}]
      \begin{scope}[shift={(0.15,-0)}]
		\tikzstyle arrowstyle=[scale=0.9]
		\draw[directed] 
		(-0.4,0.3) 
		.. controls (0.2,0.0) and (0.2,0.0) ..
		node[right=-1pt] {\color{blue} \tiny $s_3$} 
		(-0.4,-0.3);
	  \end{scope}		  
	  \end{scope}	  
	  
	  \begin{scope}[shift={(graph1)}]
      \begin{scope}[shift={(0,-0.1)}]
		\tiny
		\tikzstyle arrowstyle=[scale=0.9]
		\node (i1) at (-0.7,-0.4) {};
		\node (i2) at (-0.2,0.7) {};
		\node (o1) at (0.2,0.7) {};
		\node (o2) at (0.7,-0.4) {};
		\node [grnode, label=below:${\color{blue}t_1}$] (t1) at (0.0,0.0) {};
		\draw[directed] (i1) to node[above=-0pt] {\color{blue} \tiny $i_1$} (t1);
		\draw[directed, bend right=8] (i2) to node[left=-1pt] {\color{blue} \tiny $i_2$} (t1);
		\draw[directed, bend right=8] (t1) to node[right=-1pt] {\color{blue} \tiny $o_1$} (o1);
		\draw[directed] (t1) to node[above=-0pt] {\color{blue} \tiny $o_2$} (o2);
	  \end{scope}
	  \end{scope}

	  \begin{scope}[shift={(graph2)}]
      \begin{scope}[shift={(0.0,-0.2)}]
		\tiny
		\tikzstyle arrowstyle=[scale=0.9]
		\node (i3) at (0.75,0.2) {};
		\node (i4) at (0.6,0.6) {};
		\node (o3) at (-0.75,0.2) {};
		\node (o4) at (-0.6,0.6) {};
		\node [grnode, label=below:${\color{blue}t_2}$] (t2) at (0.0,0.0) {};
		\draw[directed, bend left=7] (i3) to node[below=-0pt] {\color{blue} \tiny $i_3$} (t2);
		\draw[directed, bend left=4] (i4) to node[above=1pt] {\color{blue} \tiny $i_4$} (t2);
		\draw[directed, bend left=7] (t2) to node[below=-0pt] {\color{blue} \tiny $o_3$} (o3);
		\draw[directed, bend left=4] (t2) to node[above=1pt] {\color{blue} \tiny $o_4$} (o4);
	  \end{scope}		  
	  \end{scope}	  

	  \begin{scope}[shift={(colim)}, 
		node distance=1.0cm,on grid,>=stealth',bend angle=22.5,auto,
	   every place/.style= {minimum size=4.2mm,thick,draw=blue!75,fill=blue!20}, 
	   every transition/.style={minimum size=3.2mm,thick,draw=black!75,fill=black!20}]
	  \footnotesize
	  \node [transition] (t1) {$t_1$};
	  \node [place](s1) [left=of t1, yshift=-7mm]  {$s_1$}; 
	  \node [place] (s2) [above=of t1] {$s_2$};
	  \node [place] (s3) [right=of t1, yshift=-7mm] {$s_3$};
	  \node [transition] (t2) [below=of t1] {$t_2$};
	  \path (s1) edge [post] node[above]  {$i_1$} (t1);
	  \path (t1) edge [post, bend right] node[right]  {$o_1$} (s2);
	  \path (s2) edge [post, bend right] node[left]  {$i_2$} (t1);
	  \path (t1) edge [post] node[above] {$o_2$} (s3);
	  \path (s3) edge [post, bend right=15] node[above, pos=0.6]  {$i_4$} (t2);
	  \path (s3) edge [post, bend left=15] node[below, pos=0.3]  {$i_3$} (t2);
	  \path (t2) edge [post, bend left=15] node[below, pos=0.6]  {$o_3$} (s1);
	  \path (t2) edge [post, bend right=15] node[above, pos=0.3]  {$o_4$} (s1);
	  \node at (-1.3,1.25) {$\pn P$};
	  \end{scope}

\end{tikzpicture}

\end{blanko}

\begin{blanko}{Marking (or state).}
  A {\em marking} (also called a {\em state}\footnote{The tendency is to use the word 
  `marking' as long as the Petri net is considered just a combinatorial or pictorial 
  gadget, whereas `state' is used when the operational interpretation is in focus.}) of a Petri net $\pn P =
  SITOS$ is a map of finite sets $M \to S$, regarded as an etale map
  from $M\emptyset\emptyset\emptyset M$ to $\pn P$. The elements of
  $M$ are called {\em tokens}.
    Throughout, we shall use the letter $M$ for a set of tokens, and 
  tacitly write $\pn M = M\emptyset\emptyset\emptyset M$ for the 
  corresponding nodeless graph.
  
  The markings of $\pn P$ naturally form a category 
  $\F \comma \pn P \simeq \F \comma S$, but we shall be more interested in the 
  corresponding groupoid $\B\comma \pn P \simeq \B \comma S$ (where $\B$ denotes the 
  groupoid of finite sets and bijections), defined by the lax pullback
  \[
  \begin{tikzcd}[column sep={31mm,between origins}]
  \B\comma \pn P \ar[d]\ar[r] & 1 \ar[d, "\name{\pn P}"]  \\
  \B \ar[ru, phantom, "\Rightarrow"]
  \ar[r, "{\tiny M\mapsto M\emptyset\emptyset\emptyset M}"'] & \kat{Petri}  .
  \end{tikzcd}
  \] 
  The objects are thus the markings, and an
  isomorphism between two markings is a bijection of sets $M
  \isopil M'$ compatible with the maps to $\pn P$.
  Two markings are isomorphic in this groupoid if and only if
  they have the same number of tokens for each place. So the isomorphism 
  classes of markings are precisely the classical multisets on $S$. More 
  categorically speaking, 
    the groupoid 
  $\B\comma S$ is the free symmetric monoidal
  category on $S$, and its set of isomorphism classes is 
  $\pi_0(\B\comma S) = C(S)$, the set of multisets on $S$.
  In traditional Petri net theory, that is the definition of marking. 
\end{blanko}

\begin{blanko}{Example.}\label{ex:state}
  To specify a marking (of the Petri net from \ref{ex:ex}) such as
  
  \small
  
  $$
  \begin{tikzcd}[row sep={14mm,between origins}, column sep={29mm,between origins}]
	\{m_1,m_2,m_3, m_3'\} \ar[d, "{\begin{array}{c}
  m_1 \mapsto s_1, m_3\mapsto s_3\\[-3pt] m_2 \mapsto s_2, m_3'\mapsto s_3
\end{array}}"] & \emptyset \ar[l]\ar[r]\ar[d]\drpullback 
	&\emptyset \ar[d] &  
	\emptyset \ar[l]\ar[r]\ar[d]\dlpullback  & \{m_1,m_2,m_3, m_3'\} \ar[d]\\
	\{s_1,s_2,s_3\}  & \{i_1,i_2,i_3,i_4\} \ar[l]\ar[r]& \{t_1,t_2\} & 
	\{o_1,o_2,o_3,o_4\} \ar[l]\ar[r] &  \{s_1,s_2,s_3\}
  \end{tikzcd}
  $$
  
  \normalsize
  
  \noindent
  it is necessary to say where each token lands, as indicated here on the left-most 
  vertical arrow (recall that the right-most vertical arrow is 
  the same map). Since any finite set $M$ can appear in a marking,
  there is generally no way to avoid specifying the map $M\to S$ element by element. However, 
  we may always choose to work with an isomorphic marking, `renaming' 
  the tokens for
  convenience. (In this example the token `names' were chosen so that the map can be
  read off the subscripts.) In the drawing
  
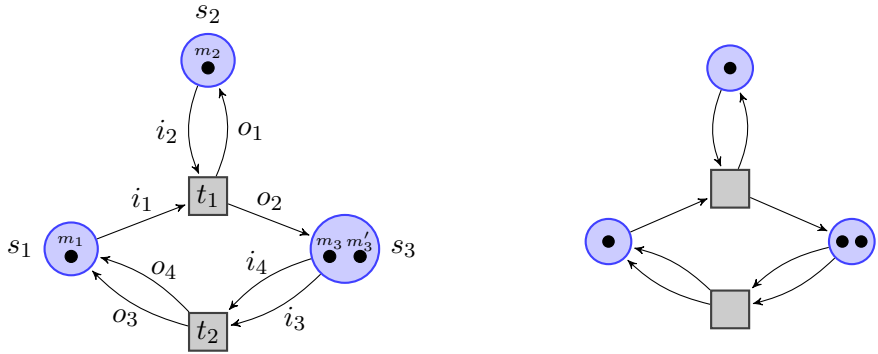
\begin{center}
	\begin{tikzpicture}[scale=0.9,
	   node distance=1.5cm,on grid,>=stealth',bend angle=22.5,auto,
	   every place/.style= {minimum size=6mm,thick,draw=blue!75,fill=blue!20}, 
	   every transition/.style={minimum size=3.5mm,thick,draw=black!75,fill=black!20}
	   every token/.style={minimum size=3.5pt, token distance=8pt}
	   ]
	  \small
	  \node [transition] (t1) {$t_1$};
	  \node [place](s1) [left=of t1, yshift=-7mm, label=left:$s_1$] {}; 
	  \node [place] (s2) [above=of t1,label=above:$s_2$] {};
	  \node [place, minimum size=9mm] (s3) [right=of t1, yshift=-7mm, label=right:$s_3$] {};
	  \node [transition] (t2) [below=of t1] {$t_2$};
	  \path (s1) edge [post] node[above]  {$i_1$} (t1);
	  \path (t1) edge [post, bend right] node[right]  {$o_1$} (s2);
	  \path (s2) edge [post, bend right] node[left]  {$i_2$} (t1);
	  \path (t1) edge [post] node[above] {$o_2$} (s3);
	  \path (s3) edge [post, bend right=15] node[above, pos=0.6]  {$i_4$} (t2);
	  \path (s3) edge [post, bend left=15] node[below, pos=0.3]  {$i_3$} (t2);
	  \path (t2) edge [post, bend left=15] node[below, pos=0.6]  {$o_3$} (s1);
	  \path (t2) edge [post, bend right=15] node[above, pos=0.3]  {$o_4$} (s1);

	  \begin{scope}[shift=(s1)]
		\filldraw[black] (0.0,-0.1)  circle (0.68mm);
		\node at (0.0,0.1) {\tiny $m_1$};
	  \end{scope}
	  \begin{scope}[shift=(s2)]
		\filldraw[black] (0.0,-0.1)  circle (0.68mm);
		\node at (0.0,0.1) {\tiny $m_2$};
	  \end{scope}
	  \begin{scope}[shift=(s3)]
		\filldraw[black] (-0.15,-0.1)  circle (0.68mm);
		\node at (-0.2,0.07) {\tiny $m_3$};
		\filldraw[black] (0.15,-0.1)  circle (0.68mm);
		\node at (0.2,0.1) {\tiny $m_3'$};
	  \end{scope}
	\end{tikzpicture}
	\hspace{18mm}\raisebox{3mm}{
	\begin{tikzpicture}[scale=0.9,
	   node distance=1.4cm,on grid,>=stealth',bend angle=22.5,auto,
	   every place/.style= {minimum size=5mm,thick,draw=blue!75,fill=blue!20}, 
	   every transition/.style={minimum size=4mm,thick,draw=black!75,fill=black!20}]
	  \small
	  \node [transition] (t1) {};
	  \node [place, tokens=1](s1) [left=of t1, yshift=-7mm] {}; 
	  \node [place,tokens=1] (s2) [above=of t1] {};
	  \node [place,tokens=2] (s3) [right=of t1, yshift=-7mm] {};
	  \node [transition] (t2) [below=of t1] {};
	  \path (s1) edge [post] (t1);
	  \path (t1) edge [post, bend right] (s2);
	  \path (s2) edge [post, bend right] (t1);
	  \path (t1) edge [post] (s3);
	  \path (s3) edge [post, bend right=15] (t2);
	  \path (s3) edge [post, bend left=15] (t2);
	  \path (t2) edge [post, bend left=15] (s1);
	  \path (t2) edge [post, bend right=15] (s1);
	\end{tikzpicture}}
  \end{center}
  the left-hand picture contains the exact same information as the diagram. 
  The right-hand picture contains only the information of the {\em isomorphism
  class} of the marked Petri net: this contains all the information about the maps in the 
  diagram, except the choices of `names' of elements. 
  In practice it may often be the case that the abstract picture (the isomorphism 
  class) exists first, for example as a result of human thought. The act of choosing 
  an explicit representative of the iso-class to work with is effectively to choose 
  the names for the elements in the picture, and use those elements as constituents 
  of the sets $S, I, T, O$ and $M$. It is thus important that {\em some} choice of 
  representing sets is made, so that there are explicit sets to work with, but it is 
  not important which choice is made.
\end{blanko}

\begin{blanko}{Individual vs.~collective tokens.} 
  (See \cite{Glabbeek-Plotkin:LICS95}, \cite{Glabbeek-Plotkin:2009},
  \cite{Bruni-Montanari:2000}, \cite{Glabbeek:2005},
  \cite{Bruni-Meseguer-Montanari-Sassone:2001} for more thorough
  discussion.) Classical Petri-net theory favours the {\em
  collective-tokens philosophy}, according to which a state is just a
  multiplicity function $S \to \N$, that is, a multiset on $S$. In the
  present formalism the tokens of a state form an explicit set, and in
  particular, each token is an individual element in a set
  (as in the individual-tokens philosophy). However, since the states form a
  groupoid, both the viewpoints are encoded simultaneously: from the
  groupoid one can pass to the individual tokens by considering the
  underlying set of the groupoid, and one can pass to the
  indistinguishable-tokens viewpoint by passing to the set of
  isomorphism classes of the groupoid.
  At this point the difference is not so big. The real difference 
  arises when it comes to tracing tokens around in processes, as we 
  shall see.
\end{blanko}

\begin{blanko}{Initial state.}
  In many applications, Petri nets have an {\em initial state}, and it is
  common to include this in the very definition of Petri net.
  We do not do so here. When required, initial states are given as
  diagrams $\pn M \to \pn P$,
  and morphisms are then required to respect this. The relevant 
  categories are then coslice categories. We shall come back to this 
  in Section~\ref{sec:unfold} in connection with unfolding.
\end{blanko}

\section{Processes}
\label{sec:processes}

\begin{blanko}{Simple firing.}\label{simplefiring}
  A {\em firing} of a transition $t\in T$ in a Petri net $\pn P = SITOS$
  in a given state $\pn M \to \pn P$ intuitively consumes a token from each of
  the places ingoing to $t$ and produces a token in each of the
  outgoing places. More precisely it consumes a token for each element
  in the pre-set $I_t$ (the restriction $S \leftarrow I_t$ tells where
  the tokens are taken from) and produces a token for each element in
  the post-set $O_t$ (the restriction $O_t \to S$ then tells where the
  new tokens are put).
  
  The minimal state in which the firing of $t\in T$ can occur is
  $$
  \begin{tikzcd}
	I_t \ar[d] & \emptyset\ar[d] \ar[l] \ar[r]\drpullback&  \emptyset\ar[d] & 
	\emptyset\ar[d] \ar[l]\ar[r]\dlpullback& I_t\ar[d]
	\\
	S & I \ar[l]\ar[r]& T & O \ar[l]\ar[r]& S,
  \end{tikzcd}
  $$
  and the state after the transition has fired will then be
  $$
  \begin{tikzcd}
	O_t \ar[d] & \emptyset\ar[d] \ar[l] \ar[r]\drpullback&  \emptyset\ar[d] & 
	\emptyset\ar[d] \ar[l]\ar[r]\dlpullback& O_t\ar[d]
	\\
	S & I \ar[l]\ar[r]& T & O \ar[l]\ar[r]& S .
  \end{tikzcd}
  $$
  The whole firing (in the minimal state enabling it) is encoded 
  geometrically by a 
  single etale map $\pn C\to \pn P$ from a corolla, namely
  $$
  \begin{tikzcd}
	I_t+O_t \ar[d] & I_t\ar[d] \ar[l] \ar[r] \drpullback &  \{t\}\ar[d] & 
	O_t\ar[d] \ar[l]\ar[r] \dlpullback & I_t+O_t\ar[d]
	\\
	S & I \ar[l]\ar[r]& T & O \ar[l]\ar[r]& S .
  \end{tikzcd}
  $$
   (Note that any other isomorphic sets could take the place of $I_t$, $O_t$ to
  constitute a corolla.)
  When reading such a $\pn C \to \pn P$ as a
  firing, the initial state is that given by the in-boundary of $\pn
  C$ and the final state is that given by the out-boundary of $\pn C$ 
  (cf.~\ref{unit} and \ref{evolution}). 
  (In the displayed case, these are $I_t$ and $O_t$.)

  A firing of the transition $t\in T$ in a general state is encoded by
  just adding more tokens. From the viewpoint of the corolla $\pn C$, this
  is to add a bunch of isolated edges, so that the general firing of
  $t\in T$ has the form
  $$
  \begin{tikzcd}
	I_t+O_t+M \ar[d] & I_t\ar[d] \ar[l] \ar[r] \drpullback &  \{t\}\ar[d] & 
	O_t\ar[d] \ar[l]\ar[r] \dlpullback & I_t+O_t+M\ar[d]
	\\
	S & I \ar[l]\ar[r]& T & O \ar[l]\ar[r]& S 
  \end{tikzcd}
  $$
  for some set $M$. The domain is then no longer a corolla, but it is 
  still a graph. The interpretation of initial and final state in
  terms of in-boundary and out-boundary of the graph is still valid,
  since the isolated edges, corresponding 
  to the tokens not changed by the firing, belong to  both the in- 
  and the out-boundary (cf.~\ref{unit}). 
\end{blanko}

\begin{blanko}{Example.}\label{firings}
   Here is a minimal firing $p:\pn C \to \pn P$ of the transition $t_1$ in
  the Petri net $\pn P$ from Example~\ref{ex:ex}:
  
  \footnotesize
  
  $$
  \begin{tikzcd}[row sep={17mm}]
 \ar[d, phantom, "p:"] & \{a_1,a_2,b_2,b_3\} \ar[d, "{\color{blue} \begin{array}{c}
  a_1\mapsto s_1\\[-4pt] a_2 \mapsto s_2\\[-4pt] b_2\mapsto s_2\\[-4pt] b_3 \mapsto s_3
\end{array}}"] & \{a_1,a_2\} \ar[l]\ar[r]
	\ar[d, "{\color{blue}\begin{array}{c}a_1\mapsto i_1\\[-4pt] a_2 \mapsto i_2\end{array}}"]\drpullback 
	&\{x_1\} \ar[d, "{\color{blue} \ x_1\mapsto t_1}"] &  
	\{b_2,b_3\} \ar[l]\ar[r]\ar[d, "{\color{blue}\begin{array}{c}b_2\mapsto o_1\\[-4pt] b_3 \mapsto o_2\end{array}}"]\dlpullback  & \{a_1,a_2,b_2,b_3\} \ar[d]
	\\
	{} &
	\{s_1,s_2,s_3\}  & \{i_1,i_2,i_3,i_4\} \ar[l]\ar[r]& \{t_1,t_2\} & 
	\{o_1,o_2,o_3,o_4\} \ar[l]\ar[r] &  \{s_1,s_2,s_3\}
  \end{tikzcd}
  $$
  
  \normalsize
    
  \noindent The top row is the corolla $\pn C$; the bottom row is the Petri net 
  $\pn P$.
  The vertical maps, constituting the etale map $p$, are specified (in blue) simply 
  by telling where each individual element goes. In the picture

  
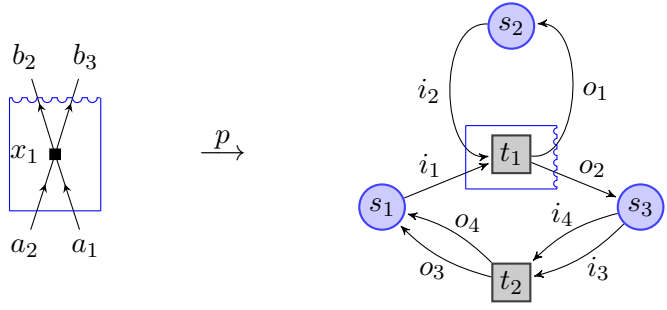
\begin{center}
	\begin{tikzpicture}[scale=0.8]
	  
	  \small
	  
	  \coordinate (C) at (0,0);
	  \coordinate (P) at (6,0);
	  \node at (2.2,0) {$\longrightarrow$};
	  \node at (2.2,0.2) {$p$};
	  
	  \begin{scope}[shift={(C)}]
		\tikzstyle arrowstyle=[scale=1]
	  
		\small
		\node [grnode, label={left:$x_1$}] (x1) at (0,0) {};
		
		\node [import] (a2) at (-0.4,-1) {$a_2$};
		\node [import] (a1) at (0.4,-1) {$a_1$};
		\node [export] (b2) at (-0.4,1) {$b_2$};
		\node [export] (b3) at (0.4,1) {$b_3$};

		\draw[directed] (a1) -- (x1);
		\draw[directed] (a2) -- (x1);
		\draw[directed] (x1) -- (b2);
		\draw[directed] (x1) -- (b3);
		
		\def\x{0.6}
		\def\y{0.75}
		\draw[blue] (-\x,-\y) to (-\x,\y);
		\draw[blue] (\x,-\y) to (\x,\y);
		\draw[blue] (-\x,-\y) to (\x,-\y);
		\draw[blue] (-\x,\y) foreach \x in {1,2,3,4,5,6} {
		  -- ++(0.035,0) 
		  arc[start angle=180, end angle=360, radius=0.064] 
		  -- ++(0.035,0) 
		  };

	  \end{scope}
	  
	  \begin{scope}[scale=0.8, shift={(P)},
	node distance=1.5cm,on grid,>=stealth',bend angle=22.5,auto,
     every place/.style= {minimum size=5mm,thick,draw=blue!75,fill=blue!20}, 
     every transition/.style={minimum size=4mm,thick,draw=black!75,fill=black!20}]
	 
	  \node [transition] (t1) {$t_1$};
	  \node [place](s1) [left=of t1, yshift=-7mm]  {$s_1$}; 
	  \node [place] (s2) [above=of t1] {$s_2$};
	  \node [place] (s3) [right=of t1, yshift=-7mm] {$s_3$};
	  \node [transition] (t2) [below=of t1] {$t_2$};

	  \path (s1) edge [post] node[above, pos=0.3]  {$i_1$} (t1);
	  \path (t1) edge [post, bend right=95] node[right]  {$o_1$} (s2);
	  \path (s2) edge [post, bend right=95] node[left]  {$i_2$} (t1);
	  \path (t1) edge [post] node[above, pos=0.7] {$o_2$} (s3);
	  \path (s3) edge [post, bend right=15] node[above, pos=0.6]  {$i_4$} (t2);
	  \path (s3) edge [post, bend left=15] node[below, pos=0.3]  {$i_3$} (t2);
	  \path (t2) edge [post, bend left=15] node[below, pos=0.6]  {$o_3$} (s1);
	  \path (t2) edge [post, bend right=15] node[above, pos=0.3]  {$o_4$} (s1);
	  
	  \begin{scope}[shift={(0,-0.04)}, scale=0.7]
		\draw[blue] (-1,-0.7) to (1,-0.7);
		\draw[blue] (-1,0.7) to (1,0.7);
		\draw[blue] (-1,-0.7) to (-1,0.7);
		\draw[blue] (1,-0.7) foreach \x in {1,2,3,4,5,6} {
		  -- ++(0,0.044) 
		  arc[start angle=-90, end angle=-270, radius=0.072] 
		  -- ++(0,0.044) 
		  };
	  \end{scope}
	  
	  \end{scope}

	\end{tikzpicture}
  \end{center}
  this information is conveyed by the `postage stamp'.
\end{blanko}

\begin{blanko}{Executions --- preliminary discussion.}\label{executions-prelim}
  An execution of a Petri net $\pn P = SITOS$ in a state $\pn M\to \pn P$
  is supposed to be just a bunch of firings taking place in sequence or
  concurrently. The fully parallel situation is given by an etale map
  $p:\pn G \to \pn P$, where $\pn G$ is a disjoint union of elementary
  graphs: the corollas in $\pn G$ then express the simultaneous firing, and
  the isolated edges are just dead weight contributing to the state. Again
  the initial state is the (restriction of $p$ to the) in-boundary of the
  graph $\pn G$, and the final state is the (restriction of $p$ to the)
  out-boundary of $\pn G$. Causal relationships between firings are
  expressed with more general graphs:
\end{blanko}

\begin{blanko}{Processes = $\pn P$-graphs.}
  A {\em process} of a Petri net $\pn P = SITOS$ is an etale map $p:\pn G \to \pn P$
  where $\pn G$ is an acyclic graph. This is also called a {\em $\pn P$-graph}.
  
  We have previously used $AINOA$ notation for graphs. From now on the symbols $I$
  and $O$ are reserved for the incidence sets of Petri nets, so for graphs we now use
  the notation $A_N$ and ${}_N A$ for the subsets of $A$ consisting of the edges that
  are incoming to some node and outgoing of some node, respectively. A process for
  $\pn P$ is thus a diagram
    $$
  \begin{tikzcd}
	A \ar[d] & A_N \ar[d] \ar[l] \ar[r]\drpullback&  N\ar[d] & 
	{}_N A \ar[d] \ar[l]\ar[r]\dlpullback& A\ar[d]
	\\
	S & I \ar[l]\ar[r]& T & O \ar[l]\ar[r]& S ,
  \end{tikzcd}
  $$
  where the top row is an acyclic graph.
  
  From the viewpoint of graphs, a process of $\pn P$ is an acyclic graph
  $\pn G$ where each edge is decorated by a place of $\pn P$ and each node
  is decorated with a transition of $\pn P$ of matching interface.
  Furthermore, the elements expressing the incidence relations of the graph
  $\pn G$ must be mapped to arcs of $\pn P$. We stress that the maps $A_N
  \to I$ and ${}_N A \to O$ must be specified too --- they are not implied
  from the maps $N \to T$ and $A\to S$. All these assignments should be
  compatible, as expressed by the commutativity of the diagram. The initial
  state of the process is the (restriction of the etale map to the)
  in-boundary of $\pn G$ and the final state is the (restriction to the)
  out-boundary of $\pn G$.
\end{blanko}

\begin{blanko}{The category of processes.}
   Define the category of processes of $\pn P$ to be the comma category
  $$
  \kat{Proc}(\pn P) := \Gr \comma \pn P .
  $$
  The morphisms are thus commutative triangles (of etale maps)
  \[\begin{tikzcd}[column sep={2em,between origins}]
  \pn G \ar[rr] \ar[rd, "p"'] && \pn G' \ar[ld, pos=0.37, "p'"] \\
   & \pn P .
  \end{tikzcd}
  \]
  
  The most important maps in $\kat{Proc}(\pn P)$ are the invertible maps: on one hand
  they allow us to use the sets we like as constituents of the graphs, and allow us
  to replace these sets by isomorphic ones, effectively `renaming' elements to our
  liking. The invertible maps keep track of these `renamings', and ensure the ability
  to distinguish elements in these sets. There may also be invertible maps from one
  graph to itself --- its symmetries.
  
  The category $\kat{Proc}(\pn P)$ also has non-invertible maps, such as in
  particular sub-graph inclusions. From the viewpoint of processes as computations,
  these represent shorter (or partial) computations. In particular it is important
  that the inclusion of the in- or out-boundary of a process is a map in
  $\kat{Proc}(\pn P)$ (representing the trivial computation, just a state): our goal
  will be to define serial composition of processes, which will be achieved (in
  Section~\ref{sec:X}) in terms of gluings along graph boundaries, in turn described
  as pushouts as in \ref{pushout}. For these pushouts (and other colimits) to make
  sense, it is essential to have non-invertible maps in $\kat{Proc}(\pn P)$. 
\end{blanko}

\begin{blankothm}{Proposition}\label{prop:PProcP}
  The assignment $\pn P \mapsto \kat{Proc}(\pn P)$ is functorial in etale 
  maps of Petri nets: an etale map of nets $f: \pn P'\to \pn P$ induces 
  canonically a functor $f\lowershriek : \kat{Proc}(\pn P') \to \kat{Proc}(\pn P)$ simply  by 
  post-composition of etale maps. Altogether, this defines a functor
  $\kat{Proc} : \kat{Petri} \to \kat{Cat}$.
\end{blankothm}

\begin{blanko}{Comparison with traditional notions of processes.}
   Modulo the differences in the definitions of Petri net and graphs, the
  above definition of process is precisely that of Goltz and
  Reisig~\cite{Goltz-Reisig:1983}. The idea of modelling processes of a Petri net
  with maps from graphs or certain posets goes back to Petri
  himself~\cite{Petri:1977} (1977).
  
  Despite this close similarity, the passage from traditional Petri nets to 
  whole-grain Petri nets implies some essential differences for processes, as 
  illustrated in the examples, and as will be important for the main theorems.
  To indicate the point briefly, consider a transition $t$: in a traditional Petri 
  net,
  $\operatorname{pre}(t)$ and $\operatorname{post}(t)$ are mere numbers, and the 
  `etale' condition on a map from a graph says that it can receive only nodes $x$
  with that many incoming and outgoing edges, but there is no way to control
  those incoming and outgoing edges, and in particular there are situations where
  they can be permuted. In the SITOS formalism, $\operatorname{pre}(t)$ and 
  $\operatorname{post}(t)$ are actual sets, and being an etale map involves explicit 
  bijections with these sets, $\operatorname{in}(x) \isopil
  \operatorname{pre}(t)$ and 
  $\operatorname{out}(x) \isopil \operatorname{post}(t)$, giving a good handle on
  symmetries. 
\end{blanko}

\begin{blanko}{Scheduling.}
  A {\em scheduling} of a process $p : \pn G\to \pn P$, also called a {\em run},
  is just a strict level function on 
  $\pn G$ (cf.~\ref{levelfunction}). The scheduling is {\em sequential} if the level function is
  bijective, and the scheduling is then called a {\em firing sequence}. 
  A strict level function $f: \pn G \to \un k$ prescribes a colimit decomposition
  of the graph as a sequence of pushouts over nodeless graphs as in 
  \ref{cut},
  \begin{equation}\label{GMGMG}
  \pn G \simeq \pn G_1 \amalg_{\pnsmall M_1} \cdots \amalg_{\pnsmall M_{k-1}} 
  \pn G_k ,
  \end{equation}
  where each layer $\pn G_i$ is a disjoint union of elementary graphs, and
  where the $\pn M_i$ are nodeless graphs. Therefore, each $\pn M_i \to \pn
  P$ is a state and each $\pn G_i$ is a simultaneous firing $\pn G_i \to
  \pn P$ with initial state $\pn M_{i-1}$ and final state $\pn M_i$ (here
  we include $\pn M_0$ defined as the in-boundary of $\pn G_1$ (which is
  also the in-boundary of $\pn G$) and $\pn M_k$ defined as the
  out-boundary of $\pn G_k$ (which is also the out-boundary of $\pn G$)).
  If the strict level function is bijective, each firing $\pn G_i \to \pn
  P$ is a simple firing (in the sense of \ref{simplefiring}).
\end{blanko}

\begin{blanko}{Example.}\label{ex:processes}
  The main purpose of this example is to show how closely tokens are kept
  track of in a process. The following diagram is a process $p : \pn G \to
  \pn P$:  

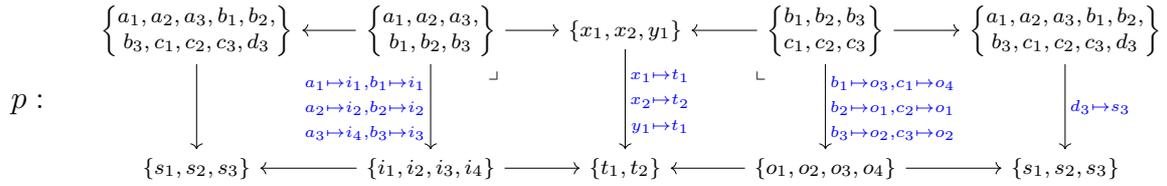
  \scriptsize
  \[
  \arraycolsep=0.0pt
  \begin{tikzcd}[row sep={52pt,between origins}]
 \ar[d, phantom, "{\text{\normalsize \(p:\)}}"] & 
 \left\{\begin{array}{c}
 a_1,a_2,a_3,b_1, b_2, \\b_3, c_1,c_2,c_3,d_3
  \end{array}\right\} 
  \ar[d] & 
  \left\{\begin{array}{c}
  a_1,a_2,a_3, \\ b_1,b_2,b_3
\end{array}\right\}
  \ar[l]\ar[r]
	\ar[d, "{\color{blue}\begin{array}{c}
  a_1 \mapsto i_1, b_1 \mapsto i_1 \\ 
  a_2 \mapsto i_2, b_2 \mapsto i_2 \\
  a_3 \mapsto i_4, b_3 \mapsto i_3
\end{array}}"']\drpullback 
	&\{x_1, x_2, y_1\} \ar[d, "{\color{blue}\begin{array}{c}
  x_1 \mapsto t_1 \\ 
  x_2 \mapsto t_2\\
  y_1 \mapsto t_1
\end{array}}"] &  
	\left\{\begin{array}{c}
	b_1,b_2,b_3 \\ c_1,c_2,c_3
	\end{array}\right\}
	\ar[l]\ar[r]\ar[d, "{\color{blue}\begin{array}{c}
  b_1 \mapsto o_3, c_1 \mapsto o_4 \\ 
  b_2 \mapsto o_1, c_2 \mapsto o_1\\
  b_3 \mapsto o_2 , c_3 \mapsto o_2
\end{array}}"]\dlpullback  & 
	 \left\{\begin{array}{c}
 a_1,a_2,a_3,b_1, b_2, \\b_3, c_1,c_2,c_3,d_3
  \end{array}\right\} 
\ar[d, "{\color{blue}d_3\mapsto s_3}"]
	\\
	{} &
	\{s_1,s_2,s_3\}  & \{i_1,i_2,i_3,i_4\} \ar[l]\ar[r]& \{t_1,t_2\} & 
	\{o_1,o_2,o_3,o_4\} \ar[l]\ar[r] &  \{s_1,s_2,s_3\}
  \end{tikzcd}
  \]
  
  \normalsize

  \noindent The bottom row is the Petri net $\pn P$ from
  Example~\ref{ex:ex}. In the top row (the graph $\pn G$), the constituent
  arrows have not been indicated, but they can be read off the following
  picture, where the graph $\pn G$ is on the left. To specify the etale map
  (the vertical arrows), it is necessary to tell where each individual
  element goes, as done with the blue annotations. For the outermost map $A
  \to S$, the assignments $a_1{\mapsto} s_1$, $a_2 {\mapsto} s_2$,
  $a_3{\mapsto} s_3$, $b_1 {\mapsto} s_1$, $b_2 {\mapsto} s_2$,
  $b_3{\mapsto} s_3$, $c_1{\mapsto} s_1$, $c_2 {\mapsto} s_2$, $c_3
  {\mapsto} s_3$ have been suppressed from the diagram to avoid clutter,
  since they can be inferred from the maps at the $I$ and $O$ levels. (Only
  the element $d_3$ is not accounted for like this, since it is an isolated
  edge.) The reason for bothering with this `economy' of annotation is that
  it is natural from the graphical viewpoint, where it is clear that the
  mapping information of $p$ is fully specified by telling where the local
  interfaces of each node are sent (and then annotating the isolated edge
  separately):

  
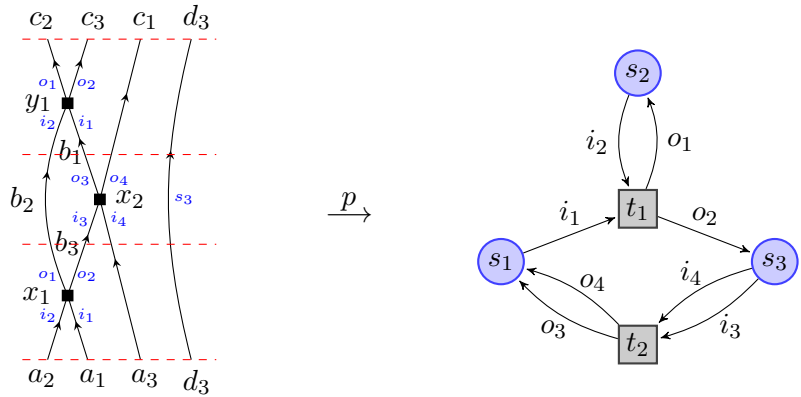
\begin{center}
	\begin{tikzpicture}[scale=0.9]
	  
	  \small
	  
	  \coordinate (G) at (-1.5,1.5 );
	  \coordinate (P) at (6,0.1);
	  \node at (2.2,0) {$\longrightarrow$};
	  \node at (2.2,0.2) {$p$};
	  
	  \begin{scope}[shift={(G)}, xscale=0.85, yscale=-0.85]
		\tikzstyle arrowstyle=[scale=1]

		\node [grnode, label=left:$x_1$] (x1) at (0,3) {};
		\node [grnode, label=right:$x_2$] (x2) at (0.5,1.5) {};
		\node [grnode, label=left:$y_1$] (y1) at (0,0) {};
		\node [import] (a1) at (0.4,4) {$a_1$};
		\node [import] (a2) at (-0.4,4) {$a_2$};
		\node [import] (a3) at (1.2,4) {$a_3$};
		\node [export] (c1) at (1.2,-1) {$c_1$};
		\node [export] (c2) at (-0.4,-1) {$c_2$};
		\node [export] (c3) at (0.4,-1) {$c_3$};
		\node [import] (d3in) at (2.0,4) {$d_3$};
		\node [export] (d3out) at (2.0,-1) {$d_3$};
		\draw[directed] (a1) -- (x1);
		\draw[directed] (a2) -- (x1);
		\draw[directed] (a3) -- (x2);
		\draw[directed] (x1) -- node[above left=-5pt, pos=0.4] {$b_3$} (x2);
		\draw[directed] (x2) -- node[below left=-7pt, pos=0.6] {$b_1$} (y1);
		\draw[directed, bend right=22pt] (x1) to  node[left] {$b_2$} (y1);
		\draw[directed] (x2) -- (c1);
		\draw[directed] (y1) -- (c2);
		\draw[directed] (y1) -- (c3);
		\draw[directed, bend right=14pt] (d3in) to (d3out);
		
		\draw[dashed, red] (-0.7,4.0) --+ (3.0,0);
		\draw[dashed, red] (-0.7,2.3) --+ (3.0,0);
		\draw[dashed, red] (-0.7,0.7) --+ (3.0,0);
		\draw[dashed, red] (-0.7,-1.0) --+ (3.0,0);
		
		\tiny
		\color{blue}
		
		\begin{scope}[shift=(y1)]
		\node at (-0.3,0.3) {$i_2$};
		\node at (-0.3,-0.3) {$o_1$};
		\node at (0.3,0.3) {$i_1$};
		\node at (0.3,-0.3) {$o_2$};
		\end{scope}
		\begin{scope}[shift=(x1)]
		\node at (-0.3,0.3) {$i_2$};
		\node at (-0.3,-0.3) {$o_1$};
		\node at (0.3,0.3) {$i_1$};
		\node at (0.3,-0.3) {$o_2$};
		\end{scope}
		\begin{scope}[shift=(x2)]
		\node at (-0.3,0.3) {$i_3$};
		\node at (-0.3,-0.3) {$o_3$};
		\node at (0.3,0.3) {$i_4$};
		\node at (0.3,-0.3) {$o_4$};
		\end{scope}
		\node at (1.8,1.5) {$s_3$};

	  \end{scope}
	  
	  \begin{scope}[shift={(P)},
	node distance=1.5cm,on grid,>=stealth',bend angle=22.5,auto,
     every place/.style= {minimum size=5mm,thick,draw=blue!75,fill=blue!20}, 
     every transition/.style={minimum size=4mm,thick,draw=black!75,fill=black!20}]
	 
	  \node [transition] (t1) {$t_1$};
	  \node [place](s1) [left=of t1, yshift=-7mm]  {$s_1$}; 
	  \node [place] (s2) [above=of t1] {$s_2$};
	  \node [place] (s3) [right=of t1, yshift=-7mm] {$s_3$};
	  \node [transition] (t2) [below=of t1] {$t_2$};

	  \path (s1) edge [post] node[above]  {$i_1$} (t1);
	  \path (t1) edge [post, bend right] node[right]  {$o_1$} (s2);
	  \path (s2) edge [post, bend right] node[left]  {$i_2$} (t1);
	  \path (t1) edge [post] node[above] {$o_2$} (s3);
	  \path (s3) edge [post, bend right=15] node[above, pos=0.6]  {$i_4$} (t2);
	  \path (s3) edge [post, bend left=15] node[below, pos=0.3]  {$i_3$} (t2);
	  \path (t2) edge [post, bend left=15] node[below, pos=0.6]  {$o_3$} (s1);
	  \path (t2) edge [post, bend right=15] node[above, pos=0.3]  {$o_4$} (s1);
	  \end{scope}

	\end{tikzpicture}
  \end{center}
  A choice of level function has been indicated with 
  the red dashed lines.

  Note that
  the maps at the $I$ and $O$ levels
  contain very precise information about token flow: we see for example that when
  $t_2$ fires in the process $p$, its $i_3$ input slot consumes the token
  that was previously produced by the $t_1$-firing (namely $b_3$) whereas the $i_4$
  input slot of $t_2$ consumes the token $a_3$ that was already there from the start.
  If the etale map were modified to have $a_3 \mapsto i_3$ instead of $a_3 \mapsto i_4$, and
  $b_3 \mapsto i_4$ instead of $b_3 \mapsto i_3$ (corresponding to interchanging 
  the two decorations $i_3$ and $i_4$ in the picture), then
  a different (and non-isomorphic) process would result, with a different flow of 
  tokens.

  A more fundamentally different $\pn P$-process $q$ (with a level function) is pictured here:
  
    
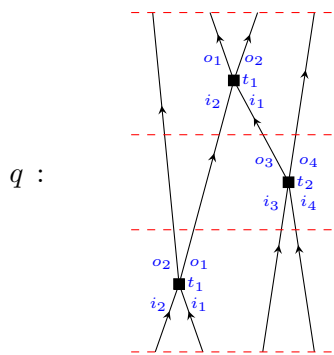
\begin{center}
	\begin{tikzpicture}[scale=0.9]
	  
	  \small
	  
	  \coordinate (H) at (11.5,1.5 );
	  \node at (9.5,0.2) {$q\,:$};
	  
	  \begin{scope}[shift={(H)}, xscale=0.9, yscale=-0.9]
		\tikzstyle arrowstyle=[scale=1]

		\node [grnode] (x1) at (0,3) {};
		\node [grnode] (x2) at (1.6,1.5) {};
		\node [grnode] (y1) at (0.8,0) {};
		\node [import] (a1) at (-0.4,4) {};
		\node [import] (a2) at (0.4,4) {};
		\node [import] (a3) at (1.2,4) {};
		\node [export] (c3) at (1.2,-1) {};
		\node [export] (b3) at (-0.4,-1) {};
		\node [export] (c2) at (0.4,-1) {};
		\node [import] (d3in) at (2.0,4) {};
		\node [export] (d3out) at (2.0,-1) {};
		\draw[directed] (a1) -- (x1);
		\draw[directed] (a2) -- (x1);
		\draw[directed] (a3) -- (x2);
		\draw[directed] (d3in) -- (x2);
		\draw[directed] (x1) -- (y1);
		\draw[directed] (x2) -- (y1);
		\draw[directed] (x1) -- (b3);
		\draw[directed] (x2) -- (d3out);
		\draw[directed] (y1) -- (c3);
		\draw[directed] (y1) -- (c2);
		\draw[dashed, red] (-0.7,4.0) --+ (3.0,0);
		\draw[dashed, red] (-0.7,2.3) --+ (3.0,0);
		\draw[dashed, red] (-0.7,0.7) --+ (3.0,0);
		\draw[dashed, red] (-0.7,-1.0) --+ (3.0,0);

	  		\tiny
		\color{blue}
		
		\begin{scope}[shift=(y1)]
		\node at (-0.3,0.3) {$i_2$};
		\node at (-0.3,-0.3) {$o_1$};
		\node at (0.35,0.3) {$i_1$};
		\node at (0.3,-0.3) {$o_2$};
		\node at (0.25,0.0) {$t_1$};
		\end{scope}
		\begin{scope}[shift=(x1)]
		\node at (-0.3,0.3) {$i_2$};
		\node at (-0.25,-0.3) {$o_2$};
		\node at (0.3,0.3) {$i_1$};
		\node at (0.3,-0.3) {$o_1$};
		\node at (0.25,0.0) {$t_1$};
		\end{scope}
		\begin{scope}[shift=(x2)]
		\node at (-0.25,0.3) {$i_3$};
		\node at (-0.35,-0.3) {$o_3$};
		\node at (0.3,0.3) {$i_4$};
		\node at (0.3,-0.3) {$o_4$};
		\node at (0.25,0.0) {$t_2$};
		\end{scope}	  
	  \end{scope}

	\end{tikzpicture}
  \end{center}
  
  \noindent
	The decorations specify precisely how the graph maps to $\pn P$, but the graph 
	itself has only been specified up to isomorphism: it remains to choose 
	representative sets (`names of elements') for the graph.
	Since we are generally only interested in structural properties of
	Petri nets and processes, it is quite reasonable to indicate only the 
	isomorphism class, leaving the implementation details such as `element names' to 
	the reader. We shall freely do this throughout the paper. 
    
  In process $p$, there is one token $d_3$ that does not participate other than
  staying put in $s_3$ all the time. In $q$, all tokens participate actively. This
  process $q$ also has the property that it could have been scheduled differently: by
  choosing a different level function one could fire $t_2$ before firing $t_1$. This
  shows that $p$ and $q$ do not have the same causal structure. In fact, it is clear
  that $p$ and $q$ are not isomorphic: they are not even isomorphic as underlying
  graphs. However, with the level functions indicated in the pictures with dashed red
  lines, the two processes both correspond the following sequence of isomorphism
  classes of firings:
  
  
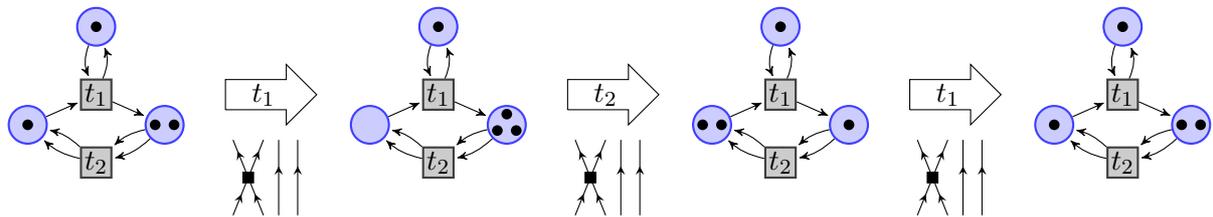
\begin{center}
	\begin{tikzpicture}[scale=0.85,
	  node distance=0.9cm,on grid,>=stealth',bend angle=22.5,auto,
	  every place/.style= {minimum size=5mm,thick,draw=blue!75,fill=blue!20}, 
	  every transition/.style={minimum 
	  size=4mm,thick,draw=black!75,fill=black!20},
	  every token/.style={minimum size=3.5pt, token distance=8pt}
	  ]
	  
	  \small
		
	  \coordinate (init) at (0,0);
	  \coordinate (aftert1) at (4.5,0);
	  \coordinate (aftert2) at (9,0);
	  \coordinate (final) at (13.5,0);

	  \coordinate (t1trans) at (2.2,0);
	  \coordinate (t1transgraph) at (2.0,-1.1);
	  \coordinate (t2trans) at (6.7,0);
	  \coordinate (t2transgraph) at (6.5,-1.1);
	  \coordinate (secondt1trans) at (11.2,0);
	  \coordinate (secondt1transgraph) at (11.0,-1.1);

  \begin{scope}[shift={(init)}]
	\node [transition] (t1) {$t_1$};
	\node [place,tokens=1](s1) [left=of t1, yshift=-4mm] {}; 
	\node [place,tokens=1] (s2) [above=of t1] {};
	\node [place,tokens=2] (s3) [right=of t1, yshift=-4mm] {};
	\node [transition] (t2) [below=of t1] {$t_2$};
	\path (s1) edge [post] (t1);
	\path (t1) edge [post, bend right] (s2);
	\path (s2) edge [post, bend right] (t1);
	\path (t1) edge [post] (s3);
	\path (s3) edge [post, bend right=15] (t2);
	\path (s3) edge [post, bend left=15] (t2);
	\path (t2) edge [post, bend left=15] (s1);
	\path (t2) edge [post, bend right=15] (s1);
  \end{scope}

  \begin{scope}[shift={(aftert1)}]
	\node [transition] (t1) {$t_1$};
	\node [place,tokens=0](s1) [left=of t1, yshift=-4mm] {}; 
	\node [place,tokens=1] (s2) [above=of t1] {};
	\node [place,tokens=3] (s3) [right=of t1, yshift=-4mm] {};
	\node [transition] (t2) [below=of t1] {$t_2$};
	\path (s1) edge [post] (t1);
	\path (t1) edge [post, bend right] (s2);
	\path (s2) edge [post, bend right] (t1);
	\path (t1) edge [post] (s3);
	\path (s3) edge [post, bend right=15] (t2);
	\path (s3) edge [post, bend left=15] (t2);
	\path (t2) edge [post, bend left=15] (s1);
	\path (t2) edge [post, bend right=15] (s1);
  \end{scope}

  \begin{scope}[shift={(aftert2)}]
	\node [transition] (t1) {$t_1$};
	\node [place,tokens=2](s1) [left=of t1, yshift=-4mm] {}; 
	\node [place,tokens=1] (s2) [above=of t1] {};
	\node [place,tokens=1] (s3) [right=of t1, yshift=-4mm] {};
	\node [transition] (t2) [below=of t1] {$t_2$};
	\path (s1) edge [post] (t1);
	\path (t1) edge [post, bend right] (s2);
	\path (s2) edge [post, bend right] (t1);
	\path (t1) edge [post] (s3);
	\path (s3) edge [post, bend right=15] (t2);
	\path (s3) edge [post, bend left=15] (t2);
	\path (t2) edge [post, bend left=15] (s1);
	\path (t2) edge [post, bend right=15] (s1);
  \end{scope}

  \begin{scope}[shift={(final)}]
	\node [transition] (t1) {$t_1$};
	\node [place,tokens=1](s1) [left=of t1, yshift=-4mm] {}; 
	\node [place,tokens=1] (s2) [above=of t1] {};
	\node [place,tokens=2] (s3) [right=of t1, yshift=-4mm] {};
	\node [transition] (t2) [below=of t1] {$t_2$};
	\path (s1) edge [post] (t1);
	\path (t1) edge [post, bend right] (s2);
	\path (s2) edge [post, bend right] (t1);
	\path (t1) edge [post] (s3);
	\path (s3) edge [post, bend right=15] (t2);
	\path (s3) edge [post, bend left=15] (t2);
	\path (t2) edge [post, bend left=15] (s1);
	\path (t2) edge [post, bend right=15] (s1);
  \end{scope}

  \begin{scope}[shift={(t1trans)}, scale=0.9]
	\draw (-0.5,-0.2) -- (0.3,-0.2) -- (0.3,-0.4) 
	-- (0.7,0.0) -- (0.3,0.4) -- (0.3,0.2) -- (-0.5,0.2)
	-- (-0.5,-0.2);
	\node (0,0) {$t_1$};
  \end{scope}
  \begin{scope}[shift={(t2trans)}, scale=0.9]
	\draw (-0.5,-0.2) -- (0.3,-0.2) -- (0.3,-0.4) 
	-- (0.7,0.0) -- (0.3,0.4) -- (0.3,0.2) -- (-0.5,0.2)
	-- (-0.5,-0.2);
	\node (0,0) {$t_2$};
  \end{scope}
  \begin{scope}[shift={(secondt1trans)}, scale=0.9]
	\draw (-0.5,-0.2) -- (0.3,-0.2) -- (0.3,-0.4) 
	-- (0.7,0.0) -- (0.3,0.4) -- (0.3,0.2) -- (-0.5,0.2)
	-- (-0.5,-0.2);
	\node (0,0) {$t_1$};
  \end{scope}

  	  \begin{scope}[shift={(t1transgraph)}, xscale=0.5, yscale=-0.5]
		\tikzstyle arrowstyle=[scale=1]

		\node [grnode] (x1) at (0,0) {};
		
		\node [port] (a1) at (-0.4,1) {};
		\node [port] (a2) at (0.4,1) {};
		\node [port] (b2) at (-0.4,-1) {};
		\node [port] (b3) at (0.4,-1) {};
		\node [port] (i3) at (0.8,1) {};
		\node [port] (i4) at (1.3,1) {};
		\node [port] (o3) at (0.8,-1) {};
		\node [port] (o4) at (1.3,-1) {};

		\draw[directed] (a1) -- (x1);
		\draw[directed] (a2) -- (x1);
		\draw[directed] (x1) -- (b2);
		\draw[directed] (x1) -- (b3);
		\draw[directed] (i3) -- (o3);
		\draw[directed] (i4) -- (o4);
	  \end{scope}
	  
  	  \begin{scope}[shift={(t2transgraph)}, xscale=0.5, yscale=-0.5]
		\tikzstyle arrowstyle=[scale=1]

		\node [grnode] (x1) at (0,0) {};
		
		\node [port] (a1) at (-0.4,1) {};
		\node [port] (a2) at (0.4,1) {};
		\node [port] (b2) at (-0.4,-1) {};
		\node [port] (b3) at (0.4,-1) {};
		\node [port] (i3) at (0.8,1) {};
		\node [port] (i4) at (1.3,1) {};
		\node [port] (o3) at (0.8,-1) {};
		\node [port] (o4) at (1.3,-1) {};

		\draw[directed] (a1) -- (x1);
		\draw[directed] (a2) -- (x1);
		\draw[directed] (x1) -- (b2);
		\draw[directed] (x1) -- (b3);
		\draw[directed] (i3) -- (o3);
		\draw[directed] (i4) -- (o4);
	  \end{scope}
	  
  	  \begin{scope}[shift={(secondt1transgraph)}, xscale=0.5, yscale=-0.5]
		\tikzstyle arrowstyle=[scale=1]

		\node [grnode] (x1) at (0,0) {};
		
		\node [port] (a1) at (-0.4,1) {};
		\node [port] (a2) at (0.4,1) {};
		\node [port] (b2) at (-0.4,-1) {};
		\node [port] (b3) at (0.4,-1) {};
		\node [port] (i3) at (0.8,1) {};
		\node [port] (i4) at (1.3,1) {};
		\node [port] (o3) at (0.8,-1) {};
		\node [port] (o4) at (1.3,-1) {};

		\draw[directed] (a1) -- (x1);
		\draw[directed] (a2) -- (x1);
		\draw[directed] (x1) -- (b2);
		\draw[directed] (x1) -- (b3);
		\draw[directed] (i3) -- (o3);
		\draw[directed] (i4) -- (o4);
	  \end{scope}

  \end{tikzpicture}
	\end{center}
\end{blanko}

\noindent This shows that the processes cannot be reconstructed, even up to
isomorphism, from knowledge of the isomorphism classes of the steps in the firing
sequence: the isomorphism classes do not retain enough information about the tokens
to be able to compose by gluing.\footnote{The fact that different graphs/posets can
correspond to the same firing sequence was stressed in the Best--Fern\'andez
book~\cite{Best-Fernandez}. Drastic quotients are required in order to make the viewpoints
match up~\cite{Best-Devillers:1987}, as we shall briefly comment on in \ref{swap}.}

\section{The symmetric monoidal Segal space of processes}

\label{sec:X}

From this point on (and up to and including Section~\ref{sec:1cats}), a little bit of
elementary homotopy theory of groupoids is involved. This is necessary since we are
interested in objects up to isomorphism, but still want to keep track of their
symmetries.  A few definitions and basic facts are recollected in
Appendix~A. The goal is to assemble processes (for a fixed Petri
net $\pn P$) into a kind of category. This will require some simplicial structures,
explained along the way; some background is given in Appendix~B.

\begin{blanko}{Towards composition of processes.}
  Fix a Petri net $\pn P$.
  Let $\ds X_0$ denote the groupoid of states of $\pn P$. Let $\ds 
  X_1 = \kat{Proc}_{\iso}(\pn P)$ denote the 
  groupoid of processes of $\pn P$.  Every state is also a process, so we 
  have a canonical map $s_0: \ds X_0 \to \ds X_1$. Every process has an 
  initial state and a final state, so we also have maps 
  $$
  \begin{tikzcd}
	\ds X_0 & \ar[l, shift left, "d_0"] \ar[l, shift right, "d_1"'] 
	\ds X_1 .
  \end{tikzcd}
  $$
   Here $d_0$ returns the {\em final} state and $d_1$ returns the {\em initial} state,
  not the other way around. Although this convention may look a bit
  counter-intuitive, the indexing is dictated by the standard simplicial formalism 
  we shall exploit,
  where an index always indicates the vertex that was {\em deleted}, as explained in
  \ref{nerve}. We think of a process $p$ as an arrow in a category, from its initial 
  state $d_1(p)$ to its final state $d_0(p)$. 

  Since states form a groupoid, whose morphisms express renaming of
  tokens, we may be more interested in composing processes that only
  match up to a specified isomorphism: given two processes $p_1:\pn
  G_1 \to \pn P$ and $p_2: \pn G_2 \to \pn P$ (that is, $p_1 \in \ds
  X_1$ and $p_2 \in \ds X_1$), we consider the situation
  where the final state of $p_1$ is isomorphic to the initial state of
  $p_2$, with a specified isomorphism $\sigma: d_0(p_1) \isopil
  d_1(p_2)$. This means precisely that the triple $(p_1,p_2,\sigma)$ is an
  element in the standard {\em homotopy pullback} of groupoids
  (see \ref{hopbk})
  \begin{equation}\label{eq:hopbk}
	\begin{tikzcd}
  \ds X_1 \times_{\ds X_0}^h \ds X_1 \drpullback \ar[rd, phantom, "\simeq" 
  description] \ar[d, "\operatorname{proj}_1"'] \ar[r, "\operatorname{proj}_2"] & \ds X_1 \ar[d, "d_1"]  \\
  \ds X_1 \ar[r, "d_0"'] & \ds X_0  .
  \end{tikzcd}
  \end{equation}

  We wish to define a weak composition law
  $$
  \ds X_1 \times_{\ds X_0}^h \ds X_1 \stackrel{\operatorname{comp.}}\longrightarrow \ds X_1 ,
  $$
  allowing to compose
  processes via connecting isomorphisms of states.\footnote{See
  Sassone~\cite{Sassone:1998} for a version of this viewpoint.} 
  To formalise this idea, we first fit $\ds X_0$ and $\ds X_1$
  into the structure of a simplicial groupoid $\ds 
  X_\bullet:\simplexcategory\op\to\Grpd$; then we show that this simplicial groupoid is a Segal space, 
  which is the standard way to encode weak categories (see
  Appendix~B for a bit of background and motivation). 
\end{blanko}

\begin{blanko}{The simplicial groupoid of processes.}
  Fix a Petri net $\pn P$. For each $k\geq 0$, denote by $\ds X_k$ the
  groupoid of $\pn P$-graphs $p : \pn G \to \pn P$ equipped with a
  level function $\pn G \to \un k$, i.e.~processes $\pn G \to \pn P$
  equipped with $k-1$ compatible cuts, as in \ref{cut}. The morphisms
  in this groupoid are isomorphisms of graphs compatible with both the
  etale map to $\pn P$ and the level function to $\un k$.
  In particular, $\ds X_1$ is just the groupoid of processes (since
  every $\pn P$-graph has a unique $1$-level function), and $\ds X_0$ is the
  groupoid of states (because a $0$-level function can exist only for
  $\pn P$-graphs with no nodes). Next, $\ds
  X_2$ is the groupoid whose objects are processes $\pn G \to \pn P$
  equipped with a cut (as in~\ref{cut}).
  
  The groupoids $\ds X_k$ assemble into a strict simplicial groupoid (see~\ref{sSet} and 
  \ref{sGrpd})
  \begin{eqnarray*}
    \ds X_\bullet : \simplexcategory\op & \longrightarrow & \Grpd  \\
    {}[k] & \longmapsto & \ds X_k .
  \end{eqnarray*}
   For this we need to describe the face and degeneracy maps (cf.~\ref{simplicial-ids}).
  
  The {\em degeneracy maps} $s_i : \ds X_k \to \ds X_{k+1}$ (for $0\leq i \leq k$) insert
  an empty layer. Formally this can be described as composing 
  level functions $\pn G \to \un k$ with injective
  monotone maps $\un k \to \un{k{+}1}$. We shall not go into details with the 
  degeneracy maps. Although they are required to define a simplicial object, they do 
  not play any role for the Segal condition. 
  
  The {\em inner face maps} $\ds X_{k-1}
  \stackrel{d_i}\longleftarrow \ds X_k$ ($0<i<k$) join adjacent 
  layers,
  or equivalently, delete cuts. Formally this is to postcompose
  level functions $\pn G \to \un k$ with surjective monotone maps $\un k \to
  \un{k{-}1}$. 
  
  So far, only the level functions are affected, whereas the underlying $\pn P$-graph
  is not changed. Finally, the {\em outer face maps} $\ds X_{k-1} 
  \overset{d_k}{\underset{d_0}{\leftleftarrows}} \ds 
  X_k$ project away the first or the
  last layer; this obviously changes the underlying $\pn P$-graph. 
  
  Having
  defined the face and degeneracy maps, we should now verify the simplicial
  identities as listed in \ref{simplicial-ids}. The way the face and degeneracy
  maps have been defined, this is straightforward. For example, the face-map identities state that
  if two cuts are deleted, it does not matter in which order, or that throwing away
  the first two layers is the same as first joining them and then throwing away the
  joined layer in one go. The following example illustrates that throwing away the
  first layer and then the last is the same as throwing away the last and then the
  first. 
\end{blanko}

\begin{blanko}{Example.} 
  The following is a picture of a $2$-layered process $p \in \ds X_2$
  (of the Petri net in Examples~\ref{ex:ex} and \ref{ex:processes}) and
  the effect of the three face maps.
  
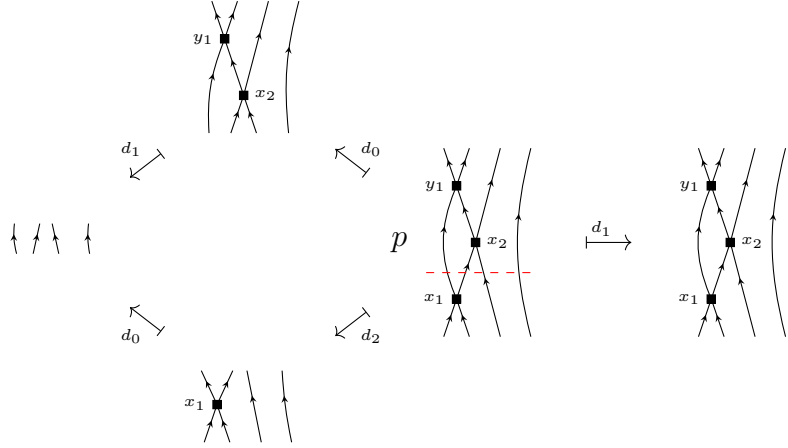
\begin{center}
	\begin{tikzpicture}
	  \coordinate (G) at (0,0);
	  \coordinate (d1) at (3.6,0);
	  \coordinate (d0) at (-2.7,2.1);
	  \coordinate (d2) at (-2.7,-2.1);
	  \coordinate (M) at (-5.4,0);

	  \node at (-0.75,0) {$p$};

	  \begin{scope}[shift={(M)}, xscale=0.5, yscale=-0.5]
	  \begin{scope}[shift={(-0.5,-2.5)}]
		\tikzstyle arrowstyle=[scale=0.8]
		\node [import] (a1) at (-0.2,2.8) {};
		\node [import] (a2) at (0.1,2.8) {};
		\node [import] (a3) at (0.9,2.8) {};
		\node [import] (i4) at (1.7,2.8) {};		
		\node [export] (c1) at (-0.3,2) {};
		\node [export] (c2) at (0.4,2) {};
		\node [export] (c3) at (0.6,2) {};
		\node [export] (o4) at (1.65,2) {};
		\draw[directed, bend right=10pt] (a1) to (c1);
		\draw[directed] (a2) -- (c2);
		\draw[directed] (a3) -- (c3);
		\draw[directed, bend right=8pt] (i4) to (o4);
	  \end{scope}
	  \end{scope}

	  \begin{scope}[shift={(d0)}, xscale=0.5, yscale=-0.5]
	  \begin{scope}[shift={(-0.7,-1.2)}]
		\tikzstyle arrowstyle=[scale=0.8]
		\tiny
		\node [grnode, label=right:$x_2$] (x2) at (0.5,1.5) {};
		\node [grnode, label=left:$y_1$] (y1) at (0,0) {};
		\node [import] (a1) at (-0.4,2.5) {};
		\node [import] (a2) at (0.1,2.5) {};
		\node [import] (a3) at (0.9,2.5) {};
		\node [export] (c1) at (1.2,-1) {};
		\node [export] (c2) at (-0.4,-1) {};
		\node [export] (c3) at (0.4,-1) {};
		\node [import] (i4) at (1.7,2.5) {};
		\node [export] (o4) at (2.0,-1) {};
		\draw[directed, bend right=12pt] (a1) to (y1);
		\draw[directed] (a2) -- (x2);
		\draw[directed] (a3) -- (x2);
		\draw[directed] (x2) -- (y1);
		\draw[directed] (x2) -- (c1);
		\draw[directed] (y1) -- (c2);
		\draw[directed] (y1) -- (c3);
		\draw[directed, bend right=10pt] (i4) to (o4);
	  \end{scope}
	  \end{scope}

	  \begin{scope}[shift={(d2)}, xscale=0.5, yscale=-0.5]
	  \begin{scope}[shift={(-0.9,-2.9)}]
		\tikzstyle arrowstyle=[scale=0.8]
		\tiny
		\node [grnode, label=left:$x_1$] (x1) at (0,3) {};
		\node [import] (a1) at (-0.4,4) {};
		\node [import] (a2) at (0.4,4) {};
		\node [import] (a3) at (1.2,4) {};
		\node [export] (o2) at (0.5,2.1) {};
		\node [export] (o1) at (-0.5,2.1) {};
		\node [export] (o3) at (0.75,2.1) {};
		\node [import] (d3in) at (2.0,4) {};
		\node [export] (d3out) at (1.7,2.1) {};
		\draw[directed] (a1) -- (x1);
		\draw[directed] (a2) -- (x1);
		\draw[directed] (a3) -- (o3);
		\draw[directed] (x1) -- (o2);
		\draw[directed] (x1) -- (o1);
		\draw[directed, bend right=4pt] (d3in) to (d3out);
	  \end{scope}
	  \end{scope}

	  \begin{scope}[shift={(G)}, xscale=0.5, yscale=-0.5]
	  \begin{scope}[shift={(-0.0,-1.5)}]
		\tikzstyle arrowstyle=[scale=0.8]
		\tiny
		\node [grnode, label=left:$x_1$] (x1) at (0,3) {};
		\node [grnode, label=right:$x_2$] (x2) at (0.5,1.5) {};
		\node [grnode, label=left:$y_1$] (y1) at (0,0) {};
		\node [import] (a1) at (-0.4,4) {};
		\node [import] (a2) at (0.4,4) {};
		\node [import] (a3) at (1.2,4) {};
		\node [export] (c1) at (1.2,-1) {};
		\node [export] (c2) at (-0.4,-1) {};
		\node [export] (c3) at (0.4,-1) {};
		\node [import] (d3in) at (2.0,4) {};
		\node [export] (d3out) at (2.0,-1) {};
		\draw[directed] (a1) -- (x1);
		\draw[directed] (a2) -- (x1);
		\draw[directed] (a3) -- (x2);
		\draw[directed] (x1) -- (x2);
		\draw[directed] (x2) -- (y1);
		\draw[directed, bend right=22pt] (x1) to (y1);
		\draw[directed] (x2) -- (c1);
		\draw[directed] (y1) -- (c2);
		\draw[directed] (y1) -- (c3);
		\draw[directed, bend right=14pt] (d3in) to (d3out);
		\draw[dashed, red] (-0.8,2.3) --+ (2.9,0);
	  \end{scope}
	  \end{scope}

	  \begin{scope}[shift={(d1)}, xscale=0.5, yscale=-0.5]
	  \begin{scope}[shift={(-0.5,-1.5)}]
		\tikzstyle arrowstyle=[scale=0.8]
		\tiny
		\node [grnode, label=left:$x_1$] (x1) at (0,3) {};
		\node [grnode, label=right:$x_2$] (x2) at (0.5,1.5) {};
		\node [grnode, label=left:$y_1$] (y1) at (0,0) {};
		\node [import] (a1) at (-0.4,4) {};
		\node [import] (a2) at (0.4,4) {};
		\node [import] (a3) at (1.2,4) {};
		\node [export] (c1) at (1.2,-1) {};
		\node [export] (c2) at (-0.4,-1) {};
		\node [export] (c3) at (0.4,-1) {};
		\node [import] (d3in) at (2.0,4) {};
		\node [export] (d3out) at (2.0,-1) {};
		\draw[directed] (a1) -- (x1);
		\draw[directed] (a2) -- (x1);
		\draw[directed] (a3) -- (x2);
		\draw[directed] (x1) -- (x2);
		\draw[directed] (x2) -- (y1);
		\draw[directed, bend right=22pt] (x1) to (y1);
		\draw[directed] (x2) -- (c1);
		\draw[directed] (y1) -- (c2);
		\draw[directed] (y1) -- (c3);
		\draw[directed, bend right=14pt] (d3in) to (d3out);
	  \end{scope}
	  \end{scope}

	  \tiny
	  
	  \draw[|-{>[width=6pt]}, 
	  shorten <=17mm, shorten >=13mm, 
	  ] (G) -- node[anchor=south, pos=0.53] {$d_1$} (d1);

	  \draw[|-{>[width=6pt]}, 
	  shorten <=15mm, shorten >=14mm, 
	  ] (G) -- node[anchor=south west] { $d_0$} (d0);

	  \draw[|-{>[width=6pt]}, 
	  shorten <=15mm, shorten >=14mm, 
	  ] (G) -- node[anchor=north west] { $d_2$} (d2);

	  \draw[|-{>[width=6pt]}, 
	  shorten <=15mm, shorten >=14mm, 
	  ] (d0) -- node[anchor=south east] { $d_1$} (M);

	  \draw[|-{>[width=6pt]}, 
	  shorten <=15mm, shorten >=14mm, 
	  ] (d2) -- node[anchor=north east] { $d_0$} (M);
	
	\end{tikzpicture}
  \end{center}
  $d_0$ applied to $p$ erases the first layer; $d_1$ joins the two layers; $d_2$ 
  erases the last layer.
  The further face maps applied give the commutative square on the left, which 
  illustrates the simplicial identity $d_0 d_2 = d_1 d_0$. 
\end{blanko}

\begin{blanko}{Remark.}
  Constructions of simplicial groupoids like this are not uncommon in combinatorics.
  In fact, this particular simplicial groupoid $\ds X_\bullet$ is almost the same as
  that of the {\em directed restriction species} of directed graphs explained in
  Example~7.13 of \cite{Galvez-Kock-Tonks:1708.02570},  where substantial details of 
  the construction can be found, although in a much more general setting. 
  The difference (apart from the
  $\pn P$-decorations) concerns isolated edges, excluded in the setting of
  restriction species, but clearly essential to keep for the present purposes.
\end{blanko}

\begin{blanko}{Segal spaces.}
  For a simplicial groupoid to work as a (weak) category, it should be a {\em Segal space}
  (see~\ref{Rezk}). First of all this means that the canonical map
  $$
  (d_2,d_0) : \ds X_2 \longrightarrow \ds X_1 \times^h_{\ds X_0} \ds X_1
  $$
  should be an equivalence of groupoids. More generally it is required that for all
  $k\geq 1$ the canonical map
    that returns the $k$ layers
  \begin{equation}\label{eq:Seg}
  \ds X_k \longrightarrow \ds X_1 \times_{\ds X_0}^h \cdots \times_{\ds 
  X_0}^h \ds X_1
  \end{equation}
  is an equivalence of groupoids.
  
  To establish this, we first establish the following important fibrancy
  condition (see~\ref{fib}), which allows to use ordinary strict pullbacks
  instead of homotopy pullbacks.
\end{blanko}

\begin{blankothm}{Lemma}\label{lem:fib}
  The face map $d_0 : \ds X_1 \to \ds X_0$ (which given a process
  returns its final state) is a fibration of groupoids
  (cf.~\ref{fib}).\footnote{By the same argument, also $d_1 :\ds X_1
  \to \ds X_0$ (initial state) is a fibration, but we shall not need
  that.}
\end{blankothm}

\begin{proof}
  Given a process $p : \pn G\to\pn P$ with final state $d_0(p) =z : \pn 
  M\to \pn P$ (meaning that $\pn M = \operatorname{out}(\pn G)$), 
  and given an isomorphism with another state $z':M'\to S$, amounting to a 
  commutative diagram
    $$\begin{tikzcd}[column sep={2em,between origins}]
  M \ar[rd, "z"'] \ar[rr, "\overset{u}{\sim}"] && M' \ar[ld, 
  "z'"] \\
   & S , &
  \end{tikzcd}
  $$
  the statement is that there exists a lift of $u$ to $p$, namely an isomorphism
  of processes $\ov u : p \isopil p'$ (for some other process $p'$), whose restriction to the 
  out-boundaries reproduces the bijection $u$.  To construct this, 
  suppose the graph $\pn G$ is given by
  $A \leftarrow A_N \to N \leftarrow {}_N A \to 
  A$, with out-boundary $\operatorname{out}(\pn G)=M$.
  By the definition of out-boundary (\ref{unit}), the set of edges thus
  splits into two disjoint subsets
  $$
  A = M + A_N .
  $$
  Now define the new graph $\pn G'$ and the morphism $\ov u$ by
  modifying as little as possible --- only the out-boundary:
  $$
  \begin{tikzcd}
	\pn G \ar[d, "\ov u"'] 
	&:&
	M+ A_N \ar[d, "u+ \id"'] & A_N\ar[d, equal] \ar[l] \ar[r] 
	\drpullback &  N\ar[d, equal] & 
	{}_N A\ar[d, equal] \ar[l]\ar[r] \dlpullback & M + A_N\ar[d, 
	"u+\id"]
	\\
	\pn G' \ar[d, "p'"'] 
	&:&	
	M'+ A_N \ar[d, "z'+ p_{\mid A_N}"'] & A_N\ar[d] \ar[l] \ar[r] 
	\drpullback &  N\ar[d] & 
	{}_N A\ar[d] \ar[l]\ar[r] \dlpullback & M' + A_N\ar[d, 
	"z'+ p_{\mid A_N}"]
	\\
	\pn P &:&
	S & I \ar[l] \ar[r] & N & O \ar[l] \ar[r] & S .
  \end{tikzcd}
  $$
  The new graph $\pn G'$ is naturally a $\pn P$-graph as indicated,
  and by construction the new map $\ov u$ is a morphism of $\pn 
  P$-graphs.
\end{proof}

The benefit of the fibrancy condition is the general fact that homotopy pullbacks
along fibrations can be computed as {\em strict} pullbacks 
(see~\ref{hopbk}--\ref{fib}).

\begin{blankothm}{Proposition}
  The simplicial groupoid $\ds X_\bullet $ is a Rezk-complete Segal space.
\end{blankothm}

\begin{proof}
  The Segal condition states that the map $\ds X_k \longrightarrow \ds X_1 \times_{\ds 
  X_0}^{h}\cdots\times_{\ds X_0}^{h} \ds X_1$ in \eqref{eq:Seg}
  is an equivalence of groupoids.
  For notational convenience we do the case $k=2$ only.\footnote{For higher $k$, one 
  can use induction, reformulating the Segal condition as $\ds X_k \simeq \ds 
  X_{k-1} \times^h_{\ds X_0} \ds X_1$.} 
  Since this homotopy pullback (which is \eqref{eq:hopbk}) is along 
  the fibration $d_0$, we can prove instead that
  $$
  \ds X_2 \stackrel{g}{\longrightarrow} \ds X_1 \times_{\ds 
  X_0}^{\operatorname{strict}} \ds X_1
  $$
  is an equivalence of groupoids. An element in this strict pullback
  is a pair $(p_1, p_2)$ of processes such that the final state $d_0(p_1)$
  of the first is literally {\em equal} to the initial state
  $d_1(p_2)$ of the second.

   The map $g$ is bijective at the level of $\pi_0$, because the pushout
  construction of \ref{pushout} provides an up-to-isomorphism
  inverse:  given processes $p_1: \pn G_1\to \pn P$ and $p_2: \pn
  G_2\to \pn P$ with $\operatorname{out}(\pn G_1) = M =
  \operatorname{in}(\pn G_2)$, the pushout produces a single $\pn
  P$-graph $\pn G_1 \amalg_M \pn G_2$ with the obvious $2$-level
  function; it is clear that this construction is inverse to $g$ up to
  isomorphism. To establish that $g$ is an equivalence it remains to
  check that it is also an isomorphism on automorphism groups
  (cf.~\ref{groupoids}). The
  automorphisms of a $2$-levelled $\pn P$-graph
  $$
  \begin{tikzcd}
  \pn G\ar[d, "p"'] \ar[r, "f"] & \un 2  \\
  \pn P & 
  \end{tikzcd}
  $$
  are the automorphisms of $\pn G$ compatible with both $p$ and $f$.
  This amounts to giving an automorphism of each layer $\pn G_1$ and 
  $\pn G_2$, whose 
  restrictions to the cut $M$ agree. But this is precisely the 
  description of the automorphism group  of the corresponding object 
  $(\pn G_1, \pn G_2)$ in $\ds X_1 \times_{\ds X_0}^{\text{strict}} 
  \ds X_1$.
  
  Finally, {\em Rezk completeness} (\ref{Rezk}) is the statement that the only
  invertible processes in $\ds X_1$ are the degenerate ones,  i.e.~in the image of $s_0 : \ds
  X_0 \to \ds X_1$. A $1$-simplex $p$ in a Segal space is called {\em invertible} if
  composition with it from either side defines an equivalence $\ds X_1 \to \ds X_1$. 
  In the present
  case, $p$ is a $\pn P$-graph and composition is gluing. But composition with a
  graph having a node will increase the number of nodes, so the only invertible 
  processes are
  the node-less graphs.  These are precisely those in the image of $s_0$, as 
  required. 
\end{proof}

\begin{blankothm}{Lemma}
  $\ds X_\bullet$ is a symmetric monoidal Segal space under disjoint
  union $+$. That is, the simplicial maps
  $$
  \begin{tikzcd}
  \ds 1_\bullet \ar[r, "\name{\emptyset}"] & \ds X_\bullet & 
  \ds X_\bullet \times \ds X_\bullet \ar[l, "+"']
  \end{tikzcd}
  $$
  satisfy the standard associative, unital, and symmetry axioms.
\end{blankothm}
Here $\ds 1_\bullet$ is the constant simplicial groupoid on the 
terminal groupoid, and the map in degree $k$ picks out the empty $k$-levelled $\pn 
P$-graph. The required coherence constraints, given separately in 
each simplicial degree, follow from
the universal properties of $\emptyset$ and $+$
as initial object and categorical sum.

\begin{blanko}{Remark.}
  The symmetric monoidal Segal space $\ds X_\bullet$ is very easy to set up, since it
  is defined in terms of {\em decomposition} instead of composition. This phenomenon,
  that decomposition is easier to achieve than composition, is quite general, and is
  one starting point of the recent theory of {\em decomposition
  spaces}~\cite{Galvez-Kock-Tonks:1512.07573}, \cite{Galvez-Kock-Tonks:1512.07577},
  \cite{Galvez-Kock-Tonks:1708.02570},
  \cite{Galvez-Kock-Tonks:1612.09225}. For the symmetric monoidal structure it must
  be appreciated that $\ds X_\bullet$ is Rezk complete. This implies that functor
  categories can be dealt with pointwise, i.e.~in each simplicial degree separately.
  The symmetric monoidal structure in each simplicial degree is obvious.
\end{blanko}

\begin{blankothm}{Proposition}\label{prop:XX'}
  An etale map of Petri nets $e:\pn P \to \pn P'$ induces a symmetric monoidal
  functor of Segal spaces
  $$
  e\lowershriek : \ds X_\bullet \to \ds X_\bullet{}\!\!' 
  $$
  by sending a layered process $\pn G \to \pn P$ to the composite 
  $\pn G \to \pn P \to \pn P'$ with the same layering.
\end{blankothm}
Indeed, the map in simplicial degree $1$ is a special case of 
Proposition~\ref{prop:PProcP}. The remaining simplicial degrees are a 
matter of level functions, and refer only to the underlying graph 
$\pn G$; they are not affected by $\pn P$- or $\pn P'$-structure.

\section{Digraphical species from Petri nets}
\label{sec:grsp}

Recall that a digraphical species is a presheaf $\gs F: \elGr\op\to\Set$
on the category of elementary graphs,
and that we call $\gs F[\triv]$ the set of {\em colours}.  The set $\gs F[\coro{m}{n}]$
is called the set of {\em $(m,n)$-ary operations}.\footnote{This terminology comes from 
the theory of operads.} 

\begin{blanko}{Digraphical species of a Petri net.}
  A Petri net $\pn P$ defines a digraphical species 
  \begin{eqnarray*}
    \gs P : \elGr\op & \longrightarrow & \Set  \\
    \pn E & \longmapsto & \Hom(\pn E,\pn P), 
  \end{eqnarray*}
  which is simply the restricted Yoneda embedding along the full inclusions
  $\elGr \subset \Gr \subset \kat{Petri}$. The set of colours of $\gs
  P$ is thus the set of places $S$, and the operations are the
  transitions $T$, symmetrised:  indeed to give an etale map from a
  corolla to $\pn P$ is first of all to say where the unique node of
  the corolla goes, say $\mapsto t\in T$. If the transition $t$
  has $m$ incoming arcs and $n$ outgoing arcs, then the domain of the map
  has to be the $[\coro{m}{n}]$-corolla, by etaleness. There are now
  are $m!n!$ different etale maps $[\coro{m}{n}] \to \pn P$ hitting $t$.
  All of these maps are valid, as there are no
  constraints on places. For each map, the corolla will acquire colours 
  on edges according to the places they are mapped to. 
\end{blanko}

From the decomposition of Petri nets into colimits of elementary 
graphs (Lemma~\ref{lem:Pcolims}), 
 we get:\footnote{by standard category theory 
\cite[Ch.~X, \S6]{MacLane:categories}} 

\begin{blankothm}{Lemma}
  (Density lemma.)\label{lem:density}
  The functor 
  \begin{eqnarray*}
  \kat{Petri} & \longrightarrow & \PrSh(\elGr)  \\
  \pn P & \longmapsto & \Hom( - , \pn P)
\end{eqnarray*}
is fully faithful. 
\end{blankothm}

 The following result provides a characterisation of Petri nets
by describing the image of this embedding. 

\begin{blankothm}{Proposition}\label{prop:flat}
  A digraphical species $\gs F:
  \elGr\op\to\Set$ is a Petri net if and only if it is flat   
  (meaning that the symmetric-group 
	actions
	$(\mathfrak S_m \times \mathfrak S_n) \times \gs F[\coro{m}{n}] \to 
	\gs F[\coro{m}{n}]$
	are free) and it takes finite values and has finite support (meaning that there 
	is an upper bound on the $m$ and $n$ for which $\gs F[\coro{m}{n}]$ is 
	non-empty).
\end{blankothm}

\begin{blanko}{Remark.}
  This result is a variation of Theorem~2.4.10 
  of \cite{Kock:0807}, which characterises polynomial endofunctors 
  among the presheaves on elementary trees.
  The terminology {\em flat} is from the theory of combinatorial 
  species~\cite{Bergeron-Labelle-Leroux}. In the theory of operads, 
  the same condition is called {\em sigma-cofibrant}. Note finally 
  that the flat digraphical species are also essentially the same 
  thing as the {\em tensor schemes} of Joyal and 
  Street~\cite{Joyal-Street:tensor-calculus}.
\end{blanko}

\begin{proof}
   Starting with a Petri net $\pn P= SITOS$, the associated graphical species is
  $\Hom( - , \pn P)$:
  \begin{eqnarray*}
    \elGr\op & \longrightarrow & \Set  \\
    {}[\triv] & \longmapsto & \Hom( [\triv], \pn P) = S \\
    {}[\coro{m}{n}] & \longmapsto & \Hom( [\coro{m}{n}], \pn P) .
  \end{eqnarray*}
  To specify an etale map $[\coro{m}{n}] \to \pn P$ is to give a transition $t\in T$
  and bijections $\beta_{\operatorname{in}} : \un m \isopil \operatorname{pre}(t)$
  and $\beta_{\operatorname{out}} : \un n \isopil \operatorname{post}(t)$. The natural
  $(\mathfrak S_m \times \mathfrak S_n)$-action
  \begin{eqnarray*}
    (\mathfrak S_m \times \mathfrak S_n) \times \Hom( [\coro{m}{n}], \pn P) & 
	\longrightarrow & \Hom( [\coro{m}{n}], \pn P)  \\
    \big((\sigma_{\operatorname{in}},\sigma_{\operatorname{out}}), 
	(t,\beta_{\operatorname{in}},\beta_{\operatorname{out}})\big) & \longmapsto & 
	(t, \beta_{\operatorname{in}}\circ 
	\sigma_{\operatorname{in}},\beta_{\operatorname{out}}\circ \sigma_{\operatorname{out}})
  \end{eqnarray*}
  is clearly free. Since each transition $t$ has finite 
  $\operatorname{pre}(t)$ and $\operatorname{post}(t)$, and since there are only
  finitely many transitions, we find that all the
  sets $\Hom([\coro{m}{n}],\pn P)$ are finite, and empty for big $m$ 
  and $n$.
  
  Conversely, given a flat digraphical species $\gs F : \elGr\op\to\Set$, we 
  construct a Petri net $\pn P=SITOS$ by putting $S:= \gs F[\triv]$ and defining
  $T$ to be the disjoint union
  $$
  T := \sum_{m,n} \gs F[\coro{m}{n}] / (\mathfrak S_m \times \mathfrak S_n) .
  $$
  This is a finite set by the finiteness assumptions on $\gs F$.
  It remains to define the sets $I$ and $O$, so as to have correct fibres over $T$ and 
  appropriate `colours'. If $t\in T$ is in the $(m,n)$-summand, then the $I$-fibre 
  should have $m$ elements and the $O$-fibre should have $n$ elements. Essentially
  $I$ and $O$ should be defined as disjoint unions over all such data, and the 
  projections to $S$ should return the corresponding colour according to the $\gs 
  F$-structure. To describe this in a natural way, we first fix $m$ and $n$ 
  and note that 
  $\gs F[\coro{m}{n}]$ can be interpreted as the set $\{ [\coro{m}{n}] \to \gs F \}$
  via the Yoneda lemma. Consider now the two fancier sets: $\{ [\triv] 
  \stackrel{\operatorname{in}}\to [\coro{m}{n}] \to \gs F \}$ where the first map 
  is any of the $m$ maps of elementary graphs that pick out an input edge, and $\{ [\triv] 
  \stackrel{\operatorname{out}}\to [\coro{m}{n}] \to \gs F \}$ where the first map 
  is any of the $n$ maps of elementary graphs that pick out an output edge.
  With the natural projections
  \[
  \begin{tikzcd}[column sep={27mm,between origins}]
   & \{ [\triv] \stackrel{\operatorname{in}}\to [\coro{m}{n}] \to \gs F \} \ar[ld] 
   \ar[rd] &  & \{ [\triv] \stackrel{\operatorname{out}}\to [\coro{m}{n}] \to \gs F \} \ar[ld] \ar[rd] & \\
  S=\{ [\triv] \to \gs F \} & & \{ [\coro{m}{n}] \to \gs F \} & & \{ [\triv] \to \gs F 
  \}=S
  \end{tikzcd}
  \]
  we get the basic building blocks for the final $SITOS$ diagram. It remains to
  divide out by the group actions, and sum up 
  all $m$ and $n$. We already have 
  the free $(\mathfrak S_m \times \mathfrak S_n)$-action on the set $\{ [\coro{m}{n}] \to 
  \gs F \}$.
  For the fancier sets, each group element $(\sigma_{\operatorname{in}}, \sigma_{\operatorname{out}}) 
  \in \mathfrak S_m \times \mathfrak S_n$ acts by permuting the 
  $[\coro{m}{n}]$ in the middle, sending the upper composite in the diagram
  \[
  \begin{tikzcd}[row sep={10mm,between origins}, column sep={22mm,between origins}]
   & {[\coro{m}{n}]} \ar[rd] &   \\
  {[\triv]}\ar[ru] \ar[rd, dotted] & & \gs F \\
     & {[\coro{m}{n}]} \ar[uu, "{(\sigma_{\operatorname{in}}, 
	 \sigma_{\operatorname{out}})}"'] \ar[ru, dotted] & 
  \end{tikzcd}
  \]
  to the lower composite. The new map 
  $\!\!
  \begin{tikzcd}[column sep={11pt}]
	{[\triv]} \ar[r, dotted] & 
  {[\coro{m}{n}]}
\end{tikzcd}
  \!\!$ is defined as the old map followed by 
  $(\sigma_{\operatorname{in}}, \sigma_{\operatorname{out}})^{-1}$, which means that
  the composite $[\triv]\to\gs F$ is not changed by the action:
  it is still the same colour of $\gs F$ picked out. This means that when we pass to 
  the quotient, the maps to $S$ are still well defined.
  
  Taking now quotients and summing over $m, n$, the final $SITOS$-diagram is 
  altogether defined as
  \[
  \begin{tikzcd}[column sep={26mm,between origins}]
   & \sum_{m,n}\frac{\{ [\triv] \stackrel{\operatorname{in}}\to [\coro{m}{n}] \to 
   \gs F \}}{\mathfrak S_m \times \mathfrak S_n} \ar[ld] 
   \ar[rd] &  & \sum_{m,n}\frac{\{ [\triv] \stackrel{\operatorname{out}}\to [\coro{m}{n}] \to 
   \gs F \}}{\mathfrak S_m \times \mathfrak S_n} \ar[ld] \ar[rd] & \\
  S=\{ [\triv] \to \gs F \} & & \sum_{m,n}\frac{\{ [\coro{m}{n}] \to \gs F \}}{\mathfrak S_m \times \mathfrak S_n} & & \{ [\triv] \to \gs F 
  \}=S
  \end{tikzcd}
  \]
  
  We omit the straightforward verification that the two constructions are inverse to 
  each other, up to natural isomorphism. 
\end{proof}

\begin{blankothm}{Lemma}\label{lem:noAuts}
  For $\pn P$ a Petri net, $\pn P$-graphs have no infinitesimal automorphisms, in the sense
  that if an automorphism fixes a node, then it is the identity on that connected 
  component.
\end{blankothm}

\begin{proof}
  Let $f:\pn G \isopil \pn G$ be an automorphism of $\pn P$-graphs that fixes a node
  $x$. Since $f$ preserves the incidences, it must map the set $\operatorname{in}(x)$
  of incoming edges of $x$ to itself and the set $\operatorname{out}(x)$ of outgoing
  edges of $x$ to itself. The $\pn P$-structure is an etale map $\pn G \to \pn P$,
  mapping $x$ to some transition $t$, and the etale condition gives bijections
  $\operatorname{in}(x) \simeq \operatorname{pre}(t)$ and $\operatorname{out}(x)
  \simeq \operatorname{post}(t)$. Since $f$ must commute with these bijections,
  it is forced to be the identity on $\operatorname{in}(x)$ and 
  $\operatorname{out}(x)$. Therefore it must also fix any incident nodes, and so on,
  forcing altogether the whole connected component to be fixed.
\end{proof}

\begin{blanko}{Remark.}
  Even when all nodes of a $\pn P$-graph are fixed, 
  it is still possible to permute
  isolated edges that map to the same place in $\pn P$. This is relevant since
  isolated edges in a process correspond to unaffected tokens (as noted in
  \ref{executions-prelim}), meaning tokens that belong to both the initial and the
  final marking.
\end{blanko}

\begin{blanko}{Example.}
  There are two possible etale maps
  
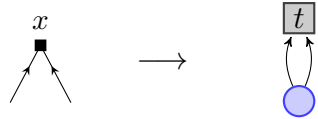
\begin{center}
	\begin{tikzpicture}
      \begin{scope}[shift={(-3.4,-0.65)}, xscale=0.7, yscale=0.7]
		\tikzstyle arrowstyle=[scale=1]
		\tikzstyle directed=[postaction={decorate,decoration={markings,
		  mark=at position .7 with {\arrow[arrowstyle]{stealth}}}}]

		\tikzstyle arrowstyle=[scale=1]
		\node [port] (i1) at (-0.5,-0.4) {};
		\node [port] (i2) at (0.5,-0.4) {};
		\node [grNode, label=above:{\small $x$}] (t1) at (0.0,0.55) {};
		\draw[directed] (i1) -- 
		(t1);
		\draw[directed] (i2) -- 
		(t1);
	  \end{scope}

	  \node at (-1.8,-0.5) {$\longrightarrow$};
	  
	  \begin{scope}[
		node distance=0.9cm,on grid,>=stealth',bend angle=22.5,auto,
		  every place/.style= {minimum size=3.5mm,thick,draw=blue!75,fill=blue!20}, 
		  every transition/.style={minimum size=3.5mm,thick,draw=black!75,fill=black!20}]
		\node [transition] (t1) {$t$};
		\node [place](s1) [below=of t1] {}; 
		\path[bend left] (s1) edge [post] 
		(t1);
		\path[bend right] (s1) edge [post] 
		(t1);
	  \end{scope}
	\end{tikzpicture}

  \end{center}
  corresponding to the two possible bijections $\operatorname{in}(x)
  \isopil \operatorname{pre}(t)$. But these two processes are isomorphic by
  the graph isomorphism that interchanges the two edges. More generally,
  there are of course infinitely many graphs corresponding to the figure,
  depending on which $2$-element set is chosen as set of edges, and so on,
  and therefore also infinitely many distinct processes of this shape. For
  any two such {\em graphs} $AINOA$ and $A'I'N'O'A'$, there are two
  possible isomorphisms between them (since there are always two distinct
  bijections $A\simeq A'$ between $2$-element sets). However, as {\em
  processes} there is precisely {\em one} possible bijection, because the
  process involves a bijection $\operatorname{in}(x) \isopil
  \operatorname{pre}(t)$, and only one bijection $A \isopil A'$ can be
  compatible with those. In conclusion, any two processes of the shape of
  the figure are uniquely isomorphic. This illustrates
  Lemma~\ref{lem:noAuts}.
  
  Here we see a crucial difference with traditional Petri net theory, where
  instead of parallel arcs there would be merely a multiplicity:
  \begin{center}
	\begin{tikzpicture}
	  	  \begin{scope}[shift={(P)},
		  node distance=0.9cm,on grid,>=stealth',bend angle=22.5,auto,
		  every place/.style= {minimum size=3.5mm,thick,draw=blue!75,fill=blue!20}, 
		  every transition/.style={minimum size=3.5mm,thick,draw=black!75,fill=black!20}]
		\node [transition] (t1) {$t$};
		\node [place](s1) [below=of t1] {}; 
		\path (s1) edge [post] node[left=-2pt]  {\footnotesize $2$} (t1);
	  \end{scope}
	\end{tikzpicture}
  \end{center}
  The domain graph of a process looks the same in traditional Petri net theory, but
  instead of the etale condition involving explicit bijections, we now have the
  condition that the number $2$ of incoming edges to the node must match the
  multiplicity $2$ in the Petri net. Again it is the case that all possible processes
  of this shape are isomorphic, but they are {\em no longer uniquely} isomorphic:
  even a fixed process admits a nontrivial automorphism, by interchanging the two
  edges of the graph. The multiplicity decoration cannot prevent this automorphism.
  This difference in the behaviour of symmetries is a key feature of whole-grain
  Petri nets compared to traditional Petri nets.
\end{blanko}

\section{Free props}
\label{sec:props}

The following description of the free-prop monad is essentially from 
\cite{Kock:1407.3744}, except that there only the free-properad monad 
is described, considering only connected graphs.

\begin{blanko}{Residue and $[\coro{m}{n}]$-graphs.}
  For an (acyclic) graph $\pn G$ (either an $\gs F$-graph or naked (i.e.~with no $\gs
  F$-structure)), the {\em residue} $\res(\pn G)$ is the naked corolla formed by
  the in-boundary and the out-boundary (and a single node).  For example,
    
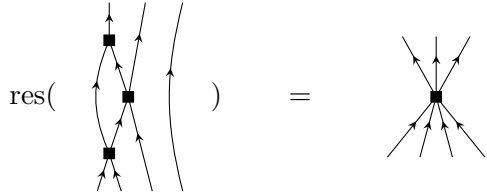
\begin{center}
	\begin{tikzpicture}

	  \small
	  
	  \coordinate (G) at (-1,0.75 );
	  \coordinate (R) at (3.3,0.0);
	  
	  \begin{scope}[shift={(G)}, xscale=0.5, yscale=-0.5]
		\tikzstyle arrowstyle=[scale=1]
		\node at (-2.0,1.5) {$\res ($};
		\node at (4.0,1.5) {$) \qquad = $};
		\node [grNode] (x1) at (0,3) {};
		\node [grNode] (x2) at (0.5,1.5) {};
		\node [grNode] (y1) at (0,0) {};
		\node [import] (a1) at (-0.4,4) {};
		\node [import] (a2) at (0.4,4) {};
		\node [import] (a3) at (1.2,4) {};
		\node [export] (c1) at (1.0,-1) {};
		\node [export] (c2) at (-0.0,-1) {};
		\node [import] (d3in) at (2.0,4) {};
		\node [export] (d3out) at (2.0,-1) {};
		\draw[directed] (a1) -- (x1);
		\draw[directed] (a2) -- (x1);
		\draw[directed] (a3) -- (x2);
		\draw[directed] (x1) -- (x2);
		\draw[directed] (x2) -- (y1);
		\draw[directed, bend right=22pt] (x1) to (y1);
		\draw[directed] (x2) -- (c1);
		\draw[directed] (y1) -- (c2);
		\draw[directed, bend right=14pt] (d3in) to (d3out);
	  \end{scope}

	  	  \begin{scope}[shift={(R)}]
		\tikzstyle arrowstyle=[scale=1]
	  
		\node [grNode] (x) at (0,0) {};
		
		\node [import] (a1) at (-0.75,-0.8) {};
		\node [import] (a2) at (-0.25,-0.8) {};
		\node [import] (a3) at (0.25,-0.8) {};
		\node [import] (a4) at (0.75,-0.8) {};
		\node [export] (b1) at (-0.52,0.8) {};
		\node [export] (b3) at (0.0,0.8) {};
		\node [export] (b4) at (0.52,0.8) {};

		\draw[directed] (a1) -- (x);
		\draw[directed] (a2) -- (x);
		\draw[directed] (a3) -- (x);
		\draw[directed] (a4) -- (x);
		\draw[directed] (x) -- (b1);
		\draw[directed] (x) -- (b3);
		\draw[directed] (x) -- (b4);
	  \end{scope}

	  \end{tikzpicture}
  \end{center} 
  This defines a functor $\res:
  \Gr_{\iso} \to \Cor$ (and for each $\gs F$ a functor $\res: \Gr_{\iso}\comma \gs F
  \to \Cor$).
  
  An {\em $[\coro{m}{n}]$-graph} is an acyclic graph $\pn G$ equipped with an
  isomorphism $\res(\pn G) \simeq [\coro{m}{n}]$. It amounts to a numbering of the
  elements in the in-boundary and a separate numbering for the out-boundary (since we
  have defined the specific corollas $[\coro{m}{n}$] in terms of natural numbers).
   More formally, the groupoid of $[\coro{m}{n}]$-graphs
  $[\coro{m}{n}]\kat{-Gr}_{\iso}$ is the homotopy fibre (see \ref{fibre}) of $\res:
  \Gr_{\iso} \to \Cor$ over $[\coro{m}{n}]$. For $\gs F$ a digraphical species, an
  {\em $[\coro{m}{n}]$-$\gs F$-graph} is an $\gs F$-graph $\pn G$ equipped with an
  isomorphism $\res(\pn G) \simeq [\coro{m}{n}]$. These are the objects of the
  groupoid $[\coro{m}{n}]\kat{-Gr}_{\iso} \comma \gs F$, the homotopy fibre over
  $[\coro{m}{n}]$ of the functor $\res:\Gr_{\iso}\comma \gs F \to \Cor$.
\end{blanko}

\begin{blanko}{Remark.}\label{not-enough-isos}
  Note that we allow non-connected graphs (in contrast to \cite{Kock:1407.3744}). In
  particular, a nodeless $\gs F$-graph $\pn U$ consisting of $m$ isolated edges is a
  $[\coro{m}{m}]$-graph in $m!m!$ ways, depending of the possible bijections
  $\operatorname{in}(\pn U)\simeq \un m$ and $\operatorname{out}(\pn U)\simeq \un m$
  (which are independent). This example also shows that $\res: \Gr_{\iso}\comma \gs F
  \to \Cor$ is not a fibration: not all automorphisms of $[\coro{m}{m}]$ admit a lift
  to $\pn U$.
\end{blanko}

\begin{blanko}{The idea of props and free props.}
   A category has objects and arrows (forming an underlying graph), and then a
  prescription for composing arrows. The free category on a graph will have not only
  the edges as arrows, but will also promote the paths in the graph to be arrows.
  Composition in a free category is just concatenation of paths.
  
  A (coloured) prop has colours and many-in/many-out operations (so as to have an
  underlying digraphical species), and then a prescription for composing operations.
  The free prop on a digraphical species $\gs F$ will have not only the $\gs
  F$-corollas as operations, but will also promote all $\gs F$-graphs to be
  operations. Composition in a free prop is just gluing of graphs. The following
  discussion, although it is a little bit technical, is just a formalisation of this
  idea. 
\end{blanko}

\begin{blanko}{Free-prop monad.}\label{free-prop-monad}
We describe the free prop\footnote{It should be remarked that the algebras for the prop monad  
described here are not exactly the same as the props of Mac Lane~\cite{MacLane:categories}, and they 
should properly be called {\em graphical props} (see 
Batanin--Berger~\cite{Batanin-Berger:1305.0086}, Remark~10.5). The 
difference is with the $(0,0)$-operations: in a Mac Lane prop, 
$\End(\mathbf{1})$ is always a commutative monoid (by the 
Eckmann--Hilton argument). In a graphical prop, $\End(\mathbf{1})$ 
can be a noncommutative monoid. The difference does not affect us 
here, as we are only concerned with {\em free} graphical props, 
and these automatically have commutative $\End(\mathbf{1})$ and are 
therefore also props in the sense of Mac Lane.} on a digraphical species.
Recall (from \ref{Sh=PrSh}) that presheaves on $\elGr$ are naturally equivalent to
  sheaves on $\Gr$, so that a presheaf on elementary graphs can be
  evaluated also on general (acyclic) graphs by the limit formula 
  $$
  \gs F[\pn G] \simeq \lim_{\pnsmall E\in \el(\pnsmall G)} \gs F[\pn E].
  $$
  The limit is over the elementary subgraphs of $\pn G$ as in \ref{colim-el}. 

  The {\em free prop monad}
  \begin{align*}
   \PrSh(\elGr) \ \longrightarrow & \ \PrSh(\elGr) \\  
   \gs F \ \longmapsto & \ \ov{\gs F} 
  \end{align*}
  is given (at the level of its underlying endofunctor) by $\ov{\gs F}[\triv] :=  
  \gs F[\triv]$ and
  \begin{align*}
  \ov{\gs F}[\coro{m}{n}] \ := & \ \colim_{\pnsmall G \in 
  [\coro{m}{n}]\kat{-Gr}_{\iso}} \gs F[\pn G] \\[6pt]
  \simeq& \ \sum_{\pnsmall G\in \pi_0([\coro{m}{n}]\kat{-Gr}_{\iso})}  
  \frac{\gs F[\pn G]}{\Aut_{[\coro{m}{n}]}(\pn G)} \\[6pt]
  \simeq& \ \pi_0 \big([\coro{m}{n}]\kat{-Gr}_{\iso} \comma \gs F \big) .
  \end{align*}
  Here the first equation follows since $[\coro{m}{n}]\kat{-Gr}_{\iso}$ is just a 
  groupoid: the sum is over isomorphism classes of $[\coro{m}{n}]$-graphs, and 
  $\Aut_{[\coro{m}{n}]}(\pn G)$ denotes
  the automorphism group of $\pn G$ in $[\coro{m}{n}]\kat{-Gr}_{\iso}$.
  The
  $(m,n)$-operations of $\ov {\gs F}$ are thus iso-classes of
  $[\coro{m}{n}]$-$\gs F$-graphs.
  
  (We omit description of the monad
  multiplication and unit (essentially gluing of graphs), although of course this is essential
  information. See \cite{Kock:1407.3744} for all details in the
  connected case, the free properad monad.)
\end{blanko}

\begin{blanko}{Free prop on a Petri net, and underlying symmetric monoidal category.}
  \sloppy The free prop on a Petri net $\pn P$ has as operations the iso-classes of processes
  $\pn G \to \pn P$ with fixed boundaries.\footnote{The fact that morphisms are $\pn
  P$-graphs equipped with specified maps from its domain and codomain onto its
  boundaries makes it an example of a cospan category with boundaries living in
  `lower dimension' than the apex. Cospans of this nature play also an important role
  in some recent approaches to open Petri nets; see for example
  \cite{Baldan-Corradini-Ehrig-Heckel:2005}, \cite{Baez-Master:1808.05415},
  \cite{Baez-Courser:1911.04630}.} An element in $\pi_0
  \big([\coro{m}{n}]\kat{-Gr}_{\iso} \comma \pn P \big)$ is a morphism in the
  underlying symmetric monoidal category of the free prop. The objects are strings of
  elements in $\pn P[\triv]$, that is maps $\un n \to \pn P[\triv]$. The domain of an
  element $p:\pn G\to \pn P$ in $\pi_0 \big([\coro{m}{n}]\kat{-Gr}_{\iso} \comma \pn
  P \big)$ is the composite $\un m \simeq \operatorname{in}(\pn G) \to \pn G
  \stackrel{p}\to \pn P , $ and similarly the codomain is the composite $\un n \simeq
  \operatorname{out}(\pn G) \to \pn G \stackrel{p}\to \pn P $. 

  \fussy Composition is given by gluing (pushout in the category $\Gr$). Formally this comes
  about from the monad multiplication, whose description we omitted.\footnote{For an
  elegant proof of the adjunction between digraphical species and
  symmetric monoidal categories, see the very recent
  Baez--Genovese--Master--Shulman~\cite{Baez-Genovese-Master-Shulman:2101.04238}.}
\end{blanko}

\begin{blanko}{Remark.}
  It is easy to see that the digraphical species $\ov{\pn P}$ is flat
  again, since the groups $\mathfrak{S}_m \times \mathfrak S_n$ act on
  $\ov{\pn P}$ only through the boundary isomorphisms. So except for
  the fact that $\ov{\pn P}$ takes infinite values, it could be
  regarded as a Petri net again, and the free-prop adjunction could
  then be interpreted directly on the category of (infinite) Petri
  nets. This viewpoint, although it has played an important role in
  Petri-net theory (from~\cite{Meseguer-Montanari:monoids}
  to~\cite{Bruni-Meseguer-Montanari-Sassone:2001}), is not given
  emphasis in the present work, where we insist on Petri nets being
  finite. While $\ov{\pn P}$ is of interest as a prop or
  symmetric monoidal category, the corresponding (infinite) Petri net
  seems to defy the purposes of Petri nets, and is hugely redundant.
  It has a transition for every process of $\pn P$ --- this includes
  an `identity' transition for every place $s\in S$ and every $n\in
  \N$, corresponding to the disjoint union of $n$ $s$-labelled unit
  graphs.
\end{blanko}

\begin{blanko}{Props in groupoids.}
  The free prop formula just described involves taking $\pi_0$. Avoiding this leads to a 
  simpler construction, that of the free prop in groupoids.
  It works for digraphical species valued in groupoids,
  $\gs F: \elGr\op\to\Grpd$.
  In particular, an ordinary graphical species, such as a Petri net,
  is a groupoid-valued graphical species via the inclusion functor 
  $\Set \to \Grpd$.

  The free prop-in-groupoid on a digraphical species 
  $\gs F : \elGr\op\to\Grpd$ is
  given by $\widetilde{\gs F}[\triv] := \gs F[\triv]$ and
  \begin{align*}
  \widetilde{\gs F}[\coro{m}{n}] \ := 
  [\coro{m}{n}]\kat{-Gr}_{\iso} \comma \gs F  ,
  \end{align*}
  without taking $\pi_0$.

  Apart from $[\triv]$ (where $\gs F\mapsto \widetilde{\gs F}$ is the identity anyway),
  $\widetilde{\gs F}$ is simply the groupoid-valued presheaf corresponding to the
  projection $\res:\Gr_{\iso}\comma \gs F \to \Cor$. (It should be noted though, that
  since $\res$ is not a fibration, it is essential to use homotopy fibres (see
  \ref{fibre}) rather than strict fibres, in order to extract the value on a given
  $[\coro{m}{n}]$.)
\end{blanko}

\begin{blanko}{Underlying Segal space of $\widetilde{\gs F}$.}
  The underlying (symmetric monoidal) Segal space $\ds Y_\bullet$
  of the free 
  prop-in-groupoids $\widetilde{\gs F}$ is given as follows: $\ds Y_1$ is the
  homotopy sum (\ref{X/G}) of all the $\widetilde{\gs F}[\coro{m}{n}]$:
  $$
  \ds Y_1 = \int^{[\coro{m}{n}]} \widetilde{\gs F}[\coro{m}{n}] 
 \ \simeq \sum_{[\coro{m}{n}]\in \pi_0(\Cor)} \widetilde{\gs 
  F}[\coro{m}{n}] /\!/ \Aut([\coro{m}{n}])
  .
  $$
  (This is the appropriate homotopical way of describing the groupoid whose homotopy
  fibre over $[\coro{m}{n}]$ is $\widetilde{\gs F}[\coro{m}{n}]$. The double bar is
  homotopy quotient (see \ref{X/G}). In degree zero we have $\ds Y_0= \B \comma \gs F
  [\triv]$, the free symmetric monoidal category on $\gs F [\triv]$. In higher
  simplicial degrees we have $\ds Y_k = \ds Y_1 \times_{\ds Y_0}^h \cdots \times_{\ds
  Y_0}^h \ds Y_1$. The face and degeneracy maps come from the monad structure: in
  particular $\ds Y_1 \stackrel{d_1}{\longleftarrow} \ds Y_2$ is the monad
  multiplication, which allows to contract a $2$-level graph to a single corolla.
  (For details, see \cite{Kock:1407.3744}, although that reference only covers the
  connected case, the case of properads.)
\end{blanko}

\begin{blankothm}{Theorem}\label{thm:Xisfree}
  For $\pn P$ a Petri net, the symmetric monoidal Segal space $\ds
  X_\bullet$ of $\pn P$-processes is the free prop-in-groupoids on the
  digraphical species $\pn P$.
\end{blankothm}

\begin{proof}
  We compare the mapping groupoids (\ref{Map(x,y)}) of the two Segal
  spaces. Fix $M$ and $N$ two states of $\pn P$ with $m$ and $n$ tokens,
  respectively. Both for $\ds X_\bullet$ and for the free prop, the mapping
  groupoids consist of processes from $M$ to $N$, and in a sense the check
  is routine; we include it to showcase the calculus of homotopy pullbacks
  (\ref{hopbk}): the two groupoids will be identified by both satisfying
  the universal property of the same homotopy pullback.
  
  The mapping groupoids appear in the following diagram of homotopy 
  pullbacks:
  \begin{equation}
	\label{Map}
	\begin{tikzcd}[row sep={43pt,between origins}, column sep={72pt,between origins}]
   \Map_{\ds X_\bullet}(M,N) \drpullback \ar[d] \ar[r, dotted] 
   \ar[rr, bend left=18pt]& \widetilde{\pn P}[\coro{m}{n}] 
   \drpullback \ar[d] \ar[r] & \ds X_1 \ar[d, "{(d_1,d_0)}"]  \\
  1 \ar[r, dotted, "{\name{(M,N)}}"'] \ar[rr, bend left=18pt, 
  crossing over, pos=0.48, "{\name{(M,N)}}"] & (\ds X_0\times \ds X_0)_{[\coro{m}{n}]} 
  \drpullback \ar[d] \ar[r] & \ds X_0 \times \ds X_0 \ar[d] \\
  & 1 \ar[r, "\name{[\coro{m}{n}]}"'] & \Cor .
  \end{tikzcd}
  \end{equation}
  The  curved homotopy-pullback rectangle is the definition of $\Map_{\ds
  X_\bullet}(M,N)$ (cf.~\ref{Map(x,y)}): homotopy pullback along the curved middle
  map $\name{(M,N)}$. The right-hand composite homotopy-pullback rectangle is the
  definition of $\widetilde {\pn P} [\coro{m}{n}]$,  realised in two steps. Note that $\Cor$
  is the groupoid of naked corollas, without reference to $\pn P$, whereas $\ds
  X_0\times \ds X_0$ is the groupoid of pairs of $\pn P$-states. Since $M$ and $N$
  are assumed to have cardinality $m$ and $n$, the middle curved arrow factors
  through $(\ds X_0\times \ds X_0)_{[\coro{m}{n}]}$, inducing first the lower dotted
  map by the universal property of the homotopy pullback $(\ds X_0\times \ds
  X_0)_{[\coro{m}{n}]}$, and inducing next the upper dotted arrow by the universal
  property of the homotopy pullback $\widetilde{\pn P}[\coro{m}{n}] $. The closure
  properties of homotopy pullbacks (\ref{lem:prism}) force the resulting upper
  left-hand square to be a homotopy pullback. This shows that $\Map_{\ds
  X_\bullet}(M,N)$  is both the mapping groupoid in $\widetilde{\pn P}$ and 
  in $\ds X_\bullet$.
  (It also exhibits $\widetilde{\pn P} [\coro{m}{n}]$ as the homotopy sum (\ref{X/G}) of its
  homotopy fibres (\ref{fibre}) over varying $(M,N)$:
  $$
  \widetilde{\pn P} [\coro{m}{n}] \simeq \int^{(M,N)\in (\ds X_0\times \ds X_0)_{[\coro{m}{n}]}}
  \Map_{\ds X_\bullet}(M,N)     .)
  $$
 
  The remaining checks (higher simplicial degrees) follow from the 
  Segal condition. The symmetric monoidal structure is clear, as it 
  is just disjoint union.
\end{proof}

\section{The symmetric monoidal category of processes}
\label{sec:1cats}

\begin{blanko}{Segal space vs.~category.}
  The symmetric monoidal Segal space $\ds X_\bullet$ is nice to work with
  for its clean combinatorial description. The fact that there is a
  groupoid of processes, $\ds X_1$, instead of just a set of them, should
  be seen as an advantage rather than a drawback. The general point is that
  since whole-grain Petri nets are configurations of finite sets, it is
  natural to admit that these configurations form groupoids, and continue
  to work with these groupoids rather than with their set quotients.
  Passing to sets of iso-classes too early can easily lead to unnecessary
  complications or even serious problems. Concretely, working in the
  groupoid of processes, we can glue processes thanks to explicit
  bijections between their boundaries (states), but it is plainly not
  possible to glue isomorphism classes of processes along isomorphism
  classes of states.

  Notwithstanding the previous paragraph, it is sometimes convenient to extract from
  $\ds X_\bullet$ a honest symmetric monoidal category $(\CC, +,\emptyset)$, which is
  then the free symmetric monoidal category on $\pn P$. (The construction is the same
  for any digraphical species $\gs F$.) The answer was already given: it is the
  symmetric monoidal category underlying the free prop on $\pn P$. However, this
  description is a bit round-about, and in fact we cheated by not describing the
  monad multiplication for the free-prop monad (since all the details were given in
  \cite{Kock:1407.3744}). Instead we can give a direct description from $\ds
  X_\bullet$, taking the opportunity to explain a few homotopy issues.
\end{blanko}

\begin{blankothm}{Theorem}\label{thm:C}
  The free symmetric monoidal category on $\pn P$, denoted $\CC$, is
  the homotopy category of $\ds X_\bullet$ (or more precisely, its
  codescent object\footnote{{\em Codescent object} means colimit of 
  $\ds X_\bullet$ in $\kat{Cat}$ weighted by $\simplexcategory \to 
  \kat{Cat}$; see Lack~\cite{Lack:1510.08925} and 
  Weber~\cite{Weber:1503.07585}.}). The objects of $\CC$ are the states $\pn M \to
  \pn P$, and its hom sets are
  $$
  \Hom_{\CC}(\pn M,\pn N) = \pi_0 \big( \Map_{\ds X_\bullet}(\pn 
  M,\pn N) \big) .
  $$
\end{blankothm}

The mapping groupoid (cf.~\ref{Map(x,y)}) already appeared in the proof of
Theorem~\ref{thm:Xisfree}: $\Map_{\ds X_\bullet}(\pn M,\pn N)$ is the
groupoid whose objects are triples $( p, \sigma_1, \sigma_2)$ consisting of
a process $p: \pn G \to \pn P$, and isomorphisms $\sigma_1 : \pn M\simeq
d_1(\pn G)$ and $\sigma_2 : \pn N \simeq d_0(\pn G)$. The arrows of the
groupoid $\Map_{\ds X_\bullet}(\pn M,\pn N)$ are isomorphisms of processes
$p \simeq p'$ compatible with the boundary isos. In conclusion, in the free
symmetric monoidal category $\CC$, the morphisms from $\pn M$ to $\pn N$
are iso-classes of processes (where the isomorphisms are required to fix
the boundary).

Before explaining this (without reference to the free prop 
construction), it is instructive to consider some 
other attempts at obtaining an ordinary symmetric monoidal 
category out of $\ds X_\bullet$.

\begin{blanko}{Brutally applying $\pi_0$ and trying to 
  cope\ldots}\label{swap}
  One could think of simply applying $\pi_0$ on $\ds X_\bullet$
  degree-wise. This does not immediately work, because $\pi_0$ does
  not preserve homotopy pullbacks and will destroy the Segal
  condition. Again, this is a well-known phenomenon, explained
  carefully in \cite{Galvez-Kock-Tonks:1612.09225} in the very similar
  example of trees.
  
  One may then try to correct the problem by taking quotients.
  Note first that $\pi_0 (\ds X_0)$ is the free commutative monoid on 
  the set of isomorphism classes of states, precisely as employed in 
  the collective-tokens approach. To correct the problem that
  $
  \pi_0( \ds X_2) \longrightarrow \pi_0(\ds X_1) \times_{\pi_0(\ds 
  X_0)} \pi_0(\ds X_1)
  $
  is not a bijection, some stuff must be quotiented out: one has to 
  identify two processes $\pn G$ and $\pn G'$ if there exist cuts
  $$
  \pn G = \pn G_1 \amalg_{\pnsmall M} \pn G_2 \qquad \text{and}
  \qquad
  \pn G' = \pn G'_1 \amalg_{\pnsmall M'} \pn G'_2
  $$
  such that $\pn M = \pn M'$, $\pn G_1 = \pn G_1'$, and $\pn G_2 = 
  \pn G_2'$ --- all
  in $\pi_0(\ds X_0)$. In other words, the
  two processes satisfy the {\em swap property} of
  Best--Devillers~\cite{Best-Devillers:1987}: one can cut and
  reconnect (a kind of surgery).  For example, the two processes $p$ and $q$ of 
  Example~\ref{ex:processes} are swap equivalent, as they have the same layers, only glued 
  together differently. 
  This is non-trivial to control, and
  the final result is the Best--Devillers category of equivalence
  classes of processes~\cite{Best-Devillers:1987}. In the end the
  building blocks are the iso-classes of simple firings. Presumably,
  this is precisely the free symmetric monoidal category of iso-classes
  of transitions, of Meseguer and
  Montanari~\cite{Meseguer-Montanari:monoids}. Note that it is in fact a
  {\em commutative} monoidal category.
  
  This is not what we want here.
\end{blanko}

\begin{blanko}{Weak-double-category viewpoint.}
  A second approach takes the viewpoint of double
  categories.\footnote{Double-categorical approaches to Petri
  nets have appeared recently in \cite{Baez-Pollard:1704.02051},
  \cite{Baez-Master:1808.05415}.} Observe that a (symmetric
  monoidal) Segal space can be regarded as a special case of a
  (symmetric monoidal) weak double category, by choosing a
  pseudo-inverse to the equivalence $\ds X_2 \to \ds X_1 \times^h_{\ds
  X_0} \ds X_1$.  In the present case, the pseudo-inverse is given by
  gluing graphs. By making a global choice of pushouts one gets a
  weak double category, which is a groupoid in the vertical direction:
  the objects are the states of $\pn P$, the vertical arrows are the
  isomorphisms of states, and the horizontal arrows are the $\pn
  P$-processes.
  
  From a weak double category one can always extract its {\em
  horizontal bicategory}, namely by disregarding the vertical arrows.
  (Getting a bicategory is an important step towards getting an
  ordinary category, because now one can take $\pi_0$; this operation
  does preserve pullbacks over discrete objects.)
  However, in the case of $\ds X_\bullet$ this construction will
  destroy the monoidal structure! The reason is that the coherence
  constraints of the monoidal structure live in the vertical
  dimension, and when they are killed the associativity is lost.
  
  It is worth mentioning this, because the horizontal-bicategory construction is a
  common way to construct symmetric monoidal bicategories (and then symmetric
  monoidal categories, by taking $\pi_0$), as explained in great detail by Hansen and
  Shulman~\cite{Hansen-Shulman:1910.09240}, and exploited in the context of Petri
  nets by Baez et al.~\cite{Baez-Master:1808.05415}, \cite{Baez-Courser:1911.04630}.
  However, the Hansen--Shulman construction specifically requires the double category
  to be {\em fibrant}, which means that $\ds X_1 \stackrel{(d_1,d_0)}\longrightarrow
  \ds X_0 \times \ds X_0$ should be a fibration.  This is not the case for
  our $\ds X_\bullet$: even though both $d_0$ and $d_1$ are fibrations individually
  (cf.~Lemma~\ref{lem:fib}), the map $(d_1,d_0)$ is not a fibration. (For example,
  let $p:\pn G \to \pn P$ be the trivial process consisting of $n$ unit graphs
  mapping to the same place, then $(d_1,d_0)$ sends $p$ to the pair of {\em two}
  copies of that state. Now $p$ has automorphism group $\mathfrak S_n$ whereas
  $(d_1,d_0)(p)$ has automorphism group $\mathfrak S_n \times 
  \mathfrak S_n$. It
  is not possible to lift automorphisms outside the diagonal. (Compare with
  Remark~\ref{not-enough-isos}.)) 
\end{blanko}

With these preliminary analyses, we are ready for the correct 
solution:

\begin{blanko}{Change of objects.}
  Instead of getting a bicategory by throwing away the vertical maps
  (isomorphisms), the passage from double category to bicategory 
  should incorporate the vertical maps into the horizontal
  part. This is achieved as an instance of the standard 
  {\em change-of-objects} construction:

  Recall the change-of-objects construction for ordinary categories
  $\CC$: given a map of sets $\phi: D \to \obj(\CC)$ one can obtain a
  new category $\CC^\phi$ with object set $D$ and hom sets
  $\Hom_{\CC^\phi}(x,y) := \Hom_{\CC}(\phi x, \phi y)$. Then by
  construction there is a fully faithful functor $\CC^\phi\to\CC$,
  which is an equivalence if $\phi$ is essentially
  surjective.
  The same construction works for Segal spaces as follows: given a
  Segal space $\ds X_\bullet$ and a map of groupoids $\phi: D \to \ds
  X_0$, one gets a new Segal space $\ds X^\phi_\bullet$ whose
  $k$-simplices are given by the homotopy pullback
  \[\begin{tikzcd}[row sep={54pt,between origins}]
  (\ds X^\phi)_k \drpullback \ar[d] \ar[r] & \ds X_k \ar[d, 
  "\parbox{40pt}{\tiny return the \\ vertices of \\ a simplex}"]
  \\
  D^{k+1} \ar[r] & (\ds X_0)^{k+1}     .
  \end{tikzcd}
  \]
  Now the bicategory associated to a Segal space $\ds X_\bullet$ is obtained by
  applying this general construction to the case $D= \obj(\ds X_0)$
  with the inclusion map $\phi: \obj(\ds X_0) \to \ds X_0$. 
  (Note that if $(d_1,d_0)$ is a fibration (as in the
  Hansen--Shulman situation), then the homotopy pullback can be taken
  to be a strict pullback, and $\ds X_\bullet^\phi$ is precisely the
  horizontal bicategory.)
  
  Since $\ds X_\bullet^\phi$ is equivalent to 
  $\ds X_\bullet$, the symmetric monoidal structure carries over.
  It should be noted that $\ds X_\bullet^\phi$ is far from being Rezk 
  complete, and the resulting symmetric monoidal structure is {\em not}
  just pointwise. For example, in simplicial degree $0$ it will appear
  that the monoidal product is not even associative, because there is
  no room for associators, but the associators (as well as the unit and 
  symmetry coherence constraints) now live in simplicial degree $1$!
  Looking into these subtleties makes one appreciate the simplicity of
  the Rezk complete $\ds X_\bullet$ itself. But $\ds X_\bullet^\phi$
  is only an intermediate step towards the:
\end{blanko}

\begin{blanko}{Codescent object.}
  Coming back to the Segal space of processes $\ds X_\bullet$, we now
  have an (equivalent) Segal space $\ds X^\phi_\bullet : 
  \simplexcategory\op\to\Grpd$
  with degree-$0$ discrete,
  and now the final step is simply to post-compose with $\pi_0:
  \Grpd\to\Set$. The functor $\pi_0$ does preserve homotopy pullbacks 
  over discrete objects, and since we have arranged for $\ds 
  X^\phi_0$ to be discrete, the composite is again a Segal space ---
  that is, a category. Since in any case $\pi_0$ preserves products, 
  the symmetric monoidal structure is preserved.
  
  The final resulting symmetric monoidal category $\CC$ is as
  stated in Theorem~\ref{thm:C}. In fact, $\CC$ is
  the {\em codescent object} of $\ds X_\bullet$, a general
  construction which in other contexts serves to compute internal-algebra classifiers
  for monads, as well as the prop envelope of an
  operad~\cite{Weber:1503.07585}. As just explained, it is rather easy
  to compute for Segal spaces (weak double categories whose vertical
  category is just a groupoid); it is considerably more involved to
  compute for general double categories \cite{Weber:1503.07585}.
\end{blanko}

\section{Hypergraphs}
\label{sec:hypergraphs}

We now change perspective a little bit, starting to work towards the unfolding 
theorem, proved in Section~\ref{sec:unfold}. 
While the process semantics of Sections~\ref{sec:processes}--\ref{sec:1cats} is
concerned with {\em gluing} $\pn P$-processes in-boundary to out-boundary,
the relevant colimits will now be {\em unions} of bigger and bigger $\pn
P$-graphs (corresponding to longer and longer computations, accumulating more and
more information), all starting from the same fixed initial state. 
Such colimits do not usually exist in the category of graphs, but
they do exist in the bigger category of (directed) hypergraphs.
In this section we set up the necessary machinery. The main task is to
characterise those hypergraphs that arise as colimits of graphs like 
this.

\begin{blanko}{Hypergraphs~\cite{Kock:1407.3744}.}
  A {\em hypergraph} $\pn H$ is a diagram of sets and maps
  $$
  \begin{tikzcd}
	A & \ar[l] I \ar[r] & N & \ar[l] O \ar[r] & A  ,
  \end{tikzcd}
  $$
  where the two spans are relations. We now allow infinite sets, but 
  insist that the maps $I \to N \leftarrow O$ be only {\em finite maps} 
  (that is, with finite fibres).
  The set $A$ is now read as the 
  set of hyper-edges. 
  We picture hypergraphs as Petri nets, with nodes represented as small 
  squares, and hyper-edges represented as small circles, like for example
  
  
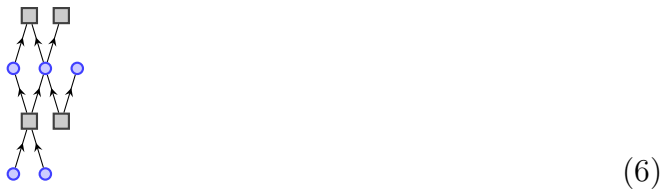
\begin{equation}\label{not-forward}
	\begin{tikzpicture}[
	  node distance=0.7cm,on grid,>=stealth',bend angle=22.5,auto,
	  ]

  	  \begin{scope}[scale=0.7,
		every place/.style= {minimum size=1.3mm,thick,draw=blue!75,fill=blue!20}, 
		every transition/.style={minimum size=1.7mm,thick,draw=black!75,fill=black!20}
		]
	  \begin{scope}
   		\tikzstyle arrowstyle=[scale=1]
		\tiny
		\node [place] (s1) at (0.0,2) {}; 
		\node [place] (s2) at (0.6,2) {};
		\node [place] (s2') at (1.2,2) {};
		\node [place] (s3) at (0.0, 0.0) {};
		\node [place] (s4) at (0.6, 0.0) {};
		\node [transition] (t1) at (0.3,1.0) {} ;
		\node [transition] (t2) at (0.9,1.0) {} ;
		\node [transition] (t3) at (0.3,3.0) {} ;
		\node [transition] (t4) at (0.9,3.0) {} ;
		\draw[directed] (s3) to (t1);
		\draw[directed]  (s4) to [post] (t1);
		\draw[directed]  (t1) to [post] (s1);
		\draw[directed]  (t1) to [post] (s2);
		\draw[directed]  (t2) to [post] (s2);
		\draw[directed]  (t2) to [post] (s2');
		\draw[directed]  (s1) to [post] (t3);
		\draw[directed]  (s2) to [post] (t3);
		\draw[directed]  (s2) to [post] (t4);
	  \end{scope}
	  \end{scope}
	\end{tikzpicture}
  \end{equation}
  
  The notion of etale map is as for graphs and Petri nets, and again an open 
  sub-hypergraph is an etale map that is injective point-wise 
  (i.e.~on $A$, $I$, $N$, $O$ components separately). Note that the
  arity of hyper-edges is not necessarily preserved by etale or open 
  maps.
  
  Denote by $\kat{Hgr}$ the category of hypergraphs and their etale maps.

\end{blanko}

\begin{blanko}{Forward hypergraphs.}
  We shall only need hypergraphs where the right-hand part is graph-like, but allow
  the left-hand part to be general-hypergraph-like.  
  We define a {\em forward
  hypergraph} to be a hypergraph where the last leg is injective:
  $$
  \begin{tikzcd}[column sep={39pt,between origins}, row sep={8pt,between origins}]
	A & \ar[l] I \ar[r] & N & \ar[l] O \ar[r, into] & A  .
	\\
	& 
	\underset{\text{\tiny relation}}{\footnotesize 
	{\underbrace{\phantom{----------}}}} &&\ar[r, phantom, 
	"\text{\tiny injective}"']&{}
  \end{tikzcd}
  $$
  This is to say: each hyper-edge is outgoing of at most one node.
  In pictures, only hyper-edges that locally look like
  
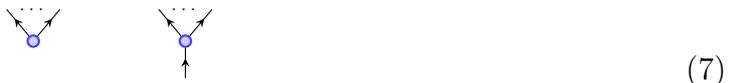
\begin{equation}\label{forward-conflict}
	\begin{tikzpicture}[
	  node distance=0.7cm,on grid,>=stealth',bend angle=22.5,auto,
	  ]
	  \begin{scope}[scale=0.7,
		every place/.style= {minimum size=1.3mm,thick,draw=blue!75,fill=blue!20}, 
		every transition/.style={minimum size=1.7mm,thick,draw=black!75,fill=black!20}
		]
   		\tikzstyle arrowstyle=[scale=1]
		\node [place] (s3) at (0.0, 0.0) {};
		\node [port] (t1) at (-0.5,0.6) {} ;
		\node at (0,0.6) {\tiny $\cdots$} ;
		\node [port] (t2) at (0.5,0.6) {} ;
		\draw[directed] (s3) to (t1);
		\draw[directed] (s3) to (t2);
	  \end{scope}
	  \begin{scope}[shift={(2,0)}, scale=0.7,
		every place/.style= {minimum size=1.3mm,thick,draw=blue!75,fill=blue!20}, 
		every transition/.style={minimum size=1.7mm,thick,draw=black!75,fill=black!20}
		]
   		\tikzstyle arrowstyle=[scale=1]
		\node [port] (i1) at (0.0, -0.7) {};
		\node [place] (s3) at (0.0, 0.0) {};
		\node at (0,0.6) {\tiny $\cdots$} ;
		\node [port] (t1) at (-0.5,0.6) {} ;
		\node [port] (t2) at (0.5,0.6) {} ;
		\draw[directed] (s3) to (t1);
		\draw[directed] (s3) to (t2);
		\draw[directed] (i1) to (s3);
	  \end{scope}
 \end{tikzpicture}
\end{equation}
  are allowed. So for example, the hypergraph above in \eqref{not-forward} is not
  forward, whereas those of \eqref{eq:Hcommay} are forward.
\end{blanko}

\begin{blanko}{Various AINOA structures.}\label{various-AINOA}
  Except for finiteness conditions, the various AINOA-shape structures introduced 
  can all be considered special cases of Petri nets. We have
    
  \small
  $$
  \text{acyc.graphs} \subset \text{occurrence hypergraphs} \subset \text{forward 
  hypergraphs} \subset \text{hypergraphs} \subset \text{Petri nets}
  $$ 
  \normalsize
  
  \medskip
  
  \noindent
  (The subtle notion of occurrence hypergraph will be introduced in \ref{occur} below.)
  From this viewpoint, hypergraphs are the Petri nets where there are no parallel 
  arcs (equivalently, in the traditional setting: all arc multiplicities are $1$).
  The forward condition prohibits what is called {\em backward conflicts} (see 
  \cite{Nielsen-Plotkin-Winskel:1981}) but still
  allows the {\em forward conflicts} pictured in \eqref{forward-conflict}. 
  Acyclic graphs are the {\em conflict-free} Petri nets,
  also called {\em causal nets}~\cite{Goltz-Reisig:1983}, \cite{Winskel:eventstructures}.

  While it is thus important that all
  these structures share the same AINOA shape, we stress this difference:
  Petri nets are required to be finite, but express infinite behaviour by not being 
  required acyclic; hypergraphs are allowed to be infinite, but should instead be required 
  to be (acyclic and) well-founded,\footnote{This contrast was stressed by 
  Scott~\cite{Scott:1971} in the context of flow diagrams.} 
  which is the next condition we impose:
\end{blanko}

\begin{blanko}{Well-founded hypergraphs.}\label{EST}
  We write $x \lessdot y$ for two nodes in a hypergraph $\pn H=AINOA$ if
  there exists a hyper-edge $a$ which is outgoing for $x$ and incoming for
  $y$: more precisely, $x$ and $a$ are $O$-related and $a$ and $y$ are
  $I$-related. Note that in a forward hypergraph the set $\{x \mid
  x\lessdot y\}$ is finite for every node $y$ (as a consequence of the
  assumption that the map $I \to N$ is finite). We denote by $<$ the
  relation on $N$ given as the transitive closure of $\lessdot$, and by
  $\leq$ the transitive and reflexive closure of $\lessdot$.
  
  A hypergraph $\pn H=AINOA$ is {\em well-founded} when the relation $<$ is
  anti-reflexive (meaning that there are no directed cycles) and for each
  node $y$ the set $\{ x \mid x \leq y\}$ is finite. In particular, the
  relation $\leq$ is then anti-symmetric, so that $(N,\leq)$ is a poset.
  The following equivalent characterisation of well-foundedness is
  technically convenient: a forward hypergraph is well-founded when the
  `earliest-start-time'
  function\footnote{Winskel~\cite{Winskel:eventstructures} calls it {\em
  depth}.} $f : N \to \N$:
  $$
  \begin{cases}
  f(y) = 1 &  \text{when }\not\exists x \lessdot y \\
  f(y) = n+1  & \text{when $f(x)$ is defined for all 
  $x\lessdot y$ and $n=\max \{f(x) \mid x\lessdot y \}$}
  \end{cases}
  $$
  is a total function. In words, $f(y)$ is the length of the longest chain ending in 
  $y$ (plus one).
\end{blanko}

\begin{blanko}{$B$-hypergraphs.}\label{BHgr}
  Since for a forward hypergraph $\pn H = AINOA$ the map $O\to A$ is injective, it
  has a well-defined complement called the {\em in-boundary} of $\pn H$. Throughout
  we shall fix a finite set $B$, with its corresponding node-less graph $\pn
  B=B\emptyset\emptyset\emptyset B$, and only consider forward hypergraphs with fixed
  in-boundary $B$. These are called {\em $B$-hypergraphs}; they are thus maps $\pn B
  \to \pn H$ for which the square
      \begin{equation}
		\label{eq:B-for-hgraphs}
  \begin{tikzcd}
	\emptyset \ar[d] \ar[r, into] \ar[rd, phantom, "{\small \Box}"]& B \ar[d] 
	\\
	O \ar[r, into]  & A
  \end{tikzcd}
  \end{equation}
  is a pushout 
  (and hence also
  a pullback).
  Note that by closure properties of pushouts, the $B$-maps $\pn H' \to \pn H$
  have the property
  that the right-most square is a pushout (and hence a pullback):
    \begin{equation}
	  \label{eq:B-map}
  \begin{tikzcd}
	A' \ar[d, "\alpha"'] & I' \ar[d] \ar[l] 
	\ar[r] \drpullback &  N' \ar[d]& 
	O' \ar[rd, phantom, 
	"{\small \Box}"]
	\ar[d] \ar[l]\ar[r, into] \dlpullback & A' \ar[d, "\alpha"] 
\\
A  & I \ar[l] \ar[r]  & N & O \ar[l]\ar[r, into]  & A
  \end{tikzcd}
  \end{equation}

  A $B$-subgraph of $\pn H$ will be called a {\em process} of $\pn H$ (from $B$).
\end{blanko}

\begin{blanko}{Lowersets.}
  Recall that a lowerset\footnote{also called {\em order ideal}, or {\em left-closed
  subset}~\cite{Winskel:eventstructures}.} of a poset $N$ is a sub-poset $L\subset N$
  such that if $x\leq y$ and $y\in L$ then also $x\in L$. A {\em lowerset} in a
  well-founded hypergraph $\pn H$ is an open sub-hypergraph (i.e., injective
  etale map) $\pn G \to \pn H$ which is a poset lowerset on nodes. Finally for a
  well-founded hypergraph $\pn H$ with in-boundary $B$, we define
  {\em $B$-lowerset} to be a lowerset whose in-boundary is $B$ again.
\end{blanko}

\begin{blankothm}{Lemma}\label{lowersetlemma}
  For a well-founded $B$-hypergraph $\pn H = AINOA$, there is a bijection
  between lowersets $L \subset N$ in the sense of posets,
  and $B$-lowersets $\pn G \subset \pn H$ as $B$-hypergraphs. 
\end{blankothm}

\begin{proof}
  From a $B$-lowerset $\pn G \subset \pn H$, just return the poset of nodes in 
  $\pn G$, which is a lowerset in $N$ by assumption. Conversely, given a lowerset 
  $L\subset N$ 
  in the poset of nodes of $\pn H$, define the hyperedges of $\pn G$ to be those
  incident to some node in $L$, together with all the hyperedges in $B$.
 \end{proof}
 
\begin{blanko}{Principal lowersets.}
  For a node $y\in N$ in a well-founded hypergraph $\pn H=AINOA$, 
  we can consider the set of 
  nodes
  $N\comma y := \{x \in N\mid x\leq y\}$, which is the lowerset in the
  sense of posets. This set of nodes spans an open sub-hypergraph of
  $\pn H$ denoted $\pn H\comma y$ called the {\em principal lowerset
  of $y$}. Note that $\pn H\comma y$ contains also all the outgoing 
  hyper-edges of all its nodes. For example, 
  
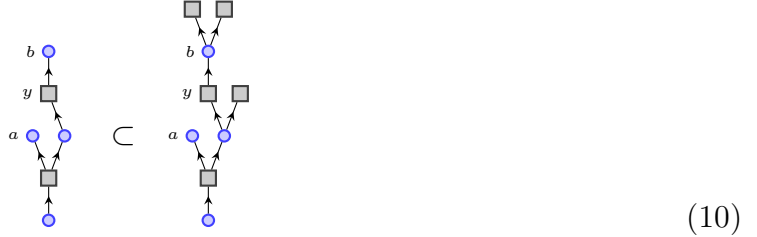
\begin{equation}\label{eq:Hcommay}
	\begin{tikzpicture}[
	  node distance=0.7cm,on grid,>=stealth',bend angle=22.5,auto,
	  ]

  	  \begin{scope}[scale=0.7,
		every place/.style= {minimum size=1.3mm,thick,draw=blue!75,fill=blue!20}, 
		every transition/.style={minimum size=1.7mm,thick,draw=black!75,fill=black!20}
		]
	  \begin{scope}
   		\tikzstyle arrowstyle=[scale=1]
		\tiny
		\node [place] (s1) at (0.0,0.0) {}; 
		\node [place, label=left:$a$] (s2) at (-0.3,1.6) {};
		\node [place] (s3) at (0.3,1.6) {};
		\node [place, label=left:$b$] (s4) at (0.0, 3.2) {};
		\node [transition] (t1) at (0.0,0.8) {} ;
		\node [transition, label=left:$y$] (t2) at (0.0,2.4) {} ;
		\draw[directed] (s1) to (t1);
		\draw[directed]  (t1) to [post] (s2);
		\draw[directed]  (t1) to [post] (s3);
		\draw[directed]  (s3) to [post] (t2);
		\draw[directed]  (t2) to [post] (s4);
	  \end{scope}
	  \node at (1.4,1.6) {$\subset$};
	  \begin{scope}[shift={(3,0)}]
   		\tikzstyle arrowstyle=[scale=1]
		\tiny
		\node [place] (s1) at (0.0,0.0) {}; 
		\node [place, label=left:$a$] (s2) at (-0.3,1.6) {};
		\node [place] (s3) at (0.3,1.6) {};
		\node [place, label=left:$b$] (s4) at (0.0, 3.2) {};
		\node [transition] (t1) at (0.0,0.8) {} ;
		\node [transition, label=left:$y$] (t2) at (0.0,2.4) {} ;
		\node [transition] (t3) at (0.6,2.4) {} ;
		\node [transition] (t4) at (-0.3,4.0) {} ;
		\node [transition] (t5) at (0.3,4.0) {} ;
		\draw[directed] (s1) to (t1);
		\draw[directed]  (t1) to [post] (s2);
		\draw[directed]  (t1) to [post] (s3);
		\draw[directed]  (s3) to [post] (t2);
		\draw[directed]  (s3) to [post] (t3);
		\draw[directed]  (t2) to [post] (s4);
		\draw[directed]  (s4) to [post] (t4);
		\draw[directed]  (s4) to [post] (t5);
	  \end{scope}
	  \end{scope}
	\end{tikzpicture}
  \end{equation}
  is the inclusion of the principal lowerset $\pn H \comma y$ in the hypergraph $\pn 
  H$. The hyperedges $a$ and $b$ have to be included even though they are not 
  below the node $y$, since otherwise there could not be an etale map.
  
  For $B$-hypergraphs, we also need the notion of {\em $B$-lowerset}, where we insist
  on including the whole in-boundary $B$ in $\pn H\comma y$, as in
  Lemma~\ref{lowersetlemma}.
  
  Note that
  in a well-founded forward hypergraph, all
  principal lowersets are {\em finite} hypergraphs, as a consequence of $I\to N$ 
  being finite.
\end{blanko}

\begin{blankothm}{Lemma}\label{lem:lowersets-preserved}
  Let $\pn H = AINOA$ be a well-founded forward $B$-hypergraph, and
  let $\pn H' \to \pn H$ be an injective etale map of $B$-hypergraphs (i.e.~an open 
  sub-$B$-hypergraph).
  Then for each node $y$ in $\pn H'$ there is a natural identification of 
  $B$-lowersets
  $\pn H' \comma y = \pn H \comma y$.
\end{blankothm}

\begin{proof}
  In view of well-foundedness, 
  it is enough to show that for $x \lessdot y$ in $\pn H$, we have also $x\lessdot 
  y$ in  $\pn H'$.
The idea is that $y$ has the same incoming hyper-edges in $\pn H'$ as in $\pn H$ 
because the map is etale, and that each such hyper-edge $a$ has the same preceding 
nodes in $\pn H'$ and in $\pn H$ by the property of $B$-maps (\ref{BHgr}, 
Equation~\eqref{eq:B-map}). The formalisation of this idea
is a pleasant exercise with pullbacks and AINOA diagrams:
  The map $f: \pn H' \to \pn H$ is a diagram
    \begin{equation}
  \begin{tikzcd}
	A' \ar[d] & I' \ar[d] \ar[l] 
	\ar[r] \drpullback &  N' \ar[d]& 
	O' \ar[d] \ar[l]\ar[r, into] \dlpullback  \ar[rd, phantom, 
	"{\small \Box}"]& A' \ar[d] 
	\\
	A  & I \ar[l] \ar[r]  & N & O \ar[l]\ar[r, into]  & A ,
  \end{tikzcd}
  \end{equation}
  and since it is a $B$-map, the last square is a pullback (cf.~\ref{BHgr}). The statement $x 
  \lessdot y \in N$ translates into the existence of a hyper-edge $a\in A$ such that $aIy$ and 
  $xOa$, which in turn means we have elements $i\in I$ and $o\in O$ like this:
  $$
  a\mapsfrom i \mapsto y   \qquad x \mapsfrom o \mapsto a .
  $$
  Furthermore $y=f(y')$. We need to construct 
  $$
  a'\mapsfrom i' \mapsto y'   \qquad x' \mapsfrom o' \mapsto a' 
  $$
  in the diagram for $\pn H'$, mapping to $x,i,o, a$. Define $i'\in I'$ to be
  the pullback of $i$ and $y'$, and 
  let $a'$ be its image in $A'$; then $a'I'y'$ by construction. Now define $o'\in O'$ to
  be the pullback of $o$ and $a'$, and let $x'$ be its image in $N'$; then 
  $x'O'a'$ by construction.
\end{proof}

\begin{blankothm}{Corollary}\label{cor:G-islower}
  Let $\pn H$ be a well-founded forward $B$-hypergraph, and
  let $\pn H' \to \pn H$ be an open sub-$B$-hypergraph.
  Then $\pn H'$ is a $B$-lowerset in $\pn H$.
\end{blankothm}

\begin{blankothm}{Corollary}\label{cor:est-agree}
  A injective etale $B$-map $\pn H' \to \pn H$ 
  of well-founded forward $B$-hypergraphs automatically preserves the
  earliest-start-time function (of \ref{EST}).
\end{blankothm}

\begin{proof}
  The value of the earliest-start-time function on a node $y$ of $\pn H$
  depends only on the principal lowerset $\pn H\comma y$, so the result follows from 
  Lemma~\ref{lem:lowersets-preserved}.
\end{proof}

\begin{blanko}{Remark/example.}
  Without the forward condition, it is not always true that $B$-sub-hyper\-graphs are lowersets,
  and it is not always true that the earliest-start-time functions agree,
  as exemplified by
  
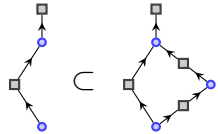
\begin{center}
  \begin{tikzpicture}[
	node distance=0.9cm,on grid,>=stealth',bend angle=22.5,auto,
	]
	
	\begin{scope}[shift={(-0.7,0)}, scale=0.45,
	  every place/.style= {minimum size=1.2mm,thick,draw=blue!75,fill=blue!20}, 
	  every transition/.style={minimum size=1.5mm,thick,draw=black!75,fill=black!20}
	  ]
	  \tikzstyle arrowstyle=[scale=0.9]
	  \node [place] (s3) at (0.0, 0.0) {};
	  \node [place] (s2) at (0.0, 2.8) {};
	  \node [transition] (t1) at (-0.9,1.4) {} ;
	  \node [transition] (t3) at (0.0,3.9) {} ;
	  \draw[directed] (s3) to (t1);
	  \draw[directed] (t1) to (s2);
	  \draw[directed] (s2) to (t3);
	\end{scope}

	\begin{scope}[shift={(0.85,0)}, scale=0.45,
	  every place/.style= {minimum size=1.2mm,thick,draw=blue!75,fill=blue!20}, 
	  every transition/.style={minimum size=1.5mm,thick,draw=black!75,fill=black!20}
	  ]
	  \tikzstyle arrowstyle=[scale=0.9]
	  \node [place] (s3) at (0.0, 0.0) {};
	  \node [place] (s2) at (0.0, 2.8) {};
	  \node [place] (sm) at (1.8, 1.4) {};
	  \node [transition] (t1) at (-0.9,1.4) {} ;
	  \node [transition] (t2) at (0.9,0.7) {} ;
	  \node [transition] (t2') at (0.9,2.1) {} ;
	  \node [transition] (t3) at (0.0,3.9) {} ;
	  \draw[directed] (s3) to (t1);
	  \draw[directed] (s3) to (t2);
	  \draw[directed] (t1) to (s2);
	  \draw[directed] (t2) to (sm);
	  \draw[directed] (sm) to (t2');
	  \draw[directed] (t2') to (s2);
	  \draw[directed] (s2) to (t3);
	\end{scope}

	\node at (-0.15,0.6) {$\subset$};
	\end{tikzpicture}
  \end{center}
\end{blanko}

\begin{blanko}{Occurrence hypergraphs.}\label{occur}
  A well-founded forward hypergraph $\pn H = AINOA$ is called a {\em occurrence
  hypergraph}\footnote{Compare~Winskel~\cite[Def.3.3.1]{Winskel:eventstructures}.} if
  every principal lowerset is a graph.\footnote{This is analogous to the following
  characterisation of forests: a poset $P$ is a forest when for each element $y$ the
  principal lowerset $P \comma y$ is a linear order.} For example, the hypergraphs in
  \eqref{eq:Hcommay} are occurrence hypergraphs, whereas the hypergraph of
  Example~\ref{ex:not-colim-of-graphs} below is not.

  We denote by $\kat{OccHgr}$ the category of occurrence hypergraphs and their etale
  maps, and by $B\kat{-OccHgr}$ the category of occurrence $B$-hypergraphs and etale
  maps that induce a bijection on in-boundaries.
\end{blanko}

\begin{blankothm}{Proposition}\label{prop:colim}
  The category of hypergraphs and etale maps has colimits of connected diagrams of
  injective maps, and they are calculated pointwise (i.e.~separately on the $A,I,N,O$
  components).

If the hypergraphs involved are 

\begin{shortlist}
  \item forward,
  
  \item $B$-hypergraphs (and the maps are 
$B$-maps),

\item well-founded,

\item occurrence hypergraphs,

  \end{shortlist}
then the same holds for the colimit hypergraph.
\end{blankothm}

\begin{proof}
  {\em General hypergraphs ---}
  We write down the colimit computed pointwise and check that the result is a 
  hypergraph again, and that the maps to it are etale and injective again.
  Connected colimits can be built from pushouts and filtered colimits, so we can 
  deal with these two cases separately. 
  
  For pushouts, 
  if $\pn G \subset \pn H'$ and $\pn G \subset \pn H''$
  are injective etale maps, then the pointwise pushout
    
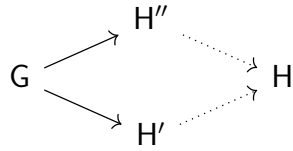
\[\begin{tikzcd}[row sep={22pt,between origins}]
  & \pn H'' \ar[rd, dotted] & \\
  \pn G \ar[ru] \ar[rd] &&  \pn H  \\
  & \pn H' \ar[ru, dotted] &
  \end{tikzcd}
  \]
  is a hypergraph again, and the dotted arrows are injective etale again.
  This is only a slight generalisation of Proposition~3.19 of 
  \cite{Kock:1407.3744}, and the proof is essentially the same.
  
  For filtered colimits, if $(\pn H_\alpha)_\alpha$ is a filtered diagram of injective
  etale maps, we need to show that
  the pointwise colimit $\pn H:=\colim_\alpha \pn H_\alpha$ is again a hypergraph, and
  the induced maps $\pn H_\alpha \to  \pn H$ are injective and etale
  again. To check that the result is a
  hypergraph again, note first that the hypergraph condition (that the two spans are
  relations) can be expressed with finite limits (namely that the pullback of $I \to
  A\times N$ with itself gives $I$ again). Now the statement follows from the fact that
  finite limits commute with filtered colimits in the category of sets. That the
  induced maps $\pn H_\alpha \to \pn H$ are injective and etale again
  can be checked separated separately on the $I \leftarrow N$ and $N \to O$
  components. Here it is precisely the fact that filtered colimits of monomorphisms
  are van Kampen in the category of sets~\cite{Heindel-Sobocinski:1101.4594}.
  Finally we check the finiteness condition:
  every node $y$ in the colimit $\pn H$ is a node in some $\pn H_\alpha$, and since the
  inclusion is etale, it follows that the pre- and post-set of $y$ can be computed
  in $\pn H_\alpha$. In particular they are finite, so that altogether the maps $I 
  \to N \leftarrow O$ are finite maps for $\pn H$, as required.

  {\em Forward hypergraphs ---}
  The condition of being forward (injectivity of the map $O \to A$) is a finite-limit
  condition, so the same reasoning as above shows that if all the hypergraphs $\pn 
  H_\alpha$ are forward, then so is the colimit $\pn H$.
  
  {\em $B$-hypergraphs ---}
  Suppose that all the hypergraphs $\pn H_\alpha$ involved are $B$-hypergraphs and 
  that all the 
  maps are $B$-maps. This means that all the right-hand squares in the AINOA 
  diagrams form pushout squares with $\emptyset\to B$, as in \eqref{eq:B-for-hgraphs}.
  But this property is 
  preserved in the colimit, since colimits commute with pushouts.

  {\em Well-founded hypergraphs ---}
  Assume that all the hypergraphs $\pn H_\alpha$ are well-founded;
   let $f_\alpha: \pn H_\alpha \to \N$ 
be the earliest-start-time functions, as in \ref{EST}. By Corollary~\ref{cor:est-agree} all the maps in 
the diagram are compatible with these functions, so we can define the
earliest-start-time function on the colimit $\pn H:=\colim_\alpha \pn H_\alpha$ by
$$
f(y) := f_\alpha(y) \quad \text{ if } y \in \pn H_\alpha .
$$
So $\pn H$ is well-founded again.

{\em Occurrence hypergraphs ---}
  Assume that all the hypergraphs $\pn H_\alpha$ are occurrence hypergraphs.
  Every node $y$ in the colimit $\pn H := \colim_\alpha \pn H_\alpha$
  belongs to some $\pn H_\alpha$, and by Lemma~\ref{lem:lowersets-preserved} the
  lowerset $\pn H \comma y$ is identified with the lowerset $\pn H_\alpha\comma y$,
  which is a graph by assumption. So $\pn H$ too is an occurrence hypergraph.
\end{proof}

\begin{blankothm}{Proposition}\label{H=colim}
  Every occurrence $B$-hypergraph $\pn H= AINOA$ is canonically the colimit of its
  $B$-processes:
  $$
  \pn H = \colim_{\pnsmall B \subset \pnsmall G \subset \pnsmall H} \pn G .
  $$
\end{blankothm}

\begin{proof}
  It is clear that $\pn H$ is the colimit of its principal $B$-lowersets, $\pn H = 
  \colim_{y\in N} \pn H\comma y$, and these lowerset are 
  all graphs by the occurrence assumption.
  There are many more $B$-graphs than principal $B$-lowersets, but each $B$-graph in 
  $\pn H$ is 
  itself a colimit of its principal $B$-lowersets, which are also principal $B$-lowersets in 
  $\pn H$ by Lemma~\ref{lem:lowersets-preserved}, so the diagram of all principal 
  $B$-lowersets in $\pn H$ is cofinal in the diagram of all $B$-graphs in $\pn H$.
\end{proof}

\begin{blanko}{Example.}\label{ex:not-colim-of-graphs}
  The result is not true without the occurrence condition, as exemplified by the 
  forward hypergraph $\pn H$
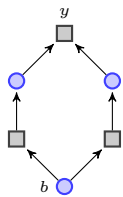
  \begin{center}
	\begin{tikzpicture}[
	  node distance=0.9cm,on grid,>=stealth',bend angle=22.5,auto,
	  every place/.style= {minimum size=5mm,thick,draw=blue!75,fill=blue!20}, 
	  every transition/.style={minimum 
	  size=4mm,thick,draw=black!75,fill=black!20},
	  every token/.style={minimum size=4pt, token distance=10pt}
	  ]
	  
	  \small
	  
	  \begin{scope}[scale=0.6,
		every place/.style= {minimum size=1.3mm,thick,draw=blue!75,fill=blue!20}, 
		every transition/.style={minimum size=1.7mm,thick,draw=black!75,fill=black!20}
		]
	  \begin{scope}[shift={(1.8,-1.9)}]
		\tikzstyle arrowstyle=[scale=0.9]
		\tiny
		\node [place, label=left:$b$] (s3) at (0.0, 0.2) {};
		\node [transition] (t1) at (-0.9,1.1) {} ;
		\node [transition] (t2) at (0.9,1.1) {} ;
		\node [place] (s1) at (-0.9,2.2) {}; 
		\node [place] (s2) at (0.9,2.2) {};
		\node [transition, label=above:$y$] (t3) at (0.0,3.1) {} ;
		\draw[directed] (s3) to (t1);
		\draw[directed] (s3) to  (t2);		
		\draw[directed] (t1) to  (s1);
		\draw[directed] (t2) to  (s2);		
		\draw[directed] (s1) to  (t3);
		\draw[directed] (s2) to  (t3);
	  \end{scope}
	  \end{scope}
  \end{tikzpicture}
  \end{center}
  where $\pn H\comma y$ is not a graph, and $y$ is not contained in any $B$-subgraph.
\end{blanko}

\section{Unfolding}
\label{sec:unfold}

An important discovery by Winskel is that the totality of all processes of a (marked)
Petri can be assembled into denotational structures; first an event structure, then a
Scott domain. The key point is the existence of a universal unfolding of a (marked)
Petri net. This was established for {\em safe} Petri nets by
Winskel~\cite{Winskel:thesis} and Nielsen, Plotkin, and
Winskel~\cite{Nielsen-Plotkin-Winskel:1981} (see
also~\cite{Winskel:eventstructures}), but did not work for non-safe nets, due to
problems with symmetries. Various workarounds were provided later
\cite{DBLP:journals/mscs/MeseguerMS97}, \cite{DBLP:journals/tcs/MeseguerMS96},
\cite{Baldan-Bruni-Montanari-DBLP:conf/wadt/BaldanBM02}, \cite{Hayman-Winskel:2008};
see \ref{unfolding-workarounds} below for further discussion.

In this section we show that in the whole-grain setting, the 
symmetry problems simply disappear: Winskel's ideas, notions and 
proof arguments in the safe case now work in full 
generality.

After the main Theorem~\ref{thm:unfolding}, we give some examples to illustrate the unfolding of some 
non-safe Petri nets to see how the symmetries are taken care of automatically. 
Finally we briefly comment on the subsequent constructions of event 
structures and domains.

\begin{blanko}{Marked Petri nets.}
  Fix a finite set $B$; denote by $\pn B$ the graph
  $B\emptyset\emptyset\emptyset B$. An {\em $B$-marked Petri net} is a
  Petri net $\pn P$ with a marking $\pn B \to \pn P$. A morphism of
  $B$-marked Petri nets is a commutative triangle of etale maps
    $$\begin{tikzcd}[column sep={2em,between origins}]
	& \pn B \ar[ld] \ar[rd] & 
	\\
	\pn P'  \ar[rr] && \pn P .
  \end{tikzcd}
  $$
  The $B$-marked Petri nets thus form the coslice category
  $B\kat{-Petri} := \pn B \comma \kat{Petri}$.
\end{blanko}

\begin{blanko}{$B$-$\pn P$-processes (= $B$-$\pn P$-graphs = $\pn P$-processes from $B$).}
  For a marked Petri net $(\pn P,B)$ we are concerned with processes that start at 
  the initial marking $B$, which is 
  to say that we only consider processes $\pn G \to \pn
  P$ for which the in-boundary of $\pn G$ agrees with $B$
  (as for hypergraphs, this means that square \eqref{eq:B-for-hgraphs} is a pushout).
  A $B$-$\pn P$-process (= 
  $\pn P$-process from $B$)
  is thus a commutative triangle of etale maps
    $$\begin{tikzcd}[column sep={2em,between origins}]
	& \pn B \ar[ld] \ar[rd] & 
	\\
	\pn G  \ar[rr] && \pn P 
  \end{tikzcd}
  $$
where the map $\pn B\to \pn G$ is a bijection onto the in-boundary of $\pn G$.
A morphism of $B$-$\pn P$-graphs, or $B$-$\pn P$-map, is 
a morphism of $\pn
P$-graphs inducing a bijection of in-boundaries.
\end{blanko}

\begin{blanko}{Grounded Petri nets.}
  A Petri net $\pn P = SITOS$ is called {\em grounded}\footnote{In the literature (see for
  example \cite{Glabbeek:2005}),
  grounded Petri nets are sometimes called `standard Petri nets'.} if the map $I\to T$ is surjective. In other words,
  the pre-sets of transitions are not allowed to be empty. The condition is
  natural from the viewpoint of unfolding, because we are interested in
  the behaviour springing from the initial marking, and without the grounding assumption a transition could fire
  spontaneously and produce an arbitrary number of tokens that there would be no control over.
  
  From a technical viewpoint, an important consequence is the following result.
\end{blanko}

\begin{blankothm}{Lemma}\label{lem:inj+noAut}
  Let $\pn P$ be a grounded Petri net $\pn P$ with initial marking $B$.
  \begin{shortlist}
    \item Every morphism of $B$-$\pn P$-processes is 
	injective.\footnote{This is analogous to the fact that a map of trees is 
	injective~\cite{Kock:0807}, and that a grounded map of forests is injective.}
    \item A $B$-$\pn P$-process has no non-trivial automorphisms.
  \end{shortlist}
\end{blankothm}

\begin{proof}
  By the etale condition, any graph $\pn G$ underlying a $\pn P$-process is
  again grounded. Let $f:\pn G \to \pn H$ be a morphism of $B$-$\pn
  P$-processes. If two nodes of $\pn G$ have the same image in $\pn H$, say
  $f(y)=f(y')$, then also some of their incoming edges have the same image
  in $\pn H$, say $f(a)=f(a')$. If $f(a)$ belongs to the in-boundary $B$ of
  $\pn H$, then so does $a$, because $f$ is a bijection on $B$, and this
  forces $a=a'$, and thereby also $y=y'$. If $f(a)$ does not belong to $B$,
  then neither do $a$ or $a'$. They are thus outgoing edges of some $x$ and
  $x'$, and since $f(a)=f(a')$ we also have $f(x)=f(x')$. Now repeat the
  argument. In a finite number of steps we arrive at $B$, where $f$ is a
  bijection by assumption. This proves (1). Statement (2) follows from a
  similar argument: we already know from Lemma~\ref{lem:noAuts} that a $\pn
  P$-process has no infinitesimal automorphisms, other than the identity.
  If an automorphism had $f(y) \neq y$, the same would be true for some
  incoming edge $a$ (which exists by the grounding condition), and so on,
  until we arrive at $B$ which is fixed by assumption, arguing as before.
\end{proof}

\begin{blanko}{Unfolding of Petri nets.}
  Let $\pn P$ be a grounded Petri net with initial state $B$. We shall continue to
  require $\pn P$ to be finite, but to accommodate the notion of unfolding, it is
  practical temporarily to enlarge the category $\kat{Petri}$ to allow infinite Petri
  nets, although we still insist on demanding the two maps $I \to T \leftarrow O$ to
  have finite fibres. An {\em unfolding} of $\pn P$ (from $B$) is a commutative triangle of
  etale maps
    $$\begin{tikzcd}[column sep={2em,between origins}]
	& \pn B \ar[ld] \ar[rd] & 
	\\
	\pn H  \ar[rr] && \pn P ,
  \end{tikzcd}
  $$
  where $\pn B \to \pn H$ is an occurrence $B$-hypergraph.
  Note that since $\pn P$ is assumed grounded, so is any unfolding (as an offending node 
  in $\pn H$ would have nowhere in $\pn P$ to land).
\end{blanko}

\begin{blankothm}{Theorem}\label{thm:unfolding}
  Fix a grounded Petri net $\pn P$, with an initial marking $\pn 
  B\to \pn P$ (given by a finite set $B$). 
  
  \begin{shortlist}
    \item There is a {\em universal unfolding} of $\pn P$ from $B$
	$$
	\pn B \to \pn U_B \pn P \stackrel{\varepsilon}\to \pn P ,
	$$
	in the sense that for every unfolding $\pn H \to \pn P$ from $B$
	there is a unique morphism of $B$-hypergraphs over $\pn P$:
    $$\begin{tikzcd}[column sep={2em,between origins}]
	& \pn B \ar[ld] \ar[rd]& \\
	\pn H \ar[rd] \ar[rr, dotted, "{\exists!}"] && \pn U_B \pn P \ar[ld, 
	"\varepsilon"] \\
	 & \pn P   . &
	\end{tikzcd}
	$$

	\item 
	The universal unfolding is constructed as the
	colimit in $B\kat{-Hgr}$ of all $\pn P$-processes
	starting at $B$:
  $$
  \pn U_B \pn P = \colim_{\pnsmall B \to \pnsmall G \to \pnsmall P}  
  \pn G  .
  $$
  \end{shortlist}
\end{blankothm}

The diagram takes place in the category $\kat{Petri}$ where we
temporarily allow infinite Petri nets. The first statement can also be
formulated like this: {\em for any grounded marked Petri net $(\pn P, B)$ and any 
occurrence $B$-hypergraph
$\pn H$, there is a canonical bijection}
$$
\Hom_{B\kat{-Petri}} (\pn H, \pn P) \simeq \Hom_{B\kat{-OccHgr}}(\pn 
H, \pn U_B \pn P) .
$$
In other words, {\em $\pn U_B \pn P$ represents
the functor}
\begin{eqnarray*}
  B\kat{-OccHgr}\op & \longrightarrow & \Set  \\
  \pn H & \longmapsto & \Hom_{B\kat{-Petri}} (\pn H, \pn P) .
\end{eqnarray*}
Note also that by restricting to graphs, we get
$$
\kat{Proc}_B(\pn P) \simeq \kat{Proc}_B(\pn U_B \pn P) .
$$

One could also adjust the categories involved so as to formulate the
statement as an adjunction: for this define $B\kat{-Petri}$ to be the
category of possibly infinite $B$-marked Petri nets, but only grounded ones,
and let $B\kat{-OccHgr}$ be the category of occurrence $B$-hypergraphs also
required to be grounded.
Then there is a forgetful functor $B\kat{-OccHgr} \to B\kat{-Petri}$,
and the statement is that this functor has a right adjoint $\pn U_B$.

\begin{proof}
  The diagram we want to take colimit of consists of injective maps, since
  morphisms of $B$-$\pn P$-processes are always injective by
  Lemma~\ref{lem:inj+noAut} (1). It is therefore of the kind that admits a colimit in
  the category of $B$-hypergraphs as in Proposition~\ref{prop:colim}, and this
  colimit is an occurrence hypergraph by \ref{prop:colim} (4).  By the
  universal property of the colimit, it comes with an etale map to $\pn P$.
  
  Given an unfolding $\pn B \to \pn H \to \pn P$, write $\pn H$ as the colimit of its
  processes: $\pn H = \colim_{\pnsmall B \subset \pnsmall G \subset 
  \pnsmall H} \pn G$, as in
  Proposition~\ref{H=colim}. By postcomposition, each $\pn G$ is also a process of
  $\pn P$ (still from $B$), and therefore it maps into the union 
  $\colim_{\pnsmall B
  \subset \pnsmall G \subset \pnsmall P} \pn G = \pn U_B \pn P$, and this map is unique since
  $\pn P$-processes under $B$ have no automorphisms (by Lemma~\ref{lem:inj+noAut}
  (2)). This collection of maps $\pn G \to \pn U_B \pn P$ defines a unique map out of
  the colimit $\pn H = \colim_{\pnsmall B \subset \pnsmall G \subset 
  \pnsmall H} \pn G$.
\end{proof}

\begin{blanko}{Remarks.}\label{unfolding-workarounds}
  In the setting of traditional Petri nets, Nielsen, Plotkin, and
  Winskel~\cite{Nielsen-Plotkin-Winskel:1981} proved Theorem~\ref{thm:unfolding} for
  safe Petri nets (see \cite[Theorems~3.3.9 and 3.3.13]{Winskel:eventstructures}). A
  marked Petri net is {\em safe} if all arcs have multiplicity $1$, and the initial
  marking as well as all reachable markings are multiplicity free. For general
  Petri nets in the traditional sense, the theorem breaks down, as the universal
  property is violated by symmetries, cf.~Example~\ref{ex:HaymanWinskel} below, which
  is also given by Hayman and Winskel~\cite{Hayman-Winskel:2008}. From the viewpoint
  of the present paper, one could say that the reason safe Petri nets do satisfy the
  theorem in the traditional case is that safe nets are already whole-grain.

  Two lines of development led to further insight into unfolding of
  non-safe Petri nets. One way to tackle the problem was devised by
  Meseguer, Montanari, and Sassone~\cite{DBLP:journals/mscs/MeseguerMS97},
  \cite{DBLP:journals/tcs/MeseguerMS96}, exploiting certain decorated
  unfoldings, which prevented the symmetries. Baldan, Bruni and
  Montanari~\cite{Baldan-Bruni-Montanari-DBLP:conf/wadt/BaldanBM02}
  established the result for pre-nets, where the symmetries do not come up,
  due to the rather restrictive notion of morphism (see also van
  Glabbeek~\cite{Glabbeek:2005}). Finally, Hayman and
  Winskel~\cite{Hayman-Winskel:2008}, rather than trying to avoid the
  symmetries, managed to embrace them, proving an unfolding theorem where
  the universal property is replaced by a universal property up to
  symmetry, a concept they formalised in terms of certain
  quasi-isomorphisms defined with spans of open
  maps~\cite{Hayman-Winskel:2009}.
    
  It must be stressed that since the theorem is a question of representability, it is
  a much easier question in the `representable' SITOS setting than in the traditional
  setting. The whole-grain formalism bypasses the difficulties, not by any miracle or
  deep insight, but simply because the tokens are now elements in specific sets, and
  this gives complete control over their symmetries.
\end{blanko}

We go through a couple of examples, in order to illustrate the workings of unfolding
in the whole-grain setting. The first example is relatively straightforward (but
it is not a safe net); the next examples are more subtle, involving multiple
markings and parallel arcs.

\begin{blanko}{Example.}\label{ex:exP}
  Consider the following Petri net $\pn P$, with its marking $B \to \pn P$
  by a $3$-element set $B$:

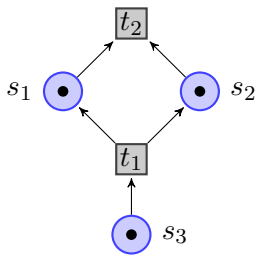
  \begin{center}
	\begin{tikzpicture}[scale=0.9,
	  node distance=0.9cm,on grid,>=stealth',bend angle=22.5,auto,
	  every place/.style= {minimum size=4mm,thick,draw=blue!75,fill=blue!20}, 
	  every transition/.style={minimum 
	  size=3.5mm,thick,draw=black!75,fill=black!20},
	  every token/.style={minimum size=3.5pt, token distance=8pt}
	  ]
	  
	  \small
		
	  \coordinate (init) at (-2,2);	  
	  
	  \begin{scope}[shift={(init)}]
		\node [transition] (t1) at (0.0,1.0) {$t_1$} ;
		\node [place,tokens=1, label=left:$s_1$] (s1) at (-0.9,1.9) {}; 
		\node [place,tokens=1, label=right:$s_2$] (s2) at (0.9,1.9) {};
		\node [place,tokens=1, label=right:$s_3$] (s3) at (0.0, 0.0) {};
		\node [transition] (t2) at (0.0,2.8) {$t_2$} ;
		\path (s3) edge [post] (t1);
		\path (t1) edge [post] (s1);
		\path (t1) edge [post] (s2);
		\path (s1) edge [post] (t2);
		\path (s2) edge [post] (t2);
	  \end{scope}
 \end{tikzpicture}
 \end{center}
	  
To compute its universal unfolding, we first list all its processes from $B$ (up to
isomorphism), arranged into a poset with the order given by $B$-$\pn P$-maps.\footnote{Note that the diagram of {\em all}
$B$-$\pn P$-graphs is hugely redundant, and includes all isomorphisms of $B$-$\pn
P$-graphs. By picking one representative from each isomorphism class we get a
skeleton to work with instead. Since $B$-$\pn P$-graphs have no nontrivial
automorphisms and all maps are injective (cf.~\ref{lem:inj+noAut}), this skeleton is
a poset, and this is the poset to work with in practice.} The small numbers in the
graph pictures indicate the $\pn P$-structure: for example, the number $1$ on an edge
indicates that it is mapped to place $s_1$, and so on. How the nodes map to
transitions is unambiguous from the drawings, and since $\pn P$ has no parallel arcs,
the effect of the etale maps on the sets of arcs is implied too. There are 9 possible
processes from $B$:

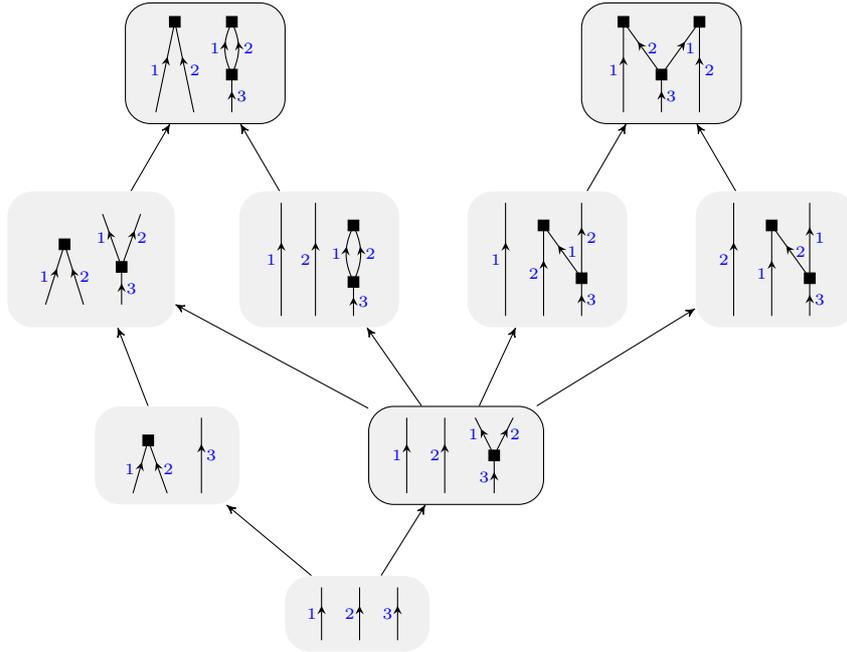
\begin{center}
	\begin{tikzpicture}[scale=0.7,
	  node distance=0.8cm,on grid,>=stealth',bend angle=22.5,auto,
	  every place/.style= {minimum size=5mm,thick,draw=blue!75,fill=blue!20}, 
	  every transition/.style={minimum 
	  size=4mm,thick,draw=black!75,fill=black!20},
	  ]
	  
	  \small
		
	  \node[rectangle, rounded corners=2ex, fill=black!06, 
		minimum width=15mm, minimum height=7mm] 
		(deg0graph) at (5.5,-1.1) {};

	  \node[rectangle, rounded corners=2ex, draw=black, fill=black!06, 
	    minimum width=16mm, minimum height=10mm]
		(deg1graph1) at (6.8,1.0) {};

	  \node[rectangle, rounded corners=2ex, fill=black!06, 
	    minimum width=13mm, minimum height=10mm] 
		(deg1graph2) at (3.0,1.0) {};

	  \node[rectangle, rounded corners=2ex, fill=black!06,
	    minimum width=16mm, minimum height=13mm]
		(deg2graph1) at (2.0,3.6) {};

	  \node[rectangle, rounded corners=2ex, fill=black!06,
	    minimum width=16mm, minimum height=13mm]
		(deg2graph2) at (5.0,3.6) {};
		
	  \node[rectangle, rounded corners=2ex, fill=black!06, 
	    minimum width=16mm, minimum height=13mm]
		(deg2graph3) at (8.0,3.6) {};
		
	  \node[rectangle, rounded corners=2ex, fill=black!06,
	    minimum width=16mm, minimum height=13mm]
		(deg2graph4) at (11.0,3.6) {};

	  \node[rectangle, rounded corners=2ex, draw=black, fill=black!06,
	    minimum width=17mm, minimum height=12mm]
		(deg3graph1) at (3.5,6.2) {};
		
	  \node[rectangle, rounded corners=2ex, draw=black, fill=black!06,
	    minimum width=17mm, minimum height=12mm]
		(deg3graph2) at (9.5,6.2) {};

	  \draw[->] (deg0graph) -- (deg1graph1);
	  \draw[->] (deg0graph) -- (deg1graph2);
	  \draw[->] (deg1graph1) -- (deg2graph1);
	  \draw[->] (deg1graph2) -- (deg2graph1);
	  \draw[->] (deg1graph1) -- (deg2graph2);
	  \draw[->] (deg1graph1) -- (deg2graph3);
	  \draw[->] (deg1graph1) -- (deg2graph4);
	  \draw[->] (deg2graph1) -- (deg3graph1);
	  \draw[->] (deg2graph2) -- (deg3graph1);
	  \draw[->] (deg2graph3) -- (deg3graph2);
	  \draw[->] (deg2graph4) -- (deg3graph2);
	  
      \begin{scope}[shift={(deg0graph)}]
      \begin{scope}[shift={(0.03,-0.35)}]
		\tikzstyle arrowstyle=[scale=0.9]
		\node [port] (i1) at (-0.5,0.0) {};
		\node [port] (i2) at (0.0,0.0) {};
		\node [port] (i3) at (0.5,0.0) {};
		\node [port] (o1) at (-0.5,0.7) {};
		\node [port] (o2) at (0.0,0.7) {};
		\node [port] (o3) at (0.5,0.7) {};
		\draw[directed] (i1) -- node[left=-1pt] {\color{blue}\tiny $1$} (o1);
		\draw[directed] (i2) -- node[left=-1pt] {\color{blue}\tiny $2$} (o2);
		\draw[directed] (i3) -- node[left=-1pt] {\color{blue}\tiny $3$} (o3);
	  \end{scope}
	  \end{scope}

	  \begin{scope}[shift={(deg1graph1)}]
      \begin{scope}[shift={(-0.25,-0.5)}]
		\tikzstyle arrowstyle=[scale=0.9]
		\node [port] (i1) at (-0.4,0.0) {};
		\node [port] (i2) at (0.1,0.0) {};
		\node [port] (i3) at (0.75,0.0) {};
		\node [port] (o1) at (-0.4,1) {};
		\node [port] (o2) at (0.1,1) {};
		\node [port] (o3) at (0.5,1) {};
		\node [port] (o4) at (1,1) {};
		\node [grnode] (x1) at (0.75,0.5) {};
		\draw[directed] (i1) -- node[left=-1pt] {\color{blue}\tiny $1$} (o1);
		\draw[directed] (i2) -- node[left=-1pt] {\color{blue}\tiny $2$} (o2);
		\draw[directed] (i3) -- node[left=-1pt] {\color{blue}\tiny $3$} (x1);
		\draw[directed] (x1) -- node[left=-1pt] {\color{blue}\tiny $1$} (o3);
		\draw[directed] (x1) -- node[right=-1pt] {\color{blue}\tiny $2$} (o4);
	  \end{scope}
	  \end{scope}

	  \begin{scope}[shift={(deg1graph2)}]
      \begin{scope}[shift={(-0.05,-0.5)}]
		\tikzstyle arrowstyle=[scale=0.9]
		\node [port] (i1) at (-0.4,0.0) {};
		\node [port] (i2) at (0.05,0.0) {};
		\node [port] (i3) at (0.5,0.0) {};
		\node [port] (o1) at (-0.4,1) {};
		\node [port] (o2) at (0.05,1) {};
		\node [port] (o3) at (0.5,1) {};
		\node [port] (o4) at (1,1) {};
		\node [grnode] (x2) at (-0.2,0.7) {};
		\draw[directed] (i1) -- node[left=-1pt] {\color{blue}\tiny $1$} (x2);
		\draw[directed] (i2) -- node[right=-1pt] {\color{blue}\tiny $2$} (x2);
		\draw[directed] (i3) -- node[right=-1pt] {\color{blue}\tiny $3$} (o3);
	  \end{scope}
	  \end{scope}

	  \begin{scope}[shift={(deg2graph1)}]
      \begin{scope}[shift={(-0.1,-0.6)}]
		\tikzstyle arrowstyle=[scale=0.9]
		\node [port] (i1) at (-0.5,0.0) {};
		\node [port] (i2) at (0.0,0.0) {};
		\node [port] (i3) at (0.5,0.0) {};
		\node [port] (o1) at (-0.5,1.2) {};
		\node [port] (o2) at (0.0,1.2) {};
		\node [port] (o3) at (0.25,1.2) {};
		\node [port] (o4) at (0.75,1.2) {};
		\node [grnode] (x2) at (-0.25,0.8) {};
		\node [grnode] (x1) at (0.5,0.5) {};
		\draw[directed] (i1) -- node[left=-1pt] {\color{blue}\tiny $1$} (x2);
		\draw[directed] (i2) -- node[right=-1pt] {\color{blue}\tiny $2$} (x2);
		\draw[directed] (i3) -- node[right=-1pt] {\color{blue}\tiny $3$} (x1);
		\draw[directed] (x1) -- node[left=-1pt] {\color{blue}\tiny $1$} (o3);
		\draw[directed] (x1) -- node[right=-1pt] {\color{blue}\tiny $2$} (o4);
	  \end{scope}
	  \end{scope}
	  
	  \begin{scope}[shift={(deg2graph2)}]
      \begin{scope}[shift={(-0.1,-0.75)}]
		\tikzstyle arrowstyle=[scale=0.9]
		\node [port] (i1) at (-0.4,0.0) {};
		\node [port] (i2) at (0.05,0.0) {};
		\node [port] (i3) at (0.55,0.0) {};
		\node [port] (o1) at (-0.4,1.5) {};
		\node [port] (o2) at (0.05,1.5) {};
		\node [port] (o3) at (0.25,1.5) {};
		\node [port] (o4) at (0.75,1.5) {};
		\node [grnode] (x2) at (0.55,1.2) {};
		\node [grnode] (x1) at (0.55,0.45) {};
		\draw[directed] (i1) -- node[left=-1pt] {\color{blue}\tiny $1$} (o1);
		\draw[directed] (i2) -- node[left=-1pt] {\color{blue}\tiny $2$} (o2);
		\draw[directed, bend left] (x1) to node[left=-1pt] {\color{blue}\tiny $1$} (x2);
		\draw[directed, bend right] (x1) to node[right=-1pt] {\color{blue}\tiny $2$} (x2);
		\draw[directed] (i3) -- node[right=-1pt] {\color{blue}\tiny $3$} (x1);
	  \end{scope}
	  \end{scope}
 
	  \begin{scope}[shift={(deg2graph3)}]
      \begin{scope}[shift={(-0.05,-0.75)}]
		\tikzstyle arrowstyle=[scale=0.9]
		\node [port] (i1) at (-0.5,0.0) {};
		\node [port] (i2) at (0.0,0.0) {};
		\node [port] (i3) at (0.5,0.0) {};
		\node [port] (o1) at (-0.5,1.5) {};
		\node [port] (o2) at (0.0,1.5) {};
		\node [port] (o3) at (0.5,1.5) {};
		\node [port] (o4) at (0.75,1.5) {};
		\node [grnode] (x2) at (0.0,1.2) {};
		\node [grnode] (x1) at (0.5,0.5) {};
		\draw[directed] (i1) -- node[left=-1pt] {\color{blue}\tiny $1$} (o1);
		\draw[directed] (i2) -- node[left=-1pt] {\color{blue}\tiny $2$} (x2);
		\draw[directed] (x1) to node[right=-1pt] {\color{blue}\tiny $1$} (x2);
		\draw[directed] (x1) to node[right=-1pt] {\color{blue}\tiny $2$} (o3);
		\draw[directed] (i3) -- node[right=-1pt] {\color{blue}\tiny $3$} (x1);
	  \end{scope}
	  \end{scope}
 
	  \begin{scope}[shift={(deg2graph4)}]
      \begin{scope}[shift={(-0.05,-0.75)}]
		\tikzstyle arrowstyle=[scale=0.9]
		\node [port] (i1) at (-0.5,0.0) {};
		\node [port] (i2) at (0.0,0.0) {};
		\node [port] (i3) at (0.5,0.0) {};
		\node [port] (o1) at (-0.5,1.5) {};
		\node [port] (o2) at (0.0,1.5) {};
		\node [port] (o3) at (0.5,1.5) {};
		\node [port] (o4) at (0.75,1.5) {};
		\node [grnode] (x2) at (0.0,1.2) {};
		\node [grnode] (x1) at (0.5,0.5) {};
		\draw[directed] (i1) -- node[left=-1pt] {\color{blue}\tiny $2$} (o1);
		\draw[directed] (i2) -- node[left=-1pt] {\color{blue}\tiny $1$} (x2);
		\draw[directed] (x1) to node[right=-1pt] {\color{blue}\tiny $2$} (x2);
		\draw[directed] (x1) to node[right=-1pt] {\color{blue}\tiny $1$} (o3);
		\draw[directed] (i3) -- node[right=-1pt] {\color{blue}\tiny $3$} (x1);
	  \end{scope}
	  \end{scope}
	    
	  \begin{scope}[shift={(deg3graph1)}]
      \begin{scope}[shift={(-0.15,-0.65)}]
		\tikzstyle arrowstyle=[scale=0.9]
		\node [port] (i1) at (-0.5,0.0) {};
		\node [port] (i2) at (0.0,0.0) {};
		\node [port] (i3) at (0.5,0.0) {};
		\node [port] (o1) at (-0.5,1.5) {};
		\node [port] (o2) at (0.0,1.5) {};
		\node [port] (o3) at (0.25,1.5) {};
		\node [port] (o4) at (0.75,1.5) {};
		\node [grnode] (x2) at (0.5,1.2) {};
		\node [grnode] (x2') at (-0.25,1.2) {};
		\node [grnode] (x1) at (0.5,0.5) {};
		\draw[directed] (i1) -- node[left=-1pt] {\color{blue}\tiny $1$} (x2');
		\draw[directed] (i2) -- node[right=-1pt] {\color{blue}\tiny $2$} (x2');
		\draw[directed, bend left] (x1) to node[left=-1pt] {\color{blue}\tiny $1$} (x2);
		\draw[directed, bend right] (x1) to node[right=-1pt] {\color{blue}\tiny $2$} (x2);
		\draw[directed] (i3) -- node[right=-1pt] {\color{blue}\tiny $3$} (x1);
	  \end{scope}		  
	  \end{scope}		  
		  
	  \begin{scope}[shift={(deg3graph2)}]
      \begin{scope}[shift={(0.0,-0.65)}]
		\tikzstyle arrowstyle=[scale=0.9]
		\node [port] (i1) at (-0.5,0.0) {};
		\node [port] (i2) at (0.0,0.0) {};
		\node [port] (i3) at (0.5,0.0) {};
		\node [port] (o1) at (-0.5,1.5) {};
		\node [port] (o2) at (0.0,1.5) {};
		\node [port] (o3) at (0.25,1.5) {};
		\node [port] (o4) at (0.75,1.5) {};
		\node [grnode] (x2) at (-0.5,1.2) {};
		\node [grnode] (x2') at (0.5,1.2) {};
		\node [grnode] (x1) at (0.0,0.5) {};
		\draw[directed] (i1) -- node[left=-1pt] {\color{blue}\tiny $1$} (x2);
		\draw[directed] (i3) -- node[right=-1pt] {\color{blue}\tiny $2$} (x2');
		\draw[directed] (x1) to node[right=-1pt] {\color{blue}\tiny $2$} (x2);
		\draw[directed] (x1) to node[right=-1pt] {\color{blue}\tiny $1$} (x2');
		\draw[directed] (i2) -- node[right=-1pt] {\color{blue}\tiny $3$} (x1);
	  \end{scope}
	  \end{scope}		  
  
\end{tikzpicture}
\end{center}
Note that the two zigzag-shaped graphs are isomorphic as abstract graphs, 
but not isomorphic as $B$-graphs.

To compute the colimit of this diagram in the category $\kat{Hgr}$, it is 
enough to compute the colimit of the subdiagram consisting of the three graphs 
pictured with black 
frame (technically: that
subdiagram is cofinal), so it amounts to computing a single pushout:

  
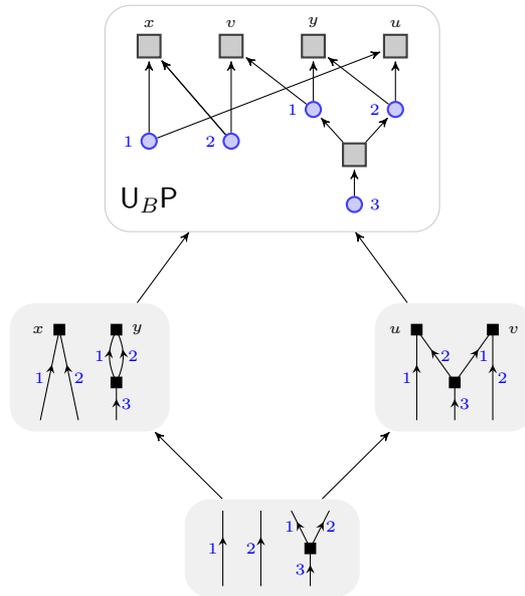
\begin{center}
	\begin{tikzpicture}[scale=0.7,
	  node distance=0.8cm,on grid,>=stealth',bend angle=22.5,auto,
	  every place/.style= {minimum size=5mm,thick,draw=blue!75,fill=blue!20}, 
	  every transition/.style={minimum 
	  size=4mm,thick,draw=black!75,fill=black!20},
	  every token/.style={minimum size=4pt, token distance=10pt}
	  ]
	  
	  \small
			  
	  \node[rectangle, rounded corners=2ex, draw=black!20, 
	    minimum width=33mm, minimum height=23mm]
		(colim) at (0.0,5.7) {};

	  \node[rectangle, rounded corners=2ex, fill=black!06, 
	    minimum width=17mm, minimum height=10mm]
		(deg1graph1) at (0.0,0.0) {};

	  \node[rectangle, rounded corners=2ex, fill=black!06,
	    minimum width=16mm, minimum height=12mm]
		(deg3graph1) at (-2.4,2.4) {};
		
	  \node[rectangle, rounded corners=2ex, fill=black!06,
	    minimum width=16mm, minimum height=12mm]
		(deg3graph2) at (2.4,2.4) {};

	  \draw[->] (deg1graph1) -- (deg3graph1);
	  \draw[->] (deg1graph1) -- (deg3graph2);
	  \draw[->] (deg3graph1) -- (colim);
	  \draw[->] (deg3graph2) -- (colim);

	  \begin{scope}[shift={(deg1graph1)}]
      \begin{scope}[shift={(-0.25,-0.5)}]
		\tikzstyle arrowstyle=[scale=0.9]
		\node [port] (i1) at (-0.4,0.0) {};
		\node [port] (i2) at (0.1,0.0) {};
		\node [port] (i3) at (0.75,0.0) {};
		\node [port] (o1) at (-0.4,1) {};
		\node [port] (o2) at (0.1,1) {};
		\node [port] (o3) at (0.5,1) {};
		\node [port] (o4) at (1,1) {};
		\node [grnode] (x1) at (0.75,0.5) {};
		\draw[directed] (i1) -- node[left=-1pt] {\color{blue} \tiny $1$} (o1);
		\draw[directed] (i2) -- node[left=-1pt] {\color{blue} \tiny $2$} (o2);
		\draw[directed] (i3) -- node[left=-1pt] {\color{blue} \tiny $3$} (x1);
		\draw[directed] (x1) -- node[left=-1pt] {\color{blue} \tiny $1$} (o3);
		\draw[directed] (x1) -- node[right=-1pt] {\color{blue} \tiny $2$} (o4);
	  \end{scope}
	  \end{scope}

	  \begin{scope}[shift={(deg3graph1)}]
      \begin{scope}[shift={(-0.15,-0.7)}]
		\tikzstyle arrowstyle=[scale=0.9]
		\node [port] (i1) at (-0.5,0.0) {};
		\node [port] (i2) at (0.0,0.0) {};
		\node [port] (i3) at (0.5,0.0) {};
		\node [port] (o1) at (-0.5,1.5) {};
		\node [port] (o2) at (0.0,1.5) {};
		\node [port] (o3) at (0.25,1.5) {};
		\node [port] (o4) at (0.75,1.5) {};
		\node [grnode, label=right:{\tiny $y$}] (x2) at (0.5,1.2) {};
		\node [grnode, label=left:{\tiny $x$}] (x2') at (-0.25,1.2) {};
		\node [grnode] (x1) at (0.5,0.5) {};
		\draw[directed] (i1) -- node[left=-1pt] {\color{blue} \tiny $1$} (x2');
		\draw[directed] (i2) -- node[right=-1pt] {\color{blue} \tiny $2$} (x2');
		\draw[directed, bend left] (x1) to node[left=-1pt] {\color{blue} \tiny $1$} (x2);
		\draw[directed, bend right] (x1) to node[right=-1pt] {\color{blue} \tiny $2$} (x2);
		\draw[directed] (i3) -- node[right=-1pt] {\color{blue} \tiny $3$} (x1);
	  \end{scope}		  
	  \end{scope}		  
		  
	  \begin{scope}[shift={(deg3graph2)}]
      \begin{scope}[shift={(-0.0,-0.7)}]
		\tikzstyle arrowstyle=[scale=0.9]
		\node [port] (i1) at (-0.5,0.0) {};
		\node [port] (i2) at (0.0,0.0) {};
		\node [port] (i3) at (0.5,0.0) {};
		\node [port] (o1) at (-0.5,1.5) {};
		\node [port] (o2) at (0.0,1.5) {};
		\node [port] (o3) at (0.25,1.5) {};
		\node [port] (o4) at (0.75,1.5) {};
		\node [grnode, label=left:{\tiny $u$}] (x2) at (-0.5,1.2) {};
		\node [grnode, label=right:{\tiny $v$}] (x2') at (0.5,1.2) {};
		\node [grnode] (x1) at (0.0,0.5) {};
		\draw[directed] (i1) -- node[left=-1pt] {\color{blue} \tiny $1$} (x2);
		\draw[directed] (i3) -- node[right=-1pt] {\color{blue} \tiny $2$} (x2');
		\draw[directed] (x1) to node[right=-1pt] {\color{blue} \tiny $2$} (x2);
		\draw[directed] (x1) to node[right=-1pt] {\color{blue} \tiny $1$} (x2');
		\draw[directed] (i2) -- node[right=-2pt] {\color{blue} \tiny $3$} (x1);
	  \end{scope}
	  \end{scope}		  
  
		\begin{scope}[shift={(colim)}, scale=0.65,
	  every place/.style= {minimum size=1.6mm,thick,draw=blue!75,fill=blue!20}, 
	  every transition/.style={minimum size=1.4mm,thick,draw=black!75,fill=black!20},
    post/.style={
        ->,
        shorten >=0.5pt, 
        >={stealth}
    },
	]
  
\begin{scope}[shift={(1.8,-1.9)}]
		\tiny
		\node [transition] (t1) at (0.0,1.1) {} ;
		\node [place, label=left:${\color{blue} 1}$] (s1) at (-4.5,1.4) {};
		\node [place, label=left:${\color{blue} 2}$] (s2) at (-2.7,1.4) {}; 
		\node [place, label=left:${\color{blue} 1}$] (s1') at (-0.9,2.1) {}; 
		\node [place, label=left:${\color{blue} 2}$] (s2') at (0.9,2.1) {};
		\node [place, label=right:${\color{blue} 3}$] (s3) at (0.0, 0.0) {};
		\node [transition, label=above:$x$] (t2i) at (-4.5,3.5) {} ;
		\node [transition, label=above:$v$] (t2ii) at (-2.7,3.5) {} ;
		\node [transition, label=above:$y$] (t2iii) at (-0.9,3.5) {} ;
		\node [transition, label=above:$u$] (t2iv) at (0.9,3.5) {} ;
		\path (s3) edge [post] (t1);
		\path (t1) edge [post] (s1');
		\path (t1) edge [post] (s2');
		
		\path (s1) edge [post] (t2i);
		\path (s2) edge [post] (t2i);
		\path (s2) edge [post] (t2ii);
		\path (s1') edge [post] (t2ii);
		\path (s1') edge [post] (t2iii);
		\path (s2') edge [post] (t2iii);
		\path (s2') edge [post] (t2iv);
		\path (s2) edge [post] (t2i);
		
		\path (s1) edge [post] (t2iv);

		\small
		
		\node at (-4.5,0.1) {$\pn U_B \pn P$};
	  \end{scope}
	  \end{scope}
 
\end{tikzpicture}
\end{center}
  In the figure, 
  the letters on the nodes are only to indicate how the graphs embed into the colimit
  hypergraph, the unfolding. As before, the numbers indicate how the four hypergraphs 
  are $\pn P$-hypergraphs.

  Intuitively, the pushout is obtained by gluing the two $3$-node graphs
  together along the overlap: the gluing locus consists of the four edges
  marked $1$ and $2$, so these become hyperedges (the blue circles) in the
  gluing, rather than plain edges. It should be stressed again that this
  computation is really formal and straightforward once the AINOA diagrams
  have been written down, and that it is nothing but computing a few
  pushouts in the category of sets. For example, all three graphs have 5
  edges, so also the colimit hypergraph has 5 hyper-edges; in each of the
  two $3$-node graphs there are 5 incoming arcs (the $I$-set in an AINOA
  diagram), and only 1 of them is in common (coming from the small graph),
  therefore the pushout $\pn U_B\pn P$ has $5+5-1=9$ incoming arcs, as
  clearly seen in the figure; and so on. It is an instructive exercise now
  to write down the processes of $\pn U_B \pn P$ from $B$. One will find
  the same poset of graphs as for $\pn P$ itself, as an illustration of
  Theorem~\ref{thm:unfolding}.
\end{blanko}

\begin{blanko}{Example.}\label{ex:HaymanWinskel}
  Consider the  Petri net $\pn P$,
  $$
  \begin{tikzcd}[column sep={18mm,between origins}]
	\{s\}  & \{u\} \ar[l] \ar[r] 
	&  \{t\} & 
	\emptyset\ar[l]\ar[r]  & \{s\} ,
  \end{tikzcd}
  $$
  with its marking $B \to \pn P$ by a $2$-element set $B=\{b_1,b_2\}$:

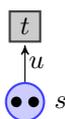
  \begin{center}
	\begin{tikzpicture}[scale=0.9,
	  node distance=1.1cm,on grid,>=stealth',bend angle=22.5,auto,
	  every place/.style= {minimum size=4.5mm,thick,draw=blue!75,fill=blue!20}, 
	  every transition/.style={minimum 
	  size=3.5mm,thick,draw=black!75,fill=black!20},
	  every token/.style={minimum size=3.5pt, token distance=8pt}
	  ]
	  
	  \small
		
	  \coordinate (init) at (-2,2);	  
	  
	  \begin{scope}[shift={(init)}]
		\footnotesize
		\node [transition] (t1) at (0.0,1.0) {$t$} ;
		\node [place,tokens=2, label=right:$s$] (s3) at (0.0, 0.0) {};
		\path (s3) edge [post] node[right=-1pt] {$u$} (t1);
	  \end{scope}
 \end{tikzpicture}
 \end{center}

  Here is the poset of processes, and the colimit, which is the universal unfolding:
  
  
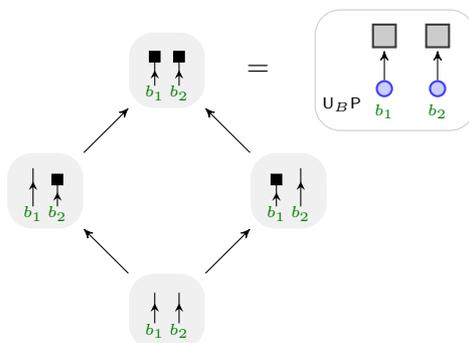
\begin{center}
	\begin{tikzpicture}[scale=0.9,
	  node distance=0.9cm,on grid,>=stealth',bend angle=22.5,auto,
	  every place/.style= {minimum size=5mm,thick,draw=blue!75,fill=blue!20}, 
	  every transition/.style={minimum size=4mm,thick,draw=black!75,fill=black!20},
	  every token/.style={minimum size=4pt, token distance=10pt}
	  ]
	  
	  \small
			  
	  \node[rectangle, rounded corners=2ex, draw=black!20, 
	    minimum width=19mm, minimum height=14mm]
		(colim) at (3.0,3.2) {};

	  \node[rectangle, rounded corners=2ex, fill=black!06, 
	    minimum width=10mm, minimum height=9mm]
		(deg2graph) at (0.0,3.2) {};

	  \node[rectangle, rounded corners=2ex, fill=black!06, 
	    minimum width=10mm, minimum height=9mm]
		(deg0graph) at (0.0,0.0) {};

	  \node[rectangle, rounded corners=2ex, fill=black!06,
	    minimum width=10mm, minimum height=9mm]
		(deg1graph1) at (-1.6,1.6) {};
		
	  \node[rectangle, rounded corners=2ex, fill=black!06,
	    minimum width=10mm, minimum height=9mm]
		(deg1graph2) at (1.6,1.6) {};

	  \draw[->] (deg0graph) -- (deg1graph1);
	  \draw[->] (deg0graph) -- (deg1graph2);
	  \draw[->] (deg1graph1) -- (deg2graph);
	  \draw[->] (deg1graph2) -- (deg2graph);
	  
	  \node at (1.2, 3.2) {$=$};

      \begin{scope}[shift={(deg0graph)}]
      \begin{scope}[shift={(0.0,-0.25)}]
		\tikzstyle arrowstyle=[scale=0.9]
		\node [port] (i1) at (-0.16,0.0) {\color{Bgreen}\tiny $b_1$};
		\node [port] (i2) at (0.16,0.0) {\color{Bgreen}\tiny $b_2$};
		\node [port] (o1) at (-0.16,0.5) {};
		\node [port] (o2) at (0.16,0.5) {};
		\draw[directed] (i1) -- node[left=-2pt] {} (o1);
		\draw[directed] (i2) -- node[left=-2pt] {} (o2);
	  \end{scope}
	  \end{scope}
	  
      \begin{scope}[shift={(deg1graph1)}]
      \begin{scope}[shift={(0.0,-0.3)}]
		\tikzstyle arrowstyle=[scale=0.9]
		\node [port] (i1) at (-0.16,0.0) {\color{Bgreen}\tiny $b_1$};
		\node [port] (i2) at (0.16,0.0) {\color{Bgreen}\tiny $b_2$};
		\node [port] (o1) at (-0.16,0.6) {};
		\node [grnode] (t2) at (0.16,0.45) {};
		\draw[directed] (i1) -- node[left=-2pt] {} (o1);
		\draw[directed] (i2) -- node[left=-2pt] {} (t2);
	  \end{scope}
	  \end{scope}
	  
      \begin{scope}[shift={(deg1graph2)}]
      \begin{scope}[shift={(0.0,-0.3)}]
		\tikzstyle arrowstyle=[scale=0.9]
		\node [port] (i1) at (-0.16,0.0) {\color{Bgreen}\tiny $b_1$};
		\node [port] (i2) at (0.16,0.0) {\color{Bgreen}\tiny $b_2$};
		\node [grnode] (t1) at (-0.16,0.45) {};
		\node [port] (o2) at (0.16,0.6) {};
		\draw[directed] (i1) -- node[left=-2pt] {} (t1);
		\draw[directed] (i2) -- node[left=-2pt] {} (o2);
	  \end{scope}
	  \end{scope}

      \begin{scope}[shift={(deg2graph)}]
      \begin{scope}[shift={(0.0,-0.3)}]
		\tikzstyle arrowstyle=[scale=0.9]
		\node [port] (i1) at (-0.16,0.0) {\color{Bgreen}\tiny $b_1$};
		\node [port] (i2) at (0.16,0.0) {\color{Bgreen}\tiny $b_2$};
		\node [grnode] (t1) at (-0.16,0.48) {};
		\node [grnode] (t2) at (0.16,0.48) {};
		\draw[directed] (i1) -- node[left=-2pt] {} (t1);
		\draw[directed] (i2) -- node[left=-2pt] {} (t2);
	  \end{scope}
	  \end{scope}

 	\begin{scope}[shift={(colim)}, scale=0.65,
	  every place/.style= {minimum size=1.6mm,thick,draw=blue!75,fill=blue!20}, 
	  every transition/.style={minimum size=1.4mm,thick,draw=black!75,fill=black!20},
    post/.style={
        ->,
        shorten >=0.5pt, 
        >={stealth}
    },
	]

	  \begin{scope}[shift={(0.35,-0.35)}]
		\tiny
		\node [place, label=below:${\color{Bgreen}b_1}$] (s1) at (-0.5,0.0) {};
		\node [place, label=below:${\color{Bgreen}b_2}$] (s2) at (0.5,0.0) {}; 
		\node [transition] (t1) at (-0.5,1.0) {} ;
		\node [transition] (t2) at (0.5,1.0) {} ;
		\path (s1) edge [post] (t1);
		\path (s2) edge [post] (t2);
				
		\node at (-1.3,-0.3) {$\pn U_B \pn P$};
	  \end{scope}
	  \end{scope}
 
\end{tikzpicture}
\end{center}
The decorations with $b_1$ and $b_2$ indicate the $B$-structure of each graph.
Note again that although the two one-node graphs are isomorphic as abstract graphs 
(and in fact isomorphic as $\pn P$-graphs), they are not isomorphic as $B$-graphs.
In this case the poset already has a terminal object, which is then of course the colimit.

This very Petri net was indicated by Hayman and
Winskel~\cite[Fig.~1]{Hayman-Winskel:2008} as an example of a Petri that does not
have a universal unfolding --- in the traditional setting. The reason is
that in the traditional setting this unfolding hypergraph (which happens to be a graph) has an automorphism
spoiling the universal property. However, this automorphism is an artefact of
thinking the marking as a mere multiplicity $2$. In the whole-grain setting, the
marking is the explicit $2$-element set $B = \{b_1, b_2\}$, which fixes the
automorphisms.
\end{blanko}

\begin{blanko}{Example.}\label{ex:uv}
  Consider the Petri net $\pn P$
  $$
  \begin{tikzcd}[column sep={18mm,between origins}]
	\{s\}  & \{u,v\} \ar[l] \ar[r] 
	&  \{t\} & 
	\emptyset\ar[l]\ar[r]  & \{s\} ,
  \end{tikzcd}
  $$
 with its marking $B \to \pn P$ by a 
$2$-element set $B =\{b_1,b_2\}$:

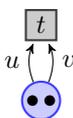
  \begin{center}
	\begin{tikzpicture}[scale=0.9,
	  node distance=1.1cm,on grid,>=stealth',bend angle=22.5,auto,
	  every place/.style= {minimum size=5mm,thick,draw=blue!75,fill=blue!20}, 
	  every transition/.style={minimum 
	  size=3.5mm,thick,draw=black!75,fill=black!20},
	  every token/.style={minimum size=3.5pt, token distance=8pt}
	  ]
	  		
	  \coordinate (init) at (-2,2);	  
	  
	  \begin{scope}[shift={(init)}]
		\footnotesize
		\node [transition] (t1) at (0.0,1.0) {$t$} ;
		\node [place,tokens=2] (s3) at (0.0, 0.0) {};
		\path[bend left] (s3) edge [post] node[left] {$u$} (t1);
		\path[bend right] (s3) edge [post] node[right] {$v$} (t1);
	  \end{scope}
 \end{tikzpicture}
 \end{center}

  Here is the poset of processes $B \to \pn G \to \pn P$, 
  and the colimit, 
  the universal unfolding:  
  
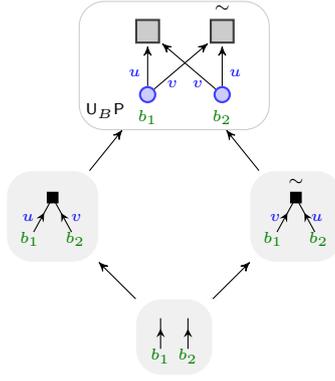
\begin{center}
	\begin{tikzpicture}[scale=0.9,
	  node distance=0.9cm,on grid,>=stealth',bend angle=22.5,auto,
	  every place/.style= {minimum size=5mm,thick,draw=blue!75,fill=blue!20}, 
	  every transition/.style={minimum size=4mm,thick,draw=black!75,fill=black!20},
	  every token/.style={minimum size=4pt, token distance=10pt}
	  ]
	  
	  \small
			  
	  \node[rectangle, rounded corners=2ex, draw=black!20, 
	    minimum width=23mm, minimum height=16mm]
		(colim) at (0.0,3.6) {};

	  \node[rectangle, rounded corners=2ex, fill=black!06, 
	    minimum width=9mm, minimum height=9mm]
		(deg0graph) at (0.0,0.0) {};

	  \node[rectangle, rounded corners=2ex, fill=black!06,
	    minimum width=11mm, minimum height=11mm]
		(deg1graph1) at (-1.6,1.6) {};
		
	  \node[rectangle, rounded corners=2ex, fill=black!06,
	    minimum width=11mm, minimum height=11mm]
		(deg1graph2) at (1.6,1.6) {};

	  \draw[->] (deg0graph) -- (deg1graph1);
	  \draw[->] (deg0graph) -- (deg1graph2);
	  \draw[->] (deg1graph1) -- (colim);
	  \draw[->] (deg1graph2) -- (colim);

      \begin{scope}[shift={(deg0graph)}]
      \begin{scope}[shift={(0.0,-0.25)}]
		\tikzstyle arrowstyle=[scale=0.9]
		\node [port] (i1) at (-0.18,0.0) {\color{Bgreen}\tiny $b_1$};
		\node [port] (i2) at (0.18,0.0) {\color{Bgreen}\tiny $b_2$};
		\node [port] (o1) at (-0.18,0.5) {};
		\node [port] (o2) at (0.18,0.5) {};
		\draw[directed] (i1) -- node[left=-2pt] {} (o1);
		\draw[directed] (i2) -- node[left=-2pt] {} (o2);
	  \end{scope}
	  \end{scope}
	  	  
      \begin{scope}[shift={(deg1graph1)}]
      \begin{scope}[shift={(0.0,-0.3)}]
		\tikzstyle arrowstyle=[scale=0.9]
		\node [port] (i1) at (-0.3,0.0) {\color{Bgreen}\tiny $b_1$};
		\node [port] (i2) at (0.3,0.0) {\color{Bgreen}\tiny $b_2$};
		\node [grnode] (t1) at (0.0,0.55) {};
		\draw[directed] (i1) -- node[left=-1pt] {\color{blue} \tiny $u$} (t1);
		\draw[directed] (i2) -- node[right=-1pt] {\color{blue} \tiny $v$} (t1);
	  \end{scope}
	  \end{scope}
      \begin{scope}[shift={(deg1graph2)}]
      \begin{scope}[shift={(0.0,-0.3)}]
		\tikzstyle arrowstyle=[scale=0.9]
		\node [port] (i1) at (-0.3,0.0) {\color{Bgreen}\tiny $b_1$};
		\node [port] (i2) at (0.3,0.0) {\color{Bgreen}\tiny $b_2$};
		\tiny
		\node [grnode, label={[label distance=-1pt]above:$\sim$}] (t1) at (0.0,0.55) {};
		\draw[directed] (i1) -- node[left=0.5pt] {\color{blue} \tiny $v$} (t1);
		\draw[directed] (i2) -- node[right=0.5pt] {\color{blue} \tiny $u$} (t1);
	  \end{scope}
	  \end{scope}

  	\begin{scope}[shift={(colim)}, scale=0.65,
	  every place/.style= {minimum size=1.6mm,thick,draw=blue!75,fill=blue!20}, 
	  every transition/.style={minimum size=1.4mm,thick,draw=black!75,fill=black!20},
    post/.style={
        ->,
        shorten >=0.5pt, 
        >={stealth}
    },
	]

\begin{scope}[shift={(0.2,-0.55)}]
		\tiny
		\node [place, label=below:${\color{Bgreen}b_1}$] (s1) at (-0.7,0.0) {};
		\node [place, label=below:${\color{Bgreen}b_2}$] (s2) at (0.7,0.0) {}; 
		\node [transition] (t1) at (-0.7,1.2) {} ;
		\node [transition, label={[label distance=-1pt]above:$\sim$}] (t2) at (0.7,1.2) {} ;
		\path (s1) edge [post] node[left=0pt, pos=0.3] {\color{blue} \tiny $u$} (t1);
		\path (s2) edge [post] node[right=0pt,pos=0.3] {\color{blue} \tiny $u$} (t2);
		\path (s1) edge [post] node[right=0pt, pos=0.1] {\color{blue} \tiny $v$} (t2);
		\path (s2) edge [post] node[left=0pt, pos=0.1] {\color{blue} \tiny $v$} (t1);
		\node at (-1.5,-0.3) {$\pn U_B \pn P$};

	  \end{scope}
	  \end{scope}
 
\end{tikzpicture}
\end{center}
  The decorations with $b_1$ and $b_2$ indicate the $B$-structure of each graph,
  whereas the decorations with $u$ and $v$ indicate the $\pn P$-structure.
  The tilde symbols only serve to remind us that the two one-node graphs are not 
  isomorphic (and how they map into the colimit): while they are isomorphic as 
  abstract graphs, they are not
  isomorphic as $\pn P$-graphs: the decorations with the letter $u$ and $v$ indicate
  the effect of the etale map on incoming arcs ($I$-sets). From an intuitive operational
  viewpoint, the difference between the two processes is which input arc receives
  which token, a distinction that would not be visible in traditional Petri nets, where 
  instead of parallel arcs there would be just a multiplicity.
\end{blanko}

With the previous example understood, we are ready to look at the following 
interesting variation of Example~\ref{ex:exP}.

\begin{blanko}{Example.}
  Consider the following Petri net $\pn Q$, with its marking $B \to \pn Q$ by a
  $3$-element set $B$:

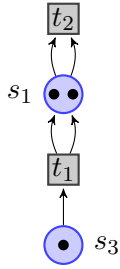
  \begin{center}
	\begin{tikzpicture}[scale=0.9,
	  node distance=1.0cm,on grid,>=stealth',bend angle=22.5,auto,
	  every place/.style= {minimum size=5mm,thick,draw=blue!75,fill=blue!20}, 
	  every transition/.style={minimum 
	  size=3.5mm,thick,draw=black!75,fill=black!20},
	  every token/.style={minimum size=3.5pt, token distance=8pt}
	  ]
	  
	  \small
		
	  \coordinate (init) at (-2,2);	  
	  
	  \begin{scope}[shift={(init)}]
		\node [transition] (t1) at (0.0,1.0) {$t_1$} ;
		\node [place,tokens=2, label=left:$s_1$] (s1) at (0.0,2) {}; 
		\node [place,tokens=1, label=right:$s_3$] (s3) at (0.0, 0.0) {};
		\node [transition] (t2) at (0.0,3) {$t_2$} ;
		\path (s3) edge [post] (t1);
		\path[bend left] (t1) edge [post] (s1);
		\path[bend right] (t1) edge [post] (s1);
		\path[bend left] (s1) edge [post] (t2);
		\path[bend right] (s1) edge [post] (t2);
	  \end{scope}
 \end{tikzpicture}
 \end{center}
Compared to the Petri net $\pn P$ from Example~\ref{ex:exP}, here two
places have been joined to one, with otherwise similar distribution of tokens and
the same pre- and post-conditions. However, since the tokens are now `all in the same
pool' they can flow in many more ways. Here is the universal unfolding:
  \begin{center}
	
	\def\mw{8.4mm}
	
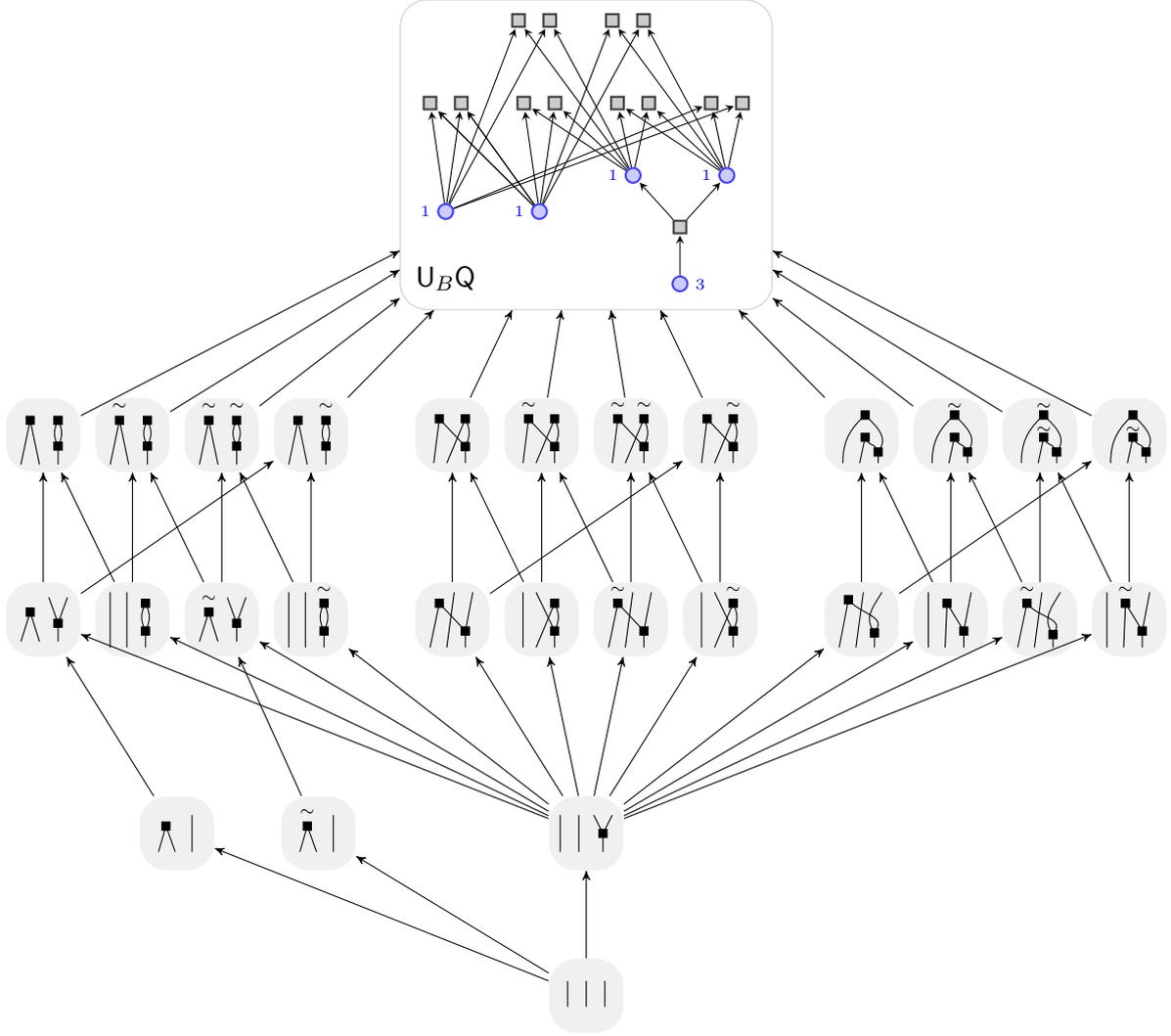
  \begin{tikzpicture}[scale=0.85,
	  node distance=0.8cm,on grid,>=stealth',bend angle=22.5,auto,
	  ]
	
	  \node[rectangle, rounded corners=2ex, draw=black!20, 
	    minimum width=40mm, minimum height=32mm]
		(colim) at (0.0,11.8) {};

	  \begin{scope}[shift={(-5.5,8)}]
		\node[rectangle, rounded corners=2ex, fill=black!06, minimum width=\mw, 
		minimum height=\mw] (deg3graph1) at (-1.8,0) {};
		\node[rectangle, rounded corners=2ex, fill=black!06, minimum width=\mw, 
		minimum height=\mw] (deg3graph2) at (-0.6,0) {};
		\node[rectangle, rounded corners=2ex, fill=black!06, minimum width=\mw, 
		minimum height=\mw] (deg3graph3) at (0.6,0) {};
		\node[rectangle, rounded corners=2ex, fill=black!06, minimum width=\mw, 
		minimum height=\mw] (deg3graph4) at (1.8,0) {};
	  \end{scope}
		
	  \begin{scope}[shift={(deg3graph1)}]
      \begin{scope}[shift={(-0.05,-0.4)}, scale=0.5]
		\node [port] (i1) at (-0.5,0.0) {};
		\node [port] (i2) at (0.0,0.0) {};
		\node [port] (i3) at (0.5,0.0) {};
		\node [port] (o1) at (-0.5,1.5) {};
		\node [port] (o2) at (0.0,1.5) {};
		\node [port] (o3) at (0.25,1.5) {};
		\node [port] (o4) at (0.75,1.5) {};
		\node [grnode] (x2) at (0.5,1.2) {};
		\node [grnode] (x2') at (-0.25,1.2) {};
		\node [grnode] (x1) at (0.5,0.5) {};
		\draw (i1) -- (x2');
		\draw (i2) -- (x2');
		\draw[bend left] (x1) to (x2);
		\draw[bend right] (x1) to (x2);
		\draw (i3) --  (x1);
	  \end{scope}		  
	  \end{scope}
	  
	  \begin{scope}[shift={(deg3graph2)}]
      \begin{scope}[shift={(-0.05,-0.4)}, scale=0.5]
		\tiny
		\node [port] (i1) at (-0.5,0.0) {};
		\node [port] (i2) at (0.0,0.0) {};
		\node [port] (i3) at (0.5,0.0) {};
		\node [port] (o1) at (-0.5,1.5) {};
		\node [port] (o2) at (0.0,1.5) {};
		\node [port] (o3) at (0.25,1.5) {};
		\node [port] (o4) at (0.75,1.5) {};
		\node [grnode] (x2) at (0.5,1.2) {};
		\node [grnode, label={[label distance=-2pt]above:$\sim$}] (x2') at (-0.25,1.2) {};
		\node [grnode] (x1) at (0.5,0.5) {};
		\draw (i1) -- (x2');
		\draw (i2) -- (x2');
		\draw[bend left] (x1) to (x2);
		\draw[bend right] (x1) to (x2);
		\draw (i3) --  (x1);
	  \end{scope}		  
	  \end{scope}
	  
	  \begin{scope}[shift={(deg3graph3)}]
      \begin{scope}[shift={(-0.05,-0.4)}, scale=0.5]
		\tiny
		\node [port] (i1) at (-0.5,0.0) {};
		\node [port] (i2) at (0.0,0.0) {};
		\node [port] (i3) at (0.5,0.0) {};
		\node [port] (o1) at (-0.5,1.5) {};
		\node [port] (o2) at (0.0,1.5) {};
		\node [port] (o3) at (0.25,1.5) {};
		\node [port] (o4) at (0.75,1.5) {};
		\node [grnode, label={[label distance=-2pt]above:$\sim$}] (x2) at (0.5,1.2) {};
		\node [grnode, label={[label distance=-2pt]above:$\sim$}] (x2') at (-0.25,1.2) {};
		\node [grnode] (x1) at (0.5,0.5) {};
		\draw (i1) -- (x2');
		\draw (i2) -- (x2');
		\draw[bend left] (x1) to (x2);
		\draw[bend right] (x1) to (x2);
		\draw (i3) --  (x1);
	  \end{scope}		  
	  \end{scope}		  

	  \begin{scope}[shift={(deg3graph4)}]
      \begin{scope}[shift={(-0.05,-0.4)}, scale=0.5]
		\tiny
		\node [port] (i1) at (-0.5,0.0) {};
		\node [port] (i2) at (0.0,0.0) {};
		\node [port] (i3) at (0.5,0.0) {};
		\node [port] (o1) at (-0.5,1.5) {};
		\node [port] (o2) at (0.0,1.5) {};
		\node [port] (o3) at (0.25,1.5) {};
		\node [port] (o4) at (0.75,1.5) {};
		\node [grnode, label={[label distance=-2pt]above:$\sim$}] (x2) at (0.5,1.2) {};
		\node [grnode] (x2') at (-0.25,1.2) {};
		\node [grnode] (x1) at (0.5,0.5) {};
		\draw (i1) -- (x2');
		\draw (i2) -- (x2');
		\draw[bend left] (x1) to (x2);
		\draw[bend right] (x1) to (x2);
		\draw (i3) --  (x1);
	  \end{scope}
	  \end{scope}
	  
	  \begin{scope}[shift={(-5.5,5.5)}]
		\node[rectangle, rounded corners=2ex, fill=black!06, minimum width=\mw, 
		minimum height=\mw] (deg2graph1) at (-1.8,0) {};
		\node[rectangle, rounded corners=2ex, fill=black!06, minimum width=\mw, 
		minimum height=\mw] (deg2graph2) at (-0.6,0) {};
		\node[rectangle, rounded corners=2ex, fill=black!06, minimum width=\mw, 
		minimum height=\mw] (deg2graph3) at (0.6,0) {};
		\node[rectangle, rounded corners=2ex, fill=black!06, minimum width=\mw, 
		minimum height=\mw] (deg2graph4) at (1.8,0) {};
	  \end{scope}
	
	  \begin{scope}[shift={(deg2graph1)}]
      \begin{scope}[shift={(-0.05,-0.3)}, scale=0.5]
		\node [port] (i1) at (-0.5,0.0) {};
		\node [port] (i2) at (0.0,0.0) {};
		\node [port] (i3) at (0.5,0.0) {};
		\node [port] (o1) at (-0.5,1.2) {};
		\node [port] (o2) at (0.0,1.2) {};
		\node [port] (o3) at (0.25,1.2) {};
		\node [port] (o4) at (0.75,1.2) {};
		\node [grnode] (x2) at (-0.25,0.8) {};
		\node [grnode] (x1) at (0.5,0.5) {};
		\draw (i1) -- (x2);
		\draw (i2) -- (x2);
		\draw (i3) -- (x1);
		\draw (x1) -- (o3);
		\draw (x1) -- (o4);
	  \end{scope}
	  \end{scope}
	  \begin{scope}[shift={(deg2graph2)}]
      \begin{scope}[shift={(-0.1,-0.38)}, scale=0.5]
		\node [port] (i1) at (-0.4,0.0) {};
		\node [port] (i2) at (0.05,0.0) {};
		\node [port] (i3) at (0.55,0.0) {};
		\node [port] (o1) at (-0.4,1.5) {};
		\node [port] (o2) at (0.05,1.5) {};
		\node [port] (o3) at (0.25,1.5) {};
		\node [port] (o4) at (0.75,1.5) {};
		\node [grnode] (x2) at (0.55,1.2) {};
		\node [grnode] (x1) at (0.55,0.45) {};
		\draw (i1) -- (o1);
		\draw (i2) --  (o2);
		\draw[bend left] (x1) to (x2);
		\draw[bend right] (x1) to (x2);
		\draw (i3) -- (x1);
	  \end{scope}
	  \end{scope}
	  \begin{scope}[shift={(deg2graph3)}]
      \begin{scope}[shift={(-0.05,-0.3)}, scale=0.5]
		\tiny
		\node [port] (i1) at (-0.5,0.0) {};
		\node [port] (i2) at (0.0,0.0) {};
		\node [port] (i3) at (0.5,0.0) {};
		\node [port] (o1) at (-0.5,1.2) {};
		\node [port] (o2) at (0.0,1.2) {};
		\node [port] (o3) at (0.25,1.2) {};
		\node [port] (o4) at (0.75,1.2) {};
		\node [grnode, label={[label distance=-2pt]above:$\sim$}] (x2) at (-0.25,0.8) {};
		\node [grnode] (x1) at (0.5,0.5) {};
		\draw (i1) -- (x2);
		\draw (i2) -- (x2);
		\draw (i3) -- (x1);
		\draw (x1) -- (o3);
		\draw (x1) -- (o4);
	  \end{scope}
	  \end{scope}
	  \begin{scope}[shift={(deg2graph4)}]
      \begin{scope}[shift={(-0.1,-0.38)}, scale=0.5]
		\tiny
		\node [port] (i1) at (-0.4,0.0) {};
		\node [port] (i2) at (0.05,0.0) {};
		\node [port] (i3) at (0.55,0.0) {};
		\node [port] (o1) at (-0.4,1.5) {};
		\node [port] (o2) at (0.05,1.5) {};
		\node [port] (o3) at (0.25,1.5) {};
		\node [port] (o4) at (0.75,1.5) {};
		\node [grnode, label={[label distance=-2pt]above:$\sim$}] (x2) at (0.55,1.2) {};
		\node [grnode] (x1) at (0.55,0.45) {};
		\draw (i1) -- (o1);
		\draw (i2) --  (o2);
		\draw[bend left] (x1) to (x2);
		\draw[bend right] (x1) to (x2);
		\draw (i3) -- (x1);
	  \end{scope}
	  \end{scope}

	  \begin{scope}[shift={(0,8)}]
		\node[rectangle, rounded corners=2ex, fill=black!06, minimum width=\mw, 
		minimum height=\mw] (deg3graph5) at (-1.8,0) {};
		\node[rectangle, rounded corners=2ex, fill=black!06, minimum width=\mw, 
		minimum height=\mw] (deg3graph6) at (-0.6,0) {};
		\node[rectangle, rounded corners=2ex, fill=black!06, minimum width=\mw, 
		minimum height=\mw] (deg3graph7) at (0.6,0) {};
		\node[rectangle, rounded corners=2ex, fill=black!06, minimum width=\mw, 
		minimum height=\mw] (deg3graph8) at (1.8,0) {};
	  \end{scope}
	  
	  \begin{scope}[shift={(deg3graph5)}]
      \begin{scope}[shift={(-0.1,-0.38)}, scale=0.5]
		\tiny
		\node [port] (i1) at (-0.3,0.0) {};
		\node [port] (i2) at (0.05,0.0) {};
		\node [port] (i3) at (0.55,0.0) {};
		\node [port] (o1) at (-0.3,1.5) {};
		\node [port] (o2) at (0.05,1.5) {};
		\node [port] (o3) at (0.25,1.5) {};
		\node [port] (o4) at (0.75,1.5) {};
		\node [grnode] (x2) at (-0.15,1.2) {};
		\node [grnode] (x2') at (0.55,1.2) {};
		\node [grnode] (x1) at (0.55,0.45) {};
		\draw (i1) -- (x2);
		\draw (i2) --  (x2');
		\draw[bend right] (x1) to (x2');
		\draw (x1) to (x2);
		\draw (i3) -- (x1);
	  \end{scope}
	  \end{scope}
	  \begin{scope}[shift={(deg3graph6)}]
      \begin{scope}[shift={(-0.1,-0.38)}, scale=0.5]
		\tiny
		\node [port] (i1) at (-0.3,0.0) {};
		\node [port] (i2) at (0.05,0.0) {};
		\node [port] (i3) at (0.55,0.0) {};
		\node [port] (o1) at (-0.3,1.5) {};
		\node [port] (o2) at (0.05,1.5) {};
		\node [port] (o3) at (0.25,1.5) {};
		\node [port] (o4) at (0.75,1.5) {};
		\node [grnode, label=above:$\sim$] (x2) at (-0.15,1.2) {};
		\node [grnode] (x2') at (0.55,1.2) {};
		\node [grnode] (x1) at (0.55,0.45) {};
		\draw (i1) -- (x2);
		\draw (i2) --  (x2');
		\draw[bend right] (x1) to (x2');
		\draw (x1) to (x2);
		\draw (i3) -- (x1);
	  \end{scope}
	  \end{scope}
	  \begin{scope}[shift={(deg3graph7)}]
      \begin{scope}[shift={(-0.1,-0.38)}, scale=0.5]
		\tiny
		\node [port] (i1) at (-0.3,0.0) {};
		\node [port] (i2) at (0.05,0.0) {};
		\node [port] (i3) at (0.55,0.0) {};
		\node [port] (o1) at (-0.3,1.5) {};
		\node [port] (o2) at (0.05,1.5) {};
		\node [port] (o3) at (0.25,1.5) {};
		\node [port] (o4) at (0.75,1.5) {};
		\node [grnode, label={[label distance=-2pt]above:$\sim$}] (x2) at (-0.15,1.2) {};
		\node [grnode, label={[label distance=-2pt]above:$\sim$}] (x2') at (0.55,1.2) {};
		\node [grnode] (x1) at (0.55,0.45) {};
		\draw (i1) -- (x2);
		\draw (i2) --  (x2');
		\draw[bend right] (x1) to (x2');
		\draw (x1) to (x2);
		\draw (i3) -- (x1);
	  \end{scope}
	  \end{scope}
	  \begin{scope}[shift={(deg3graph8)}]
      \begin{scope}[shift={(-0.1,-0.38)}, scale=0.5]
		\tiny
		\node [port] (i1) at (-0.3,0.0) {};
		\node [port] (i2) at (0.05,0.0) {};
		\node [port] (i3) at (0.55,0.0) {};
		\node [port] (o1) at (-0.3,1.5) {};
		\node [port] (o2) at (0.05,1.5) {};
		\node [port] (o3) at (0.25,1.5) {};
		\node [port] (o4) at (0.75,1.5) {};
		\node [grnode] (x2) at (-0.15,1.2) {};
		\node [grnode, label={[label distance=-2pt]above:$\sim$}] (x2') at (0.55,1.2) {};
		\node [grnode] (x1) at (0.55,0.45) {};
		\draw (i1) -- (x2);
		\draw (i2) --  (x2');
		\draw[bend right] (x1) to (x2');
		\draw (x1) to (x2);
		\draw (i3) -- (x1);
	  \end{scope}
	  \end{scope}

	  \begin{scope}[shift={(0,5.5)}]
		\node[rectangle, rounded corners=2ex, fill=black!06, minimum width=\mw, 
		minimum height=\mw] (deg2graph5) at (-1.8,0) {};
		\node[rectangle, rounded corners=2ex, fill=black!06, minimum width=\mw, 
		minimum height=\mw] (deg2graph6) at (-0.6,0) {};
		\node[rectangle, rounded corners=2ex, fill=black!06, minimum width=\mw, 
		minimum height=\mw] (deg2graph7) at (0.6,0) {};
		\node[rectangle, rounded corners=2ex, fill=black!06, minimum width=\mw, 
		minimum height=\mw] (deg2graph8) at (1.8,0) {};
	  \end{scope}
	  
	  \begin{scope}[shift={(deg2graph5)}]
      \begin{scope}[shift={(-0.1,-0.38)}, scale=0.5]
		\node [port] (i1) at (-0.4,0.0) {};
		\node [port] (i2) at (0.05,0.0) {};
		\node [port] (i3) at (0.55,0.0) {};
		\node [port] (o1) at (-0.4,1.5) {};
		\node [port] (o2) at (0.05,1.5) {};
		\node [port] (o3) at (0.25,1.5) {};
		\node [port] (o4) at (0.75,1.5) {};
		\node [grnode] (x2) at (-0.15,1.2) {};
		\node [grnode] (x1) at (0.55,0.45) {};
		\draw (i1) -- (x2);
		\draw (i2) --  (o3);
		\draw (x1) to (x2);
		\draw (x1) to (o4);
		\draw (i3) -- (x1);
	  \end{scope}
	  \end{scope}
	  \begin{scope}[shift={(deg2graph6)}]
      \begin{scope}[shift={(-0.1,-0.38)}, scale=0.5]
		\node [port] (i1) at (-0.3,0.0) {};
		\node [port] (i2) at (0.05,0.0) {};
		\node [port] (i3) at (0.55,0.0) {};
		\node [port] (o1) at (-0.3,1.5) {};
		\node [port] (o2) at (0.05,1.5) {};
		\node [port] (o3) at (0.25,1.5) {};
		\node [port] (o4) at (0.75,1.5) {};
		\node [grnode] (x2) at (0.55,1.2) {};
		\node [grnode] (x1) at (0.55,0.45) {};
		\draw (i1) -- (o1);
		\draw (i2) --  (x2);
		\draw[bend right] (x1) to (x2);
		\draw (x1) to (o2);
		\draw (i3) -- (x1);
	  \end{scope}
	  \end{scope}
	  \begin{scope}[shift={(deg2graph7)}]
      \begin{scope}[shift={(-0.1,-0.38)}, scale=0.5]
		\tiny
		\node [port] (i1) at (-0.4,0.0) {};
		\node [port] (i2) at (0.05,0.0) {};
		\node [port] (i3) at (0.55,0.0) {};
		\node [port] (o1) at (-0.4,1.5) {};
		\node [port] (o2) at (0.05,1.5) {};
		\node [port] (o3) at (0.25,1.5) {};
		\node [port] (o4) at (0.75,1.5) {};
		\node [grnode, label={[label distance=-2pt]above:$\sim$}] (x2) at (-0.15,1.2) {};
		\node [grnode] (x1) at (0.55,0.45) {};
		\draw (i1) -- (x2);
		\draw (i2) --  (o3);
		\draw (x1) to (x2);
		\draw (x1) to (o4);
		\draw (i3) -- (x1);
	  \end{scope}
	  \end{scope}
	  \begin{scope}[shift={(deg2graph8)}]
      \begin{scope}[shift={(-0.1,-0.38)}, scale=0.5]
		\tiny
		\node [port] (i1) at (-0.3,0.0) {};
		\node [port] (i2) at (0.05,0.0) {};
		\node [port] (i3) at (0.55,0.0) {};
		\node [port] (o1) at (-0.3,1.5) {};
		\node [port] (o2) at (0.05,1.5) {};
		\node [port] (o3) at (0.25,1.5) {};
		\node [port] (o4) at (0.75,1.5) {};
		\node [grnode, label={[label distance=-2pt]above:$\sim$}] (x2) at (0.55,1.2) {};
		\node [grnode] (x1) at (0.55,0.45) {};
		\draw (i1) -- (o1);
		\draw (i2) --  (x2);
		\draw[bend right] (x1) to (x2);
		\draw (x1) to (o2);
		\draw (i3) -- (x1);
	  \end{scope}
	  \end{scope}

	  \begin{scope}[shift={(5.5,8)}]
		\node[rectangle, rounded corners=2ex, fill=black!06, minimum width=\mw, 
		minimum height=\mw] (deg3graph9) at (-1.8,0) {};
		\node[rectangle, rounded corners=2ex, fill=black!06, minimum width=\mw, 
		minimum height=\mw] (deg3graph10) at (-0.6,0) {};
		\node[rectangle, rounded corners=2ex, fill=black!06, minimum width=\mw, 
		minimum height=\mw] (deg3graph11) at (0.6,0) {};
		\node[rectangle, rounded corners=2ex, fill=black!06, minimum width=\mw, 
		minimum height=\mw] (deg3graph12) at (1.8,0) {};
	  \end{scope}
	  
	  \begin{scope}[shift={(deg3graph9)}]
      \begin{scope}[shift={(-0.05,-0.38)}, scale=0.5]
		\node [port] (i1) at (-0.4,0.0) {};
		\node [port] (i2) at (0.05,0.0) {};
		\node [port] (i3) at (0.55,0.0) {};
		\node [port] (o1) at (-0.4,1.5) {};
		\node [port] (o2) at (0.05,1.5) {};
		\node [port] (o3) at (0.25,1.5) {};
		\node [port] (o4) at (0.75,1.5) {};
		\node [grnode] (x2) at (0.2,1.3) {};
		\node [grnode] (x2') at (0.2,0.7) {};
		\node [grnode] (x1) at (0.55,0.3) {};
		\draw[bend left] (i1) to (x2);
		\draw (i2) --  (x2');
		\draw (x1) to[out=60, in=-45] (x2);
		\draw (x1) to[out=120, in=300]  (x2');
		\draw (i3) -- (x1);
	  \end{scope}
	  \end{scope}
	  \begin{scope}[shift={(deg3graph10)}]
      \begin{scope}[shift={(-0.05,-0.38)}, scale=0.5]
		\tiny
		\node [port] (i1) at (-0.4,0.0) {};
		\node [port] (i2) at (0.05,0.0) {};
		\node [port] (i3) at (0.55,0.0) {};
		\node [port] (o1) at (-0.4,1.5) {};
		\node [port] (o2) at (0.05,1.5) {};
		\node [port] (o3) at (0.25,1.5) {};
		\node [port] (o4) at (0.75,1.5) {};
		\node [grnode] (x2) at (0.2,1.3) {};
		\node at (0.2, 1.55) {$\sim$};
		\node [grnode] (x2') at (0.2,0.7) {};
		\node [grnode] (x1) at (0.55,0.3) {};
		\draw[bend left] (i1) to (x2);
		\draw (i2) --  (x2');
		\draw (x1) to[out=60, in=-45] (x2);
		\draw (x1) to[out=120, in=300]  (x2');
		\draw (i3) -- (x1);
	  \end{scope}
	  \end{scope}
	  \begin{scope}[shift={(deg3graph11)}]
      \begin{scope}[shift={(-0.05,-0.38)}, scale=0.5]
		\tiny
		\node [port] (i1) at (-0.4,0.0) {};
		\node [port] (i2) at (0.05,0.0) {};
		\node [port] (i3) at (0.55,0.0) {};
		\node [port] (o1) at (-0.4,1.5) {};
		\node [port] (o2) at (0.05,1.5) {};
		\node [port] (o3) at (0.25,1.5) {};
		\node [port] (o4) at (0.75,1.5) {};
		\node [grnode] (x2) at (0.2,1.3) {};
		\node at (0.2, 1.55) {$\sim$};
		\node [grnode] (x2') at (0.2,0.7) {};
		\node at (0.2, 0.9) {$\sim$};
		\node [grnode] (x1) at (0.55,0.3) {};
		\draw[bend left] (i1) to (x2);
		\draw (i2) --  (x2');
		\draw (x1) to[out=60, in=-45] (x2);
		\draw (x1) to[out=120, in=300]  (x2');
		\draw (i3) -- (x1);
	  \end{scope}
	  \end{scope}
	  \begin{scope}[shift={(deg3graph12)}]
      \begin{scope}[shift={(-0.05,-0.38)}, scale=0.5]
		\tiny
		\node [port] (i1) at (-0.4,0.0) {};
		\node [port] (i2) at (0.05,0.0) {};
		\node [port] (i3) at (0.55,0.0) {};
		\node [port] (o1) at (-0.4,1.5) {};
		\node [port] (o2) at (0.05,1.5) {};
		\node [port] (o3) at (0.25,1.5) {};
		\node [port] (o4) at (0.75,1.5) {};
		\node [grnode] (x2) at (0.2,1.3) {};
		\node [grnode] (x2') at (0.2,0.7) {};
		\node at (0.2, 0.9) {$\sim$};
		\node [grnode] (x1) at (0.55,0.3) {};
		\draw[bend left] (i1) to (x2);
		\draw (i2) --  (x2');
		\draw (x1) to[out=60, in=-45] (x2);
		\draw (x1) to[out=120, in=300]  (x2');
		\draw (i3) -- (x1);
	  \end{scope}
	  \end{scope}

	  \begin{scope}[shift={(5.5,5.5)}]
		\node[rectangle, rounded corners=2ex, fill=black!06, minimum width=\mw, 
		minimum height=\mw] (deg2graph9) at (-1.8,0) {};
		\node[rectangle, rounded corners=2ex, fill=black!06, minimum width=\mw, 
		minimum height=\mw] (deg2graph10) at (-0.6,0) {};
		\node[rectangle, rounded corners=2ex, fill=black!06, minimum width=\mw, 
		minimum height=\mw] (deg2graph11) at (0.6,0) {};
		\node[rectangle, rounded corners=2ex, fill=black!06, minimum width=\mw, 
		minimum height=\mw] (deg2graph12) at (1.8,0) {};
	  \end{scope}
	  
	  \begin{scope}[shift={(deg2graph9)}]
      \begin{scope}[shift={(-0.1,-0.38)}, scale=0.5]
		\node [port] (i1) at (-0.4,0.0) {};
		\node [port] (i2) at (0.05,0.0) {};
		\node [port] (i3) at (0.55,0.0) {};
		\node [port] (o1) at (-0.4,1.5) {};
		\node [port] (o2) at (0.05,1.5) {};
		\node [port] (o3) at (0.25,1.5) {};
		\node [port] (o4) at (0.75,1.5) {};
		\node [grnode] (x2) at (-0.15,1.3) {};
		\node [grnode] (x1) at (0.55,0.4) {};
		\draw (i1) -- (x2);
		\draw (i2) --  (o3);
		\draw (x1) to[out=60, in=-45] (x2);
		\draw (x1) to[out=120, in=240]  (o4);
		\draw (i3) -- (x1);
	  \end{scope}
	  \end{scope}

	  \begin{scope}[shift={(deg2graph10)}]
      \begin{scope}[shift={(-0.1,-0.38)}, scale=0.5]
		\node [port] (i1) at (-0.4,0.0) {};
		\node [port] (i2) at (0.05,0.0) {};
		\node [port] (i3) at (0.55,0.0) {};
		\node [port] (o1) at (-0.4,1.5) {};
		\node [port] (o2) at (0.05,1.5) {};
		\node [port] (o3) at (0.25,1.5) {};
		\node [port] (o4) at (0.75,1.5) {};
		\node [grnode] (x2) at (0.1,1.2) {};
		\node [grnode] (x1) at (0.55,0.45) {};
		\draw (i1) -- (o1);
		\draw (i2) --  (x2);
		\draw (x1) to (x2);
		\draw (x1) to (o4);
		\draw (i3) -- (x1);
	  \end{scope}
	  \end{scope}
	  \begin{scope}[shift={(deg2graph11)}]
      \begin{scope}[shift={(-0.1,-0.38)}, scale=0.5]
		\tiny
		\node [port] (i1) at (-0.4,0.0) {};
		\node [port] (i2) at (0.05,0.0) {};
		\node [port] (i3) at (0.55,0.0) {};
		\node [port] (o1) at (-0.4,1.5) {};
		\node [port] (o2) at (0.05,1.5) {};
		\node [port] (o3) at (0.25,1.5) {};
		\node [port] (o4) at (0.75,1.5) {};
		\node [grnode, label={[label distance=-2pt]above:$\sim$}] (x2) at (-0.15,1.2) {};
		\node [grnode] (x1) at (0.55,0.35) {};
		\draw (i1) -- (x2);
		\draw (i2) --  (o3);
		\draw (x1) to[out=60, in=-45] (x2);
		\draw (x1) to[out=120, in=240]  (o4);
		\draw (i3) -- (x1);
	  \end{scope}
	  \end{scope}

	  \begin{scope}[shift={(deg2graph12)}]
      \begin{scope}[shift={(-0.1,-0.38)}, scale=0.5]
		\tiny
		\node [port] (i1) at (-0.4,0.0) {};
		\node [port] (i2) at (0.05,0.0) {};
		\node [port] (i3) at (0.55,0.0) {};
		\node [port] (o1) at (-0.4,1.5) {};
		\node [port] (o2) at (0.05,1.5) {};
		\node [port] (o3) at (0.25,1.5) {};
		\node [port] (o4) at (0.75,1.5) {};
		\node [grnode, label={[label distance=-2pt]above:$\sim$}] (x2) at (0.1,1.2) {};
		\node [grnode] (x1) at (0.55,0.45) {};
		\draw (i1) -- (o1);
		\draw (i2) --  (x2);
		\draw (x1) to (x2);
		\draw (x1) to (o4);
		\draw (i3) -- (x1);
	  \end{scope}
	  \end{scope}

	  \node[rectangle, rounded corners=2ex, fill=black!06, 
	    minimum width=\mw, minimum height=\mw]
		(deg1graph1) at (0.0,2.6) {};
	  \node[rectangle, rounded corners=2ex, fill=black!06, 
	    minimum width=\mw, minimum height=\mw]
		(deg1graph1') at (-3.6,2.6) {};
	  \node[rectangle, rounded corners=2ex, fill=black!06, 
	    minimum width=\mw, minimum height=\mw]
		(deg1graph1'') at (-5.5,2.6) {};
		
	  \begin{scope}[shift={(deg1graph1)}]]
      \begin{scope}[shift={(-0.15,-0.25)}, scale=0.5]
		\node [port] (i1) at (-0.4,0.0) {};
		\node [port] (i2) at (0.1,0.0) {};
		\node [port] (i3) at (0.75,0.0) {};
		\node [port] (o1) at (-0.4,1) {};
		\node [port] (o2) at (0.1,1) {};
		\node [port] (o3) at (0.5,1) {};
		\node [port] (o4) at (1,1) {};
		\node [grnode] (x1) at (0.75,0.5) {};
		\draw (i1) -- (o1);
		\draw (i2) -- (o2);
		\draw (i3) -- (x1);
		\draw (x1) -- (o3);
		\draw (x1) -- (o4);
	  \end{scope}
	  \end{scope}
	  \begin{scope}[shift={(deg1graph1')}]
      \begin{scope}[shift={(-0.05,-0.25)}, scale=0.5]
		\tiny
		\node [port] (i1) at (-0.4,0.0) {};
		\node [port] (i2) at (0.05,0.0) {};
		\node [port] (i3) at (0.5,0.0) {};
		\node [port] (o1) at (-0.4,1) {};
		\node [port] (o2) at (0.05,1) {};
		\node [port] (o3) at (0.5,1) {};
		\node [port] (o4) at (1,1) {};
		\node [grnode, label={[label distance=-2pt]above:$\sim$}] (x2) at (-0.2,0.7) {};
		\draw (i1) -- (x2);
		\draw (i2) -- (x2);
		\draw (i3) -- (o3);
	  \end{scope}
	  \end{scope}
	  \begin{scope}[shift={(deg1graph1'')}]
      \begin{scope}[shift={(-0.05,-0.25)}, scale=0.5]
		\tiny
		\node [port] (i1) at (-0.4,0.0) {};
		\node [port] (i2) at (0.05,0.0) {};
		\node [port] (i3) at (0.5,0.0) {};
		\node [port] (o1) at (-0.4,1) {};
		\node [port] (o2) at (0.05,1) {};
		\node [port] (o3) at (0.5,1) {};
		\node [port] (o4) at (1,1) {};
		\node [grnode] (x2) at (-0.2,0.7) {};
		\draw (i1) -- (x2);
		\draw (i2) -- (x2);
		\draw (i3) -- (o3);
	  \end{scope}
	  \end{scope}

	  \node[rectangle, rounded corners=2ex, fill=black!06, 
	    minimum width=\mw, minimum height=\mw]
		(deg0graph) at (0.0,0.4) {};
      \begin{scope}[shift={(deg0graph)}]
      \begin{scope}[shift={(0.0,-0.15)}, scale=0.5]
		\node [port] (i1) at (-0.5,0.0) {};
		\node [port] (i2) at (0.0,0.0) {};
		\node [port] (i3) at (0.5,0.0) {};
		\node [port] (o1) at (-0.5,0.7) {};
		\node [port] (o2) at (0.0,0.7) {};
		\node [port] (o3) at (0.5,0.7) {};
		\draw (i1) -- node[left=-2pt] {} (o1);
		\draw (i2) -- node[left=-2pt] {} (o2);
		\draw (i3) -- node[left=-2pt] {} (o3);
	  \end{scope}
	  \end{scope}

	  \draw[->] (deg3graph1) -- (colim);
	  \draw[->] (deg3graph2) -- (colim);
	  \draw[->] (deg3graph3) -- (colim);
	  \draw[->] (deg3graph4) -- (colim);
	  \draw[->] (deg3graph5) -- (colim);
	  \draw[->] (deg3graph6) -- (colim);
	  \draw[->] (deg3graph7) -- (colim);
	  \draw[->] (deg3graph8) -- (colim);
	  \draw[->] (deg3graph9) -- (colim);
	  \draw[->] (deg3graph10) -- (colim);
	  \draw[->] (deg3graph11) -- (colim);
	  \draw[->] (deg3graph12) -- (colim);
	  \draw[->] (deg0graph) -- (deg1graph1);
	  \draw[->] (deg0graph) -- (deg1graph1');
	  \draw[->] (deg0graph) -- (deg1graph1'');
	  \draw[->] (deg1graph1) -- (deg2graph1);
	  \draw[->] (deg1graph1') -- (deg2graph3);
	  \draw[->] (deg1graph1) -- (deg2graph2);
	  \draw[->] (deg1graph1) -- (deg2graph3);
	  \draw[->] (deg1graph1) -- (deg2graph4);
	  \draw[->] (deg1graph1) -- (deg2graph5);
	  \draw[->] (deg1graph1) -- (deg2graph6);
	  \draw[->] (deg1graph1) -- (deg2graph7);
	  \draw[->] (deg1graph1) -- (deg2graph8);
	  \draw[->] (deg1graph1) -- (deg2graph9);
	  \draw[->] (deg1graph1) -- (deg2graph10);
	  \draw[->] (deg1graph1) -- (deg2graph11);
	  \draw[->] (deg1graph1) -- (deg2graph12);
	  \draw[->] (deg1graph1'') -- (deg2graph1);

	  \draw[->] (deg2graph1) -- (deg3graph1);
	  \draw[->] (deg2graph2) -- (deg3graph1);
	  \draw[->] (deg2graph2) -- (deg3graph2);
	  \draw[->] (deg2graph3) -- (deg3graph2);
	  \draw[->] (deg2graph3) -- (deg3graph3);
	  \draw[->] (deg2graph4) -- (deg3graph3);
	  \draw[->] (deg2graph4) -- (deg3graph4);
	  \draw[->] (deg2graph1) -- (deg3graph4);

	  \draw[->] (deg2graph5) -- (deg3graph5);
	  \draw[->] (deg2graph6) -- (deg3graph5);
	  \draw[->] (deg2graph6) -- (deg3graph6);
	  \draw[->] (deg2graph7) -- (deg3graph6);
	  \draw[->] (deg2graph7) -- (deg3graph7);
	  \draw[->] (deg2graph8) -- (deg3graph7);
	  \draw[->] (deg2graph8) -- (deg3graph8);
	  \draw[->] (deg2graph5) -- (deg3graph8);

	  \draw[->] (deg2graph9) -- (deg3graph9);
	  \draw[->] (deg2graph10) -- (deg3graph9);
	  \draw[->] (deg2graph10) -- (deg3graph10);
	  \draw[->] (deg2graph11) -- (deg3graph10);
	  \draw[->] (deg2graph11) -- (deg3graph11);
	  \draw[->] (deg2graph12) -- (deg3graph11);
	  \draw[->] (deg2graph12) -- (deg3graph12);
	  \draw[->] (deg2graph9) -- (deg3graph12);

	\begin{scope}[shift={(colim)}, scale=0.65,
	  every place/.style= {minimum size=1.6mm,thick,draw=blue!75,fill=blue!20}, 
	  every transition/.style={minimum size=1.4mm,thick,draw=black!75,fill=black!20},
    post/.style={
        ->,
        shorten >=0.5pt, 
        >={stealth}
    },
	]

	\begin{scope}[shift={(1.8,-2.5)}]
		\tiny
		\node [transition] (t1) at (0.0,1.1) {} ;
		\node [place, label=left:${\color{blue} 1}$] (s1) at (-4.5,1.4) {};
		\node [place, label=left:${\color{blue} 1}$] (s2) at (-2.7,1.4) {}; 
		\node [place, label=left:${\color{blue} 1}$] (s1') at (-0.9,2.1) {}; 
		\node [place, label=left:${\color{blue} 1}$] (s2') at (0.9,2.1) {};
		\node [place, label=right:${\color{blue} 3}$] (s3) at (0.0, 0.0) {};
		\node [transition] (t2i) at (-4.8,3.5) {} ;
		\node [transition] (t2i') at (-4.2,3.5) {} ;
		\node [transition] (t2ii) at (-3.0,3.5) {} ;
		\node [transition] (t2ii') at (-2.4,3.5) {} ;
		\node [transition] (t2iii) at (-1.2,3.5) {} ;
		\node [transition] (t2iii') at (-0.6,3.5) {} ;
		\node [transition] (t2iv) at (1.2,3.5) {} ;
		\node [transition] (t2iv') at (0.6,3.5) {} ;

		\node [transition] (t2v) at (-3.1,5.1) {} ;
		\node [transition] (t2v') at (-2.5,5.1) {} ;
		\node [transition] (t2vi) at (-1.3,5.1) {} ;
		\node [transition] (t2vi') at (-0.7,5.1) {} ;

		\path (s3) edge [post] (t1);
		\path (t1) edge [post] (s1');
		\path (t1) edge [post] (s2');
		
		\path (s1) edge [post] (t2i);
		\path (s1) edge [post] (t2i');
		\path (s2) edge [post] (t2i);
		\path (s2) edge [post] (t2i');
		\path (s2) edge [post] (t2ii);
		\path (s2) edge [post] (t2ii');
		\path (s1') edge [post] (t2ii);
		\path (s1') edge [post] (t2ii');
		\path (s1') edge [post] (t2iii);
		\path (s1') edge [post] (t2iii');
		\path (s2') edge [post] (t2iii);
		\path (s2') edge [post] (t2iii');
		\path (s2') edge [post] (t2iv);
		\path (s2') edge [post] (t2iv');
		\path (s2) edge [post] (t2i);
		\path (s2) edge [post] (t2i');
		
		\path (s1) edge [post] (t2iv);
		\path (s1) edge [post] (t2iv');

		\path (s1) edge [post] (t2v);
		\path (s1) edge [post] (t2v');
		\path (s1') edge [post] (t2v);
		\path (s1') edge [post] (t2v');

		\path (s2) edge [post] (t2vi);
		\path (s2) edge [post] (t2vi');
		\path (s2') edge [post] (t2vi);
		\path (s2') edge [post] (t2vi');

		\small
		
		\node at (-4.5,0.1) {$\pn U_B \pn Q$};
	  \end{scope}
	  \end{scope}
	  
  \end{tikzpicture}

\end{center}
Note that after $t_1$ has fired, there
are $4$ tokens in $s_1$, and that above that point, the universal unfolding consists 
of ${4 \choose 2} = 6$ small figures \raisebox{-4pt}{
  \begin{tikzpicture}[
	  node distance=0.7cm,on grid,>=stealth',bend angle=22.5,auto,
	      post/.style={
        ->,
        shorten >=0.5pt, 
        >={stealth}
    },
	  ]
	  \begin{scope}[shift={(colim)}, scale=0.35,
	    every place/.style= {minimum size=1.2mm,thick,draw=blue!75,fill=blue!20}, 
		every transition/.style={minimum size=1.2mm,thick,draw=black!75,fill=black!20}
	  ]
	  \begin{scope}[shift={(0.0,-0.5)}]
		\tiny
		\node [place] (s1) at (-0.6,0.0) {};
		\node [place] (s2) at (0.6,0.0) {}; 
		\node [transition] (t1) at (-0.6,1.2) {} ;
		\node [transition] (t2) at (0.6,1.2) {} ;
		\path (s1) edge [post] (t1);
		\path (s2) edge [post] (t2);
		\path (s1) edge [post] (t2);
		\path (s2) edge [post] (t1);
	  \end{scope}
	  \end{scope}
\end{tikzpicture}
}
  as in Example~\ref{ex:uv}, corresponding to which two of the four tokens in $s_1$ are
  consumed (and how).
\end{blanko}

\begin{blanko}{Event structures and Scott domains.}
  The universal unfolding of a Petri net is the crucial step towards denotational
  semantics in terms of Scott domains. These discoveries and insights are due to
  Winskel~\cite{Winskel:thesis} and Nielsen, Plotkin, and
  Winskel~\cite{Nielsen-Plotkin-Winskel:1981}. We briefly explain how the steps from
  unfoldings to event structures and domains look in the SITOS formalism. Recall
  (from Winskel~\cite{Winskel:eventstructures}) that a {\em prime event structure} is
  a poset $(P,\sqsubseteq)$ equipped with a {\em conflict relation} $\#$: it is an
  symmetric, anti-reflexive relation compatible with the partial order $\sqsubseteq$
  in the sense that if $x \# y$ and $y\sqsubseteq z$ then also $x \# z$. The elements
  of $P$ are called {\em events}, and $\sqsubseteq$ expresses causality. The
  compatibility axiom says that if two events are in conflict, then any event that
  causally depend on the second event will again be in conflict with the first (and
  conversely, by symmetry).

  On the the other hand, a {\em finitary prime algebraic domain} (the same thing as
  the dI-domains of Berry~\cite{Berry:thesis}) are certain Scott domains which are
  distributive and where each finite element has finite lowerset. We shall not need
  the precise definition here; we refer instead to
  Winskel~\cite{Winskel:eventstructures}, both for the definition and for the
  equivalence of the category of these domains with the category
  of prime event structures. It is sufficient here to say that the domain associated
  to a prime event structure is the poset of conflict-free lowersets.
\end{blanko}

For any occurrence hypergraph $\pn H=AINOA$, consider the poset $N$ with
$\sqsubseteq$ defined as the transitive and reflexive closure of the
relation $\lessdot$. As in \cite{Winskel:eventstructures}, the conflict
relation is defined in two steps, whereby it comes to obey the conflict
axiom by construction: we first declare two nodes $x$ and $y$ to be in
immediate conflict, written $x \#_m y$, if $x\neq y$ and there exists a
hyper-edge $a$ such that $aIx$ and $aIy$. Then we define $\#$ to be the
closure of $\#_m$ under $\sqsubseteq$, meaning that $x \# z$ holds if there
exists $y \leq z$ with $x \#_m y$. By construction, this is a prime event
structure.

The associated domain has as its elements the conflict-free lowersets of
this event structure~\cite{Winskel:eventstructures}. Now the bijection in
Lemma~\ref{lowersetlemma} yields a one-to-one correspondence between the
conflict-free lowersets of the event structure $(N,\sqsubseteq, \#)$ and
the conflict-free lowersets of the occurrence hypergraph that have the
whole $B$ as in-boundary. In other words, the elements of the domain are
precisely the (possibly infinite) $B$-processes of the occurrence
hypergraph. So far the reasoning concerns any occurrence hypergraph. In the
special case of the universal unfolding $\pn U_B \pn P$ of a Petri net $\pn
P$, the main theorem \ref{thm:unfolding} tells us that the (possibly
infinite) processes of $\pn U_B \pn P$ from $B$ are in bijection with the
(possibly infinite) processes of $\pn P$ from $B$. It is clear that these
bijections are also isomorphisms of posets (as in all three cases, the
order relation is given by inclusion). All told, the domain associated to a
Petri net $(\pn P, B)$ is canonically identified with the set of its
(possibly infinite) $B$-processes, with poset structure given by
$B$-preserving inclusions.

\section{Rational maps (of Petri nets)}
\label{sec:generalmap}

We are going to introduce more general morphisms of Petri nets,
corresponding to notions studied in the literature, generally 
motivated by bisimulation. Winskel's 
papers~\cite{Winskel:1984}, \cite{Winskel:1987} seem to be the most complete studies.
His notions of morphism are defined in terms of multi-relations.
In the whole-grain setting, multi-relations become spans.
The general morphisms of Petri nets we consider will be spans 
\begin{equation}\label{eq:generalmaps}
\begin{tikzcd}
  \pn P & \ar[l, "\text{special}"'] \pn P' \ar[r, "\text{etale}"] & 
  \pn Q
 \end{tikzcd}
\end{equation}
consisting of a backwards map of a certain special type, followed by 
the `main part' which is an etale map.
Where the etale part preserves the interfaces of transitions, the
special backwards maps will serve to make a kind of `place-respecting
correction' prior to the etale map.  In the most general situation,
`special' will simply mean `place-etale', meaning respecting the
interfaces of places.  A {\em place-etale map} is thus a diagram of
the form
  $$
  \begin{tikzcd}
	S' \ar[d, "\beta"'] & I' \ar[d] \ar[l] 
	\ar[r] \dlpullback &  T' \ar[d]& 
	O' \ar[d] \ar[l]\ar[r] \drpullback & S' \ar[d, "\beta"] 
\\
S  & I \ar[l] \ar[r]  & T & O \ar[l]\ar[r]  & S   ,
  \end{tikzcd}
  $$  
and when combined with the etale maps as in \eqref{eq:generalmaps}, 
the general kind of morphism will thus be diagrams like the following,
to be read from top to bottom:
  \begin{equation}\label{eq:generalmapspelledout}
  \begin{tikzcd}
	\pn P   &:& S  & I \ar[l] \ar[r]  &  
	T & 
	O \ar[l]\ar[r]  & S
	\\
	\pn P' \ar[u, "\text{place-etale}"]\ar[d, "\text{etale}"'] &:& S' \ar[u, "\beta"] \ar[d, "\alpha"'] & I' \ar[u] \ar[d] \ar[l] 
	\ar[r] \drpullback\ulpullback &  T'\ar[d] \ar[u]& 
	O' \ar[u] \ar[d] \ar[l]\ar[r] \dlpullback\urpullback & S' \ar[u, "\beta"'] \ar[d, "\alpha"]
	\\
	\pn Q &:& R & J \ar[l]\ar[r]& Z & U \ar[l]\ar[r]& R 
  \end{tikzcd}
  \end{equation}
This idea does not come out of the blue: composite squares with pullbacks like the
ones from $T$ to $R$ are precisely the
commutative squares in the (bi)category of spans (see
\cite{Galvez-Kock-Tonks:1602.05082},
\cite{Galvez-Kock-Tonks:1512.07573}): the pullback conditions express
that there is a natural isomorphism between the composite of $T
\leftarrow O \to S$ with $S \leftarrow S' \to R$ and the composite of
$T\leftarrow T'\to Z$ with $Z \leftarrow U\to R$ (and similarly with the left-hand 
composite square).

Spans are matrices of sets, and they are the categorical counterpart of
linear maps, such as maps given by multi-relations~\cite{Winskel:1984}, or
maps of commutative monoids~\cite{Meseguer-Montanari:monoids}. In the
latter two notions, there is a matrix of natural-number multiplicities
instead of a matrix of sets. This is analogous to the difference in the
very definitions of Petri nets: in the traditional definition of Petri
nets, instead of spans $S \leftarrow I \to T$ and $T \leftarrow O \to S$,
there are multi-relations \cite{Goltz-Reisig:1983}, \cite{Winskel:1984}, or
--- equivalently --- Kleisli maps for the commutative-monoid monad:
$$
C(S) \leftarrow T \quad \text{ and } \quad T \to C(S)  ,
$$
as introduced by Meseguer and 
Montanari~\cite{Meseguer-Montanari:monoids}.

The general maps of \eqref{eq:generalmapspelledout} correspond 
to the most general multi-relation maps considered by
Winskel~\cite{Winskel:1984}, \cite{Winskel:1987}.
It may not be entirely clear what the Petri-net meaning of such a 
general map should be, and Winskel himself concentrates on two more 
restrictive notions of morphism of Petri nets. A first restriction is
to require the transition-level multi-relation to be a partial 
map; these are simply called {\em morphisms} by Winskel~\cite{Winskel:1984}.
In the present setting, this amounts to demanding the map $T' \to 
T$ to be injective. A second restriction consists in actually 
demanding it to be a total function, which is to say that $T'\to T$ is 
the identity. These are for Winskel~\cite{Winskel:1984} the {\em 
synchronous} maps, and they are (modulo the difference in definitions)
precisely the Petri-net morphisms considered by Meseguer 
and Montanari~\cite{Meseguer-Montanari:monoids}, which in the 
free-commutative-monoid formalism take the form
\begin{equation}\label{eq:CMmaps}
\begin{tikzcd}
T \ar[d, "f"']\ar[r, shift left] \ar[r, shift right] & 
C(S)  \ar[d, "g"]\\
Z \ar[r, shift left] \ar[r, shift right] & C(R) 
\end{tikzcd}
\end{equation}
(with notation as in \eqref{eq:generalmapspelledout}).
A third restriction asks also the multi-relation at the place-level to 
be a total function; these are called {\em folding maps} by 
Winskel. In the monoid formalism, as in \eqref{eq:CMmaps},
this amounts to demanding $g$ to 
be free on a set map. These correspond to the etale maps.

\bigskip

In the following, we could work with any of these three classes of
special maps, yielding three classes of morphisms of Petri nets beyond
the etale maps. For simplicity, we choose to cover only the case of
`synchronous' maps, in the sense of Winskel~\cite{Winskel:1984}, which
are also the main class of morphisms studied by Meseguer and
Montanari~\cite{Meseguer-Montanari:monoids} (modulo the difference in the 
definition of Petri net). The definition will come in \ref{rational}.

\begin{blanko}{Cabling maps.}
  For (graphs or) Petri nets, a {\em cabling map} is a diagram of the 
  form
  $$
  \begin{tikzcd}
	S' \ar[d, "\beta"'] & I' \ar[d] \ar[l] 
	\ar[r] \dlpullback &  T \ar[d, equal]& 
	O' \ar[d] \ar[l]\ar[r] \drpullback & S' \ar[d, "\beta"] 
\\
S  & I \ar[l] \ar[r]  & T & O \ar[l]\ar[r]  & S  .
  \end{tikzcd}
  $$  
  They are thus bijective on transitions, and arity preserving on 
  places.
  
  They can do two things: 
  
  (1)
  They can add a place to a Petri net,
and connect that new place to existing transitions in any way.

(2) Or they can take a set of parallel  places (i.e.~all having the same interface
(pre-sets and post-sets) and all connecting to the same transitions)
and `cable' them into a single place with that same interface with 
those same transitions.

Note that the interfaces of transitions are not preserved by these changes. 

  It may also be helpful to interpret the morphism in the opposite
  direction: the moves one can do are then (1) delete a place and all
  its arcs (but not the transitions). (2) pick a place, and refine it 
  into $k$ parallel copies with the same arcs (a cable with $k$ wires), 
  linking in the same way to the same transitions.
  
  This shows that given a Petri net $\pn P = SITOS$, the possible cabling 
  maps to $\pn P$ are given by keeping $T$ fixed, and by pullback along 
  any map $\beta:S'\to S$ like this:
  $$
  \begin{tikzcd}
	S' \ar[d, "\beta"'] & \cdot \ar[d, dotted] \ar[l, dotted] 
	\ar[r, dashed] \dlpullback &  T \ar[d, equal]& 
	\cdot \ar[d, dotted] \ar[l, dashed]\ar[r, dotted] \drpullback & S' \ar[d, "\beta"] 
	\\
	S  & I \ar[l] \ar[r]  & T & O \ar[l]\ar[r]  & S   .
  \end{tikzcd}
  $$
\end{blanko}

Since injections are stable under pullback, we get:

\begin{blankothm}{Lemma}\label{lem:cableG}
  Given a cabling map of Petri nets $\pn Q \to \pn G$ where $\pn G$ is a graph,
  then also $\pn Q$ is a graph.
\end{blankothm}

Let $\kat{Petri}_{\operatorname{cabl}}$ denote the category of Petri 
nets and cabling maps.
There is a pseudo-functor
\begin{eqnarray*}
  \kat{Petri}_{\operatorname{cabl}}\op & \longrightarrow & \kat{Cat}  \\
  \pn P & \longmapsto & \kat{Proc}(\pn P)  .
\end{eqnarray*}
It takes a cabling map $b: \pn P' \to \pn P$ to the functor
\begin{eqnarray*}
  \kat{Proc}(\pn P) & \longrightarrow & \kat{Proc}(\pn P')  \\
  p & \longmapsto & b\upperstar (p)
\end{eqnarray*}
defined by the pointwise pullback
\[
\begin{tikzcd}
\pn G' \drpullback \ar[d, "b\upperstar p"'] \ar[r, "\text{cabling}"] & \pn G \ar[d, "p"]  \\
\pn P' \ar[r, "\text{cabling}", "b"'] & \pn P    .
\end{tikzcd}
\]
It follows easily from closure properties of pullbacks that the map $b\upperstar p$
defined by this pointwise pullback is etale again, and that the map $\pn G' \to \pn
G$ is a cabling again. Therefore (by Lemma~\ref{lem:cableG}), $\pn G'$ is a graph,
and altogether $p':=b\upperstar p$ is a process. The upshot is that $\kat{Proc}$ is
contravariantly pseudo-functorial in cabling maps.

\begin{blanko}{Cablings of graphs.}
  The notion of cabling maps is the same for graphs. It should also be noted that
  cabling interacts well with level functions. Given a level function $f: \pn G \to
  \un k$, and given a cabling map $\pn G' \to \pn G$, there is induced an obvious
  level function $f' : \pn G' \to \un k$. Also the induced colimit decompositions
  restrict along cabling maps (for simplicity we only treat the case $\un k=\un 2$):
  if
  $$
  \pn G = \pn G_1 \amalg_{\pnsmall M} \pn G_2
  $$
  is the colimit decomposition induced by a $2$-level function $\pn G\to\un 2$, and
  if $b:\pn G' \to \pn G$ is a cabling map, then the colimit decomposition $\pn G' =
  \pn G'_1 \amalg_{\pnsmall M'} \pn G'_2$ resulting from the level function $\pn G' \to\pn
  G \to \un 2$ can also be described as the pullback of the original decomposition.
  In other words, $b$ restricts to cabling maps between the subgraphs involved. From
  these observations, the next result follows readily:
\end{blanko}

\begin{blankothm}{Lemma}\label{lem:cabling XX'}
  A cabling map of Petri nets $b : \pn P' \to \pn P$ induces a 
  symmetric monoidal simplicial map $\ds X_\bullet \to \ds X_\bullet{}\!\!'$
\end{blankothm}

\begin{blanko}{Rational maps of Petri nets.}\label{rational}
  A {\em rational map} of Petri nets from $\pn P$ to $\pn Q$ is by 
  definition a diagram
$$
\begin{tikzcd}
  \pn P & \ar[l, "\text{cabling}"'] \pn P' \ar[r, "\text{etale}"] & 
  \pn Q
 \end{tikzcd}
 $$
--- a cabling backwards followed by an etale map. Spelled out, these 
are thus diagrams
  $$
  \begin{tikzcd}
	\pn P   &:& S  & I \ar[l] \ar[r]  &  
	T\ar[d, equal] & 
	O \ar[l]\ar[r]  & S
	\\
	\pn P' \ar[u, "\text{cabling}"]\ar[d, "\text{etale}"'] &:& S' \ar[u, "b"] \ar[d, "\alpha"'] & I' \ar[u] \ar[d] \ar[l] 
	\ar[r] \drpullback\ulpullback &  T\ar[d] & 
	O' \ar[u] \ar[d] \ar[l]\ar[r] \dlpullback\urpullback & S' \ar[u, "b"'] \ar[d, "\alpha"]
	\\
	\pn Q &:& R & J \ar[l]\ar[r]& Z & U \ar[l]\ar[r]& R .
  \end{tikzcd}
  $$

  Rational maps are composed by pointwise span composition, i.e.~by
  pullback in the category of sets. It is a routine exercise in
  pullbacks to see that in the result the required squares are
  pullbacks. Pullback is only defined up to isomorphism, so the 
  composition law is only weak, and the result is not a category but 
  a bicategory.
  The $2$-cells are isomorphisms of spans
  \[\begin{tikzcd}[row sep={2em,between origins}]
   & \pn P' \ar[ld] \ar[rd] \ar[dd, "\simeq"] &  \\
  \pn P  & & \pn Q   .\\
   & \pn P'' \ar[lu] \ar[ru] &
  \end{tikzcd}
  \]
  The universal property of pullbacks ensures all the required coherences
  in the usual way.\footnote{All this is very similar to the situation for polynomial
endofunctors~\cite{Gambino-Kock:0906.4931}: there the cartesian natural
transformations play the role of etale maps, whereas general natural
transformations uniquely decompose into something `backward' followed
by cartesian.}
\end{blanko}

All these constructions restrict to the subcategory of graphs. The only thing to
check is that the injectivity condition is preserved under these pullback
manipulations, which is because injective maps are stable under pullback.

\begin{blanko}{Functoriality in rational maps.}
  If $\pn P'\to \pn P$ is a rational map of Petri nets (backward cabling 
  followed by forward etale), there is induced a functor 
  $$\kat{Proc}(\pn P') \to \kat{Proc}(\pn P)$$
  from the 
  category of processes of $\pn P'$ to the category of processes of $\pn P$.
  (Recall that $\kat{Proc}(\pn P)=\Gr\comma \pn P$ is the category of 
  etale maps from graphs.)
  This is just a question of combining the functorialities already 
  established.
  This works by refactorisation: a process is itself an etale map
  $\pn G' \to \pn P'$.
  So now we have altogether an etale map followed by a backward
  cabling, followed by an etale map
  $$
  \pn G' \stackrel{\operatorname{etale}}\to \pn P' \stackrel{\operatorname{cabling}}\leftarrow \pn Q 
  \stackrel{\operatorname{etale}}\to \pn P.
  $$
  Just refactor the first pair
  into backward cabling followed by etale:
  $$
  \pn G' \stackrel{\operatorname{cabling}} \leftarrow \pn G 
  \stackrel{\operatorname{etale}}\to \pn Q 
  \stackrel{\operatorname{etale}}\to \pn P .
  $$
  The middle object $\pn G$ appearing here is a graph by 
  Lemma~\ref{lem:cableG}. Then throw away $\pn G'$.
  
  In other words, the process will have a different underlying graph, 
  but with the same nodes.
  The way $\pn G'$ is modified is the minimal way to ensure an etale map 
  to $\pn P$.
\end{blanko}

Combining the functorialities already established 
(\ref{prop:XX'} and \ref{lem:cabling XX'}), we find

\begin{blankothm}{Proposition}
  The construction of the symmetric monoidal Segal space $\ds 
  X_\bullet$ is functorial in rational maps.
\end{blankothm}

\section{Notes}
\label{sec:outlook}

\begin{blanko}{Geometric vs.~algebraic process semantics.}
  The geometric process semantics was pioneered by Goltz and
  Reisig~\cite{Goltz-Reisig:1983}, in the language of posets. As noted in
  Section~\ref{sec:processes}, this cannot immediately be linked with firing
  sequences and the idea of a categorical composition law, but Best and
  Devillers~\cite{Best-Devillers:1987} actually figured out the equivalence relations
  required both on processes and on `occurrence sequences' (certain free categories)
  in order to make them match up. Meanwhile, purely algebraic approaches were
  introduced by Winskel~\cite{Winskel:1987} and
  Meseguer--Montanari~\cite{Meseguer-Montanari:monoids}, giving symmetric monoidal
  categories, but without clear connection to the Goltz--Reisig processes.

  It took a longer journey, begun perhaps by Degano, Meseguer, and
  Montanari~\cite{Degano-Meseguer-Montanari:1996}, to find a reconciliation of the
  two viewpoints; this inevitably involved a gradual shift from collective- to
  individual-tokens viewpoints. Roughly, the ideas were to introduce book-keeping
  devices in ingenious ways, in terms of numbering schemes, in order to compensate
  for the slack of multisets, to get enough control over the symmetries of states and
  processes to be able to compose them in a meaningful way. The short survey of
  Meseguer--Montanari--Sassone~\cite{Meseguer-Montanari-Sassone:birthday97} describes
  this: first {\em concatenable processes}~\cite{Degano-Meseguer-Montanari:1996},
  introducing numberings of subsets of in- and out-boundaries of processes mapping to
  the same place; then {\em strongly concatenable processes}~\cite{Sassone:1998},
  with full numbering of in- and out-boundaries, in terms of `strings as explicit
  representatives of multisets'. Finally, these developments culminated with the
  clean idea of pre-nets of Bruni, Meseguer, Montanari, and
  Sassone~\cite{Bruni-Meseguer-Montanari-Sassone:CTCS99},
  \cite{Bruni-Meseguer-Montanari-Sassone:2001}, where `everything' is numbered: here
  multisets are finally discarded, and replaced by lists: a {\em pre-net} is the data
  of
  $$
  T \rightrightarrows M(S)
  $$
  where $M$ is the free-monoid monad, instead of the free-commutative-monoid monad $C$
  of the definition of Meseguer and Montanari~\cite{Meseguer-Montanari:monoids}.
  In combination with the idea
  \cite{Baldan-Bruni-Montanari-DBLP:conf/wadt/BaldanBM02} of working with states as
  words (or lists) of places instead of multisets of places, this gives finally
  full-blown individual-tokens semantics. The individuality of a token in a given
  state is encoded as its position in a word, and the individuality of token
  occurrences is encoded through its history, as in unfolding semantics.
\end{blanko}

\begin{blanko}{Symmetry problems in unfolding semantics.}
  The unfolding semantics is not directly affected by the dichotomy 
  geometric/algebraic, as it is not concerned with composition of processes end to 
  end. It may be considered geometric in that the processes are defined as maps from
  `causal nets', a viewpoint that goes back to Petri himself~\cite{Petri:1977}. 
  Nevertheless, the symmetry problems turn out to be of very similar nature, and as
  we have seen, the problems can be given the diagnosis that traditional Petri nets
  are not structured enough to control symmetries of processes.
  
  Although the developments from concatenable processes to pre-nets were motivated by
  monoidal-category semantics, the ideas were also applied to unfolding, where again
  the rigidity imposed by decorations and numberings served to work around the
  symmetry problems. Important contributions in this direction were
  Meseguer--Montanari--Sassone~\cite{DBLP:journals/tcs/MeseguerMS96},
  \cite{DBLP:journals/mscs/MeseguerMS97} and
  Baldan--Bruni--Montanari~\cite{Baldan-Bruni-Montanari-DBLP:conf/wadt/BaldanBM02}.
\end{blanko}

\begin{blanko}{Pre-nets vs.~whole-grain Petri nets.}
  \sloppy Some of the successive key insights of the
  Bruni--Meseguer--Montanari--Sassone line of work are subsumed in the
  SITOS formalism. Note first that traditional Petri nets too induce
  digraphical species, but not generally flat ones. The
  strongly-concatenable-processes insight of numbering boundaries is
  subsumed in the general idea of the free-prop monad, and more
  particularly in the notion of $[\coro{m}{n}]$-graph, an inherent
  ingredient in the theory of symmetric monoidal categories, present since
  the notion of tensor scheme of Joyal and
  Street~\cite{Joyal-Street:tensor-calculus}. 

  \fussy The ultimate consequence of those developments, replacing Petri nets
  with pre-nets for the sake of handling symmetries, is a
  question of imposing structure enough to land in the subcategory of
  {\em flat} digraphical species. Indeed,
  {\em  Every (finite) flat digraphical species is the symmetrisation of a pre-net}
  --- simply, because finite sets (as in SITOS diagrams)
  {\em admit} a linear ordering.
  However, the functor from pre-nets to digraphical species is not
  full. Its full image is equivalent to the category of whole-grain Petri 
  nets (by \ref{lem:density} and \ref{prop:flat}).
  If, to correct this deficiency,
  morphisms of pre-nets were defined to be morphisms of their associated
  flat digraphical species, then in the end 
  it would amount essentially to 
  working with whole-grain Petri nets, but with the numberings imposed on top.

  Compared to the Bruni--Meseguer--Montanari--Sassone line of
  development, the present work takes a different approach. Instead of
  numbering schemes and linear orders, the individuality of elements
  is encoded simply by taking seriously the graphical rendition of
  Petri nets. The elements of the sets
  $$
  S \leftarrow I \to T \leftarrow O \to S  
  $$
  are precisely the elements seen in a picture of a Petri net: places, transitions,
  and arcs; the notion is {\em very} close to the traditional definition of Petri nets,
  and does not involve numberings. But having sets of arcs instead of just numbers of
  arcs makes a big difference, leading to the representability feature formalised
  through the identification (Proposition~\ref{prop:flat}) of SITOS Petri nets with
  flat digraphical species.
  
  The SITOS Petri nets contain the pre-nets, as the special case where
  the maps
  $$
  I \to T \leftarrow O
  $$
  are forced to be of the special kind
  $$
  \sum_{t\in T} \un m_t \longrightarrow T \longleftarrow \sum_{t\in T} 
  \un n_t
  $$
  where all $\un m_t$ and $\un n_t$ are standard linear orders.
   But while every Petri net is isomorphic to one of this kind, by 
  appropriate choices of representing sets, the general 
  morphisms will not preserve these choices. 
\end{blanko}

\begin{blanko}{Comparison functors.}
  Baez, Genovese, Master, and Shulman~\cite{Baez-Genovese-Master-Shulman:2101.04238}
  have furthered the theory of whole-grain Petri nets and digraphical species (which 
  they call $\Sigma$-nets), and provide in particular the following system of 
  adjunctions which clarifies the relationship between the various notions of net:
  
\newsavebox{\adju}
\sbox{\adju}{\tiny $\top$}
\newcommand{\adj}{\usebox{\adju}}

  \[
  \begin{tikzcd}[row sep=3em]
  \kat{PreNet} \ar[rd, bend right] \ar[r, shift right=2] \ar[r, phantom, 
  "\adj" description]
  &  \PrSh(\elGr) \ar[l, shift right=2] \ar[r, shift right=2] \ar[r, phantom, 
  "\adj" description]  
  & 
  \kat{TradPetri}  \ar[l, dotted, shift right=2] \\
   & \kat{Petri} \ar[u] \ar[ru, bend right] & {}
  \end{tikzcd}
  \]
  Note that the right-hand triangle from whole-grain to traditional to digraphical 
  species does not commute --- a key difference between whole-grain 
  Petri nets and traditional Petri nets is the fact that they
  embed differently into digraphical species.  Note that in all these categories,
  the morphisms are only the etale maps. Further work will be required to perform
  similar comparison with fancier notions of morphisms, as in 
  Section~\ref{sec:generalmap}, and in particular it is not clear if embedding into
  digraphical species would work.
\end{blanko}

\begin{blanko}{Which monad?}\label{whichmonad}
  Classical Petri nets are diagrams $T \rightrightarrows C(S)$ for $C$
  the free-commutative-monoid monad. Pre-nets are diagrams $T
  \rightrightarrows M(S)$ for $M$ the free-monoid monad.
  Master~\cite{Master:1904.09091} has begun the study of notions of
  Petri nets relative to other monads than these two, in fact relative
  to Lawvere theories. This generality covers also other interesting flavours of 
  Petri nets.
  
  The free-commutative-monoid monad $C$ does not provide sufficient
  grip on the symmetries. The free-monoid monad $M$ introduces artificial linear 
  orders that are not present
  in the intuitive picture we have of Petri nets, and it also does not
  lead to reasonable notions of morphism. In between the
  options $C$ and $M$ there is the symmetric-monoidal-category monad.
  It has the commutative flavour of the free-commutative-monoid monad
  (which indeed is its $\pi_0$), but at the same time has the
  important property shared with the free-monoid monad that it is {\em
  cartesian}. The only issue with the free-symmetric-monoidal-category
  monad --- at first sight a serious blow to the 
  idea --- is that it does not exist on the category of sets! It
   requires at least groupoids: the free symmetric monoidal category on a set gives a 
  groupoid.  Sassone, in the outlook section of 
  \cite{Sassone:1998}, suggests that the symmetry issues related to
  the strongly-concatenable-processes formalism should be overcome by
  stepping up to $2$-categories. The free-symmetric-monoidal-category
  monad on $\Grpd$ fits this idea perfectly. The short expository 
  paper by
  Bruni--Meseguer--Montanari--Sassone~\cite{Meseguer-Montanari-Sassone:birthday97}
  also hints at this, describing {\em the role of the 2-structure
  being to carry information about multisets, thus making an explicit
  quotient construction unnecessary.} This viewpoint, however, did not
  find its way into the final paper 
  \cite{Bruni-Meseguer-Montanari-Sassone:2001}, and apparently it was 
  not pursued.
  The symmetric monoidal Segal space $\ds
  X_\bullet$ can be seen as fleshing out this idea, while insisting
  that Petri nets are just configurations of sets.
\end{blanko}
  
\begin{blanko}{SITOS formalism and symmetric-monoidal-category monad.}\label{S}
  The difference between traditional multisets and representable
  multisets constitutes the passage from the free-commutative-monoid
  monad to the free-symmetric-monoidal-category monad, in the
  following sense, giving an interpretation of the SITOS formalism in
  terms of the free-symmetric-monoidal-category monad.
  
  In its strictest version, the free-symmetric-monoidal-category 
  monad sends a set $S$ (regarded as a discrete groupoid)
  to the  groupoids whose underlying set is the set
  of words in $S$, and whose morphisms are the permutations. 
  A word is just a set map
  $\un{n} \to S$ and a permutation is a commutative diagram
  $$\begin{tikzcd}[column sep={2em,between origins}]
  \un n \ar[rd] \ar[rr, "\sim"] && \un n \ar[ld] \\
   & S  .  &
  \end{tikzcd}
  $$
  A more invariant presentation uses {\em arbitrary} finite sets 
  instead of insisting on the sets $\un n$. The groupoid is
  then $\B\comma S$. In this view on multisets, to assign to every transition $t$ its 
  post-set 
  is to specify a map $O_t \to S$. All these maps can 
  be conveniently bundled into the configuration $T \leftarrow O \to 
  S$ (and similarly of course with pre-sets and $S\leftarrow I \to T$). The SITOS formalism is therefore almost unavoidable.
\end{blanko}

\begin{blanko}{Symmetric operads vs.~non-symmetric operads vs.~polynomial monads.}
  The use of set-con\-fig\-urations instead of numberings was advocated in the
  polynomial formalism in operad theory~\cite{Kock:0807}. Symmetric operads
  can be seen as the many-in/one-out case of props. Their symmetry issues
  are subtle, as expressed by the fact that the corresponding monads on the
  category of sets are generally not cartesian but only weakly cartesian. Every
  non-symmetric operad defines a symmetric operad by symmetrisation, and
  the result is sigma-cofibrant by construction. Every sigma-cofibrant
  operad {\em admits} the structure of non-symmetric operad, but does not
  come with one.\footnote{The distinction is subtle, and it is the origin
  of some mistakes in the literature on operads and Lawvere theories, as
  explained in detail in Leinster's book~\cite{Leinster:0305049}.} It turns
  out that sigma-cofibrant operads are the same thing as finitary
  polynomial monads. This hinges on the crucial representability feature of
  polynomial endofunctors~\cite{Gambino-Kock:0906.4931}: they are
  represented by diagrams
  $$
  I \leftarrow E \to B \to I   .
  $$
  The theory of polynomial functors~\cite{Gambino-Kock:0906.4931}, 
  \cite{Kock:0807} is the original inspiration for
  both \cite{Kock:1407.3744} and the present work. 
  
  It is interesting to note that the distinction between (finitary)
  polynomial monads and operads, which is purely a question of symmetries,
  goes away in the world of $\infty$-categories, by a theorem
  of~\cite{Gepner-Haugseng-Kock:1712.06469} (in fact a groupoid-enriched
  setting is enough). By the same arguments, the distinction between
  (whole-grain) Petri nets and digraphical species would vanish if Petri
  nets were allowed to have a groupoid of transitions instead of just a set
  of them. While this might seem wild from the perspectives of current
  Petri-net theory, it would look natural from the viewpoint of
  implementing Petri nets in (intensional) dependent type
  theory~\cite{HoTT-book}, with places and transitions being types instead
  of sets. The recent work of Baez, Genovese, Master, and
  Shulman~\cite{Baez-Genovese-Master-Shulman:2101.04238} can also be seen
  to point in that direction, their definition of digraphical species
  (there called $\Sigma$-net) having very much the flavour of a dependent
  type.
\end{blanko}

\appendix

\stepcounter{section}
\section*{Appendix A: Groupoids and homotopy pullbacks}
\label{app:groupoids}


We recall here a few basic notions regarding groupoids. 

\begin{blanko}{Groupoids.}\label{groupoids}
  A groupoid is a category in which all arrows are invertible; a map 
  of groupoids is just a functor. For
  their use in combinatorics, it is rather their topological-space
  aspects that are important. If $X$ is a groupoid, $\pi_0(X)$ denotes
  the set of connected components, i.e.~the set of iso-classes. For
  each $x\in X$, we denote by $\pi_1(X,x)=\Aut_X(x)$ the group of
  automorphisms. A map of groupoids is an equivalence iff it is
  bijective on $\pi_0$ and invertible on each $\pi_1$ (Whitehead's
  theorem). 
  
  We are interested in groupoids up to equivalence, and since the
  ordinary categorical constructions with groupoids --- such as
  pullbacks and fibres --- are not invariant under equivalence, they
  should be replaced by their homotopy analogues, which are
  characterised by universal properties up to equivalence. If just
  these homotopy notions are used consistently, they behave very much
  like the ordinary notions do for sets.
\end{blanko} 
  
\begin{blanko}{Homotopy pullbacks.}\label{hopbk}
  A {\em homotopy pullback} is an up-to-isomorphism commutative square
  \[
  \begin{tikzcd}
  P \ar[d] \ar[r] & Y \ar[d, "q"]  \\
  X \ar[r, "p"'] \ar[ru, phantom, "\simeq" description]& S
  \end{tikzcd}
  \]
  satisfying a universal property among all such squares with common $p$ and $q$. As
  such it is determined uniquely up to equivalence. There are different (but
  equivalent) models for homotopy pullback. The {\em standard homotopy pullback} $P =
  X \times_S^h Y$ has as objects triples $(x,y,\sigma)$ consisting of $x\in X$, $y\in
  Y$, and $\sigma:px\to qy$ in $S$; its arrows $(x,y,\sigma)\to (x',y',\sigma')$ are
  pairs $(\phi,\psi)\in \Hom_X(x,x') \times \Hom_Y(y,y')$ such that $\sigma' \circ
  p(\phi)=q(\psi)\circ \sigma$.  (From the viewpoint of category theory, the
  standard homotopy pullback is thus the comma category $p \comma q$.)  While
  the standard homotopy pullback is always correct, it can be a bit
  cumbersome to use. It is sometimes possible to be more economical by exploiting
  fibrations: 
\end{blanko}

\begin{blanko}{Fibrations.}\label{fib}
  A map of groupoids $p:X\to B$ is a {\em fibration} when it satisfies the 
  {\em path lifting property}: for each $x\in X$ and $\beta: 
  p(x) \isopil b$ in $B$, there exists an arrow $\phi: 
  x\to x'$ such 
  that $p(\phi)=\beta$. The benefit of this notion is that
  an ordinary (strict) pullback
    \[
  \begin{tikzcd}[cramped]
  X \times_S^{\text{strict}} Y \ar[d] \ar[r] & Y \ar[d, "q"]  \\
  X \ar[r, "p"'] & S
  \end{tikzcd}
  \]
  is also a homotopy pullback whenever
  one of the two maps $p$ and $q$ is a fibration.
\end{blanko}

\begin{blankothm}{Lemma}\label{lem:prism}
  (Prism Lemma.)
  Given a prism diagram of groupoids
  \begin{center}
    \begin{tikzcd}
        X''  \ar[dr, phantom, "\simeq" description]  \ar[r, ""] \ar[d, ""]
          & X' \ar[r, ""]\ar[d, ""] \ar[dr, phantom, "\simeq" description]  \drpullback
          & X\ar[d, ""]\\
        Y'' \ar[r, ""'] & Y' \ar[r, ""] & Y
    \end{tikzcd}
  \end{center}
	in which the right-hand square is a homotopy pullback, then the outer
	rectangle is a homotopy pullback if and only of the left-hand square
	is a homotopy pullback.
\end{blankothm}

\begin{blanko}{Homotopy fibre.}\label{fibre}
  Given a map of groupoids $p: X \to S$ and an element
  $s \in S$, the \emph{homotopy fibre} $X_s$ of $p$ over $s$ is the 
  homotopy pullback
\[
    \begin{tikzcd}
        X_s \ar[r, ""] 
            \ar[d, ""'] 
            \drpullback \ar[dr, phantom, "\simeq" description]
             & X \ar[d, "p"]\\
        1 \ar[r, "\name{s}"'] & S.
    \end{tikzcd}
\]
\end{blanko}

\begin{blanko}{Homotopy quotient and homotopy sum.}\label{X/G}
  Given a group action $G \times X \to X$ (for
  $G$ a group and $X$ a set or groupoid), instead of the naive
  quotient (set of orbits), the {\em homotopy quotient} $X/\!/G$ is obtained
  from $X$ by sewing in a path from $x$ to $g.x$ for each $x\in X$ and
  $g\in G$. Note that it is a groupoid, not a set.
   (See Baez--Dolan~\cite{Baez-Dolan:finset-feynman} for a very enjoyable account of 
  this.) 
  
  A {\em homotopy sum} is a (homotopy) colimit indexed by a groupoid, just 
  like an ordinary sum is a colimit indexed by a set. It can be 
  computed as an ordinary sum of homotopy quotients: if $B$ is a 
  groupoid and $F: B \to \Grpd$ is a diagram of groupoids indexed by 
  $B$, then the homotopy sum is
  $$
  \int^{b\in B} F(b) \simeq \sum_{b\in \pi_0(B)} F(b) /\!/ \Aut(b) .
  $$
\end{blanko}

\stepcounter{section}

\section*{Appendix B: Simplicial groupoids and Segal spaces}
\label{app:simplicial}


Simplicial methods are fundamental in topology and algebra. Further information can
be found in May's little 1967 book {\em Simplicial Methods in Algebraic
Topology}~\cite{May:ss}. In this paper we use simplicial groupoids to encode `weak
categories', a standard idea in higher category theory~\cite{Leinster:0305049}, 
\cite{Bergner:0610239}. For their role 
in combinatorics, see
\cite{Galvez-Kock-Tonks:1612.09225}.

\begin{blanko}{The idea of simplicial sets.}\label{simplicial-ids}
  Where a directed graph $X_0 
  \leftleftarrows X_1$ (in the ordinary sense, without the open-ended edges of the 
  graphs in Section~\ref{sec:graphs}) has vertices and edges, a {\em simplicial set} 
  $X_\bullet$ has (directed)
cells in all dimensions:\footnote{The bullet in $X_\bullet$ is there to remind that it is
  not a single set, but a whole configuration of sets.} after vertices and edges come a set $X_2$ of triangles, a
set $X_3$ of tetrahedra, a set $X_4$ of $4$-dimensional simplices, and so on. Just
like an edge has two endpoints which are vertices, a triangle has three vertices,
numbered $0,1,2$, but it also has three sides which are edges, so there should be
three maps $\begin{tikzcd}[column sep={16mm,between origins}]X_1 &  X_2 \ar[l, shift 
left=2]\ar[l]\ar[l, 
shift right=2]\end{tikzcd}$. Similarly, a tetrahedron has four vertices, numbered
$0,1,2,3$, and four faces which are triangles (so as to have four maps from $X_4$ to
$X_3$), and so on. 
Altogether, a simplicial set $X_\bullet$ has a 
sequence of sets $X_0$, $X_1$, $X_2$, $X_3$, $X_4$, $\ldots$ with {\em face maps}
\begin{equation}
  \begin{tikzcd}[column sep={12mm,between origins}]
	{X_0} && {X_1} && {X_2} && {X_3} & \cdots
	\arrow["{d_{0}}"{description, pos=0.6}, shift left=1.65, shorten <=4pt, from=1-3, to=1-1]
	\arrow["{d_{1}}"{description, pos=0.6}, shift right=1.65, shorten <=4pt, from=1-3, to=1-1]
	\arrow["{d_{1}}"{description, pos=0.6}, shorten <=4pt, from=1-5, to=1-3]
	\arrow["{d_{2}}"{description, pos=0.6}, shift right=3.3, shorten <=4pt, from=1-5, to=1-3]
	\arrow["{d_{0}}"{description, pos=0.6}, shift left=3.3, shorten <=4pt, from=1-5, to=1-3]
	\arrow["{d_{3}}"{description, pos=0.6}, shift right=4.95, shorten <=4pt, from=1-7, to=1-5]
	\arrow["{d_{2}}"{description, pos=0.6}, shift right=1.65, shorten <=4pt, from=1-7, to=1-5]
	\arrow["{d_{1}}"{description, pos=0.6}, shift left=1.65, shorten <=4pt, from=1-7, to=1-5]
	\arrow["{d_{0}}"{description, pos=0.6}, shift left=4.95, shorten <=4pt, from=1-7, to=1-5]
\end{tikzcd}
\end{equation}
subject to the identities given below.
The most systematic way
of numbering the sides of a triangle, the faces of a tetrahedron, and so on, is to
record the vertex {\em removed} to get it. So, for example,
face $2$ of a tetrahedron is the triangle opposite vertex
$2$. 
It is thus important to index the face maps such that $d_i$ removes vertex 
number $i$ to get a simplex of one dimension lower. 
With these conventions (which have proven optimal through many decades of
work in topology and algebra), the {\em face map identities} are
$$
  d_i d_j=d_{j-1} d_i \qquad \text{ for all } i<j \text{, and in all dimensions},
$$
which essentially say that for two vertices, it does not matter in which order they 
are removed.
(It turns out to be harmless to use the 
same name for face maps in all dimensions, although of course they are different maps.)

Furthermore, a simplicial set should have {\em degeneracy maps}, whereby a vertex can be regarded as a 
degenerate edge, and an edge can be regarded as a degenerate triangle (in two ways),
and so on. The degeneracy maps are always denoted $s_i$.
With this further structure we arrive at the full picture of a simplicial 
set:
it is a diagram
\begin{equation}
  \begin{tikzcd}
	{X_0} && {X_1} && {X_2} && {X_3} & \cdots
	\arrow["{d_{0}}"{description, pos=0.7}, shift left=2.2, shorten <=4pt, from=1-3, to=1-1]
	\arrow["{d_{1}}"{description, pos=0.7}, shift right=2.2, shorten <=4pt, from=1-3, to=1-1]
	\arrow["{s_{0}}"{description, pos=0.7}, color={rgb,255:red,214;green,92;blue,92}, shorten <=4pt, from=1-1, to=1-3]
	\arrow["{d_{1}}"{description, pos=0.7}, shorten <=4pt, from=1-5, to=1-3]
	\arrow["{d_{2}}"{description, pos=0.7}, shift right=4.4, shorten <=4pt, from=1-5, to=1-3]
	\arrow["{d_{0}}"{description, pos=0.7}, shift left=4.4, shorten <=4pt, from=1-5, to=1-3]
	\arrow["{s_{1}}"{description, pos=0.7}, shift left=2.2, color={rgb,255:red,214;green,92;blue,92}, shorten <=4pt, from=1-3, to=1-5]
	\arrow["{s_{0}}"{description, pos=0.7}, shift right=2.2, color={rgb,255:red,214;green,92;blue,92}, shorten <=4pt, from=1-3, to=1-5]
	\arrow["{d_{3}}"{description, pos=0.7}, shift right=6.6, shorten <=4pt, from=1-7, to=1-5]
	\arrow["{d_{2}}"{description, pos=0.7}, shift right=2.2, shorten <=4pt, from=1-7, to=1-5]
	\arrow["{d_{1}}"{description, pos=0.7}, shift left=2.2, shorten <=4pt, from=1-7, to=1-5]
	\arrow["{d_{0}}"{description, pos=0.7}, shift left=6.6, shorten <=4pt, from=1-7, to=1-5]
	\arrow["{s_{2}}"{description, pos=0.7}, shift left=4.4, color={rgb,255:red,214;green,92;blue,92}, shorten <=4pt, from=1-5, to=1-7]
	\arrow["{s_{1}}"{description, pos=0.7}, color={rgb,255:red,214;green,92;blue,92}, shorten <=4pt, from=1-5, to=1-7]
	\arrow["{s_{0}}"{description, pos=0.7}, shift right=4.4, color={rgb,255:red,214;green,92;blue,92}, shorten <=4pt, from=1-5, to=1-7]
\end{tikzcd}
\end{equation}
subject to the {\em simplicial identities}:
$$
d_is_i=d_{i+1}s_i=1, \quad (\forall i)
$$
$$
  d_id_j=d_{j-1}d_i,\quad
  d_{j+1}s_i=s_id_j,\quad
  d_is_j=s_{j-1}d_i,\quad
  s_js_i=s_is_{j-1},\quad
  \qquad(i<j).
  $$

While these identities express clear geometric intuition,
it takes some practice to get familiar with the 
indices. Luckily there is a categorical way to
describe the structure more conceptually:
\end{blanko}

\begin{blanko}{Definition (of simplicial set).}\label{sSet}
  Let $\simplexcategory$ denote the category whose objects are the nonempty
  finite total orders
  $$[k] := \{0 \leq 1 \leq \cdots \leq k\}$$
  and whose arrows are the monotone maps. A {\em simplicial set} is a functor
  $X_\bullet: \simplexcategory\op\to\Set$. The value of $X_\bullet$ on
  $[k]$ is denoted $X_k$, and its elements are called {\em $k$-simplices}.
  
  A simplicial map $X_\bullet \to Y_\bullet$ between simplicial sets is a natural 
  transformation of functors $\simplexcategory\op\to\Set$.
\end{blanko} 

The simplicial identities are now accounted for by noting that the category
$\simplexcategory$ is generated by the injections $d^i : [k{-}1]\to [k]$ that skip
the value $i$ and the surjections $s^i : [k{+}1] \to [k]$
that repeat $i$. The relations holding between these
maps are obvious and induce the simplicial identities by functoriality.

\bigskip

The reason
simplicial sets are fundamental objects in topology is that they provide a combinatorial 
model for spaces up to homotopy~\cite{May:ss}.
In the present context they are important for their
role in category theory, which is via the nerve construction:

\begin{blanko}{The nerve of a small category.}\label{nerve}
  The
  {\em nerve} of a small category $\CC$ is the simplicial set
  $$
  N\CC : \simplexcategory\op\to\Set
  $$
  whose set of $k$-simplices is the set of sequences of $k$ composable
  arrows in $\CC$ (allowing identity arrows). 
  
  --- The degeneracy maps are given by inserting an identity map in the
  sequence.
  
  --- The inner face maps are given by composing adjacent arrows.
  
  --- The outer face maps discard arrows at the
  beginning or the end of the sequence.
  
  By regarding the total order $[k]$
  as a category, we see that 
  a sequence of $k$ composable arrows in $\CC$ is the same thing
  as a functor $[k]\to \CC$, and more formally the set of $k$-simplices can be 
  described as
  $$
  (N\CC)_k = \Fun([k],\CC).
  $$
  
  Note that $d_0:X_1 \to X_0$ assigns to an arrow its codomain, and
  $d_1: X_1 \to X_0$ assigns to an arrow its domain. 
  Note further that $d_1: X_2 \to X_1$ is the composition map.  Also, $d_0:X_2 \to
  X_1$ assigns to a composable pair the second arrow, and $d_2 : X_2 \to X_1$
  assigns to a composable pair the first arrow (cf.~the standard index convention of 
  \ref{simplicial-ids}).
\end{blanko}

\begin{blanko}{Nerve theorem.}
  The nerve functor is fully faithful, and it image is characterised by the {\em
  Segal condition}, which states that the canonical projection map that returns the
  edges of a $k$-simplex
  \begin{equation}\label{Segalmaps}
  X_k \longrightarrow  X_1 \times_{X_0} X_1\times_{X_0} \dots \times_{X_0} X_1
  \end{equation}
  is a bijection for each $k\geq 1$. The fibre products express that the edges match 
  up at the vertices.
  Extracting the edges of a simplex can be done 
  with iterated use of face maps. The easiest and most important case of 
  the Segal condition is $k=2$. Here the condition says that the map
  $(d_2,d_0) : X_2 \to X_1 
  \times_{X_0} X_1$ is a bijection, which is equivalent to saying that
  this square is a pullback:
    \[
  \begin{tikzcd}
     X_2\drpullback \ar[r, "d_0"]\ar[d, "d_2"'] & X_1 \ar[d, "d_1"] \\
     X_1 \ar[r, "d_0"'] & X_0 .
   \end{tikzcd}
  \]
  The condition thus says that the $2$-simplices are precisely the composable pairs.
\end{blanko}

\begin{blanko}{Simplicial groupoids.}\label{sGrpd}
  One can talk about simplicial objects in any category $\mathcal{E}$: they are
  functors $X_\bullet :\simplexcategory\op\to\mathcal{E}$. In particular a {\em simplicial
  groupoid}
  is a functor
  $$
  X_\bullet : \simplexcategory\op\to\Grpd ,
  $$
  and it thus amounts to a sequence of groupoids
  $X_k$, ($k\geq 0$), and face and degeneracy map
  $d_i:X_k\to X_{k-1}$, $s_i:X_k\to X_{k+1}$, $(0\leq i\leq k)$, subject to the
  simplicial identities above.
\end{blanko}

From the viewpoint of simplicial groupoids, it is now possible (and natural) to
weaken the notion of category, by relaxing the requirement that the 
maps~\eqref{Segalmaps} should
be isomorphisms to just demanding they be equivalences of groupoids, with the caveat
that we should now refer to the homotopy pullback instead of the strict pullback.
Additionally there is one more technical condition called Rezk completeness:

\begin{blanko}{Rezk-complete Segal spaces.}\label{Rezk}
  A simplicial groupoid is a {\em Segal space}\footnote{The full-blown notion used in 
  homotopy theory and $\infty$-category theory uses $\infty$-groupoids instead of 
  just ordinary groupoids. Rezk-complete Segal 
  spaces are an important model for weak categories in higher
  category theory and homotopy theory; see
  Bergner~\cite{Bergner:0610239} for a survey.
  } if $X_2 \simeq X_1\times^h_{X_0}
  X_1$, and in general the canonical projection map
  $$
  X_k \longrightarrow  X_1 \times^h_{X_0} X_1\times^h_{X_0} \dots \times^h_{X_0} X_1
  $$ 
  is an equivalence of groupoids for each $k\geq 1$.

  Composition in a Segal space is up to equivalence: given a pair of arrows that are
  composable up to equivalence, their {\em composite} is given by
  $$
   X_1\times^h_{X_0} X_1 \underset{\sim}{\stackrel{(d_2,d_0)}\longleftarrow} X_2 
   \stackrel{d_1}\longrightarrow X_1 ,
  $$
  where we first have to go backwards along the equivalence to $X_2$:
  the Segal condition tells us that a composable pair is equivalent to 
  a whole $2$-simplex, and 
  now we can return its `long edge', using $d_1$. (Note that $d_1$ deletes the 
  middle vertex of a triangle, so as to leave us with the `long edge' from vertex 
  $0$ to vertex $2$.)  
  
  An arrow $f\in X_1$ is called {\em invertible} if composition with it
  from either side defines an equivalence of groupoids $X_1 \to X_1$.
  A Segal space $X_\bullet$ is {\em Rezk complete} if $s_0: X_0 \to X_1$ is fully faithful 
  and has as its essential image the invertible arrows.
\end{blanko}

\begin{blanko}{Mapping groupoids.}\label{Map(x,y)}
  Just as an ordinary category has hom sets, a Segal space has {\em mapping 
  groupoids} (often called {\em mapping spaces}). Given two objects $x,y$ in a Segal space 
  $X_\bullet$, the mapping groupoid $\Map_{X_\bullet}(x,y)$ is defined as the homotopy pullback
  \[
  \begin{tikzcd}
  \Map_{X_\bullet}(x,y) \drpullback \ar[dr, phantom, "\simeq" description] \ar[d] \ar[r] & X_1 \ar[d, "{(d_1,d_0)}"]  \\
  1 \ar[r, "\name{(x,y)}"'] & X_0 \times X_0 ,
  \end{tikzcd}
  \]
  which is the appropriate way of picking out those arrows that start in 
  $x$ and end in $y$ --- up to homotopy.
\end{blanko}

From the viewpoint of homotopy theory, it is quite natural to consider that if some
combinatorial structure with symmetries is hard to assemble into a category, then it
should perhaps be a weak category instead --- a (Rezk-complete) Segal space. This is
what happens in Section~\ref{sec:X}.

\bigskip

\noindent {\bf Acknowledgments.} I wish to thank Pawe\l\ Soboci\'nski, 
Filippo Bonchi, Steve Lack, and Mike Shulman for feedback and help. 
This work was presented as a plenary keynote address at the 2020 International 
Conference on Applied Category Theory (ACT 2020). I am grateful to the
programme committee for their confidence and for the opportunity.
Finally I am thankful to various anonymous referees for their feedback, which led to 
many improvements.


\small

\begin{thebibliography}{10}

\bibitem{AlgebraicPetri}
{\rm Andrew Baas, James Fairbanks, Micah Halter, and Evan Patterson}.
2020.
\newblock {\em Algebraic Petri}. Julia library for building Petri net models compositionally.
\newblock \url{https://github.com/AlgebraicJulia/AlgebraicPetri.jl}

\bibitem{Baez-Courser:1911.04630}
{\rm John~C. Baez and Kenny Courser}.
2020.
\newblock {\em Structured cospans}.
Theory Appl. Categ. {\bf 35} (2020), 1771--1822,
\newblock ArXiv:1911.04630.

\bibitem{Baez-Dolan:finset-feynman}
{\rm John~C. Baez and James Dolan}.
2001.
\newblock {\em From finite sets to {F}eynman diagrams}.
\newblock In B.~Eng\-quist and W.~Schmid, editors, {\em Mathematics
  unlimited---2001 and beyond}, pp. 29--50. Springer-Verlag, Berlin, 2001.
\newblock ArXiv:math.QA/0004133.

\bibitem{Baez-Genovese-Master-Shulman:2101.04238}
{\rm John C. Baez, Fabrizio Genovese, Jade Master, and Michael Shulman}.
2021.
\newblock {\em Categories of nets}.
\newblock In {\em Proceedings of the 36th Annual ACM/IEEE Symposium on Logic in Computer 
Science}, art. no. 16, pp.13. Association for Computing Machinery, New York, 
NY, USA, 2021.
\newblock ArXiv:2101.04238.

\bibitem{Baez-Master:1808.05415}
{\rm John~C. Baez and Jade Master}.
2020.
\newblock {\em Open {P}etri nets}.
\newblock Math. Structures Comput. Sci. {\bf 30} (2020), 314--341.
\newblock ArXiv:1808.05415.

\bibitem{Baez-Pollard:1704.02051}
{\rm John~C. Baez and Blake~S. Pollard}.
2017.
\newblock {\em A compositional framework for reaction networks}.
\newblock Rev. Math. Phys. {\bf 29} (2017), 1750028, 41pp.
\newblock ArXiv:1704.02051.

\bibitem{Baldan-Bruni-Montanari-DBLP:conf/wadt/BaldanBM02}
{\rm Paolo Baldan, Roberto Bruni, and Ugo Montanari}.
2002.
\newblock {\em Pre-nets, read arcs and unfolding: A functorial presentation}.
\newblock In M.~Wirsing, D.~Pattinson, R.~Hennicker, editors,
  {\em Recent Trends in Algebraic Development Techniques, 16th International
  Workshop, {WADT} 2002, Frauenchiemsee, Germany, September 24-27, 2002,
  Revised Selected Papers}, vol. 2755 of Lecture Notes in Computer Science, pp.
  145--164. Springer, 2002.

\bibitem{Baldan-Corradini-Ehrig-Heckel:2005}
{\rm Paolo Baldan, Andrea Corradini, Hartmut Ehrig, and Reiko Heckel}.
2005.
\newblock {\em Compositional semantics for open {P}etri nets based on
  deterministic processes}.
\newblock Math. Structures Comput. Sci. {\bf 15} (2005), 1--35.

\bibitem{Batanin-Berger:1305.0086}
{\rm Michael Batanin and Clemens Berger}.
2017.
\newblock {\em Homotopy theory for algebras over polynomial monads}.
\newblock Theory Appl. Categ. {\bf 32} (2017), 148--253.

\bibitem{Bergeron-Labelle-Leroux}
{\rm Fran{\c{c}}ois Bergeron, Gilbert Labelle, and Pierre Leroux}.
1998.
\newblock {\em Combinatorial species and tree-like structures}, vol.~67 of
  Encyclopedia of Mathematics and its Applications.
\newblock Cambridge University Press, Cambridge, 1998.
\newblock Translated from the 1994 French original by Margaret Readdy, with a
  foreword by Gian-Carlo Rota.

\bibitem{Bergner:0610239}
{\rm Julia~E. Bergner}.
2010.
\newblock {\em A survey of {$(\infty,1)$}-categories}.
\newblock In {\em Towards higher categories}, vol. 152 of IMA Vol. Math. Appl.,
  pp. 69--83. Springer, New York, 2010.
\newblock ArXiv:math.AT/0610239.

\bibitem{Berry:thesis}
{\rm G{\'e}rard Berry}.
1979.
\newblock {\em Mod{\`e}les comple{\`e}tement ade{\'e}quats et stables des
  lambda-calculs typ{\'e}s}.
\newblock PhD thesis, Th{\`e}se de Doctorat d'Etat, Universit{\'e} Paris VII,
  1979.

\bibitem{Best-Devillers:1987}
{\rm Eike Best and Raymond Devillers}.
1987.
\newblock {\em Sequential and concurrent behaviour in {P}etri net theory}.
\newblock Theoret. Comput. Sci. {\bf 55} (1987), 87--136.

\bibitem{Best-Fernandez}
{\rm Eike Best and C\'{e}sar Fern\'{a}ndez}.
1988.
\newblock {\em Nonsequential processes --- A Petri net view}, vol.~13 of EATCS
  Monographs on Theoretical Computer Science.
\newblock Springer-Verlag, Berlin, 1988.


\bibitem{Bezanson-Karpinski-Shah-Edelman:1209.5145}
{\rm Jeff Bezanson, Stefan Karpinski, Viral B. Shah, and Alan Edelman}.
2012.
\newblock {\em Julia: a fast dynamic language for technical computing}.
\newblock ArXiv:1209.5145.

\bibitem{Bruni-Meseguer-Montanari-Sassone:CTCS99}
{\rm Roberto Bruni, Jos\'{e} Meseguer, Ugo Montanari, and Vladimiro Sassone}.
1999.
\newblock {\em Functorial semantics for {P}etri nets under the individual token
  philosophy}.
\newblock In {\em C{TCS} '99: {C}onference on {C}ategory {T}heory and
  {C}omputer {S}cience ({E}dinburgh)}, Electron. Notes Theor.
  Comput. Sci. {\bf 29} (1999), 29003, 18pp.

\bibitem{Bruni-Meseguer-Montanari-Sassone:2001}
{\rm Roberto Bruni, Jos\'{e} Meseguer, Ugo Montanari, and Vladimiro Sassone}.
2001.
\newblock {\em Functorial models for {P}etri nets}.
\newblock Inform. and Comput. {\bf 170} (2001), 207--236.

\bibitem{Bruni-Montanari:2000}
{\rm Roberto Bruni and Ugo Montanari}.
2000.
\newblock {\em Zero-safe nets: comparing the collective and individual token
  approaches}.
\newblock Inform. and Comput. {\bf 156} (2000), 46--89.

\bibitem{Degano-Meseguer-Montanari:1996}
{\rm Pierpaolo Degano, Jos\'{e} Meseguer, and Ugo Montanari}.
1996.
\newblock {\em Axiomatizing the algebra of net computations and processes}.
\newblock Acta Inform. {\bf 33} (1996), 641--667.

\bibitem{Galvez-Kock-Tonks:1602.05082}
{\rm Imma G{\'a}lvez-Carrillo, Joachim Kock, and Andrew Tonks}.
2018.
\newblock {\em Homotopy linear algebra}.
\newblock Proc. Royal Soc. Edinburgh A {\bf 148} (2018), 293--325.
\newblock ArXiv:1602.05082.

\bibitem{Galvez-Kock-Tonks:1512.07573}
{\rm Imma G{\'a}lvez-Carrillo, Joachim Kock, and Andrew Tonks}.
2018.
\newblock {\em Decomposition spaces, incidence algebras and {M}\"{o}bius
  inversion {I}: {B}asic theory}.
\newblock Adv. Math. {\bf 331} (2018), 952--1015.
\newblock ArXiv:1512.07573.

\bibitem{Galvez-Kock-Tonks:1512.07577}
{\rm Imma G{\'a}lvez-Carrillo, Joachim Kock, and Andrew Tonks}.
2018.
\newblock {\em Decomposition spaces, incidence algebras and {M}\"{o}bius
  inversion {II}: {C}ompleteness, length filtration, and finiteness}.
\newblock Adv. Math. {\bf 333} (2018), 1242--1292.
\newblock ArXiv:1512.07577.

\bibitem{Galvez-Kock-Tonks:1708.02570}
{\rm Imma G{\'a}lvez-Carrillo, Joachim Kock, and Andrew Tonks}.
2020.
\newblock {\em Decomposition spaces and restriction species}.
\newblock Int. Math. Res. Notices {\bf 2020} (2020), 7558--7616
\newblock ArXiv:1708.02570.

\bibitem{Galvez-Kock-Tonks:1612.09225}
{\rm Imma G{\'a}lvez-Carrillo, Joachim Kock, and Andrew Tonks}.
2016.
\newblock {\em Decomposition spaces in combinatorics}.
\newblock Preprint, arXiv:1612.09225.

\bibitem{Gambino-Kock:0906.4931}
{\rm Nicola Gambino and Joachim Kock}.
2013.
\newblock {\em Polynomial functors and polynomial monads}.
\newblock Math. Proc. Cambridge Phil. Soc. {\bf 154} (2013), 153--192.
\newblock ArXiv:0906.4931.

\bibitem{Genovese-Gryzlov-Herold-Perone-Post-Videla:1904.12974}
{\rm Fabrizio Genovese, Alex Gryzlov, Jelle Herold, Marco Perone, Erik Post,
  and Andr{\'e} Videla}.
  2019.
\newblock {\em Computational Petri nets: adjunctions considered harmful}.
\newblock Preprint, arXiv:1904.12974.

\bibitem{Genrich-StankiewiczWiechno}
{\rm Hartmann~J. Genrich and Ewa Stankiewicz-Wiechno}.
1980.
\newblock {\em A dictionary of some basic notions of net theory}.
\newblock In {\em Net theory and applications ({P}roc. {A}dv. {C}ourse
  {G}eneral {N}et {T}heory {P}rocesses {S}ystems, {H}amburg, 1979)}, vol.~84 of
  Lecture Notes in Computer Science, pp. 519--535. Springer, Berlin-New York, 1980.

\bibitem{Gepner-Haugseng-Kock:1712.06469}
{\rm David Gepner, Rune Haugseng, and Joachim Kock}.
2022.
\newblock {\em $\infty$-operads as analytic monads}.
Int. Math. Res. Notices {\bf 2022} (2022), 12516--12624.
DOI: http://doi.org/10.1093/imrn/rnaa332.
\newblock ArXiv:1712.06469.

  \bibitem{Glabbeek:2005}
{\rm Robert~Jan van Glabbeek}.
2005.
\newblock {\em The individual and collective token interpretations of {P}etri
  nets}.
\newblock In {\em C{ONCUR} 2005---concurrency theory}, vol. 3653 of Lecture
  Notes in Computer Science, pp. 323--337. Springer, Berlin, 2005.

  \bibitem{Glabbeek-Plotkin:LICS95}
{\rm Rob~J. van Glabbeek and Gordon~D. Plotkin}.
1995.
\newblock {\em Configuration structures}.
\newblock In {\em Proceedings, 10th Annual {IEEE} Symposium on Logic in
  Computer Science, San Diego CA, 1995}, pp. 199--209. {IEEE} Computer Society,
  1995.

\bibitem{Glabbeek-Plotkin:2009}
{\rm Rob~J. van Glabbeek and Gordon~D. Plotkin}.
2009.
\newblock {\em Configuration structures, event structures and {P}etri nets}.
\newblock Theoret. Comput. Sci. {\bf 410} (2009), 4111--4159.

\bibitem{Goltz-Reisig:1983}
{\rm Ursula Goltz and Wolfgang Reisig}.
1983.
\newblock {\em The nonsequential behaviour of {P}etri nets}.
\newblock Inform. and Control {\bf 57} (1983), 125--147.

\bibitem{Hack:PhD}
{\rm Michel Hack}.
1976.
\newblock {\em Decidability questions for Petri nets}.
\newblock PhD thesis, Massachusetts Institute of Technology, Cambridge, MA,
  {USA}, 1976.

\bibitem{Halter-Patterson:epidem}
{\rm Micah Halter and Evan Patterson}.
2020.
\newblock {\em Compositional epidemiological modeling using structured cospans}.
October 2020.
\newblock 
\url{https://www.algebraicjulia.org/blog/post/2020/10/structured-cospans}.


\bibitem{Hansen-Shulman:1910.09240}
{\rm Linde~Wester Hansen and Michael Shulman}.
2019.
\newblock {\em Constructing symmetric monoidal bicategories functorially}.
\newblock Preprint, arXiv:1910.09240.

\bibitem{Hayman-Winskel:2008}
{\rm Jonathan Hayman and Glynn Winskel}.
2008.
\newblock {\em The unfolding of general {P}etri nets}.
\newblock In {\em F{STTCS} 2008: {IARCS} {A}nnual {C}onference on {F}oundations
  of {S}oftware {T}echnology and {T}heoretical {C}omputer {S}cience}, vol.~2 of
  LIPIcs. Leibniz Int. Proc. Inform., pp. 223--234. Schloss Dagstuhl.
  Leibniz-Zent. Inform., Wadern, 2008.

\bibitem{Hayman-Winskel:2009}
{\rm Jonathan Hayman and Glynn Winskel}.
2009.
\newblock {\em Symmetry in {P}etri nets}.
\newblock In {\em Perspectives in concurrency theory}, pp. 231--263. Univ.
  Press, Hyderabad, 2009.

\bibitem{Heindel-Sobocinski:1101.4594}
{\rm Tobias Heindel and Pawel Soboci\'{n}ski}.
2011.
\newblock {\em Being {V}an {K}ampen is a universal property}.
\newblock Log. Meth. Comput. Sci. {\bf 7} (2011).
\newblock ArXiv:1101.4594.

\bibitem{Joyal:1981}
{\rm Andr{\'e} Joyal}.
1981.
\newblock {\em Une th\'eorie combinatoire des s\'eries formelles}.
\newblock Adv. Math. {\bf 42} (1981), 1--82.

\bibitem{Joyal-Kock:0908.2675}
{\rm Andr{\'e} Joyal and Joachim Kock}.
2011.
\newblock {\em {F}eynman graphs, and nerve theorem for compact symmetric
  multicategories (extended abstract)}.
\newblock In {\em Proceedings of the 6th International Workshop on Quantum
  Physics and Logic (Oxford 2009)},
  Electron. Notes Theor.
  Comput. Sci. {\bf 270} (2011), 105--113.
\newblock ArXiv:0908.2675.

\bibitem{Joyal-Moerdijk:openmaps}
{\rm Andr{\'e} Joyal and Ieke Moerdijk}.
1994.
\newblock {\em A completeness theorem for open maps}.
\newblock Ann. Pure Appl. Logic {\bf 70} (1994), 51--86.

\bibitem{Joyal-Nielsen-Winskel}
{\rm Andr{\'e} Joyal, Mogens Nielsen, and Glynn Winskel}.
1996.
\newblock {\em Bisimulation from open maps}.
\newblock Inform. and Comput. {\bf 127} (1996), 164--185.

\bibitem{Joyal-Street:tensor-calculus}
{\rm Andr{\'e} Joyal and Ross Street}.
1991.
\newblock {\em The geometry of tensor calculus. {I}}.
\newblock Adv. Math. {\bf 88} (1991), 55--112.

\bibitem{Julia}
{\em The Julia Programming Language}.
\url{https://julialang.org/}.

\bibitem{Kock:0807}
{\rm Joachim Kock}.
2011.
\newblock {\em Polynomial functors and trees}.
\newblock Int. Math. Res. Notices {\bf 2011} (2011), 609--673.
\newblock ArXiv:0807.2874.

\bibitem{Kock:1407.3744}
{\rm Joachim Kock}.
2016.
\newblock {\em Graphs, hypergraphs, and properads}.
\newblock Collect. Math. {\bf 67} (2016), 155--190.
\newblock ArXiv:1407.3744.

\bibitem{Kock:1611.10342}
{\rm Joachim Kock}.
2018.
\newblock {\em Cospan construction of the graph category of {B}orisov and
  {M}anin}.
\newblock Publ. Mat. {\bf 62} (2018), 331--353.
\newblock ArXiv:1611.10342.

\bibitem{Lack:1510.08925}
{\rm Stephen Lack}.
2002.
\newblock {\em Codescent objects and coherence}.
\newblock J. Pure Appl. Algebra {\bf 175} (2002), 223--241.
\newblock ArXiv:1510.08925.

\bibitem{Leinster:0305049}
{\rm Tom Leinster}.
2004.
\newblock {\em Higher Operads, Higher Categories}.
\newblock London Math. Soc. Lecture Note Series. Cambridge University Press,
  Cambridge, 2004.
\newblock ArXiv:math.CT/0305049.

\bibitem{Libkind-Baas-Halter-Patterson-Fairbanks:2203.16345}
{\rm Sophie Libkind, Andrew Baas, Micah Halter, Evan Patterson, and James 
Fairbanks.}
2022.
\newblock {\em An algebraic framework for structured epidemic modeling}.
\newblock Philos. Trans. A Math. Phys. Eng. Sci. {\bf 380} (2022), 20210309.
DOI: 10.1098/rsta.2021.0309
\newblock Preprint, arXiv:2203.16345.

\bibitem{MacLane:categories}
{\rm Saunders {Mac~Lane}}.
1998.
\newblock {\em Categories for the working mathematician, second edition}.
\newblock No.~5 in Graduate Texts in Mathematics. Springer-Verlag, New York,
  1998.

\bibitem{Master:1904.09091}
{\rm Jade Master}.
2020.
\newblock {\em Petri nets based on Lawvere theories}.
Math. Structures Comput. Sci. {\bf 30} (2020), 833--864.
(Previously circulated under the title 
{\em Generalized Petri nets}.)
\newblock ArXiv:1904.09091.

\bibitem{May:ss}
{\rm J.~Peter May}.
1967.
\newblock {\em Simplicial objects in algebraic topology}.
\newblock Van Nostrand Mathematical Studies, No. 11. D. Van Nostrand Co., Inc.,
  Princeton, N.J.-Toronto, Ont.-London, 1967.

\bibitem{Meseguer-Montanari:monoids}
{\rm Jos\'{e} Meseguer and Ugo Montanari}.
1990.
\newblock {\em Petri nets are monoids}.
\newblock Inform. and Comput. {\bf 88} (1990), 105--155.

\bibitem{DBLP:journals/tcs/MeseguerMS96}
{\rm Jos{\'{e}} Meseguer, Ugo Montanari, and Vladimiro Sassone}.
1996.
\newblock {\em Process versus unfolding semantics for place/transition Petri
  nets}.
\newblock Theor. Comput. Sci. {\bf 153} (1996), 171--210.

\bibitem{DBLP:journals/mscs/MeseguerMS97}
{\rm Jos{\'{e}} Meseguer, Ugo Montanari, and Vladimiro Sassone}.
1997.
\newblock {\em On the semantics of place/transition Petri nets}.
\newblock Math. Structures Comput. Sci. {\bf 7} (1997), 359--397.

\bibitem{Meseguer-Montanari-Sassone:birthday97}
{\rm Jos{\'{e}} Meseguer, Ugo Montanari, and Vladimiro Sassone}.
1997.
\newblock {\em Representation theorems for Petri nets}.
\newblock In C.~Freksa, M.~Jantzen, R.~Valk,
  editors, {\em Foundations of Computer Science: Potential - Theory -
  Cognition, to Wilfried Brauer on the occasion of his 60th birthday}, vol.
  1337 of Lecture Notes in Computer Science, pp. 239--249. Springer, 1997.

\bibitem{Nielsen-Plotkin-Winskel:1981}
{\rm Mogens Nielsen, Gordon Plotkin, and Glynn Winskel}.
1981.
\newblock {\em Petri nets, event structures and domains. {I}}.
\newblock Theoret. Comput. Sci. {\bf 13} (1981), 85--108.

\bibitem{Patterson-Lynch-Fairbanks:2106.04703}
{\rm Evan Patterson, Owen Lynch, and James Fairbanks}.
2022.
\newblock {\em Categorical data structures for technical computing}.
\newblock Compositionality {\bf 4} (2022).
DOI: 10.32408/compositionality-4-5.
\newblock ArXiv:2106.04703.

\bibitem{Peterson:1977}
{\rm James~L. Peterson}.
1977.
\newblock {\em Petri nets}.
\newblock {ACM} Comput. Surv. {\bf 9} (1977), 223--252.

\bibitem{Petri:1977}
{\rm Carl~A. Petri}.
1977.
\newblock {\em Nicht-sequentielle {P}rozesse.}
\newblock Technical report, Universit{\"a}t Erlangen-N{\"u}rnberg,
  Arbeitsberichte des IMMD, Vol.9, Nr.8, pp. 57-82 (1976) also: Gesellschaft
  f{\"u}r Mathematik und Datenverarbeitung Bonn, ISF-76-6, 3.
  revidierte und erg{\"a}nzte Auflage. Translation {\em Non-Sequential
  Processes} by P.~Krause, J.~Low. Gesellschaft f{\"u}r
  Mathematik und Datenverarbeitung Bonn, ISF-77-5,
  1977.

\bibitem{DBLP:books/sp/Reisig85a}
{\rm Wolfgang Reisig}.
1985.
\newblock {\em Petri Nets: An Introduction}, vol.~4 of {EATCS} Monographs on
Theoretical Computer Science.
\newblock Springer, 1985.

\bibitem{Sassone:1998}
{\rm Vladimiro Sassone}.
1998.
\newblock {\em An axiomatization of the category of {P}etri net computations}.
\newblock Math. Structures Comput. Sci. {\bf 8} (1998), 117--151.

\bibitem{Scott:1971}
{\rm Dana~S. Scott}.
1971.
\newblock {\em The lattice of flow diagrams}.
\newblock In E.~Engeler, editor, {\em Symposium on Semantics of Algorithmic
  Languages}, vol. 188 of Lecture Notes in Mathematics, pp. 311--366. Springer,
  1971.

\bibitem{HoTT-book}
{\rm The {Univalent Foundations Program}}.
2013.
\newblock {\em Homotopy type theory---univalent foundations of mathematics}.
\newblock The Univalent Foundations Program, Princeton, NJ; Institute for
  Advanced Study (IAS), Princeton, NJ, 2013.
\newblock Available from \url{http://homotopytypetheory.org/book}.

\bibitem{Weber:1503.07585}
{\rm Mark Weber}.
2015.
\newblock {\em Internal algebra classifiers as codescent objects of crossed
  internal categories}.
\newblock Theory Appl. Categ. {\bf 30} (2015), 1713--1792.

\bibitem{Winskel:thesis}
{\rm Glynn Winskel}.
1980.
\newblock {\em Events in computation}.
\newblock PhD thesis, Department of Computer Science, University of Edinburgh,
  1980.

\bibitem{Winskel:1984}
{\rm Glynn Winskel}.
1984.
\newblock {\em A new definition of morphism on {P}etri nets}.
\newblock In {\em S{TACS} 84 ({P}aris, 1984)}, vol. 166 of Lecture Notes in
  Computer Science, pp. 140--150. Springer, Berlin, 1984.

\bibitem{Winskel:eventstructures}
{\rm Glynn Winskel}.
1986.
\newblock {\em Event structures}.
\newblock In W.~Brauer, W.~Reisig, G.~Rozenberg,
  editors, {\em Petri Nets: Central Models and Their Properties, Advances in
  Petri Nets, Part II, Proceedings of an Advanced Course, Bad Honnef,
  Germany, 8-19 September 1986}, vol. 255 of Lecture Notes in Computer Science,
  pp. 325--392. Springer, 1986.

\bibitem{Winskel:1987}
{\rm Glynn Winskel}.
1987.
\newblock {\em Petri nets, algebras, morphisms, and compositionality}.
\newblock Inform. and Comput. {\bf 72} (1987), 197--238.

\end{thebibliography}
\end{document}